\documentclass{LMCS}

\usepackage{enumerate}
\usepackage[dvips]{epsfig}
\usepackage{latexsym,amsfonts, amssymb,amsmath,amsthm}
\usepackage{bussproofs}
\usepackage{cmll}
\usepackage{hyperref}
\usepackage{pstricks,pst-node,pst-tree}
\usepackage{proof}

\newcommand{\rt}[1]{\underline{\sigma}(#1)}

\newcommand{\single}[1]{\langle #1 \rangle}

\newcommand{\view}[1]{\ulcorner #1\urcorner }

\newcommand{\crocl}{\hbox{[\hskip -1.5pt [}}
\newcommand{\crocr}{\hbox{]\hskip -1.5pt ]}}

\newcommand{\cut}[1]{\crocl#1\crocr}

\renewcommand{\b}{{}^{\bot}}

 \newcommand{\pref}{\sqsubseteq}

\newcommand{\bG}{\mathbf{G}}

\newcommand{\bF}{\mathbf{F}}
\newcommand{\bE}{\mathbf{E}}
\newcommand{\bA}{\mathbf{A}}

\newcommand{\bB}{\mathbf{B}}
\newcommand{\bC}{\mathbf{C}}
\newcommand{\bP}{\mathbf{P}}
\newcommand{\bQ}{\mathbf{Q}}

\newcommand{\bS}{\mathbf{S}}

\newcommand{\bN}{\mathbf{N}}

\newcommand{\One}{\pmb{1}}
\newcommand{\Zero}{\pmb{0}}
\newcommand{\Toppo}{\pmb{\top}}
\newcommand{\Botto}{\pmb{\bot}}

\newcommand{\semone}{\posm\pmb{1}}
\newcommand{\sembot}{\negm\pmb{\bot}}

\newcommand{\posm}{\wn_{\!\textsl{P}}\,}
\newcommand{\negm}{\oc_{\!\!\!\,\,\,\textsl{N}}\,}

\newcommand{\uno}{\mathsf{I}\,}
\newcommand{\ult}{\ensuremath{\rotatebox[origin=c]{180}{\textsf{T}}\,}}
\newcommand{\vero}{\mathsf{T}\,}
\newcommand{\falso}{\mathsf{0}\,}

\newcommand{\pl}{\textup{\ensuremath{\crocl}}}
\newcommand{\pr}{\textup{\ensuremath{\crocr}}}

 \newcommand{\dai}{\dagger}
\newcommand{\senza}{\setminus}

\newcommand{\coslj}{\mathsf{\top}}

\def\C{\mathcal C}
\newcommand{\D}{\mathcal D}
\newcommand{\A}{\mathcal A}
\newcommand{\F}{\mathcal F}
\newcommand{\B}{\mathcal B}
\newcommand{\E}{\mathcal E}

\def\R{\mathcal R}
\def\X{\mathcal X}

\def\GG{\Gamma}

\def\DD{\Delta}

\def\ss{\sigma}

\newcommand{\bGG}{\boldsymbol{\GG}}
\newcommand{\bDD}{\boldsymbol{\DD}}
\newcommand{\bPP}{\boldsymbol{\Pi}}

\newcommand{\Dai}{\mathfrak{Dai}}
\newcommand{\Fid}{\mathfrak{Fid}}

\newcommand{\HS}{\mathbf{MELLS}}

\newcommand{\freccia}{\blacktriangleleft}

\theoremstyle{plain}
\def\eg{{\em e.g.}, }
\def\cf{{\em cf.} }
\def\ie{{\em i.e.}, }

\def\wrt{{\em w.r.t.} }
\def\wloge{{\em W.l.o.g.}}

\newcommand{\sA }{{\mathsf{A}}}
\newcommand{\sB }{{\mathsf{B}}}
\newcommand{\sC }{{\mathsf{C}}}

\newcommand{\sF }{{\mathsf{F}}}

\newcommand{\sN }{{\mathsf{N}}}
\newcommand{\sM }{{\mathsf{M}}}

\newcommand{\sP }{{\mathsf{P}}}
\newcommand{\sQ }{{\mathsf{Q}}}

\newcommand{\sGamma}{{\mathsf{\Gamma}}}

\newcommand{\sDelta }{{\mathsf{\Delta}}}
\newcommand{\sPi }{{\mathsf{\Pi}}}
\newcommand{\sPl }{{\mathsf{p}}}
\newcommand{\sNl }{{\mathsf{n}}}

\newcommand{\sumP}[1]{  {\bigoplus_{#1}}^\tau}
\newcommand{\sumN}[1]{  {\sum_{#1}}^\tau}

\def\doi{7 (2:13) 2011}
\lmcsheading%
{\doi}
{1--85}
{}
{}
{Jan.~\phantom05, 2010}
{May.~17, 2011}
{}

\begin{document}

\title[Ludics with repetitions]{Ludics  with repetitions \\
{\normalsize(Exponentials, Interactive types and Completeness)}\rsuper*}

\author[M.~Basaldella]{Michele Basaldella\rsuper a}	
\address{{\lsuper a}Research Institute for Mathematical Sciences, Kyoto University, Kitashirakawa Oiwakecho, Sakyo-ku, Kyoto 606-8502, Japan.}	
\email{mbasalde@kurims.kyoto-u.ac.jp}  

\author[C.~Faggian]{Claudia Faggian\rsuper b}	
\address{{\lsuper b}\'{E}quipe PPS, CNRS and Universit\'{e} Paris 7, 2 place Jussieu, case
7017, 75251 Paris cedex 05, France.}
\email{faggian@pps.jussieu.fr}

\keywords{linear logic,
ludics,
game semantics, internal completeness}
\subjclass{  F.4.1 Mathematical Logic (Proof theory), F.3 Logics and Meanings of Programs}
\titlecomment{{\lsuper*}This paper is a completely revised and extended version of ~\cite{BasFag}.}

\begin{abstract}

Ludics is peculiar in the panorama of game semantics: we first have
the definition of interaction-composition and then we have semantical
types, as a set of strategies which ``behave well'' and react in the
same way to a set of tests.  The semantical types which are
interpretations of logical formulas enjoy a fundamental property,
called internal completeness, which characterizes ludics and sets it
apart also from realizability.  Internal completeness entails standard
full completeness as a consequence.

A growing body of work start to explore the potential of this specific
interactive approach.  However, ludics has some limitations, which are
consequence of the fact that in the original formulation, strategies
are abstractions of $\mathbf{MALL}$ proofs. On one side, no
repetitions are allowed.  On the other side, the proofs tend to rely
on the very specific properties of the $\mathbf{MALL}$ proof-like
strategies, making it difficult to transfer the approach to semantical
types into different settings.

In this paper, we provide an extension of ludics which allows
repetitions and show that one can still have interactive types and
internal completeness. From this, we obtain full completeness \wrt a
polarized version of $\mathbf{MELL}$.  In our extension, we use less
properties than in the original formulation, which we believe is of
independent interest. We hope this may open the way to applications of
ludics approach to larger domains and different settings.

\end{abstract}

\maketitle

\section{Introduction}

Ludics is a  research program started by  Girard
  ~\cite{GirLoc} with the aim of  providing a foundation
for logic based on interaction.
It can be seen as  a form of game semantics
where {\em first} we have the definition of {\em interaction} (equivalently called composition, normalization),
and {\em then} we have semantical {\em types}, as sets of strategies which ``behave well'' with respect to
composition. This role of interaction in the definition of types is where lies the specificity of ludics in the panorama of game semantics.

Recently, a growing body of work is starting to explore and  to develop
 the potential of this specific approach, and to
 put at work the more general notion of type offered by ludics:
 the notion of  \emph{type defined through interaction}.
We mention in particular work by Saurin on interactive proof-search as a logic  programming paradigm
~\cite{alexis}, and work by Terui on computability
~\cite{Terui}.
Terui gives an especially interesting use of the notion of \emph{orthogonality} (``to interact well''): if  the {\em strategy}
$\D$ describes an automaton, $\{\D\}\b$ (the set of all strategies which ``interact well'' with it) is  the language accepted by that automaton.
In ~\cite{BasalTLCA09} Basaldella and Terui have studied the traditional
logical
duality between proofs and models
in the setting of computational ludics ~\cite{Terui}
enriched with exponentials (following our approach
to exponentials ~\cite{BasFag}, this paper). Both proofs and models live in an homogeneous
setting, both  are strategies,
which are related by orthogonality. Finally, we observe that
 interactive types seem  to be very natural also in   process calculi;  a bridge between
process calculi and ludics has already been established in
Faggian and Piccolo  ~\cite{FagPic}, which shows a close correspondence between the strategies of ludics
 and the terms of the linear $\pi$-calculus ~\cite{yobeho} | from this one can hope to transfer the whole approach  of ludics  to that
setting.

There are also other lines of work
in the literature which use orthogonality
to define semantical types.
We mention  work by Pitts on parametricity
~\cite{Pittspoly},  work by
Krivine  on realizability ~\cite{KriRea}, work
by Hyland and Schalk on categorical models of linear logic ~\cite{HylScha},
 work by
Melli{\`e}s and
Vouillon
on recursive types ~\cite{MelvouRea}
and  work by Paolini
on parametric $\lambda$-calculus ~\cite{paolini08tcs}.

\subsubsection*{Interactive types} The computational objects of ludics | \emph{designs} | can be seen as a linear form of
Hyland-Ong (HO) innocent strategies (as shown in ~\cite{FagHyl}) or as Curien's abstract
B\"ohm trees ~\cite{CurAbs,CurHerb}.

 However, in  game semantics,
we first define the types (\emph{arenas}, \emph{games}), and then
strategies on a given type.
 The type information
 guarantees that
strategies compose well.
In ludics,
{\em strategies are untyped}, in the
sense that all strategies are given on a universal arena (the arena of all possible moves);
strategies can always interact with each other, and the interaction may {\em terminate well} (the two strategies ``accept each other'', and are said {to be} {\em orthogonal}) or not (they {\em deadlock}).
An {\em interactive type}
 is a  set of strategies which ``compose well'', and  reacts in the same way to a set of tests
(see Section  ~\ref{ludics}). More concretely,
 a semantical type $\bG$ is any set of strategies which reacts well to the
same set of tests $\bE$, which are themselves strategies (counter-strategies), that is $\bG=\bE\b$.

\subsubsection*{Internal completeness} With ludics, Girard also introduces a new notion of completeness, which
is called {\em internal completeness} (see
Section  ~\ref{internal completeness}).
This is a key | \emph{really defining} | element of ludics. We have already said that a
 semantical type is  a set of strategies closed by biorthogonal ($\bG = \bG\b\b$).
Internal completeness {(in ~\cite{GirLoc})} is the property which {essentially} says that  {\em the constructions on semantical types do not require
any closure operation},
\ie they are already complete.

 For instance, {in ~\cite{GirLoc}} the interpretation of $A\oplus B$ (where
   $\oplus$ denotes
  the additive disjunction of linear logic)
   {is defined as} $(\bA \cup \bB) \b\b$. This set of terms
 could be in general strictly greater than $\bA \cup \bB$.
 But {under certain conditions}, it is possible to \emph{prove} that $\bA \cup \bB$ is  equal to $(\bA \cup \bB)\b\b$, and
 since the closure by  biorthogonal does not introduce new terms,
  $\bA \cup \bB$ already gives a simple and {\em complete} description of what inhabits the semantical type, \ie we have \emph{internal completeness}.

While it is standard in realizability that a semantical type is a set $\bS$ of terms closed
by biorthogonal ($\bS=\bS\b\b$), when interpreting types one has to perform some kind of closure, and this
operation {might} {\em introduce new terms}. Such
  new terms  do not pose any essential problem when  we are only interested in proving
the
\emph{soundness}
of a calculus \wrt a realizability model,
that is,
roughly,
the property which states that
 if   $\pi$ is a proof  of a  formula $A$, then
we can construct from $\pi$ a strategy $\D$ which realizes the
 semantical interpretation $\bA$ of $A$, \ie
 $\D \in \bA$. However, introducing new terms does make a difference when we  are also interested in  the other direction, namely (full) \emph{completeness}:
given a ``good" strategy $\D \in \bA$,
we want to  effectively associate to $\D$ a proof $\pi$ of $A$.
Now the new terms which might have been introduced by the closure operation have to be taken into account.  What internal completeness guarantees, is that we actually have a {\em complete}  description of all the terms of a semantical types which correspond to logical proofs  (or at least a subset which includes the ``good'' strategies).

In Girard's paper on ludics, the semantical types which are interpretations of propositional formulas enjoy internal completeness.
This is really the key property (and the one used in ~\cite{alexis,Terui}).
Full completeness (for multiplicative-additive-linear logic $\mathbf{MALL}$, in the case of
~\cite{GirLoc}) directly  {\em follows} from it.

\subsection{Contributions of the paper} The purpose of this paper is two-fold.

On the
one hand, we  show that it is possible to overcome the main limitation of ludics,
namely the constraint of linearity, hence the lack of exponentials: we show that
 internal completeness (and from that full completeness) can be obtained also when having repetitions, if one  extends in a rather natural way the setting of ludics.

On the other hand, we provide proofs which make use of less properties than the original ones given by Girard. Not only  we do believe this improves the understanding of the results, but | more fundamentally | we hope this opens the way to the application of the approach of ludics to a larger domain.

We now give more details on the content of the paper.

\subsubsection{Ludics architecture}
A difficulty  in ~\cite{GirLoc} is that there is a huge amount of structure (see the ``analytical theorems''), without a clear distinction
between what are properties observed in the specific objects, and what is necessary to the construction.
Strategies are an abstraction of $\mathbf{MALL}$ proofs, and enjoy many good properties.
In ~\cite{GirLoc}, all proofs of the high level structure of ludics
 make essential use of these properties.
Since some of those properties are very specific to the particular nature of
the objects,
this makes it difficult in principle to extend the | very interesting | approach of ludics
to a different setting, or build the interactive types on different
 computational objects.

Ludics, as introduced in ~\cite{GirLoc}, is composed of several layers.

\begin{enumerate}[$\bullet$]
\item  At the {\em low level}, there is the definition of the {\em untyped computational
structures} (strategies, there called \emph{designs}) and their {\em dynamics} (interaction).
Interaction allows the definition of {\em orthogonality.}
\begin{enumerate}[$-$]

\item  The computational objects satisfy certain remarkable properties,
called {\em analytical theorems}, in particular {\bf separation property}, the ludics analogue of B\"ohm theorem for $\lambda$-calculus:
two strategies $\A,\B$ are syntactically equal if and only if
they are observationally equal (\ie for any
counter-strategy $\C$, the strategies $\A,\B$ react in the same way  to $\C$).

\end{enumerate}

\item At the {\em high level}, there is the definition of {\em interactive types},
which satisfy {\em internal completeness.}
\end{enumerate}

By relying on less structure, we show that the high level architecture of ludics
 is somehow independent from the low level entities (strategies), and in
fact could be built on other | more general | computational objects.

In particular,
separation  is a strong property. It is a great property, but it is not a common one to have.
However, the fact that computational
objects do not enjoy separation does not mean
that it is not possible to build the ``high level architecture'' of ludics. In fact,
we show  (Section  ~\ref{int_compl})
that the proofs of internal and full completeness rely on much less structure, namely  operational
properties of
the interaction.

We believe that discriminating between internal completeness and the properties which are  specific to the objects
 is important both to
 improve understanding
of the results, and to  make it possible to build the same construction on different entities.

In particular, strategies with repetitions have weaker properties with respect to the original | linear |
ones. We show that
it is still possible to have interactive types, internal completeness, and
from  this full completeness for a polarized version of
the constant-only fragment of
$\mathbf{MELL}$
(multiplicative-exponential-linear logic) that we call $\HS$
(Section  ~\ref{HS}).

The reason  we restrict our attention to the constant-only
fragment is that the treatment of propositional variables
is rather complicated in ludics (see ~\cite{GirLoc} and also
~\cite{introunif}) and not strictly related to our purposes:
the analysis of ludics with repetitions of actions.
On the other hand,
the extension of our framework to additives
is straightforward.

\subsubsection{Exponentials in ludics}
The treatment of exponentials has been the main open problem in ludics since ~\cite{GirLoc}.
Maurel ~\cite{MauTh}  has been the first one to propose a solution
(a summary  of this solution can { also} be found  in
 ~\cite{CurHerb,GirBook}).
The focus of Maurel's work is  to recover a form of separation when having repetitions;
for this purpose, he develops a sophisticated setting, which is   based
on the use of \emph{probabilistic strategies}: two probabilistic strategies
``compose well'' with a certain probability.
This approach is however limited by
 its technical complexity; this is the main obstacle which stops Maurel from going further, and studying    interpretation and full completeness issues.

In this work, we do not analyze the issue of  separation, while we focus exactly into interactive types and internal completeness,  and develop a fully complete interpretation from it.

Maurel   also explores a simpler solution in order to
introduce exponentials,
but he does not pursue it further
 because of  the failure of  the separation property.
Our work starts from an analysis of  this simpler solution, and builds on it.

\subsubsection{Our approach}
In the literature, there are two standard branches
of game semantics which have been extensively used
to build denotational models of various fragments
of linear logic.  On the one hand, we have Abramsky-Jagadeesan-Malacaria style game semantics (AJM)
 ~\cite{AbrJagMac} which is essentially inspired by  Girard's
geometry of interaction  ~\cite{GirGoI1}.    On the other hand, we have  Hyland-Ong
 style game semantics (HO)  ~\cite{Hy-On}, introducing {\em innocent strategies}.
 The main difference between those two game models is how
 the semantical structures corresponding to exponential modalities are
built. In AJM, given a game $A$,
$\oc A$ is treated as an infinite tensor product
of $A$, where each copy of  $A$ receives a different labeling index.
Two strategies in $\oc A$ which only differ by a different labeling
of moves  are identified.
By contrast, in HO the notion of justification pointer substitutes that of index. The games $A$ and $\oc A$  share the  same arena.
Informally, a strategy in $\oc A$ is a kind of ``juxtaposition'' of strategies
of $A$ such that by following the  pointer structure, we can  \emph{unambiguously} decompose it
as a set of strategies of $A$.

Girard's designs  ~\cite{GirLoc}  are a linear form of HO innocent strategies
 ~\cite{FagHyl}.
Hence, the most natural solution to extend ludics to the  exponentials
 is to consider as strategies, \emph{standard HO innocent strategies} (on an universal arena).
But in order to do so, there is a new kind of difficulty, which we deal with in this paper: we need to have enough tests.

More precisely, as we illustrate in Section  ~\ref{repetitions}, we need  \emph{non-uniform}
counter-strategies. We implement and concretely realize this idea
 of non-uniform (non-deterministic) tests
 by introducing a \emph{non-deterministic sum} of
strategies,  which  builds on and refines work
by Faggian and Piccolo  ~\cite{FP09}.
More  precise motivations and a sketch of the solution are detailed in Section
 ~\ref{NUtests}.

 \subsection{Plan of the paper}

 In Section  ~\ref{HS}, we introduce the
 polarized fragment of linear logic $\mathbf{MELLS}$
 for which we will show a fully complete model in
 Section  ~\ref{completeness}.

  In Section
  ~\ref{Game Semantics}, we recall the basic notions
 of HO innocent game semantics, which we then use  in Section ~\ref{ludics}
 to present Girard's ludics.

 In Section  ~\ref{int_compl} we review
 the  results
 of internal completeness for linear strategies and
  outline a direct  proof of full completeness.

 In Section  ~\ref{repetitions}, we   provide
 the motivations and an informal description of
 non-uniform strategies, and in Section  ~\ref{lwR section II}
 we give  the formal constructions.

In Section  ~\ref{VAM} we describe in detail the
composition of non-uniform strategies and in Section
~\ref{orthog_sec} we revise the notion of orthogonality
in the non-uniform setting.

 In Section  ~\ref{type sez} we introduce
  semantical types for $\mathbf{MELLS}$, and
 we extend internal completeness to non-linear strategies.
Full completeness is developed in  Section  ~\ref{completeness}.

In Section ~\ref{conclusion}
we  discuss   related work
and  conclude the paper.



\section{Calculus} \label{HS}
We start by introducing  a calculus that we call $\HS$, which will be our ``underlying'' syntax; in
 Section  ~\ref{completeness}, we prove that our model is
fully complete for this calculus.

$\HS$ is  a polarized variant of the constant-only propositional fragment of multipli\-cative-exponential linear logic  $\mathbf{MELL}$ ~\cite{LinearLogic}  based on   synthetic connectives ~\cite{GirMeanII}.
Polarization, which we discuss in  Section ~\ref{polarization},  is fundamental to
Girard's computational analysis  of   classical logic
~\cite{GirLC} in which the system $\mathbf{LC}$
was introduced, and more recently to the design of   Laurent's polarized linear logic $\mathbf{LLP}$  ~\cite{LaurentThesis,LauPol}.

\subsection{\texorpdfstring{$\mathbf{MELL}$}{MELL}} \label{MELL subsection}
Formulas of the constant-only, propositional, multiplicative-exponential linear logic $\mathbf{MELL}$ ~\cite{LinearLogic} are finitely generated
by the following grammar:
$$F::=
 \uno  \ | \ \ult \ | \ F \otimes F \ |  \ F \parr F \ | \  \oc F \ | \
  \wn F.$$

The involutive linear negation $^\bot$ is defined as follows:
$$
\uno \ ^\bot  :=  \ult;  \qquad
F \otimes G \ ^\bot  :=  F\b \parr G\b; \qquad
\oc F \ ^\bot  :=  \wn (F\b).
$$

A sequent, written $\vdash \Gamma$,  consists of a (possibly empty) multi-set of formulas $\Gamma = F_1,\ldots,F_n$.
Given two multi-sets $\Gamma$ and $\Delta$ the expression
$\Gamma,\Delta$ denotes their multi-set union.
Given a multi-set  $\Gamma =  F_1,\ldots,F_n$
we write $\wn \Gamma$ for $\wn F_1,\ldots,\wn F_n$.

Sequent  calculus rules  are given in Table  ~\ref{mell_fig}.

\begin{table}[h]
  \centering

\fbox{
\begin{tabular}{ccc}
&&\\
Multiplicative rules &
\AxiomC{$\vdash \Gamma,F$}
\AxiomC{$\vdash \Delta,G$}
\RightLabel{\scriptsize{$\otimes$}}
\BinaryInfC{$\vdash \Gamma,\Delta, F \otimes G $}
\DisplayProof
\qquad
\AxiomC{}
\RightLabel{\scriptsize{$\uno$}}
\UnaryInfC{$\vdash \uno$}
\DisplayProof
&
\AxiomC{$\vdash \Gamma, F,G$}
\RightLabel{\scriptsize{$\parr$}}
\UnaryInfC{$\vdash \Gamma, F \parr G$}
\DisplayProof
\qquad
\AxiomC{$\vdash \Gamma$}
\RightLabel{\scriptsize{$\ult$}}
\UnaryInfC{$\vdash \Gamma,\ult$}
\DisplayProof \\

&&\\
Exponential rules &

\AxiomC{$\vdash \wn \Gamma, F$}
\RightLabel{\scriptsize{$\oc$}}
\UnaryInfC{$\vdash \wn \Gamma,\oc F$}
\DisplayProof
\qquad
\AxiomC{$\vdash  \Gamma, F$}
\RightLabel{\scriptsize{$\wn$}}
\UnaryInfC{$\vdash  \Gamma,\wn F$}
\DisplayProof
&
\AxiomC{$\vdash  \Gamma$}
\RightLabel{\scriptsize{W}}
\UnaryInfC{$\vdash  \Gamma,\wn F$}
\DisplayProof
\qquad
\AxiomC{$\vdash  \Gamma,\wn F, \wn F$}
\RightLabel{\scriptsize{C}}
\UnaryInfC{$\vdash \Gamma,\wn F$}
\DisplayProof \\

&&\\
Cut-rule &
\AxiomC{$\vdash \Gamma,F$}
\AxiomC{$\vdash \Delta,F\b$}
\RightLabel{\scriptsize{Cut}}
\BinaryInfC{$\vdash \Gamma,\Delta$}
\DisplayProof &\\
&&\\
\end{tabular}
}
 \caption{$\mathbf{MELL}$}
  \label{mell_fig}
\end{table}

A key feature of linear logic  is the distinction between:
\begin{enumerate}[$\bullet$]
\item
\emph{linear formulas}:
$ \uno,  \ult,    F \otimes F, F \parr  F$;
\item\emph{exponential formulas}: $\wn F,\oc F$.
\end{enumerate}
Linear formulas can only be used once, while the
 modalities $\oc, \wn$ allow \emph{formulas} and sequent calculus \emph{derivations} to be \emph{erased} or \emph{duplicated} in the sense we now make precise.

The possibility of discarding  formulas is
expressed in the sequent calculus by the {\em weakening} rule  (or \emph{erase}, in the bottom-up reading of a derivation)  on $\wn F$ formulas:
\begin{prooftree}
\AxiomC{$\vdash  \Gamma$}
\RightLabel{\scriptsize{W}}
\UnaryInfC{$\vdash  \Gamma,\wn F$}
\end{prooftree}
The possibility of  repeating formulas is taken into account
by the {\em contraction} rule  (or \emph{duplication}, in the bottom-up reading)  on $\wn F$ formulas:
\begin{prooftree}
\AxiomC{$\vdash  \Gamma,\wn F, \wn F$}
\RightLabel{\scriptsize{C}}
\UnaryInfC{$\vdash \Gamma,\wn F$}
\end{prooftree}
In a dual sense, the modality $\oc$ allows   derivations to be erased or duplicated  during \emph{cut-elimination procedure}. Namely, (recall that  $\wn F\b$ is defined as $\oc (F\b)$)  we have:

\textbf{Weakening}: a cut
\AxiomC{$\vdots$ $\pi$}
\noLine
\UnaryInfC{$\vdash  \Gamma$}
\RightLabel{\scriptsize{W}}
\UnaryInfC{$\vdash \Gamma, \wn F$}
\AxiomC{$\vdots$ $\rho$}
\noLine
\UnaryInfC{$\vdash \wn \Delta,\oc (F \b)$}
\RightLabel{\scriptsize{Cut}}
\BinaryInfC{$\vdash \Gamma,\wn \Delta$} \DisplayProof
reduces to
\AxiomC{$\vdots$ $\pi$}
\noLine
\UnaryInfC{$\vdash  \Gamma$}
\UnaryInfC{$\vdots $ weakenings $ \vdots$}
\UnaryInfC{$\vdash \Gamma,\wn \Delta$} \DisplayProof \hspace{0.2cm} :\\

\noindent the derivation  of  $\vdash \wn \Delta,\oc (F \b)$ is erased.

More important (for the purposes of this paper) is the case
of contraction rule:

\textbf{Contraction}: a cut
\AxiomC{$\vdots$ $\pi$}
\noLine
\UnaryInfC{$\vdash  \Gamma,\wn F, \wn F$}
\RightLabel{\scriptsize{C}}
\UnaryInfC{$\vdash\Gamma, \wn F$}
\AxiomC{$\vdots$ $\rho$}
\noLine
\UnaryInfC{$\vdash \wn \Delta,\oc (F \b)$}
\RightLabel{\scriptsize{Cut}}
\BinaryInfC{$\vdash \Gamma,\wn \Delta$} \DisplayProof
reduces to\\
\begin{center}
\AxiomC{$\vdots$ $\pi$}
\noLine
\UnaryInfC{$\vdash  \Gamma,\wn F, \wn F$}
\AxiomC{$\vdots$ $\rho$}
\noLine
\UnaryInfC{$\vdash  \wn \Delta,\oc (F \b)$}
\RightLabel{\scriptsize{Cut}}
\BinaryInfC{$\vdash  \Gamma,\wn \Delta,\wn F$}
\AxiomC{$\vdots$ $\rho$}
\noLine
\UnaryInfC{$\vdash \wn \Delta,\oc (F \b)$}
\RightLabel{\scriptsize{Cut}}
\BinaryInfC{$\vdash \Gamma,\wn \Delta,\wn \Delta$}
\UnaryInfC{$\vdots $ contractions $ \vdots$}
\UnaryInfC{$\vdash \Gamma,\wn \Delta$}
\DisplayProof
\hspace{0.2cm}:
\end{center}

\noindent the derivation $\rho$ of  $\vdash \wn \Delta,\oc (F \b)$ can be used  several times, once for each duplication of $\wn F$.

\subsection{Polarities, focalization and synthetic connectives}
\label{polarization}

The connectives and constants of  linear logic are split into two families according to their {\em polarity} (positive or negative).
Let us discuss first the multiplicative-additive fragment $\mathbf{MALL}$, where the distinction is clearly highlighted also typographically.\\

\begin{tabular}{rcclc}
\textbf{
Positive multiplicative}\ : & $\uno,\otimes$, &  \qquad & \textbf{
Positive additive}\ : &  $\falso,\oplus, $\\
&&&&\\
\textbf{
Negative  multiplicative} \ : & $\ult,\parr$, & \qquad & \textbf{Negative additive} \ : & $\vero, \&$.\\
&&&&\\
\end{tabular}

\noindent The significance of polarities in linear logic was made explicit by Andreoli's seminal
work on {\em focalization}  ~\cite{AndreoliLP}.
The distinction into positive and negative corresponds in fact to properties  of the connectives  in proof construction ~\cite{GirMeanII,CurLLL2}.
If a sequent is provable in  linear logic, then it is provable with a
proof which satisfies the proof-search strategy which we recall next
(we recall that the polarity  of a formula is the polarity of the outermost connective).

{\em In the bottom-up construction of a proof:
\begin{enumerate}[\em(1)]

\item If there is a negative formula, keep on decomposing it until
we get to atoms or positive subformulas.

\item If there are not negative formulas,  choose a positive formula, and keep on decomposing it until
we get to atoms or negative
subformulas.
\end{enumerate}
}

\noindent For the exponential modalities $\wn,\oc$, the situation is however a bit more complex (see \eg ~\cite{AndreoliLP} and ~\cite{LaurentThesis}). This has lead Girard to analyze the exponentials via the following decomposition ~\cite{meaning1,LauPol}:
$$ \oc F = \shpos \sharp F,  \qquad \wn F = \shneg \flat F,$$
where $\sharp$ is the negative modality, $\flat$ is the positive modality, and $\shpos$, $\shneg$ are operators which change the polarity. The connectives
 $\sharp$ and $\flat$ are the ``true'' modalities,
 responsible of duplicative features. Hence,
{after the previous decomposition,} the contraction rule { would} become:
\begin{prooftree}
\AxiomC{$\vdash \Gamma, \flat P, \flat P$}
\RightLabel{\scriptsize{C}}
\UnaryInfC{$\vdash \Gamma, \flat P$}
\end{prooftree}

\noindent In this paper, we decided to use  symbols which are more familiar,
and we simply
 write $\negm $ (instead of $\sharp$) for the {negative} modality,
and $\posm$ (instead of $\flat$) for the {positive} modality.

Polarities allow us to have  \emph{synthetic connectives} ~\cite{GirMeanII,GirBook}
\ie  maximal clusters of connectives
of the same polarity.
The key ingredient that allows for the definition of
synthetic  connectives is precisely
 { the focalization property}.
In fact, from the point of view of logic, focalization (see the proof-search strategy above) means that each cluster of formulas with the same polarity can be introduced by a single
logical  rule (with several premises).
By using synthetic connectives,  formulas
are in a  canonical form, where immediate subformulas
have opposite polarity. This means that
in a (cut-free) proof, there is a  \emph{positive/negative  alternation
of  rules}, which matches  the standard
Player (positive)/Opponent (negative) alternation of moves in a strategy (see Section  ~\ref{Game Semantics}).

\subsection{\texorpdfstring{$\HS$}{HS}} \label{MELLS formulas and rules}
We now introduce in detail our calculus.\medskip

\emph{Formulas of $\mathbf{\HS}$} split into positive $P$ and negative $N$ formulas, and they are
inductively generated by the following grammar:
$$
\begin{array}{rcccc}
\mbox{ Positive formulas : } & P & ::= &  \posm(N_1\otimes \dots\otimes N_n) & (n \geq 0) \\
\mbox{ Negative formulas : } & N & ::= &  \negm(P_1\parr \dots\parr P_n) & (n \geq 0) \\
\end{array}
$$
When $n=0$, we
write $\posm \uno$ and $\negm \ult$ for the positive
and negative formula  respectively.
They are the only \emph{ground} formulas of our calculus.

We will use $F$ as a variable for formulas and indicate the polarity also
by writing  $F^+$ or $F^-$. We use $P,Q,R,\ldots$ (resp.\ $N,M,L,\ldots$)
for positive (resp.\ negative) formulas. To stress the immediate subformulas of some formula $F$, we often write
 $F^+(N_1, \dots, N_n)$ and  $F^-(P_1,\dots, P_n)$.

The involutive \emph{linear negation} $\b$  is defined in the natural way:
$$
\posm(N_1\otimes \dots\otimes N_n) \ ^\bot :=
 \negm(N_1^\bot\parr \dots\parr N_n^\bot).$$
In particular, $\posm \uno \ \b = \negm \ult$.

A sequent of $\HS$ is a (possibly empty) multi-set of formulas $\Gamma$,   written
 $\vdash \Gamma$, such that $\Gamma$  contains at most one (occurrence of) negative formula. In the sequel,
 $\Pi$ always stands for a (possibly empty) multi-set consisting of
 positive formulas only.\medskip

\emph{Rules of $\HS$} are given in Table  ~\ref{mells_fig}.

\begin{table}[h]
  \centering

\fbox{
\begin{tabular}{cc}
&\\
Positive rules : &
\AxiomC{$\vdash \Pi,P,N_1$}
\AxiomC{$\ldots$} \AxiomC{$\vdash \Pi,P,N_n$}
\RightLabel{\scriptsize{Pos$_n$}}
\TrinaryInfC{$ \vdash  \Pi, P$} \DisplayProof

\\
$P =
\posm(N_1\otimes \dots\otimes N_n)$ and  $n\geq 0 $ & \\
& \\

Negative rules : &

\AxiomC{$\vdash  \Pi,P_1,\ldots,P_n$}
\RightLabel{\scriptsize{Neg$_n$}}
\UnaryInfC{$\vdash \Pi,N$} \DisplayProof

\\

$N =
\negm(P_1\parr \cdots\parr P_n)$ and $n\geq 0$ &
\\

&\\
\end{tabular}
}

 \caption{$\mathbf{MELLS}$}
  \label{mells_fig}
\end{table}
\noindent In particular, when  $n=0$ (and
hence $P=\posm \uno$ and $N=\negm \ult$) we have:

\[\begin{tabular}{ccc}
\AxiomC{}
\RightLabel{\scriptsize{Pos$_0$}}
\UnaryInfC{$ \vdash  \Pi, \posm \uno$} \DisplayProof &
\AxiomC{$\vdash \Pi$}
\RightLabel{\scriptsize{Neg$_0$}}
\UnaryInfC{$\vdash \Pi, \negm \ult$} \DisplayProof
\end{tabular}
\]\medskip

\noindent In the sequel we use variables $\pi,\rho,\psi,\theta\ldots$
for derivations of sequents in $\HS$.

Structural rules (weakening and contraction)
are on  \emph{positive} formulas only, and
given implicitly in the positive rules.

\subsection{Expressivity of \texorpdfstring{$\HS$}{HS}}

In Appendix  ~\ref{calcolo_app} we  discuss the expressivity of
 $\HS$ by relating it to more standard systems.
In Appendix   ~\ref{intuition} we give  a
correspondence between
 $\HS$
and a ``focalized and synthesized" version
of the  $\neg ,\wedge$
fragment of the sequent calculus for { propositional}
intuitionistic logic  $\mathbf{LJ}$.
In Appendix
  ~\ref{corrispondenza}
we relate $\HS$ to a more standard
polarized version of $\mathbf{MELL}$ called $\mathbf{MELL}_{\mathsf{pol}}$ ~\cite{LaurentThesis}.

\subsection{ Cut-rule} \label{mellscut}

The cut-rule for $\HS$ is the following one:

\begin{prooftree}
\AxiomC{$\vdash \Xi,\Pi,P$}
\AxiomC{$\vdash \Delta, P\b$}
\RightLabel{\scriptsize{Cut}}
\BinaryInfC{$\vdash \Xi,\Pi,\Delta$}
\end{prooftree}
where the multi-set $\Xi$ is either empty or it consists of exactly one
(occurrence of) negative formula and $\Delta$ is a multi-set
of positive formulas.

This \textrm{cut} is \emph{admissible}   in  $\HS$ (Theorem ~\ref{principal}), \ie
it does not improve the derivability of sequents
\wrt  the cut-free formulation of $\HS$.
A proof of this fact is given in Appendix  ~\ref{cut adm_app}.

\section{HO innocent game semantics} \label{Game Semantics}

An {\em innocent
 strategy} ~\cite{Hy-On} can be described either in terms of all possible interactions for the player | {\em strategy as set of plays} | or in a more compact way, which provides only the minimal information for Player to move | {\em strategy as set of views} (see \eg ~\cite{Nickau,harthesis,CurGS}).
It is standard that the two presentations are equivalent: from a play one can extract the views, and from the views one can calculate the play.

 In this paper we use the ``strategy as set of views'' description. Our presentation of innocent strategies adapts to our needs  the presentations   by Harmer  ~\cite{harGS} and Laurent
~\cite{LauPol}.

Before introducing the formal notions, let us use an image. A strategy tells the player how to respond to a counter-player move. The dialogue between two players | let us call them \textit{P} (Player) and \textit{O} (Opponent) | will produce an \emph{interaction} (a play).
The ``universe of moves'' which can be played is set by the \emph{arena}.
Each move belongs to only one of the players, hence there are \textit{P}-moves and \textit{O}-moves. For \textit{P}, the moves which \textit{P} plays are positive (active, output), while the moves  played by \textit{O} are negative (passive, input), to which \textit{P} has to respond.

\subsubsection*{Polarities}
 Let $Pol :=\{+,-\}$ be the set of polarities:  positive (for Player) and negative (for Opponent).
We use the symbol $\epsilon$ as a variable to range over polarities.

\subsubsection*{Arenas}
An arena is given by  {\em a directed acyclic graph}, d.a.g. for short,
 which describes a dependency relation between moves
 and {\em a polarity function}, which assigns a polarity to the moves.
\begin{defi}[Arena]\label{arenaGS}
An \textbf{arena} $A=(M_A,\vdash_A,  \lambda_A)$
is given by:
\begin{enumerate}[$\bullet$]
\item a {\bf directed acyclic graph}  $(M_A,\vdash_A)$ where:

\begin{enumerate}[$-$]
\item $M_A$ (nodes of the d.a.g.) is the set of {\bf moves};
\item $\vdash_A$ (edges of the d.a.g.)
 is a well founded, binary {\bf enabling relation} on $A$. If there is an edge from $m$
to $n$,  we write $m\vdash_A n$. We call {\bf initial} each move $m$ such that no other move
enables it, and we write this as $\vdash_A n$.
\end{enumerate}
 \item
  a {\bf function}  $\lambda_A: M_A \to Pol$ which labels each element with a polarity $\epsilon$.
\end{enumerate}\medskip

\noindent Enabling relation and polarity have to satisfy the following property of {\bf alternation}:
\begin{center}
if $n\vdash_A m$,  they have opposite polarity.
\end{center}
 A {\em non-empty} arena whose initial moves have all the same polarity $\epsilon$, is said to be \textbf{polarized}  ~\cite{LauPol}.
 In such a case, if $\epsilon$ is positive (resp.\ negative), we say that the arena is  {\bf positive} (resp.\  {\bf negative}).

With a slight abuse of notation, we will write $m\in A$ for $m\in M_A$.

\end{defi}

\subsubsection*{Strategies}

A {\bf pointing string} over a set $X$   is a string $s\in X^*$ with pointers between the occurrences of $s=s_1.\dots s_n$
 such that, if $s_i$
points to $s_j$ then $j < i$, \ie pointers always point back to earlier occurrences, and we have at most one pointer from any given occurrence of $s_i$.

\begin{defi}[Justified sequence]
Let $A$ be an arena.
A {\bf justified sequence} $s$
on $A$ is  a pointing string $s=s_1.\dots s_n \in A^*$
 which satisfies the following properties.
\begin{enumerate}[$\bullet$]

\item
{\em Justification.} For each non-initial move $s_i$ of $s$, there is
 a unique pointer to an earlier occurrence of move $s_j$, called the justifier of $s_i$, such that $s_j\vdash_A s_i$.

\end{enumerate}
\end{defi}\medskip

\noindent The {\bf polarity} of a move in a justified sequence  is given by the arena.
We sometimes put in evidence the polarity of a move $x$ by writing $x^+$ or $x^-$.

\begin{defi}[View] \label{viewdef}

 A {\bf view}   on $A$  
 is a justified sequence  on $A$ which   satisfies:
\begin{enumerate}[$\bullet$]
\item
{\em Alternation.} No two following moves have the same polarity.
\item {\em View.} If $s=s_1.\dots s_n$,
for each  negative (Opponent) move $s_i$ such that $i>1$,  $s_i$ is justified
by its immediate predecessor $s_{i-1}$.
\end{enumerate}

\end{defi}

\begin{defi}[Strategy] \label{HOst}

A {\bf strategy} $\D$ on $A$, denoted by $\D:A$
is a prefix-closed set of {\em non-empty} views, such that:
\begin{enumerate}[(1)]
\item {\em Coherence.}  If $s.m,s.n\in \D$ and $m \not= n$ then $m,n$ are
negative.

\item {\em Positivity.}  If $s.m$ is maximal in $\D$ (\ie no other view extends it), then $m$ is positive.

\end{enumerate}

\end{defi}

In case that the arena is polarized |which is the case for all arenas we deal with in this paper|
we call {\em positive}  (resp.\ {\em negative}) a strategy on a positive
(resp.\ negative) arena.

\begin{rem}
The choice of defining a strategy as a set of non-empty views is not standard, but is coherent with the setting of ludics. Following
~\cite{GirLoc}, a strategy {\em can} be an empty set (of views), whereas it {\em never} contains the empty view.
\end{rem}

As a consequence, condition (1) (Coherence) in Definition ~\ref{HOst} implies that  in a {\em positive strategy}, all views have the {\em same first move}.  On the other hand, a negative strategy may also be a forest, { rather than a tree (see Section ~\ref{conventions})}.

\subsection{Constructions on arenas}
We  give some constructions on arenas, which  we will need
in Section  ~\ref{ludics}.\\

Let $A_1, A_2$ be {positive} arenas  such that
the sets of moves are {\em disjoint} (\ie $M_{A_1} \cap M_{A_2}=\emptyset$).\\
The arena $A_1 \parallel A_2$ is defined as follows:
\begin{enumerate}[$\bullet$]
\item $M_{A_1 \parallel A_2}\  := \ M_{A_1} \cup  M_{A_2}$;
\item
$\lambda_{A_1 \parallel A_2}(m)  \ :=  \ \lambda_{A_1}(m)$ if $m\in M_{A_1}$, \ \
 $\lambda_{A_1 \parallel A_2}(m)  \ :=  \ \lambda_{A_2}(m)$ if $m\in M_{A_2}$;
\item $\vdash_{A_1 \parallel A_2} \ := \ \vdash_{A_1} \cup  \vdash_{A_2}$.

\end{enumerate}

Observe that the structure of $A_1 \parallel A_2$ is inherited from its constituent arenas; we place the arenas side-by-side.
The construction immediately generalizes to the $n$-ary case: if $A_1 \ldots,A_n$ ($n\geq 1$) are { positive} arenas  such that
 for any $ 1\leq i,j \leq n$
the sets of moves of $A_i$ and $A_j$ are pairwise disjoint, we obtain  $A_1 \parallel \cdots \parallel A_n$.
It is immediate that $A_1 \parallel \cdots \parallel A_n$ is a positive arena.

Let $C$ be a negative and $A$ be a positive arena,  with {\em disjoint} sets of moves.\\  The \ arena
 $C\freccia  A$ is defined as follows.
 \begin{enumerate}[$\bullet$]
\item $M_{C\freccia  A}\  := \ M_C \cup M_A$;
\item $\lambda_{C\freccia  A}(m)  \ :=  \ \lambda_C(m)$ if $m\in M_C$, \ \ $\lambda_{C\freccia  A}(m)\ :=  \ \lambda_A(m)$ if $m\in M_A$;
\item$m\vdash_{C\freccia  A} n$ holds if
 \begin{enumerate}[$-$]
 \item[] $m\vdash_{C} n$ , or
  \item[] $m\vdash_{A} n$ , or
 \item[] $m$ is {\em initial} in $C$ and $n$ is {\em initial} in $A$.
\end{enumerate}
\end{enumerate}

\noindent In words, the arenas $C$ is lifted on top of the roots of $A$, by adding each root of $A$ as child to each root of $C$.

\subsection{Composition of strategies}

Composition of strategies as sets of views
has been studied in particular by Curien and Herbelin. They  have introduced the
View-Abstract-Machine (VAM) ~\cite{CurAbs,CurHerb} by elaborating
 Coquand's Debates machine ~\cite{Coq}.
We will give more details in section  ~\ref{VAM}.

\subsection{Conventions and notation} \label{conventions}
It is now convenient to fix
some conventions and notation
we will employ in all the rest of this paper.

We will  often deal with sets of $n\geq 0$ elements, which we write
$\{L_i: 1\leq i \leq n\}$ or $\{L_1, \dots, L_n\}$. The case $n=0$ always corresponds to the empty set.

Given $n\geq 0$, $I_n$ \emph{always stands} for the set
given as follows:\\
\begin{center}
$I_n :=$ $\left\{
  \begin{array}{ll}
    \{1,\dots,n\}, & \hbox{if $n>0$;} \\
    \emptyset, & \hbox{if $n=0$.} \\
  \end{array}
\right.
$
\end{center}
Given an index  set $S$, we will often use  (as in ~\cite{GirLoc})  the notation $(L_s), s\in S$, to indicate a family of elements, indexed by $S$.
When $S$ is clear from the context, we just write $(L_s)$.\smallskip

Let $\X$ be a set of views. We call {\bf root} each occurrence of move $a$ such that
$a.s\in \X$.

 Emphasizing the arborescent structure of
a strategy, it is  convenient to write a strategy with a single root  $a$ as
$\D=a.\X$, where $\X$  is the set of pointing strings $\{s : a.s \in \X\}$. Notice that
$\X$ does not need to be a strategy. In case
 $\X$ is empty  we just
 write $\D=a$.

 We will come back on this in  Section ~\ref{linearity} and in Section ~\ref{compound types}, Lemma  ~\ref{projections}.\smallskip

To better grasp the intuitions, we will draw strategies as  trees
whose nodes are labeled by moves.  Nodes which are labeled by positive moves are circled.

\begin{exa}
Let $\D$ be  the strategy given by the  closure under non-empty prefix
of the set of views $\{ a^+.b^-.d^+ \ , \ a^+.c^-.e^+ \}$ (less formally, we could also write
$\D= a^+.\{b^-.d^+, c^-.e^+ \}$).
We represent $\D$
by the following tree:

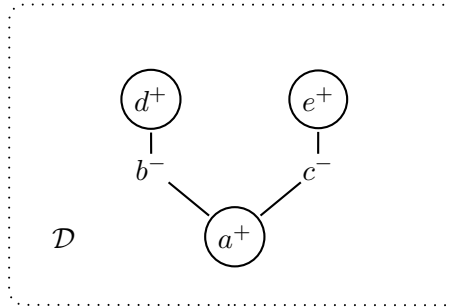
\begin{center}
\psmatrix[rowsep=9pt,colsep=0.3cm]
&&&&&&&\\
&&&\rnode{up}{}&&&&\\
&&\circlenode{leaf1}{$d^+$}&& \circlenode{leaf2}{$e^+$}&& \\
&& \rnode{x1m}{$b^-$} && \rnode{x2m}{$c^-$}&& \\
$\D$&&&\circlenode{xp}{$a^+$}&&&&\\
\endpsmatrix
 \psset{shortput=nab,arrows=->,labelsep=6pt}
 \small
\ncline[nodesep=1pt]{-}{leaf1}{x1m}
\ncline[nodesep=1pt]{-}{leaf2}{x2m}
\ncline[nodesep=1pt]{-}{x1m}{xp}
\ncline[nodesep=1pt]{-}{x2m}{xp}
\ncbox[nodesep=.5cm,boxsize=3,linearc=.2,
linestyle=dotted]{xp}{up}
\end{center}
\end{exa}

\section{Ludics, the linear case}\label{ludics}
In this and next section we give a compact but complete presentation of ludics ~\cite{GirLoc}, introducing
 all definitions and technical results which are relevant to our approach, including
internal completeness and full completeness.
Our choice here is to give a presentation which fits into the language of game semantics.

Let us first stress again the peculiarity of ludics in the panorama of game semantics.
In game semantics, one defines constructions on arenas which correspond to the
interpretation of types.  A strategy is always ``typed'', in the sense
that it is a strategy on a specific arena: first we have the ``semantical type'' (the arena), and then the strategy on that arena. When strategies are opportunely typed, they
interact (compose) well.

In the approach of ludics, there is only one arena (up to renaming): the universal arena of all possible moves.
Strategies
are ``untyped'', in the sense that all strategies are defined on the universal arena. Strategies then interact with each other, and the interaction can
{\em terminate well} (the two strategies ``accept'' each other)  or not ({\em deadlock}).

Two opposite strategies $\D,\E$ whose interaction terminates well, are said to be
{\bf orthogonal}, written $\D\bot\E$.
Orthogonality allows us to define interactive types.
A semantical type $\bG$ is any set of strategies which react well to the
same set of tests $\bE$, which are themselves strategies (counter-strategies), that is $\bG=\bE\b$.

\subsubsection*{Daimon} One of the goals in
the program of ludics is to overcome the distinction between syntax (the formal
 system) on one side and semantics (its interpretation) on the other side.
Rather then having two separate worlds,  proofs are interpreted via
 proofs. To determine and test properties, a {\em proof} of $A$ should
be tested with {\em proofs} of $ A\b$. Ludics provides a setting in which
proofs of $A$ interact with proofs of  $ A\b$; to this aim, it
generalizes the notion of proof.

A proof should be thought  in the sense of   ``proof-search'' or
``proof-construction'': we start from the conclusion, and guess a last rule,
then the rule above.  What if we cannot apply any rule?
 A new rule is introduced, called
 \textbf{daimon}:
$$\infer[\dai]{\vdash \Pi}{}
$$
Such a rule allows us to  assume any conclusion, or said in other words,
it allows to close any open branch in the proof-search tree
of a sequent.

In the semantics, the daimon is a special action which acts as a
{\em termination signal}.

\subsection{Strategies on a universal arena} \label{strategy on uni}
We now introduce the notion of strategy as defined in ludics.
The role of the arena becomes somehow secondary, while the central notion, is that of {\em name}.
Strategies communicate on names. We can think of names as process algebras channels, which can be used to send outputs (if positive) or to receive inputs
(if negative). Each strategy $\D$ will have an \emph{interface}, which
provides the names on which  $\D$  can communicate
with the rest of the world, and the use (input/output) of each name.\\

 A {\bf name} (called {\em locus} in  ~\cite{GirLoc})
is a string of natural numbers. We use the
 variables $\xi,\ss,\alpha,\dots$
to range over names.
Two names are {\bf disjoint} if  neither is a prefix of the other one.

 An {\bf action} $x$ is either the symbol $\dai$ (called \textbf{daimon})
or a pair
$(\xi,I)$,
where $\xi$ is a name, and $I$ is a finite subset  of  $\mathbb{N}$. In this paper, we will always assume $I$ of the form $I_n=\{1,\dots,n\}$, for some $n\geq 0$ (see Section ~\ref{conventions})\footnote{ We
 can think of the set
 $I_n$ only just as an ``arity provider.'' We use the notation $I_n$ first of all for compatibility (and comparability) with ~\cite{GirLoc}, in which the general case is needed to deal with the additives. Secondly, this choice leaves open the possibility to extend our work with the additive structure of ludics, without essential modifications.}.
Given an action $(\xi, I)$ on the name $\xi$, the set $I$ is actually a shortcut for the set of the
 names $\{\xi i: i\in I\}$ which are generated from $\xi$ by this action. We call {\bf proper} an action of the form $(\xi,I)$ (to contrast with a $\dai$ action, which has a different function). The role of proper actions and of the $\dai$ actions will be very different when defining interaction between strategies.\\

The prefix relation  (written $\xi\pref \sigma$)
induces a natural relation of dependency on names, and hence on the proper actions, which
generates an arena.

\begin{defi}[Universal arena (on a name)] \label{A_name}
 Given a name $\xi$ and a polarity $\epsilon\in \{+,-\}$, the
\textbf{universal  arena}   $A(\xi,\epsilon)$
is the tuple $(M,\vdash, \lambda)$ defined as follows:

\begin{enumerate}
\item
The {\bf set of moves} $M$ is the
 set of all actions of the form $(\xi',I)$,
where
 $\xi \pref \xi'$  and $I$ is a finite subset of $\mathbb{N}$.

\item The {\bf polarity} of the initial actions is $\lambda((\xi,I))=\epsilon$ for each $I$;
the polarity of any  other action is the one induced by alternation.

The {\bf enabling relation} $x\vdash y$ is defined as follows:

\begin{center}
$x\vdash y$ if $x=(\xi',I)$ and $y= (\xi' i, J)$, with $  i\in I$.
\end{center}
\end{enumerate}
\end{defi}\medskip

\noindent The choice of a name ``delocalizes'' the universal arena on a specific name. Of course, all universal arenas are isomorphic up to   the choice of some  name $\xi$.

\begin{defi}[Interface]
 An {\bf  interface} $\Gamma$ (called \emph{base} in ~\cite{GirLoc}) is a  (possibly empty) finite
 set of pairwise disjoint names, together with a polarity for each name, such that
{\em at most one name is negative}. If a name $\xi$ has polarity $\epsilon$, we write $\xi^{\epsilon}\in \Gamma$.

With an abuse of notation, in the sequel we often write
$\Gamma=\xi_1^{\epsilon_1},\ldots,\xi_n^{\epsilon_n}$
instead of $\Gamma=\{\xi_1^{\epsilon_1},\ldots,\xi_n^{\epsilon_n}\}$.
When no confusion arises, we  also omit the polarities and simply write
$\Gamma=\xi_1,\ldots,\xi_n$.
\end{defi}

An interface $\Gamma$ is {\em negative} if it contains a negative name,
{\em positive} otherwise. In particular, the empty interface is positive.

\begin{defi}[Universal arena on an interface] \label{Uarenaonint}
We denote by $A_\dai$ the  arena whose set of moves is  $\{\dai\}$, with the polarity of $\dai$ being positive.

Let $\Pi = \xi_1,\ldots,\xi_n$ be a (possibly empty) {\em positive} interface.

The {\bf universal arena on the interface $\Pi$} is the arena
 $$U(\Pi) := \  A(\xi_1,+) \parallel \cdots \parallel A(\xi_n,+)\parallel A_\dai.$$

The  {\bf universal arena
on a (negative) interface $\{\sigma^-\} \cup \Pi$} is the arena
$$U(\Gamma) := \ A(\sigma,-)\freccia  U(\Pi).$$

\end{defi}

\begin{rem}We observe that, according to Definition  ~\ref{arenaGS}, a universal arena $U(\GG)$ is a polarized arena; the polarity of the arena is that of its initial actions,  which results the same as the polarity of the interface $\GG$.
\end{rem}

\begin{exa}
The universal arena $U(\xi^+)$  can be pictured as in Figure ~\ref{fig:universal}. An arrows from $b$ to $a$ stands for
an enabling
$a \vdash b$;
 the polarity of the actions is given as follows: actions
 lying on even (resp.\ odd) layers have positive (resp.\ negative)
 polarity.

\begin{figure}[h!]
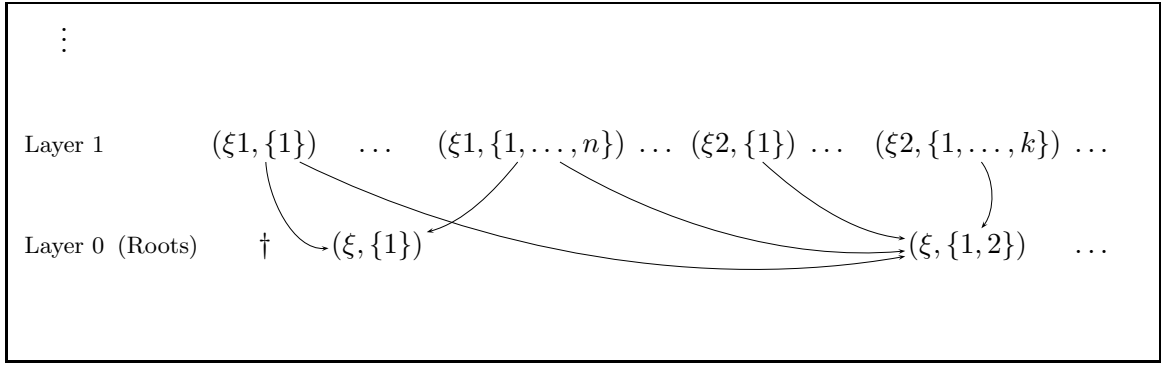

  \centering
  \fbox{
\psmatrix[rowsep=25pt,colsep=0.17cm]

\rnode{alto}{$\vdots$} & & & & & & & & & &  & & &  \\

\footnotesize{Layer $1$}  & & \rnode{xi1K}{$(\xi 1,\{1\})$} & \dots & \rnode{xi1N}{$(\xi 1,\{1,\dots,n\})$} & \dots & \rnode{xi2K}{$(\xi 2,\{1\})$} & \dots &  & \rnode{xi2N}{$(\xi 2,\{1,\dots,k\})$} & \dots \\

\footnotesize{Layer $0$}  & \footnotesize{(Roots)} & $\dai$   & \rnode{xiI}{$(\xi,\{1\})$} & & &  & &  & \rnode{xiJ}{$(\xi,\{1,2\})$}& \dots& & & \\

 \endpsmatrix
 \psset{shortput=nab,arrows=->,labelsep=6pt}
 \small
\ncarc[linewidth=0.2pt,nodesep=1pt,arcangle=-45]{xi1K}{xiI}
\ncarc[linewidth=0.2pt,nodesep=1pt,arcangle=-18]{xi1K}{xiJ}
\ncarc[linewidth=0.2pt,nodesep=1pt,arcangle=18]{xi1N}{xiI}
\ncarc[linewidth=0.2pt,nodesep=1pt,arcangle=-18]{xi1N}{xiJ}
\ncarc[linewidth=0.2pt,nodesep=1pt,arcangle=-18]{xi2K}{xiJ}
\ncarc[linewidth=0.2pt,nodesep=1pt,arcangle=38]{xi2N}{xiJ}
}
  \caption{The universal arena $U(\xi^+)$}
  \label{fig:universal}
\end{figure}

\end{exa}

\begin{defi}[Untyped strategies] \label{untyped}  Let $\Gamma$
be an interface.
 A {\bf strategy $\D$ on $\Gamma$,} also written $\D: \Gamma$,
is a strategy (in the sense of Definition  ~\ref{HOst}) on the universal arena $U(\Gamma)$.
\end{defi}

We point out that  here we call \emph{strategies} and \emph{views} (following the language of game semantics) what  in  ~\cite{GirLoc}
is indicated as \emph{designs} and \emph{chronicles}
respectively.

\begin{exas}[Basic strategies: $\Dai$, $\Fid$, $\emptyset$] \label{daimon ex}
Let us point out a few relevant strategies.

\begin{enumerate}[$\bullet$]
\item There are two {\em positive} strategies which play a key role in ludics: $\Dai$ and $\Fid$.
Both are defined on any {\em positive} interface, and in particular also on the empty interface. In fact,
 they are {\em
the only possible strategies on the empty interface.}

\begin{enumerate}[$-$]
\item $\Dai$ is  the  strategy which consists  of only one action
$\{\dai\}$; it is called \emph{daimon}.

\item $\Fid$ is  the {\em empty} strategy;
it is called \emph{faith}.
\end{enumerate}

\item
We highlight also a simple example of negative strategy: the {\em empty} strategy on a negative interface.
We will denote this strategy simply by $\emptyset$.
\end{enumerate}
  \end{exas}

\subsubsection{Totality} \label{tot section}
  \begin{defi}[Totality]
We say that an untyped strategy is {\bf total} when it is not $\Fid$.
\end{defi}

 The definition of totality deserves some explanations.
 First, it is not the usual totality condition
of game semantics.
 Totality, in ludics, is closely connected with the definition of orthogonality, hence it will become clear only  after Section ~\ref{ortho}.
 Let us however anticipate some remarks, to justify why $\Fid$ is not a total strategy (while the empty negative strategy $\emptyset$ is total).
The definition of orthogonality is based on the fact that there are only two possible outcomes when we make interact two strategies on dual interfaces: either $\Fid$ or $\Dai$. The former is interpreted as failure of the interaction process (no output/deadlock), while  $\Dai$ is interpreted as
 success of the interaction process. In such a case, the strategies are said to be orthogonal.
We will  see  (Example \ref{DaiFid}) that, for the way in which the interaction is defined, when we have a negative strategy $\E$, its interaction with $\Dai$ always succeeds, while its interaction  with $\Fid$ always fails (\ie there is no output).

Moreover, the fact that  the interaction with $\Dai$ succeeds for
 any negative strategy, including the case  of the empty  negative strategy  allows us also
to understand why  the {\em negative} empty strategy is total. Such a strategy
 is in fact  the smallest strategy to be orthogonal to $\Dai$.

 In terms of linear logic,
 the
  empty negative strategy $\emptyset$ interprets the  rule for $\vero$, while $\Dai$ would correspond to a  ``proof"  of $\falso$ (here
  $\vero$ and $\falso$ denote the additive constants \emph{top} and \emph{zero}
  respectively).
But there is no corresponding ``proof-object" for $\Fid$.
We want a strategy which corresponds to a proof to be total.

\subsubsection{Linearity.}\label{linearity}
\begin{defi}[Linearity] \label{linearity def}
Given a strategy $\D:\Gamma$, we say that an occurrence of  action $(\xi,I)$ in $\D$ is {\bf linear}
if the name $\xi$ is only used by that occurrence of action.
We say that  $\D:\Gamma$ is {\bf linear} if each occurrence of  (proper)  action in $\D$ is linear.
\end{defi}
Linearity has as consequence that all {\em pointers} are trivial (each move has only one possible justifier and
the prefix relation between names  univocally tells us which is), and then can be forgotten.

\begin{rem}
Linear strategies are essentially the strategies introduced in ~\cite{GirLoc}.
Our condition is actually more strict than the condition in ~\cite{GirLoc}, and allows us a simplification of some details, by  restricting the setting only to multiplicatives.
The linearity condition in ~\cite{GirLoc}  is slightly more
complex, in order
 to take into account also the additive structures (additive duplication
is allowed), but
for our discussion
it is enough to ask that in a strategy
 {\em each name is only used once.}
\end{rem}

Observe that
a relevant consequence of our strict  condition of linearity  is that {\em both positive and negative strategies have a single root}
 (\emph{\ie they are trees, and not forests}). Hence, we can always write a negative strategy of root $x$ as $x^-.\C$.  It is immediate to check  that
 $\C$ is a positive strategy.
 As for positive strategies, let
 $\D$ have root $(\sigma, I_n)^+ $, \ie  $\D = (\sigma, I_n)^+.\E$.
  All views in $\E$ have a first action  of the form  $(\sigma i, K)$, with $i\in I_n$.
  We partition $\E$ into maximal subsets  $\E_i, 1\leq i \leq n $
  of views which start with the same action.
  The linearity condition implies that each $\E_i$  is a negative strategy
which either  has a unique root of the form $(\sigma i,K)^-$ or
 is the empty negative strategy (\ie $\E_i= \emptyset$).
We will also write $\D = (\sigma, I_n)^+. \{\E_1,\dots,\E_n\}$,
 to emphasize the tree structure.\\

{\em From now on, and till the end of Section  ~\ref{int_compl}, strategies are always linear strategies.}

\subsection{Dynamics in the linear case}\label{lin_norm} The composition of untyped strategies can be described via the VAM machine (see Section  ~\ref{VAM}).
For the moment, we only describe normalization in the linear case (see ~\cite{GirLoc,FagTra}). This case is
simpler, but has all the key ingredients
to follow most of the examples of  this paper.\smallskip

The central idea beyond names, is that we can compose two strategies   when their interfaces
have a common name, which appears in the two interfaces with opposite polarity.
The key case is the following: $\D_1:\sigma^+,\Gamma$  and $\D_2:\sigma^-,\Delta$, with $\GG\cap \DD=\emptyset$.

$\D_1$ can communicate with $\D_2$ through the  name $\sigma$.
The shared name $\ss$ is called a {\bf cut}.

 Rather than define the dynamics for pairs of strategies, it is
more convenient to define it for an arbitrary finite number of strategies at the same time.

\begin{defi}[Cuts and cut-nets] \label{cut-net}

Two interfaces $\GG_1, \GG_2$ are said to be {\bf compatible} if all names are pairwise disjoint or equal,
and in the latter case any name which appears in both the interfaces, appears with opposite polarity.
If $\xi^+ \in \GG_i$ and $\xi^-\in \GG_j$, the name $\xi$ is said to be a {\bf cut}.

A non-empty finite set of interfaces $\{\Gamma_1,\ldots,\Gamma_n\}$
is said to be {\bf valid} if:
\begin{enumerate}[$\bullet$]
\item  $\Gamma_1,\dots, \Gamma_n$ are \emph{pairwise compatible} interfaces;
\item the graph defined here below is  {\em connected}
and {\em acyclic}:
\begin{enumerate}[$-$]
\item \emph{nodes}:   $\Gamma_1,\dots, \Gamma_n$;
 \item \emph{edges}:
 there is an edge
 between $\Gamma_i$ and $\Gamma_j$ for each cut  $\xi$ such that $\xi^+ \in \GG_i$ and $\xi^-\in \GG_j$.
\end{enumerate}
\end{enumerate}
A  non-empty finite set of strategies $\R = \{\D_1 : \Gamma_1,\ldots, \D_n : \Gamma_n \}$ is said to be
a {\bf cut-net} if $\{\Gamma_1,\ldots,\Gamma_n\}$ is a valid set of interfaces.

We also write $\R = \{\D_1,\ldots, \D_n \}$ when
the interfaces of the strategies are clear from the contexts
or irrelevant for our purposes.

\end{defi}

\begin{rem}

The acyclicity condition implies that any two interfaces in a valid set have at most one name in common.
Observe  that if $\{\Gamma_1,\Gamma_2\}$ is a valid set of interfaces, then  there is exactly one cut (at most one by acyclicity, at least one by connectedness).
\end{rem}

\begin{exa} \label{exa cut-net}
The following
 three sets of interfaces are all pairwise compatible, but only the third one (3) is valid: (1) is not acyclic, an (2) is not connected.
\footnote{In the picture, we orient the edges for future uses
of this example (Proposition ~\ref{main strategy prop}).
But recall that the
edges of the graph  as given in Definition ~\ref{cut-net} are not oriented.}

 \begin{center}
\begin{tabular}{l c}
&\\
(1)$^{\phantom{A^{A^{AV}}}}$ &
\psmatrix[rowsep=10pt,colsep=1cm]

 \rnode{p1}{$\xi^-,\sigma^+$}
 & \rnode{p2}{$\sigma^-,\alpha^+,\beta^+$}
 & \rnode{p3}{$\beta^-,\xi^+$}
\endpsmatrix
 \psset{shortput=nab,arrows=->,labelsep=2pt}
 \small
\ncarc[linewidth=0.2pt,nodesep=3pt,arcangle=25]{p1}{p2}^{\scriptsize{$\sigma$}}
\ncarc[linewidth=0.2pt,nodesep=3pt,arcangle=25]{p2}{p3}^{\scriptsize{$\beta$}}
\ncarc[linewidth=0.2pt,nodesep=3pt,arcangle=25]{p3}{p1}_{\scriptsize{$\xi$}} \\
& \\
& \\
(2) &
\psmatrix[rowsep=10pt,colsep=1cm]

 \rnode{q1}{$\beta^-,\xi^+$}
 & \rnode{q2}{$\alpha^-, \gamma^+$}
\endpsmatrix \\
& \\
& \\
(3)$^{\phantom{A^{A}}}$ &
\psmatrix[rowsep=10pt,colsep=1cm]

 \rnode{r1}{$\sigma^-,\alpha^+,\beta^+$}
 & \rnode{r2}{$\beta^-,\xi^+$}
 & \rnode{r3}{$\alpha^-\gamma^+$}
\endpsmatrix
 \psset{shortput=nab,arrows=->,labelsep=2pt}
 \small
\ncarc[linewidth=0.2pt,nodesep=3pt,arcangle=25]{r1}{r2}^{\scriptsize{$\beta$}}
\ncarc[linewidth=0.2pt,nodesep=3pt,arcangle=-25]{r1}{r3}_{\scriptsize{$\alpha$}}\\
& \\
& \\
\end{tabular}
\end{center}

 \end{exa}

\begin{defi}[Closed cut-net]
A cut-net  $\R$ is  {\bf closed}  if all names in the interfaces are cuts.
\end{defi}

\begin{exa}
We have a typical example of closed cut-net  when we have strategies on opposite
 interface, such as $ \D:\xi^+$ and $\E: \xi^-$.
 \end{exa}

\begin{rem} \label{remark only one}
 Closed cut-nets contain exactly one positive strategy
 (see also Remark~\ref{main_rem}).
\end{rem}

Given a cut-net $\R = \{\D_1:\GG_1,\dots,\D_n:\GG_n\}$, the result of the
composition is called {\bf normal form}, and denoted by $\cut{\R}$.
Composition (normalization) follows the standard paradigm of {\em parallel composition} (\emph{the interaction}) \emph{plus hiding} of internal communication: $\cut \R$
is obtained from the result of the interaction  by hiding all the actions on internal names.
   The result is
 a strategy on the interface which is obtained from $\bigcup \GG_i$ by hiding all cut names.

Given the cut-net $\R$, if
 $\Theta$ is the set of the  cuts, then
 $(\bigcup_{1 \leq i \leq n} \Gamma_i) \senza \Theta$ is an interface,
which is called \textbf{the interface of $\R$}.\smallskip

Composition of the strategies in a cut-net can be described in several equivalent ways.
The {\em merging of orders} ~\cite{GirLoc,FP09} is the most compact and mathematically pleasant | but it is specific to the
linear case.
In this paper, we prefer to describe composition via an {\em abstract machine} (the LAM ~\cite{GirBook,FagTra}), because this will serve as an introduction to the machine which performs the composition of strategies with repetitions.

Let us first introduce with a small example  the basic ideas on the way in which strategies in a cut-net interact with each other to produce the normal form.
Since each action  appears only once, the dynamics  is extremely simple: we match actions of opposite polarity.

\begin{exa} \label{esempio}
Let us consider the following  strategies:
\begin{enumerate}[$\bullet$]
\item  $\D=a^+.b^-.\dai$;
\item $\E=a^+ $;
\item     $\F=  a^-.b^+$;
\end{enumerate}
where $a=(\xi,I)$ and
$b=(\xi 1,K)$, so that
$\D,\E : \xi^+$ and $\F :\xi^-$. Notice that
 $b$ is enabled by $a$ in both the underlying universal arenas.
We can draw these strategies as follows:
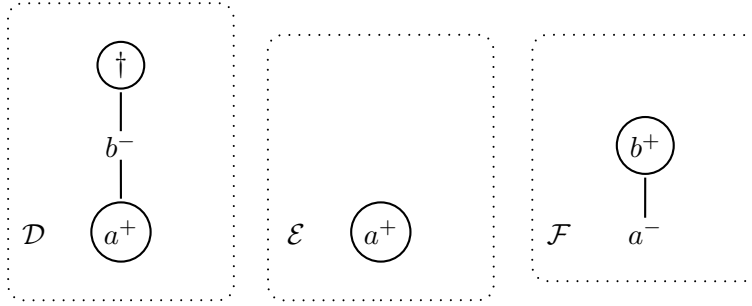
\vspace{1cm}

\begin{center}
\psmatrix[rowsep=9pt,colsep=0.3cm]

&&&&&&&&
&&&&&&&&&&&&&&&&
\\

&&\circlenode{leaf1}{$\dai$}&&&&&&
&&\rnode{fake2}{}&&&&&&&&\rnode{fake3}{}&&&&&&
\\
&&\rnode{node1}{$b^-$}&&&&&&
&&&&&&&&
&&\circlenode{node3}{$b^+$}&&&&&&
\\
$\D$&&\circlenode{root1}{$a^+$}&&&&&&
 $\E$&&\circlenode{root2}{$a^+$} &&&&&&
$\F$&&\rnode{root3}{$a^-$}&&&&&&
\\
&&&&&&&&
&&&&&&&&&&&&&&&&
\\
\endpsmatrix
 \psset{shortput=nab,arrows=->,labelsep=6pt}
 \small
\ncline[nodesep=1pt]{-}{root1}{node1}
\ncline[nodesep=1pt]{-}{leaf1}{node1}
\ncline[nodesep=1pt]{-}{root3}{node3}

\ncbox[nodesep=.5cm,boxsize=1.5,linearc=.2,
linestyle=dotted]{root1}{leaf1}
\ncbox[nodesep=.5cm,boxsize=1.5,linearc=.2,
linestyle=dotted]{root2}{fake2}
\ncbox[nodesep=.5cm,boxsize=1.5,linearc=.2,
linestyle=dotted]{root3}{fake3}
\end{center}

\noindent Let us have $\D$ interact with $\F$. Remember the intuition  that ``$+$" corresponds to an output, and ``$-$" to an input.
$\D$ starts by playing the  move $a^+$, $\F$ receives $a$ and checks its answer to that move, which is $b^+$. If $\D$ receives input $b$, its answer is $\dai$, which terminates the interaction. Summing up, the interaction produces $a^+.a^-.b^+.b^-.\dai$.
If we hide the internal communication, we get $\dai$, \ie
$\pl \D, \F \pr =\Dai$.

If we have $\E$ interacting with $\F$, we again match $a^+$ with $a^-$.
Then $\F$ plays $b$,
but $\E$ has not considered the action $b$.
Here we have a deadlock \ie $\pl \E,\F \pr = \Fid$.
\end{exa}

\subsubsection{Linear composition: \emph{LAM} (Loci-Abstract-Machine)}\label{LAMsec}

To formally define  normalization, we need to introduce a notion of order on the strategies of a cut-net; such a definition
 makes explicit the  order in which normalization accesses the strategies.\smallskip

Up to the end of this section,
let us fix a cut-net
$\R = \{\D_1 : \Gamma_1,\dots,\D_n : \Gamma_n\}$.

\begin{prop}[Main strategy] \label{main strategy prop} Given a cut-net $\R$, we define the following \emph{precedence relation} $<$
 on the strategies:
 \begin{center}

$\D_i:\GG_i < \D_j:\GG_j$ \  if  there is a cut $\xi$ such that \ $\xi^+ \in \GG_i$ and $\xi^-\in \GG_j$.

 \end{center}
The order induced by this precedence relation has a minimal element, which
 is called the  {\bf main strategy} of $\R$.
\end{prop}

\proof Since the graph
induced by the valid set of interfaces is acyclic (Definition  ~\ref{cut-net}),  the order is arborescent, because each interface contains at most one negative
  name, hence each strategy has at most one immediate predecessor.

  To verify that there is a unique  minimal element, we observe that
 the precedence relation induces also  an orientation on the edges of the graph:
for each cut  $\xi$ such that $\xi^+ \in \GG_i$ and $\xi^-\in \GG_j$, the  edge is oriented
 {\em from} $\Gamma_i$ {\em to} $\Gamma_j$ (see {Example ~\ref{exa cut-net}}).  Since the graph is connected,
 the oriented graph is a tree, and not a forest. Hence
   there is a unique  minimal
 strategy.
\qed

\begin{rem}\label{main_rem} \label{rem poscut} In a cut-net,  there is at most one  positive strategy, which necessarily is the main one. The argument is the same as in the previous proof.

\end{rem}

A cut name $\xi$ and all the names hereditarily generated from $\xi$,
are said to be {\bf internal}. We call internal an action on an internal name, otherwise the action is said to be \textbf{visible}.

Observe that if $\R$ is a  {closed cut-net}, then all the names are internal   (and  its interface is empty). The only possible visible actions are $\dai$ actions.

The following lemma is also an easy preliminary observation.

\begin{lem}
Let $\R$ be a cut-net. Given a proper  action $x^{\epsilon}$, it occurs in at most one  $\D\in \R$. \qed
\end{lem}

\begin{defi}[LAM ~\cite{FagTra,GirBook}] \label{LAMdef} Given a cut-net $\R $,
the set $LAM(\R)$
 is the
 set of sequences of actions defined as follows.

\begin{enumerate}[(1)]

\item {\bf (Initialization)}
If the main strategy
of $\R$ is  empty, we set $LAM(\R):= \emptyset$. Otherwise, if $a$ is the root
 of the main strategy, we set
$a \in LAM(\R)$.

\item
 Let $ p=x_1\dots x_n \in LAM(\R)$. We have the following two cases:
\begin{enumerate}[a.]
\item[a.] {\bf (Continuation)}

The action $x_n$ is either a {\bf negative action} or a {\bf positive
  visible action}. If $x_n$ is proper, by the previous lemma there
exists a unique strategy $\D \in \R$ such that a view $s.x_n \in \D$.
In such a case, for each action $a$ which extends $s.x_n$ in $\D$ we
set $p.a \in LAM(\R)$.

\item[b.] {\bf (Jump)}  The action $x_n= a^+$ is an \textbf{internal    positive}
action.
If there is  $\E\in \R$ such that a view $s.a^-\in \E$, we
set $ p.a^- \in LAM(\R)$.  We say that  $a^-$ matches $a^+$.
\end{enumerate}
\end{enumerate}

\end{defi}\medskip

\noindent Let us  informally explain how the  machine
calculates the interaction of a cut-net $\R$. The machine visits actions of the strategies in $\R$
and  collects the sequences of visited actions, proceeding as follows:

\begin{enumerate}[$\bullet$]

\item
We start on
 the root of
the main strategy of a cut-net $\R$.

 \item If we visit a visible action $a$ occurring in some  $\D \in \R$,
we continue to explore the current strategy $\D$. The process
  branches when $a$ is a branching node of $\D$.

\item
If we visit an
 internal action $a^+$ occurring in $\D$ we match it
with its opposite
 $a^-$ occurring in $\E \in \R$, then we continue to collect actions  in $\E$ (this is a \emph{jump}
of the machine).

\item We may eventually stop when either we reach a maximal action
or  an internal action which has no match.
\end{enumerate}

\begin{defi}[Hiding]
Given a cut-net $\R$,  and   $p=x_1,\dots, x_n \in LAM(\R)$, we define $hide(p)$ as the sequence obtained from $p$ by deleting all the internal actions.
\end{defi}

\begin{defi}[Normal form] 
Let $\R$ be a cut-net.
 We define the set $I(\R)$ of the {\bf interactions} of $\R$ as the
closure by  non-empty prefix of $$ \{q.c\in LAM(\R):  c \mbox{ visible and
positive}\}.$$
The \textbf{normal form}
 of  $\R$, denoted by $\pl \R \pr$ is defined as
 $$\pl \R \pr = \{hide(p):  p\in I(\R)  \mbox{ and }  hide(p)   \mbox{ non-empty} \} .$$

\end{defi}\medskip

\noindent The normal form of a cut-net is a strategy ~\cite{GirLoc,CurLLL2,GirBook}.

\begin{thm} \label{nor_lin}
If $\R$ is a cut-net of interface $\GG$, then $\pl\R\pr$ is a strategy on $\GG$. \qed
\end{thm}

Composition satisfies associativity,  in the following sense
\cite{GirLoc,CurLLL2,GirBook}.
\begin{thm}[Associativity] \label{ass_lineare}
Let $\R$  be a cut-net which can be partitioned into
cut-nets $\R= \R_1 \cup \ldots \cup \R_n$. We have:
$$ \pl \R  \pr =  \pl  \pl \R_1 \pr, \ldots, \pl \R_n \pr  \pr.$$
\qed
\end{thm}

\begin{rem}
The standard  associativity of game semantics  (in terms of ``morphisms")
can be  expressed in this setting,  considering  the universal arena on an interface
of the form $\alpha^-,\beta^+$ as
an arena ``of the form $A \to B$." Let us write for now
$\alpha \to \beta$ instead of $\alpha^-,\beta^+$.

Let $\alpha,\beta,\gamma,\delta$ be disjoint names and consider $\D : \alpha \to \beta$,
$\E : \beta \to \gamma$   and $\F :  \gamma \to \delta$.
By Theorems  ~\ref{nor_lin} and   ~\ref{ass_lineare},  we have that
 $$ \pl \pl \D,\E \pr,\F  \pr  = \pl\D,\E,\F \pr =  \pl \D, \pl  \E, \F\pr \pr $$
 is a strategy on interface $\alpha \to \delta$.
\end{rem}

\begin{exa}[Interaction with $\Dai$ or $\Fid$]\label{DaiFid} We can now examine how $\Fid$ and $\Dai$ behave in the normalization, and so make precise the discussion in Section \ref{tot section}.
 Let $\R =\{\D, \E_1,\ldots,\E_n\}$ be a positive cut-net, where
   $\D$ is the unique positive strategy (see Remark \ref{rem poscut}).
Let us see what happens when $\D= \Dai$ or $\D=\Fid$.

\begin{enumerate}[$\bullet$]

\item The interaction starts from $\D$ (because it is the main strategy).

\item If $\D$ is $\Dai$, the interaction reaches $\dai$ at the first step, and terminates immediately, whatever the other (negative) strategies
    in $\E_1,\ldots,\E_n$ are.

\item If $\D$ is $\Fid$, nothing can happen. Since $\D$ is empty, the output of the interaction is also empty.
\end{enumerate}
Summing up,
we always have that:
$$\cut {\Dai, \E_1,\ldots,\E_n}= \Dai  \quad \mbox{ and } \quad
\cut{\Fid,  \E_1,\ldots,\E_n} = \Fid.$$
\end{exa}

\subsubsection{A notation to describe the interaction}

In the sequel, given two strategies $\D$ and $\F$ we often describe their
interaction  in the following graphical way:
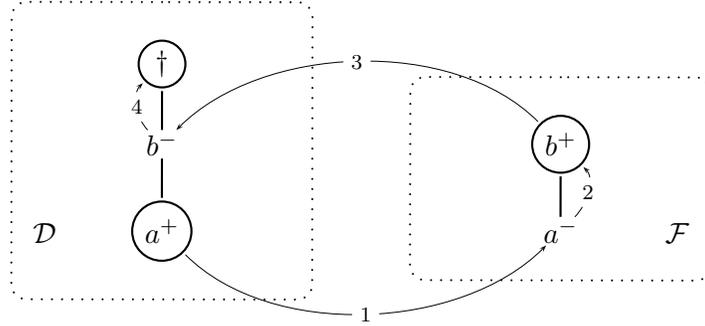
\vspace{1cm}
\begin{center}
\psmatrix[rowsep=9pt,colsep=0.5cm]
&&\circlenode{daimon}{$\dai$}&&&&&&&&&  && \\
&&\rnode{x1m}{$b^-$} &&&&&&&&& \circlenode{x1p}{$b^+$}&& \\
$\D$&&\circlenode{xp}{$a^+$}&&&&&&&&&  \rnode{xm}{$a^-$} && $\F$ \\
\endpsmatrix
 \psset{shortput=nab,arrows=->,labelsep=6pt}
 \small
\ncarc[linewidth=0.2pt,nodesep=1pt,arcangle=-45]{xp}{xm}
\ncput*{\scriptsize 1}
\ncarc[linewidth=0.2pt,nodesep=1pt,arcangle=-45]{xm}{x1p}
\ncput*{\scriptsize 2}
\ncarc[linewidth=0.2pt,nodesep=1pt,arcangle=-45]{x1p}{x1m}
\ncput*{\scriptsize 3}
\ncarc[linewidth=0.2pt,nodesep=1pt,arcangle=45]{x1m}{daimon}
\ncput*{\scriptsize 4}

\ncline[nodesep=1pt]{-}{daimon}{x1m}
\ncline[nodesep=1pt]{-}{x1m}{xp}
\ncline[nodesep=1pt]{-}{x1p}{xm}

\ncbox[nodesep=.5cm,boxsize=2,linearc=.2,
linestyle=dotted]{daimon}{xp}

\ncbox[nodesep=.5cm,boxsize=2,linearc=.2,
linestyle=dotted]{x1p}{xm}

\end{center}

\noindent Here, we have taken $\D$ and $\F$ as in Example  ~\ref{esempio}.
We draw tagged arrows to denote the matching of
actions (\eg $a^+$ matches $a^-$ at step $1$) and the
(unique) positive action (the ``answer'') above a reached
negative action (\eg, $b^+$ after $a^-$).
The tags $1,2,\dots$ are only needed to record
the chronological order in which actions are visited.
Following the arrows with this order, we retrieve
the sequence of actions $a^+.a^-.b^+.b^-.\dai$ which corresponds to the interaction of $\D$ and $\F$
of Example  ~\ref{esempio}.

\subsection{Orthogonality}\label{ortho} \label{types}

The most important case of composition in ludics is the closed case, the typical example being a cut-net
given by two strategies on opposite interfaces
 $\D:\xi^+$ and $\E:\xi^-$.

We already observed that in this case, all names are internal,
and the interface of the cut-net is empty. Since we know that  there are only two possible strategies which have
empty interface: $\Fid$ and $\Dai$,  we only have {\em two possible values} as normal form.
 $\Fid$ and $\Dai$ are respectively interpreted as failure (deadlock) or success of the interaction process. More precisely,
 either normalization gives no output at all | and in this case the result is the empty strategy, \ie $\Fid$ | or it succeeds by reaching the action $\dai$, which
signals termination | and in this case the result is the strategy $\Dai$.
 In the  case of success, we say that the strategies are  {\em orthogonal}.

The orthogonality relation is defined only on   {\em total} strategies. In fact, we already know (Example \ref{DaiFid}) that if  $\Fid\in \R$ then
 $\pl \R\pr=\Fid$, whatever are the other strategies.\smallskip

Let us first anticipate the definition in the key case of strategies on a unary interface,
 and then give the general definition of being orthogonal.

Given two total strategies  $\D:\xi^+$ and $\E:\xi^-$, they are said to be {\em orthogonal}, written
 $\D\bot \E$ (or equivalently $\E \bot \D$),  if $\pl \D, \E \pr=\Dai$.

Orthogonality  means that  at each step of the interaction, the  positive action
$x^+$ {\em is matched with} its negative dual action $x^-$, and the computation terminates by reaching
  a $\dai$ action.

We can then define the orthogonal set of a set of total strategies $\bS$ on the same interface $\xi^\epsilon$ in a standard way as
$$\bS\b=\{\E :  ~ \E \mbox{ is a total strategy on the interface } \xi^{\overline{\epsilon}} \mbox{ and } \E \bot \D \mbox{ for any } \D\in \bS\}$$
Notice that (as in ~\cite{GirLoc}) the partial strategy $\Fid$ is ruled
out  in the definition of the orthogonal sets.
For instance,
if $\bS$ is the empty set of negative strategies
on $\xi^-$, then $\bS^\bot$ is the set
of all \emph{total} strategies on $\xi^+$
(in particular, $\Fid \notin \bS\b$).

\begin{exa}
 In  example  ~\ref{esempio}, $\D \bot \F$, while $\E$ and $\F$ are not orthogonal.
\end{exa}

We now make this notion general.

\begin{defi}[Counter-strategies]\label{counter-strategies}
Given an interface  $\GG$,
we call {\bf family of counter-strategies} (\wrt $\GG$)  any family of total strategies
$(\E_\xi : \xi^{\overline{\epsilon}})_{\xi^{\epsilon} \in \GG}$.

With a slight abuse of notation, we will write simply $(\E_\xi)_{\xi \in \GG}$, by omitting the indication of the polarity (when clear from the context, we will also omit the indication of the indexing set $\GG$).

If $\D:\GG$ is a strategy, we will  use the notation $\{\D, (\E_\xi) \}$ for the cut-net $\{\D\} \cup \{\E_\xi:\xi\in \GG\}$ they induce. Observe that $\{\D, (\E_\xi) \}$ is a closed cut-net.
\end{defi}

\begin{exa}
If $\GG=\xi^+$, a counter-strategy has the form $\E:\xi^-$.

 If $ \GG=\xi^-,\alpha^+,\beta^+$, then  $\{\E_1: \xi^+, \quad \E_2:\alpha^-,\quad \E_3:\beta^-\}$ is a family of counter-strategies.
\end{exa}

We can now define the orthogonality relation and
orthogonal sets. We use
the same notation of Definition ~\ref{counter-strategies}.

\begin{defi}[Orthogonality, orthogonal set] \label{ortho def}  Let $\D:\GG$ be a total strategy and
$(\E_\xi)$ be  a family of counter-strategies \wrt $\Gamma$.
$\D$ and  $(\E_\xi)$
 are said to be \textbf{orthogonal}, written  $\D \bot (\E_\xi)$
 (or equivalently $(\E_\xi) \bot \D$),
if $\pl \D, (\E_\xi) \pr$ is total.
In other words,
 we have orthogonality if $\pl \D, (\E_\xi) \pr = \Dai$.

Given  a set $\bS$ of total strategies on the same  interface $\Gamma$, its \textbf{orthogonal set} is defined as $$\bS\b:=\{(\E_\xi) : ~ (\E_\xi)  \mbox{ is a family of counter-strategies \wrt $\Gamma$ and } (\E_\xi)\bot \D \mbox{ for any } \D \in \bS \}.$$
Similarly, given a set $\bC$ of families of counter-strategies \wrt $\Gamma$,
its \textbf{orthogonal set} is defined as $$\bC\b:=\{\D :   \D \mbox{ is total  on interface } \Gamma \mbox{ and } \D \bot (\E_\xi)  \mbox{ for any } (\E_\xi) \in \bC \}.$$
\end{defi}\medskip

\noindent Orthogonality satisfies the usual
closure properties:
 if  $\bE,\bF$ are sets of total strategies  on the same  interface (resp.\ sets of families of counter-strategies
\wrt the same interface), then
\begin{enumerate}[$\bullet$]
\item   $\bE \subseteq \bF$ implies $\bF\b \subseteq \bE\b$
(and thus $\bE\b\b \subseteq \bF\b\b$);
\item  $\bE \subseteq \bE\b\b$;
    \item $\bE\b = \bE\b\b\b$.
\end{enumerate}

In terms of games,
 orthogonality allows the players to agree (or not), without this being guaranteed in advance by the type:
$\{\D\}\b$ is the set of the families of counter-strategies which are consensual (\ie well interact) with $\D$.

\subsection{Interactive types (behaviours)}

\begin{defi}[Behaviour]
A {\bf behaviour} (or  \textbf{interactive type}) on the interface $\GG$
is a set $\bG$ of   strategies $\D:\GG$ such that $\bG\b\b=\bG$
(\ie it is closed by bi-orthogonal).
We say that a behaviour $\bG$ is positive or negative according to its interface.
\end{defi}
We observe that (by the definition of orthogonal set) an interactive type only contains total strategies.

When is useful to  emphasize that
 $\bG$ is a set of strategies on the name $\xi$, we may annotate the name $\xi$
as a subscript: $\bG_{\xi}$.

\begin{exa}[$\Zero, \Toppo$] \label{zero,top} Given a name $\xi$, the \emph{minimal positive behaviour} on $\xi^+$ is the one
 generated by the empty set of total strategies on $\xi^+$.
 Its
 orthogonal set consists of all the negative strategies
 on interface $\xi^-$. Let us call $\Toppo$ this set,
 which is the \emph{maximal negative behaviour} on $\xi$.
 By the definition of orthogonality,
 a strategy in $\Toppo \b$ must
 be consensual to all negative strategies on interface $\xi^-$.
 But the only strategy which can do this is $\Dai$.
 Hence the closure by bi-orthogonal of the empty set of
 positive total strategies on $\xi^+$ is $\{\Dai\}$. Let us call
 $\Zero$ this  behaviour.

 In ~\cite{GirLoc} $\Toppo$ and $\Zero$ are the interpretations
 of the additive constants $\vero$ and $\falso$
 of linear logic respectively, whence their name.
 \end{exa}

\begin{rem}
 In our setting | but not in ~\cite{GirLoc}| behaviours can be empty.
  While a positive behaviour is never empty (because
  it contains at least $\Dai$),
  a negative behaviour can be empty:  if $\bS$ is the empty set of strategies on a {\em negative} interface, $\bS=\bS\b\b$.
 This difference with ~\cite{GirLoc}  is a limit of our choice of ``strict linearity'' (Section ~\ref{linearity}), and will disappear in
 the non-linear setting.

 This mismatch  deserves some discussion.
 The purpose of this and next section is to give a compact but complete presentation of the construction of ludics and of internal completeness in the linear case. In particular, we want to
  show  (in this setting which is easier to grasp)
 how full completeness follows from internal completeness, and provide a proof which  will then be possible to generalize to a full setting in the second part of the paper.
For this reason, we will restrict our attention to the multiplicatives. To have additive structure would
make the   behaviours never empty, but
require some more
technical definitions, without adding anything substantial to our purposes. The small price is the explicit request
for the behaviours to be ``non-empty'' in the various constructions below.
\end{rem}

\subsection{Linear types constructors} \label{typecons}

In this section we consider the behaviours which will interpret logical formulas, more precisely multiplicative formulas.

\subsubsection{Constructions on strategies}

Let  $\D_1:\xi1^-$, \dots, \ $\D_n:\xi n^-$ with $n \geq 0$  be  negative strategies. We obtain a new
positive strategy on the interface $\xi^+$, denoted by $ \D_1\bullet \dots \bullet \D_n$,
by adding to the union of the strategies the positive root $(\xi,I_n)^+$,
\ie
$$ \D_1\bullet \dots\bullet \D_n:= (\xi,I_n)^+. \{\D_1, \dots, \D_n\}.$$
Recall that as we stipulated in  Section ~\ref{conventions},
$I_n$ denotes either $\{1,\ldots,n\}$ if $n >0$, or
$\emptyset$ if $n=0$.

\subsubsection{Constructions on behaviours}\label{regular_behaviours}

\begin{defi}[Tensor/Par of behaviours]\label{mult beh}
Let $\bN_1\ldots,\bN_n$  be {\em non-empty} {\em negative} behaviours on interface  $\xi 1^-$, \ldots, $\xi n^-$ respectively.

We define the set of strategies:
$$
\bN_1\bullet \dots\bullet \bN_n  :=  \{\D_1\bullet \dots \bullet \D_n : \D_i\in \bN_i \mbox{ for any } 1 \leq i \leq n \}$$
and the behaviours
$$
\begin{array}{cclc}
\bN_1 \otimes \dots \otimes\bN_n  & := & (\bN_1\bullet \dots\bullet \bN_n)\b\b & \mbox{ positive behaviour on } \xi^+; \\

\bN_1^\bot \parr \dots \parr \bN_n^\bot & := & (\bN_1\bullet \dots\bullet \bN_n)\b &
\mbox{ negative behaviour on } \xi^-
\end{array}
$$
We call {\bf multiplicative} a behaviour which is inductively generated by these constructions.
\end{defi}

\begin{rem}[0-ary case] \label{0 ary costr}
In the case that $n=0$, the set  $\bN_1\bullet \dots\bullet \bN_n$ consists of a unique strategy $\D= (\xi,\emptyset)^+$. We have that
$\bN_1^\bot \parr \dots \parr \bN_n^\bot$ only contains the strategy $\E= (\xi,\emptyset)^-.\Dai$ , since there is no other possibility to be orthogonal to $\D$. Finally,
$\bN_1 \otimes \dots \otimes\bN_n = \{\D, \Dai\}$, again because there are no other possibilities.

 We denote by $\One$ the behaviour $\bN_1 \otimes \dots \otimes\bN_n$ and by $\Botto$ the behaviour $\bN_1^\bot \parr \dots \parr \bN_n^\bot$ in the case  $n=0$. Of course, both behaviours are multiplicative.
\end{rem}

\begin{prop}\label{nonempty_prop} Let    $\bP = \bN_1 \otimes \dots \otimes\bN_n$ and  $\bN=\bN_1^\bot \parr \dots \parr \bN_n^\bot$.

\begin{enumerate}[$\bullet$]
\item $\bP,\bN$ are non-empty;
\item $\bP,\bN$ contain only non-empty  strategies.
\end{enumerate}

In particular, multiplicative behaviours are never empty.
\end{prop}
\proof A positive behaviour  is never empty, because it contains at least $\Dai$. Moreover, $\Fid \notin \bP$, by definition of orthogonality.
For negatives behaviours generated
by our construction, observe that since
$\bN_1,\ldots,\bN_n$ are non-empty,
$\bN_1\bullet \dots \bullet  \bN_n$ is non-empty too and
by construction all the strategies have the same root $(\xi, I_n)$.
A negative behaviour $\bN$ of the form
$(\bN_1\bullet \dots \bullet  \bN_n)\b$  is also never empty, because it contains    $\E :=(\xi, I_n)^-.\dai$.
The empty strategy cannot belong to $\bN$, because it is not orthogonal to any strategy in $\bN_1\bullet \dots \bullet  \bN_n$.
\qed

\subsection{Sequent of behaviours}
As a behaviour on a unary interface  corresponds in ludics to a logical formula, the  notion
of sequent of behaviours corresponds
to the notion of sequent.

\begin{defi}[Sequent of behaviours]\label{seq_deflin}
Let   $\Gamma=\xi_1^{\epsilon_1}, \dots,\xi_n^{\epsilon_n} $ ($n\geq 0$) be an interface, and  let
 $\bGG = \bG_{\xi_1},\dots,
\bG_{\xi_n}$    ($n \geq 0$)   behaviours of respective polarities $\epsilon_1, \dots, \epsilon_n$.

We define a new behaviour  on the same interface $\Gamma$,
which we call {\bf sequent of behaviours} and denote by $\vdash \bGG$,
as follows:
$$ \vdash \bGG := \{ \D :\D \mbox{ is total on interface }
\Gamma \mbox{ and }
 \D \bot \{\E_1,\ldots,\E_n\}  \mbox{ for all  } \E_1 \in \bG_{\xi_1}^\bot, \ldots,\E_n \in \bG_{\xi_n}^\bot  \}.$$
\end{defi}\medskip

\noindent It is clear that a sequent of behaviours is itself a behaviour, since
it is the orthogonal set of the set of families of counter-strategies
$(\E_1,\ldots,\E_n)$ such that $\E_1 \in \bG_{\xi_1}^\bot, \ldots,\E_n \in \bG_{\xi_n}^\bot$.

Observe  that:
\begin{enumerate}[$\bullet$]
\item  $\vdash \bP \  =  \ \bP$ and $\vdash \bN \ = \ \bN$.
\item if $\Pi$ is empty, $\vdash \bPP = \{\Dai\}$.
\end{enumerate}\smallskip

When moving to full completeness
(and thus consider sequents of multiplicative behaviours), we will use also the following property, which is immediate by associativity (Theorem ~\ref{ass_lineare}).

\begin{prop} \label{closure} Let $\bA, \bG_1, \dots \bG_n$ ($n\geq 0$) be a sequence of  multiplicative
behaviours, and $\bGG= \bG_1, \dots \bG_n$.  We have that:
\begin{enumerate}[$\bullet$]
\item
$\D\in \ \vdash \bGG,\bA $ if and only if  for each
$\F \in \bA\b$, $\pl \D,\F \pr \in \ \vdash \bGG$.

\item $\D\in \ \vdash \bGG,\bA $ if and only if
 $\pl \D,(\E_i)  \pr \in \ \vdash \bA$, for each family
$(\E_i)$ such that
$\E_1 \in \bG_{1}^\bot, \ldots,$ $ \E_n \in \bG_{n}^\bot$.
\end{enumerate}
\end{prop}
\proof
 Let us abbreviate the set $\{(\E_i) : ~
\E_1 \in \bG_{1}^\bot, \ldots,\E_n \in \bG_{n}^\bot\}$
by $\bC$.
We then have:

$$\begin{array}{rcll}
\D \in \ \vdash \bGG,\bA  & \Leftrightarrow  & \forall \F \in \bA\b \ \forall (\E_i) \in \bC,   \
 \D \bot \{\F,(\E_i)\}
  & \\
 & \Leftrightarrow  & \forall \F \in \bA\b \ \forall (\E_i) \in \bC, \
 \pl \D, \F,  (\E_i) \pr \mbox{ is total }
  & \\
   & \Leftrightarrow  &\forall \F \in \bA\b \ \forall (\E_i) \in \bC, \
 \pl \pl \D,  \F \pr, (\E_i) \pr \mbox{ is total }
  & \mbox {(associativity)} \\
   & \Leftrightarrow  &\forall \F \in \bA\b \ \forall (\E_i) \in \bC, \
  \pl \D,  \F \pr \bot (\E_i)
  &  \\
   & \Leftrightarrow  &\forall \F \in \bA\b, \ \pl \D,  \F \pr \in \ \vdash \bGG.
  &  \\
\end{array}
$$
The second claim is obtained by iterating the first one.
\qed

 \section{Ludics in the linear case: internal and full completeness}
 \label{int_compl} \label{internal completeness}
We  introduce the notion of internal completeness
and give a direct proof of  internal completeness, as well as full completeness,
without relying on separation.\smallskip

In ~\cite{GirLoc}, the set of strategies which interprets
 $\mathbf{MALL}$ formulas satisfies a remarkable
closure property, called {\em internal completeness}: the set $\bS$
of strategies produced by the construction is {\em essentially}
equal to its biorthogonal ($\bS=\bS\b\b$).  Since the biorthogonal
 does not introduce new objects, we have a complete description of all
strategies in the behaviour.\smallskip

The best example is the interpretation $\bN_1 \otimes \bN_2:=
(\bN_1 \bullet \bN_2)\b\b$
 of a tensor formula.
One proves that $(\bN_1\bullet\bN_2) \cup \{\Dai\}=(\bN_1\bullet\bN_2)\b\b$, \ie
 we do not add new
objects when closing
by biorthogonal: our description | the one which generates the set | is already complete.

 From this, full completeness follows. In fact,
because of internal completeness, if $\D\in \bN_1 \otimes \bN_2$
{ and $\D \neq \Dai$} we know we can
decompose it as $\D_1 \bullet \D_2$, with
 $\D_1\in\bN_1$ and   $\D_2\in\bN_2$.
This  corresponds to writing the rule:
\begin{prooftree} \AxiomC{$\vdots$}
 \noLine
\UnaryInfC{$\vdash N_1 $}
 \AxiomC{$\vdots$}
 \noLine
\UnaryInfC{$\vdash N_2 $}
\RightLabel{\scriptsize{$\otimes$}}
\BinaryInfC{$\vdash N_1 \otimes N_2$}
\end{prooftree}
\ie if  each $\D_i$
corresponds to a proof of $N_i$, and
 $\D$ corresponds to a proof of $N_1\otimes N_2$.

\subsection{Internal completeness}
Until the end of Section ~\ref{int_compl} , we assume the following:

\begin{enumerate}[$\bullet$]
\item we fix a name $\xi$

\item  $\bN_1, \dots \bN_n $ $(n \geq 0)$ are  {\em non-empty} negative behaviours, respectively   on $\xi 1^-$, $\dots$, $\xi n^-$.

\end{enumerate}

Let us examine the form of the strategies which inhabit a behaviour of the type we have introduced in the last section.
Let us first consider
$\bN_1\bullet\bN_2$. By construction,
each strategy in $\bN_1\bullet\bN_2$  is on  $\xi^+$ and
has $x^+=(\xi,\{1,2\})^+$ as root.

What is  $(\bN_1\bullet\bN_2)\b$? By definition of linear normalization, each strategy
has as root
the action $x^-=(\xi,\{1,2\})^-$ (otherwise, normalization would fail immediately). In particular we have the strategy $x^-.\dai$.

What is $(\bN_1\bullet\bN_2)\b\b$? All strategies have a positive root, which,
to normalize against $(\bN_1\bullet\bN_2)\b$,  must be either $\dai$, or
 $x^+$.
Hence, we know that a strategy   $\D\in \bN_1\otimes\bN_2$ is either $\Dai$ or has the form
$x^+.\{\D_1,\D_2\}$, where $\D_1:\xi 1$ and $\D_2:\xi 2$.
The following picture represents this.
\begin{center}
\psmatrix[rowsep=9pt,colsep=0.3cm]
&&&&&&&&& &&&& && \\
&&&&\rnode{up}{}&&&&& &&&& && \\
&\rnode{leaf1a}{}&\rnode{leaf1}{} & \rnode{leaf1b}{}&&
  \rnode{leaf2b}{} & \rnode{leaf2}{} & \rnode{leaf2a}{}&& &&&& && &\\
&&& &  & && & &&&& \rnode{yml}{}  &\rnode{ym1}{} & \rnode{ymr}{}  \\
&\rnode{x1ma}{\footnotesize{$\D_1$}}& \rnode{x1m}{$(\xi 1, J)^-$} &&&& \rnode{x2m}{$(\xi 2, K)^-$}& \rnode{x2ma}{\footnotesize{$\D_2$}}& &&&& & &  &    \\
&&&&\circlenode{xp}{$x^+$}&&&&& &&&& \rnode{ym}{$x^-$} &&   \\
&&&&\rnode{xz}{$\D \in (\bN_1\bullet\bN_2)\b\b$}&&&&& &&&& \rnode{zz}{$\E \in (\bN_1\bullet\bN_2)\b$} && \\
&&&&&&&&& &&&& && \\
\endpsmatrix
 \psset{shortput=nab,arrows=->,labelsep=6pt}
 \small
\ncline[nodesep=1pt]{-}{x1m}{xp}
\ncline[nodesep=1pt]{-}{x2m}{xp}
\ncline[nodesep=0pt,linewidth=.01]{-}{yml}{ym}
\ncline[nodesep=0pt,linewidth=.01]{-}{ymr}{ym}
\ncline[nodesep=0pt,linewidth=.01]{-}{ymr}{yml}
\ncline[nodesep=0pt,linewidth=.01]{-}{leaf1a}{leaf1b}
\ncline[nodesep=0pt,linewidth=.01]{-}{x1m}{leaf1b}
\ncline[nodesep=0pt,linewidth=.01]{-}{leaf1a}{x1m}
\ncline[nodesep=0pt,linewidth=.01]{-}{leaf2a}{leaf2b}
\ncline[nodesep=0pt,linewidth=.01]{-}{x2m}{leaf2b}
\ncline[nodesep=0pt,linewidth=.01]{-}{leaf2a}{x2m}
\ncbox[nodesep=.5cm,boxsize=4.8,linearc=.2,
linestyle=dotted]{xz}{up}
\ncbox[nodesep=.5cm,boxsize=1.7,linearc=.15,
linestyle=dotted]{x2m}{leaf2}
\ncbox[nodesep=.5cm,boxsize=1.7,linearc=.15,
linestyle=dotted]{x1m}{leaf1}
\ncbox[nodesep=.5cm,boxsize=2,linearc=.2,
linestyle=dotted]{zz}{ym1}
\end{center}

\noindent We now want to prove that if $\D\in (\bN_1\otimes\bN_2)$
(and $\D \neq \Dai$) then $\D_1\in \bN_1$ and $\D_2\in \bN_2$, which
means that $(\bN_1\bullet\bN_2)$ was already {\em complete}, \ie
closed by biorthogonal.

\begin{prop}[Internal completeness of tensor]\label{compl+}
Let $\bN_1, \dots, \bN_n$ be $n \geq 0$  non-empty negative behaviours, respectively
on $\xi 1^-$, $\dots$,
$\xi n^-$.
We have that $$\bN_1\otimes \dots \otimes \bN_n =\bN_1\bullet \dots \bullet \bN_n\cup \{\Dai\}.$$
\end{prop}

\proof
We have already shown that this is true in the case $n=0$ (Remark ~\ref{0 ary costr}).
\wloge, we prove the claim in the case $n=2$.

Let $\D\in \bN_1\otimes \bN_2$, $\D \not= \Dai$.  We know (by the discussion above) that $\D$ has the form $x^+.\{\D_1,\D_2\}$, with $x=(\xi, \{1,2\})$. We now
 prove  that $\D_1\in \bN_1$
and $\D_2\in \bN_2$.

\begin{enumerate}[(i)]
 \item Given any  $\E:\xi 1^+\in \bN\b_1$, we obtain the strategy  $\E':\xi^-=x^-.\E$ by adding
the root $x^-$.  We have that
\begin{equation} \label{eq int comp}
\cut{x^-.\E, x^+.\{\D_1,\D_2\}} = \cut{\E,\D_1}  \tag{$\star$}\end{equation}
by definition of normalization, and by the fact that since in $\E$ there are
only names generated by $\xi 1$. Hence, $\E:\xi1^+$ only interact with the subtree $\D_1:\xi 1^-$. No action in $\D_2$ is ever used.

 \item $\E'\in  (\bN_1\bullet\bN_2)\b$, because
by using (\ref{eq int comp})  we deduce that
$\E'\bot\D$, for any $\D\in  (\bN_1\bullet\bN_2)$.
 \item[(iii)] Given any $\D\in \bN_1\otimes\bN_2$, by definition we have that  $\D\bot\E$ for each $\E\in (\bN_1\bullet\bN_2)\b$. Hence in particular, for each $\E\in\bN\b_1  $, we have $\D\bot\E'$ ($\E'$ defined as above). Again because of (\ref{eq int comp}),  $\D_1\bot \E$. This says that
$\D_1 \in (\bN\b_1)\b=\bN_1$. \qed
\end{enumerate}

\begin{rem}[Important]\label{importante}
The key to extend this argument  to the case of ludics with repetition, is that here we only use two properties of the strategies:
 the dynamics (normalization), and the fact that
{\em the root is the only action on the name $\xi$}
(to say at point (i) in the proof above that occurrences of $\xi 1$ only appear inside $\D_1$).
\end{rem}

\begin{prop}[Internal completeness of par]\label{compl-}
Let $\bN_1, \dots, \bN_n$  be as in ~\ref{compl+} (hence
 $\bN_1^\bot, \dots, \bN\b_n$ are  positive behaviours respectively on $\xi 1^+, \dots, \xi n^+$), and let
$x=(\xi, I_n)$. We have:
$$
x^-.\E\in \bN_1^\bot\parr \dots \parr \bN_n^\bot \ \Leftrightarrow \
\E\in \ \vdash \bN_1^\bot, \dots, \bN_n^\bot.$$
\end{prop}

\proof
The case $n=0$ is
actually given by Remark ~\ref{0 ary costr}, observing
that the empty sequent of behaviours contains
only $\Dai$.

Let us give the proof in the case $n=2$.
A strategy $x^-.\E$ belongs to $\bN_1^\bot\parr\bN_2^\bot$ if and only if
for any $x^+.\{\D_1,\D_2\} \in  \bN_1 \bullet \bN_2$, we have that
$ x^-.\E ~\bot~ x^+.\{\D_1,\D_2\}$. By definition of normalization,
 $ \pl x^-.\E, x^+.\{\D_1,\D_2\} \pr =\pl \E, \D_1,\D_2 \pr$, and from this and the
  definition of sequent of behaviours
(Definition ~\ref{seq_deflin}) the claim follows  immediately.
\qed


\subsection{Full completeness}\label{fullc_lin}

We are now ready to show how full completeness is obtained from internal completeness, in the linear case. More precisely, we are going to introduce the calculus  $\mathbf{MLLS}$, which is the multiplicative fragment of the calculus $\mathbf{HS}$ introduced by Girard in ~\cite{GirLoc,GirBook}.
We then show that full completeness can be derived  from internal completeness of tensor and par
and Proposition ~\ref{closure}.

While the proof in ~\cite{GirLoc} relies on separation, we  give a direct proof which uses only the properties of the dynamics.
The choice of omitting the additive here only simplifies (and shortens) the presentation; the proof can be extended to the additive structure without problems  (but in that case, we would need to add one more ``winning conditions'', as we will discuss below).

\subsubsection{$\mathbf{\mathbf{MLLS}}$}
As it is the case for $\mathbf{HS}$ ~\cite{GirLoc,GirBook}, the calculus $\mathbf{MLLS}$ is \emph{affine}, \ie we have  weakening  (but restricted to positive formulas).

The calculus
$\mathbf{MLLS}$ can be seen as
 the \emph{affine}  (with implicit weakening but without implicit contraction for positive formulas) restriction of the calculus $\HS$
 given in Section ~\ref{HS}
\footnote{We do not consider the cut-rule here, but we
do it in Section ~\ref{intF}.}.\smallskip

\emph{Formulas of $\mathbf{\mathbf{MLLS}}$} are inductively given by:
$$
P  ::=   N_1\otimes \dots\otimes N_n   \qquad
N  ::=   P_1\parr \dots\parr P_n \qquad  (n \geq 0).
$$
When $n=0$, we
write $ \uno$ and $ \ult$ for the positive
and negative formula respectively.\smallskip

\emph{Rules of $\mathbf{MLLS}$} are given in Table  ~\ref{mlls_fig}.

\begin{table}[h]
  \centering

\fbox{
\begin{tabular}{cc}
&\\
Positive rules : &
\AxiomC{$\vdash \Pi_1,N_1$}
\AxiomC{$\ldots$} \AxiomC{$\vdash \Pi_n,N_n$}
\RightLabel{\scriptsize{Pos$_n$}}
\TrinaryInfC{$ \vdash  \Pi_0,\Pi_1,\ldots,\Pi_n, P$} \DisplayProof

\\
$P =
N_1\otimes \dots\otimes N_n$ and  $n\geq 0 $ & \\
& \\

Negative rules : &

\AxiomC{$\vdash  \Pi,P_1,\ldots,P_n$}
\RightLabel{\scriptsize{Neg$_n$}}
\UnaryInfC{$\vdash \Pi,N$} \DisplayProof

\\

$N =
P_1\parr \cdots\parr P_n$ and $n\geq 0$ &
\\

&\\
\end{tabular}
}
 \caption{$\mathbf{MLLS}$}
  \label{mlls_fig}
\end{table}


\noindent Notice the (implicit) weakening on
 occurrences of positive formulas $\Pi_0$ in the positive rule.

\subsubsection{Interpretation}

We only give an outline  of the interpretation of formulas
and sequents.
We will discuss it  in full detail the interpretation of derivations in the setting
 which also includes exponentials in Section ~\ref{intF}.\smallskip

 Given a formula $F$ of $\mathbf{MLLS}$ and an arbitrary name
$\xi$ we associate a  multiplicative behaviour of the same polarity $\bF$ on interface $\xi$
inductively as follows.
$$
 \begin{array}{rcl}
N_1\otimes \cdots \otimes N_n \ ^\xi & := & N_1 \ ^{\xi1} \otimes \cdots \otimes N_n \ ^{\xi n}; \\ P_1\parr \cdots \parr P_n \ ^\xi & := &  P_1 \ ^{\xi1} \parr \cdots \parr P_n \ ^{\xi n}.  \\
 \end{array}
$$
Given a  positive sequent  $\vdash P_1,\ldots,P_n$,
and a positive interface $\xi_1^+ \ldots,\xi_n^+$
we associate the
 sequent of behaviours $\vdash P_1 \ ^{\xi_1},\ldots,P_{n}
 \ ^{\xi_n}$.
Given a  negative sequent  $\vdash N, P_1,\ldots,P_n$
and a negative interface $\sigma^-,\xi_1^+ \ldots,\xi_n^+$
we associate the
 sequent of behaviours $\vdash N \ ^{\sigma}, P_1 \ ^{\xi_1},\ldots,P_{n}
 \ ^{\xi_n}$.

The interpretation of a proof $\pi$ of $\vdash \Gamma$ will be a daimon-free strategy  in the sequent of behaviours $\vdash
\bGG$.
Once given the interpretation of proofs (which we only do in Section ~\ref{full completeness}) one can establish the following theorem.
\begin{thm}
Let $\pi$ be a proof of a sequent $\vdash \Gamma$ in ${\mathbf{MLLS}}$.
There exists a daimon-free  strategy  $ \D \in \ \vdash \bGG$ such that $\D$ is interpretation of $\pi$. \qed
\end{thm}

\subsubsection{Full  completeness} \label{full linear}
We now show the following in detail:

\begin{thm}[Full completeness]\label{fullcompl_lin} Let $\vdash \bGG$ be a  sequent of behaviours which is interpretation of the sequent $\vdash \Gamma$ in ${\mathbf{MLLS}}$.
If $\D$ is a daimon-free strategy
in   $\vdash \bGG$
then
$\D$ is the interpretation of a  proof $\pi$ of the sequent $\vdash \Gamma$ in ${\mathbf{MLLS}}$. \qed
\end{thm}

We notice that we associate proofs only to daimon-free strategies. Being daimon-free is what is called  a ``winning condition.''
Since we are only concerned with the multiplicative structure, this is the only condition necessary for   linear strategies.

When working with additive structure, or with exponentials, one also needs  the notion of materiality, which is introduced in ~\cite{GirLoc}
and that we will discuss in  Section
~\ref{completeness}. However, since we restrict out attention to the multiplicative fragment,
 we can overlook materiality for the moment.\smallskip

Let $\vdash \bGG$ be the interpretation of the sequent $\vdash \Gamma$, and $\D\in \ \vdash \bGG$
 a daimon-free strategy.
Our purpose is
 to associate to $\D$ a derivation $\D^\star$
of   $\vdash \Gamma$
 in
$\mathbf{MLLS}$ by progressively decomposing $\D$, \ie
inductively writing ``the last rule.''
To be able to use internal completeness, which is defined on behaviours on unary interfaces (and not on sequents of behaviours),  we will use | back and forth | the definition  of sequent of behaviours and in particular Proposition  ~\ref{closure}.

The formula on which the last rule is applied is indicated by the name of the root action.
For example, let us assume that the root of $\D$ is $(\xi,I)$;
then if  $\D\in \ \vdash \bG_\alpha,\bF_\xi$,
the behaviour which corresponds to the last  rule is the one on $\xi$, that is  $\bF_\xi$.

The proof is by induction on
the number of logical symbols occurring in the sequent $\vdash \Gamma$ we have interpreted in $\vdash \bGG$.

In the sequel,  we  consider sequents
of behaviours of the form $\vdash \bPP,\bF_\xi$, which are interpretations of  sequents $\vdash  \Pi, F$ of $\mathbf{MLLS}$.
$\bF$ is the interpretation of a formula $F$
and $\bPP$ is a sequence of $m$ behaviours  $\bQ_{\alpha_1},\ldots,\bQ_{\alpha_m}$  which respectively interpret formulas
$Q_1, \dots, Q_m$.
Observe that by
the shape of the rules
of $\mathbf{MLLS}$ and the fact that the interpretation
of formulas preserves the polarity,
  $\bPP$  always consists of positive behaviours only.

We also remark the following facts:
\begin{enumerate}[$\bullet$]

\item Since we consider daimon-free strategies,
all the actions in $\D $ are proper actions.
\item By linearity, both positive and negative strategies have a single root (\ie they are trees, and not forests), and hence are of the form
$\D=x^+.\{\D_1, \dots, \D_n\}$ or $\D=x^-.\D'$.
\item Suppose that $\D$ belongs to some negative behaviour $\bF$
which is interpretation of a formula of $\mathbf{MLLS}$. Since  $\bF$ has to be of the form $(\bN_1\bullet \dots \bullet  \bN_n)\b$, then $\D$ is not empty (see Proposition ~\ref{nonempty_prop}).
\end{enumerate}
Below we use the following convention:
we write $(\E)$ and
$(\E) \in \bC$ for $\E_1,\ldots,\E_m$
and
 $\E_1 \in \bQ_{\alpha_1}^\bot,\ldots,
 \E_m \in \bQ_{\alpha_m}^\bot$
 respectively.

We have two cases.

\subsubsection*{Positive case}
Let  $\D = (\xi,I_n)^+.\{\D_1,\dots, \D_n\}$ be a {\em positive} daimon-free
strategy which belongs to $\vdash \bPP,\bF_\xi$.
Let $\bF_\xi=\bN_1\otimes \dots \otimes  \bN_k$.
 By  Proposition ~\ref{closure},  for any $(\E) \in \bC$, we have that  $\cut{\D,(\E)}\in \bF_\xi$,
 hence its root is   $(\xi,I_k)$.
 By definition of normalization, the root  of
 $\pl \D,(\E) \pr$ is still
  $(\xi,I_n)$ (because it is a visible action), hence $k=n$.

We proceed as follows, using internal completeness.
\begin{enumerate}[(1)]

\item Let $\C={\pl \D,(\E) \pr}$. By  internal completeness of the tensor (Proposition ~\ref{compl+}), we have that
$\C$ can be written as $\C_1\bullet \dots\bullet  \C_n,$
for some  $\C_i\in \ \vdash \bN_i$.
\item By linearity, observe that
a name $\alpha \in \{\alpha_1,\ldots,\alpha_m\}$ of the context $\bPP = \bQ_{\alpha_1},\ldots,\bQ_{\alpha_m}$ { either appears in  one of the  $\D_i$ | and there is only one
such a $\D_i$ | or it does not appear at all}.
This allows us to univocally split the context
$\bPP$ into \emph{disjoint} subsets as follows. We define:
\begin{enumerate}[a.]
\item $\bPP_i := \{ \bQ_{\alpha} \in \bPP : \mbox{ the name } \alpha \mbox { occurs in }  \D_i \}$, for $1\leq i \leq n$;
\item $\bPP_0 := \bPP \setminus \ \bigcup_{1 \leq i \leq n} \bPP_i$.
\end{enumerate}

Let us split  $(\E) \in \bC$
into $(\E)_0 \in \bC_0,(\E)_1 \in \bC_1 \ldots, (\E)_n \in \bC_n$.
By definition of normalization
we have:
$$\begin{array}{rcl} {\C \ = \ }
\pl x^+.\{\D_1,\dots,\D_n\}  ,(\E) \pr & = &
x^+.\{\pl \D_1,(\E)_1 \pr ,\dots, \pl \D_n,(\E)_n \pr \} \\
& = &  \pl \D_1,(\E)_1 \pr \bullet \cdots \bullet \pl \D_n,(\E)_n \pr \\ & = & \C_1\bullet \cdots\bullet  \C_n. \\
\end{array}
$$
\end{enumerate}
 From this, we conclude that $\pl \D_1,(\E)_1 \pr \in \ \vdash
\bN_1$, \ldots, $\pl \D_n,(\E)_n \pr \in \ \vdash
\bN_n$.
By applying
Proposition ~\ref{closure} again, we have that
$\D_1\in \ \vdash \bPP_1,\bN_1$, \ldots,
$\D_n\in \ \vdash \bPP_n,\bN_n$.
The size has decreased for all sequents, and
$\D_1,\ldots,\D_n$ are obviously all daimon-free. We can then apply the inductive hypothesis and
  write the derivation:
\begin{center}
\AxiomC{$\vdots \ \D_1^\star$}\noLine
\UnaryInfC{$\vdash \Pi_1,N_1$}
\AxiomC{$\ldots$}
\AxiomC{$\vdots \ \D_n^\star$}\noLine
\UnaryInfC{$\vdash \Pi_n,N_n$}
\RightLabel{\scriptsize{Pos$_n$}}
\TrinaryInfC{$ \vdash   \Pi_0,\Pi_1\ldots,\Pi_n, N_1\otimes \dots\otimes N_n$} \DisplayProof
\end{center}

Notice that in case  $\bF_\xi = \One$ then $\D=(\xi,\emptyset)^+$ and
the procedure above gives the rule
 \begin{center}
 \AxiomC{}
\RightLabel{\scriptsize{Pos$_0$}}
\UnaryInfC{$ \vdash  \Pi_0, \uno$} \DisplayProof
 \end{center}

\subsubsection*{Negative case}  The negative case is an immediate application
 of the internal completeness for par (Proposition ~\ref{compl-})  and Proposition ~\ref{closure}.

 Let $\D = x^-.\D'$, and  $x=(\xi,I_n)$.
 Assume
$\D \in \ \vdash  \bPP, \bF_\xi$
and $\D$ daimon-free, where $\bF_\xi=\bP_1\parr \dots \parr\bP_n$ is the interpretation
of the formula  $P_1\parr \dots \parr P_n$ and
 $\bPP$ is the interpretation of the context $\Pi$
of a negative sequent $\vdash  \Pi, P_1\parr \dots \parr P_n$ of $\mathbf{MLLS}$.

For any family $(\E) \in \bC$, we have
\begin{enumerate}[(1)]
\item
$\pl \D,(\E)\pr \in \bP_1\parr \dots \parr\bP_n$, and the root is still $x^-$. This
allows us
to use internal completeness.

\item By internal completeness  (Proposition~\ref{compl-}), we conclude that
    $\pl \D,(\E)\pr$ is of the form $x^-.\D''$
    with $\D'' \in \ \vdash \bP_{ 1},\dots, \bP_{ n}$.

    \item By the definition of normalization,
    $$ \pl \D,(\E) \pr = \pl x^-.\D', (\E) \pr =
    x^-.\pl \D',(\E)\pr.$$
From this, we have that
 $\D'' =
\pl \D',(\E)\pr$ and hence
 $\pl \D',(\E)\pr \in
 \ \vdash \bP_{ 1},\dots, \bP_{ n}$.

  \end{enumerate}
By applying
Proposition ~\ref{closure} again, we have that
 $ \D'\in
 \ \vdash \bPP, \bP_{\xi 1},\dots, \bP_{\xi n}$.   By applying the inductive hypothesis, we can  write the derivation:
\begin{center}
\AxiomC{$\vdots \ \D^{' \star}$}\noLine
\UnaryInfC{$\vdash \Pi, P_1,\dots,P_n,$}
\RightLabel{\scriptsize{Neg$_n$}}
\UnaryInfC{$ \vdash   \Pi, P_1 \parr\dots \parr P_n$} \DisplayProof
\end{center}


\section{Ludics with repetitions: what, how, why} \label{repetitions}

In the previous section, we assumed linearity of the strategies to prove internal completeness.
 From now on, we go back to the general definition of strategy
(on an universal arena)  as in {the beginning of} Section ~\ref{ludics}, without any
hypothesis of linearity.
This means that strategies now allow repeated actions.

In this section, we mainly discuss the difficulties in extending the approach of ludics to this setting, and introduce our solution, which will be technically developed in
Section ~\ref{lwR section II}.\smallskip

First, let us introduce  some operations which we will use in this section to
 deal with repeated actions and describe the composition.

\subsection{Copies and renaming}\label{renaming}

\subsubsection*{Renaming} Given a strategy $\D:\xi$ of arbitrary polarity, let us indicate
by $\D[\sigma/\xi]$ the strategy obtained from $\D$ by renaming, in all occurrences of action, the prefix
 $\xi$ into $\sigma$, \ie each name  $\xi.\alpha$ becomes $\sigma.\alpha$.
Obviously, if $\D:\xi$, then $\D[\sigma/\xi]:\sigma$.

\subsubsection*{Renaming of the root}  Given a positive strategy $\D:\xi^+$, let us indicate by $\rt{\D}$ the strategy obtained by
renaming the prefix $\xi$ into $\sigma$ in the root, and in all actions which are hereditarily
justified by the root. If $\D:\xi^+$,
 we obtain a new strategy $\rt{\D}: \sigma^+,
\xi^+$.\smallskip

\begin{figure}[htbp]
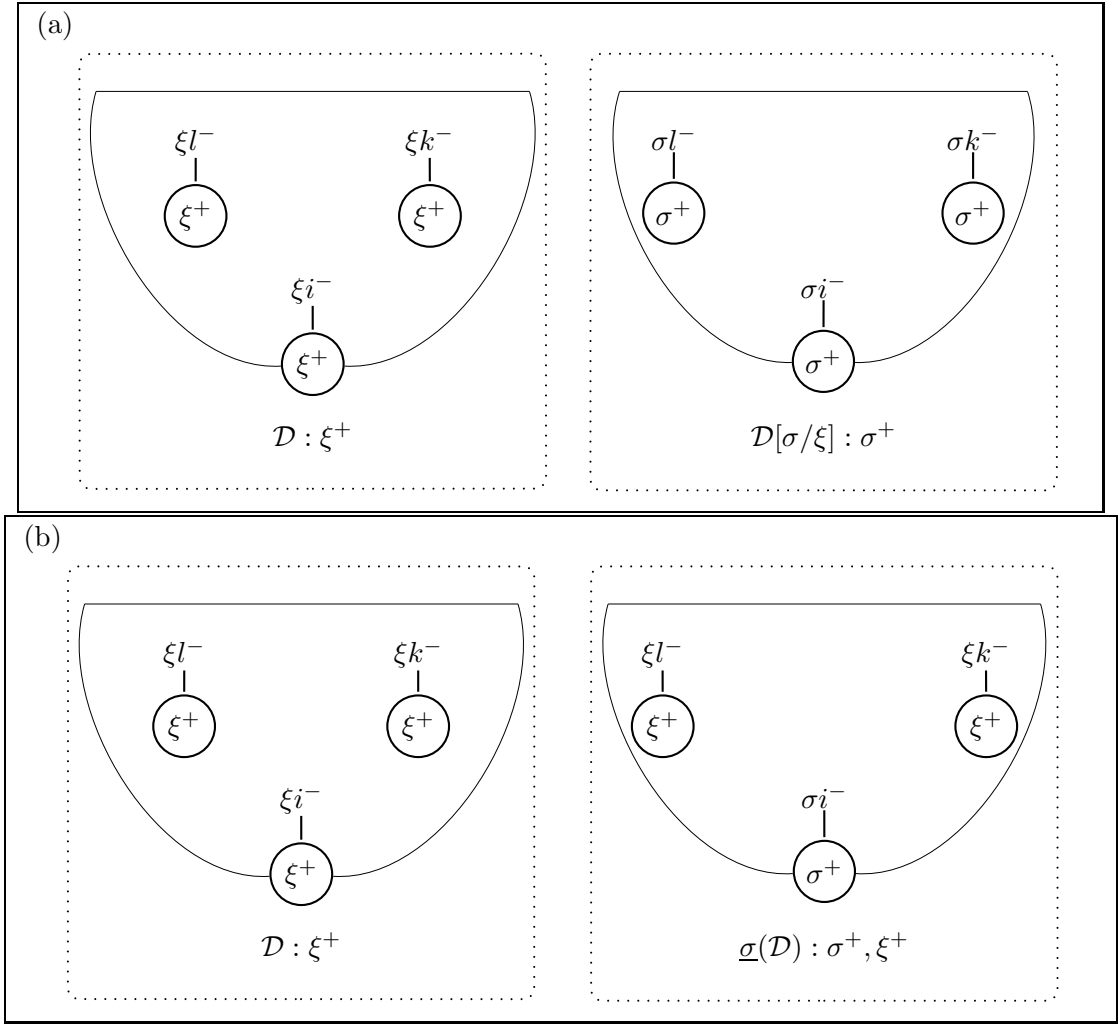


\centering
\fbox{
\psmatrix[rowsep=9pt,colsep=0.3cm]
(a) &&& & &  & && &
&&&&&&
 & &  &  & & &
\\
& \rnode{l3}{} &&
 &  & & \rnode{up1}{} & &  &

 &&
 \rnode{l4}{}
 &&&&

 \rnode{al3}{}  &  & & \rnode{aup1}{} & &  & \rnode{al4}{}

 &&&
 \\

 &&&

\rnode{l1}{} & \rnode{xil}{$\xi l^-$} & \rnode{l1bis}{} &  & \rnode{l2bis}{} & \rnode{xik}{$\xi k^-$} & \rnode{l2}{}

 &&&&&&

\rnode{al1}{} & \rnode{axil}{$\sigma l^-$} & \rnode{al1bis}{} &  & \rnode{al2bis}{} & \rnode{axik}{$\sigma k^-$} & \rnode{al2}{}

\\

&&&

& \circlenode{xia}{$\xi^+$} & & \rnode{fa}{} & & \circlenode{xib}{$\xi^+$} &

 &&&&&&

& \circlenode{axia}{$\sigma^+$} & & \rnode{afa}{} & & \circlenode{axib}{$\sigma^+$} &

 \\
 &&&

& &  \rnode{node1}{} & \rnode{xii}{$\xi i^-$} &
 \rnode{node2}{}
 & &

 &&&&&&

& &  \rnode{anode1}{} & \rnode{axii}{$\sigma i^-$} &
 \rnode{anode2}{}
 & &

 \\

 &&&

 & &  & \circlenode{root}{$\xi^+$}  &

 & &

  &&&&&&

 & &  & \circlenode{aroot}{$\sigma^+$}  &

 & &

 \\

 &&&

  & &  & \rnode{name}{$\D : \xi^+$}  &

 & &

   &&&&&&

   & &  & \rnode{aname}{$\D[\sigma/\xi] : \sigma^+$}  &

 & &

\\

 &&& & &  & && &
&&&&&&
 & &  &  & & &
\\
\endpsmatrix
 \psset{shortput=nab,arrows=->,labelsep=6pt}
 \small
\ncline[nodesep=1pt]{-}{xia}{xil}
\ncline[nodesep=1pt]{-}{xib}{xik}
\ncline[nodesep=1pt]{-}{root}{xii}
\ncline[nodesep=1pt]{-}{axia}{axil}
\ncline[nodesep=1pt]{-}{axib}{axik}
\ncline[nodesep=1pt]{-}{aroot}{axii}
\ncline[nodesep=0pt,linewidth=.01]{-}{l3}{l4}
\ncarc[nodesep=0pt,linewidth=.01,arcangle=-55]{-}{l3}{root}
\ncarc[nodesep=0pt,linewidth=.01,arcangle=55]{-}{l4}{root}

\ncline[nodesep=0pt,linewidth=.01]{-}{al3}{al4}
\ncarc[nodesep=0pt,linewidth=.01,arcangle=-55]{-}{al3}{aroot}
\ncarc[nodesep=0pt,linewidth=.01,arcangle=55]{-}{al4}{aroot}

\ncbox[nodesep=.5cm,boxsize=3.1,linearc=.15,
linestyle=dotted]{name}{up1}

\ncbox[nodesep=.5cm,boxsize=3.1,linearc=.15,
linestyle=dotted]{aname}{aup1}

}

\fbox{
\psmatrix[rowsep=9pt,colsep=0.3cm]
(b) &&& & &  & && &
&&&&&&
 & &  &  & & &
\\
& \rnode{l3}{} &&
 &  & & \rnode{up1}{} & &  &

 &&
 \rnode{l4}{}
 &&&&

 \rnode{al3}{}  &  & & \rnode{aup1}{} & &  & \rnode{al4}{}

 &&&
 \\

 &&&

\rnode{l1}{} & \rnode{xil}{$\xi l^-$} & \rnode{l1bis}{} &  & \rnode{l2bis}{} & \rnode{xik}{$\xi k^-$} & \rnode{l2}{}

 &&&&&&

\rnode{al1}{} & \rnode{axil}{$\xi l^-$} & \rnode{al1bis}{} &  & \rnode{al2bis}{} & \rnode{axik}{$\xi k^-$} & \rnode{al2}{}

\\

&&&

& \circlenode{xia}{$\xi^+$} & & \rnode{fa}{} & & \circlenode{xib}{$\xi^+$} &

 &&&&&&

& \circlenode{axia}{$\xi^+$} & & \rnode{afa}{} & & \circlenode{axib}{$\xi^+$} &

 \\
 &&&

& &  \rnode{node1}{} & \rnode{xii}{$\xi i^-$} &
 \rnode{node2}{}
 & &

 &&&&&&

& &  \rnode{anode1}{} & \rnode{axii}{$\sigma i^-$} &
 \rnode{anode2}{}
 & &

 \\

 &&&

 & &  & \circlenode{root}{$\xi^+$}  &

 & &

  &&&&&&

 & &  & \circlenode{aroot}{$\sigma^+$}  &

 & &

 \\

 &&&

  & &  & \rnode{name}{$\D : \xi^+$}  &

 & &

   &&&&&&

   & &  & \rnode{aname}{$\rt\D : \sigma^+,\xi^+$}  &

 & &

\\

 &&& & &  & && &
&&&&&&
 & &  &  & & &
\\
\endpsmatrix
 \psset{shortput=nab,arrows=->,labelsep=6pt}
 \small
\ncline[nodesep=1pt]{-}{xia}{xil}
\ncline[nodesep=1pt]{-}{xib}{xik}
\ncline[nodesep=1pt]{-}{root}{xii}
\ncline[nodesep=1pt]{-}{axia}{axil}
\ncline[nodesep=1pt]{-}{axib}{axik}
\ncline[nodesep=1pt]{-}{aroot}{axii}
\ncline[nodesep=0pt,linewidth=.01]{-}{l3}{l4}
\ncarc[nodesep=0pt,linewidth=.01,arcangle=-55]{-}{l3}{root}
\ncarc[nodesep=0pt,linewidth=.01,arcangle=55]{-}{l4}{root}

\ncline[nodesep=0pt,linewidth=.01]{-}{al3}{al4}
\ncarc[nodesep=0pt,linewidth=.01,arcangle=-55]{-}{al3}{aroot}
\ncarc[nodesep=0pt,linewidth=.01,arcangle=55]{-}{al4}{aroot}

\ncbox[nodesep=.5cm,boxsize=3.1,linearc=.15,
linestyle=dotted]{name}{up1}

\ncbox[nodesep=.5cm,boxsize=3.1,linearc=.15,
linestyle=dotted]{aname}{aup1}

}

  \caption{Renaming (a) and renaming of the root (b)}
  \label{fig:D'}
\end{figure}

We  picture both the operations  in Figures ~\ref{fig:D'}(a) and (b).
For readability,
we indicate an action on $\xi$ simply with the name $\xi$.

\subsubsection*{Copies of a behaviour} We remind that to  emphasize that
 $\bA$ is a set of strategies on interface $\xi$, we annotate the name $\xi$
as a subscript: $\bA_{\xi}$.
If $\bA_\xi$ is a set of strategies on the name $\xi$,
we write $\bA_\sigma $ for $\{\D[\sigma/\xi] : \D\in \bA_\xi\}$. $\bA_\sigma $ is a copy of
$\bA_\xi$: they are \emph{equal up to renaming}.

\subsection{Composition (normalization)} \label{composition}
 In a strategy, actions can now be repeated.
Composition of strategies as sets of views can be described via the VAM
 abstract machines
introduced  in ~\cite{CurHerb}. We describe composition in details
in Section ~\ref{VAM}.

We now give an example of composition of strategies using
the graphical notation introduce before.

However, what we will really need is only that
 composition has a fundamental property,
expressed  by the following equation:

\begin{equation} \label{Copies}
\pl \D, \E \pr = \pl \rt{\D}, \E, \E[\sigma/\xi] \pr
\end{equation}
for any $\D : \xi^+$ and $\E : \xi^-$ such that the operations ``$\rt\_$" and ``$[\sigma/\xi]$"  make sense.
This property will also hold for strategies
with silent actions we introduce later. The  proof for a more
 general equation  is given in Section ~\ref{VAM}
(Proposition ~\ref{propo copies1}).

\begin{rem}
   Equation ~\ref{Copies} is reminiscent of the
separation of head occurrence property of ~\cite{AbrJagMac}
and the linearization of the head occurrence property of
\cite{Abr} (see also the discussion therein).
\end{rem}

From Equation (\ref{Copies}), we have in particular:

\begin{cor}\label{Copies_cor}
 $ \D \bot \E$   if and only if  $ \rt{\D} \bot \{ \E, \E[\sigma/\xi]\}$. \qed
\end{cor}

Let us see how  Equation (\ref{Copies}) works by giving a description of  the
composition.

\begin{figure}[htbp]
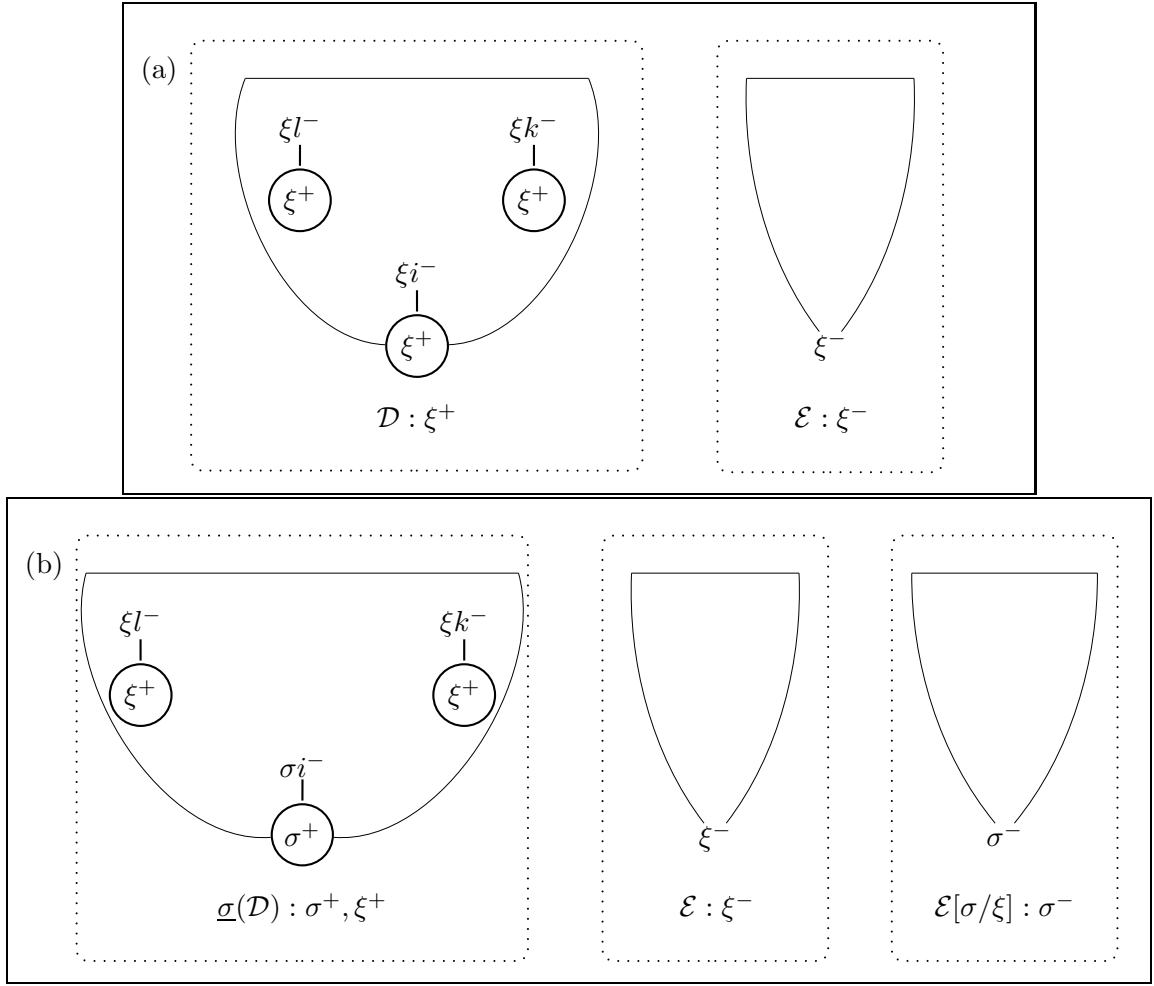

  \centering
\fbox{
\psmatrix[rowsep=9pt,colsep=0.3cm]
 &&& & &  & &&
&&&&&&
 & &  &  & & &
\\
(a)&&&
\rnode{l3}{} &  & & \rnode{up1}{} & & & \rnode{l4}{}

 &&&&&

  &  & \rnode{al3}{} & \rnode{aup1}{} & \rnode{al4}{} &  &

 &&&
 \\

 &&&

\rnode{l1}{} & \rnode{xil}{$\xi l^-$} & \rnode{l1bis}{} &  & \rnode{l2bis}{} & \rnode{xik}{$\xi k^-$} & \rnode{l2}{}

 &&&&&

\rnode{al1}{} &  & \rnode{al1bis}{} &  & \rnode{al2bis}{} &  & \rnode{al2}{}

\\

&&&

& \circlenode{xia}{$\xi^+$} & & \rnode{fa}{} & & \circlenode{xib}{$\xi^+$} &

 &&&&&

&  & &  & & &

 \\
 &&&

& &  \rnode{node1}{} & \rnode{xii}{$\xi i^-$} &
 \rnode{node2}{}
 & &

 &&&&&

& &  \rnode{anode1}{} & \rnode{axii}{} &
 \rnode{anode2}{}
 & &

 \\

 &&&

 & &  & \circlenode{root}{$\xi^+$}  &

 & &

  &&&&&

 & &  & \rnode{aroot}{$\xi^-$}  &

 & &

 \\

 &&&

  & &  & \rnode{name}{$\D : \xi^+$}  &

 & &

   &&&&&

   & &  & \rnode{aname}{$\phantom{[\sigma} \E : \xi^-\phantom{\xi]}$}  &

 & &

\\

 &&& & &  & &&
&&&&&&
 & &  &  & & &
\\
\endpsmatrix
 \psset{shortput=nab,arrows=->,labelsep=6pt}
 \small
\ncline[nodesep=1pt]{-}{xia}{xil}
\ncline[nodesep=1pt]{-}{xib}{xik}
\ncline[nodesep=1pt]{-}{root}{xii}

\ncline[nodesep=0pt,linewidth=.01]{-}{l3}{l4}
\ncarc[nodesep=0pt,linewidth=.01,arcangle=-55]{-}{l3}{root}
\ncarc[nodesep=0pt,linewidth=.01,arcangle=55]{-}{l4}{root}

\ncline[nodesep=0pt,linewidth=.01]{-}{al3}{al4}
\ncarc[nodesep=0pt,linewidth=.01,arcangle=-20]{-}{al3}{aroot}
\ncarc[nodesep=0pt,linewidth=.01,arcangle=20]{-}{al4}{aroot}

\ncbox[nodesep=.5cm,boxsize=3,linearc=.15,
linestyle=dotted]{name}{up1}

\ncbox[nodesep=.5cm,boxsize=1.5,linearc=.15,
linestyle=dotted]{aname}{aup1}

}

\fbox{
\psmatrix[rowsep=9pt,colsep=0.3cm]

&&
 & & &

 &&&

  &  & & & &

 &&&

 &&&&&&&
 \\

(b)& \rnode{l3}{}
 &  & & \rnode{up1}{} & &  & \rnode{l4}{}

 &&&

  &  & \rnode{al3}{} & \rnode{aup1}{} & \rnode{al4}{} &  &

 &&&\rnode{bl3}{}&\rnode{bup1}{}&\rnode{bl4}{}& &
 \\

 &

\rnode{l1}{} & \rnode{xil}{$\xi l^-$} & \rnode{l1bis}{} &  & \rnode{l2bis}{} & \rnode{xik}{$\xi k^-$} & \rnode{l2}{}

 &&&

\rnode{al1}{} &  & \rnode{al1bis}{} &  & \rnode{al2bis}{} &  & \rnode{al2}{}

&&&&&&&

\\

&

& \circlenode{xia}{$\xi^+$} & & \rnode{fa}{} & & \circlenode{xib}{$\xi^+$} &

 &&&

&  & &

&&&&&&&&&&

 \\
 &

& &  \rnode{node1}{} & \rnode{xii}{$\sigma i^-$} &
 \rnode{node2}{}
 & &

 &&&

& &  \rnode{anode1}{} & \rnode{axii}{} &
 \rnode{anode2}{}
 &  &&&&  \rnode{bnode1}{} & \rnode{bxii}{} &
 \rnode{bnode2}{} &&

 \\

 &

 & &  & \circlenode{root}{$\sigma^+$}  &

 & &

  &&&

 & &  & \rnode{aroot}{$\xi^-$}  &

 & &

 &&&   & \rnode{broot}{$\sigma^-$} &
 &&

 \\

 &

  & &  & \rnode{name}{$\rt\D : \sigma^+,\xi^+$}  &

 & &

   &&&

   & &  & \rnode{aname}{$\phantom{[\sigma} \E : \xi^-\phantom{\xi]}$}  &

 & &

 &&&   & \rnode{bname}{$\E[\sigma/\xi] : \sigma^-$} &
 &&

\\

 &&&   & && &
&&&&&&
 & &  &  & & &
\\
\endpsmatrix
 \psset{shortput=nab,arrows=->,labelsep=6pt}
 \small
\ncline[nodesep=1pt]{-}{xia}{xil}
\ncline[nodesep=1pt]{-}{xib}{xik}
\ncline[nodesep=1pt]{-}{root}{xii}

\ncline[nodesep=0pt,linewidth=.01]{-}{l3}{l4}
\ncarc[nodesep=0pt,linewidth=.01,arcangle=-55]{-}{l3}{root}
\ncarc[nodesep=0pt,linewidth=.01,arcangle=55]{-}{l4}{root}

\ncline[nodesep=0pt,linewidth=.01]{-}{al3}{al4}
\ncarc[nodesep=0pt,linewidth=.01,arcangle=-20]{-}{al3}{aroot}
\ncarc[nodesep=0pt,linewidth=.01,arcangle=20]{-}{al4}{aroot}

\ncline[nodesep=0pt,linewidth=.01]{-}{bl3}{bl4}
\ncarc[nodesep=0pt,linewidth=.01,arcangle=-20]{-}{bl3}{broot}
\ncarc[nodesep=0pt,linewidth=.01,arcangle=20]{-}{bl4}{broot}

\ncbox[nodesep=.5cm,boxsize=3,linearc=.15,
linestyle=dotted]{name}{up1}

\ncbox[nodesep=.5cm,boxsize=1.5,linearc=.15,
linestyle=dotted]{aname}{aup1}

\ncbox[nodesep=.5cm,boxsize=1.5,linearc=.15,
linestyle=dotted]{bname}{bup1}

}

  \caption{Composition (with repeated actions)}
  \label{fig:copies}

\end{figure}

  Let $\D:\xi^+$ and $\E:\xi^-$ be two strategies,
which we represent
 in Figure ~\ref{fig:copies} (a) (again, we indicate an  action $x$ on $\xi$ simply with the name $\xi$).
The idea behind the abstract machine in ~\cite{CurHerb} is
that, when  the two strategies $\D$ and $\E$
interact, every time  $\D$ plays an action
$x$ on $\xi$, a copy of $\E$ is created; \ie composition works as if we had a copy of $\E$ for each occurrence of $x$ in $\D$. It is rather intuitive
that the result of normalization is the same if we make this explicit, by  renaming
 one occurrence of $x$ (namely the root),
and making  an explicit copy of $\E$, as illustrated in
Figure ~\ref{fig:copies} (b).

\begin{exa} \label{ex intrep}
Let us consider the  strategies $\D$ and $\E$ in Figure
~\ref{fig:intrep}, where
we indicate an  action $x$ on $\xi$ simply with the name $\xi$.
Observe that we explicitly need to draw a pointer from
$\xi 11^+$ to the right occurrence of $\xi 1^-$ (the lowermost
one in our case) which justifies it. The other pointers
can be univocally determined.
The interaction is the sequence  given by following the arrows and
the normal form is $\Dai$.

\begin{figure}[htbp]
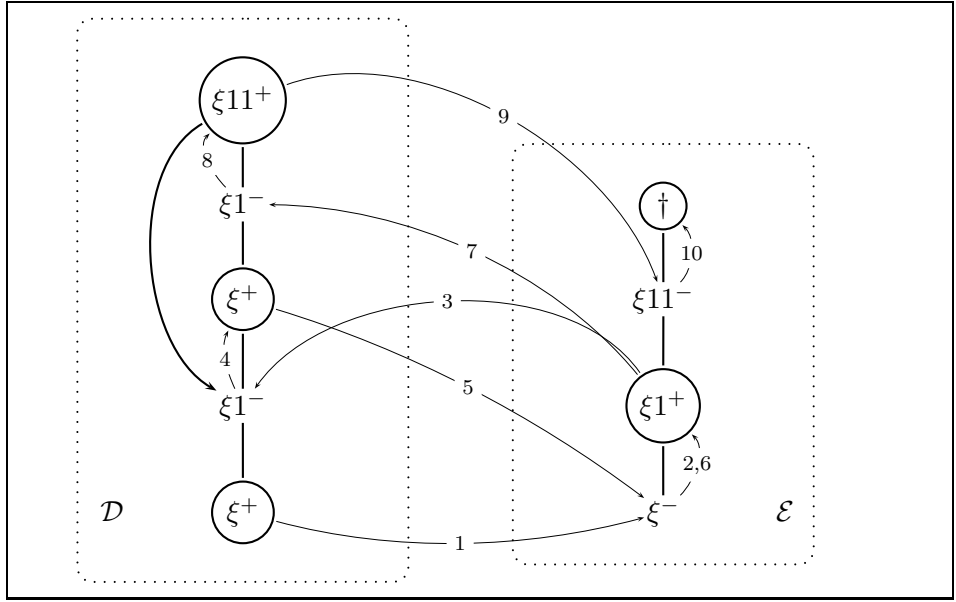

  \centering
\fbox{
\psmatrix[rowsep=13pt,colsep=0.5cm]
&&&&&&&&&&&&&&&&& &&\\
&&&&\circlenode{x11pa}{$\xi 11^+$}&&&&&&&&&&& &&\\
&&&&\rnode{x1mb}{$\xi 1^-$}&&&&&&&&& \circlenode{daimon}{$\dai$}&& &&\\
&&&&\circlenode{xpb}{$\xi^+$}&&&&&&&&& \rnode{x11m}{$\xi 11^-$} && &&\\
&&&&\rnode{x1ma}{$\xi 1^-$} &&&&&&&&& \circlenode{x1p}{$\xi 1^+$}&& &&\\
&&$\D$&&\circlenode{xpa}{$\xi^+$}&&&&&&&&&  \rnode{xm}{$\xi^-$} && $\E$ &&\\
&&&&&&&&&&&&&&& &&\\
\endpsmatrix
 \psset{shortput=nab,arrows=->,labelsep=6pt}
 \small
 \ncarc[nodesep=1pt,arcangle=-60]{x11pa}{x1ma}

\ncarc[linewidth=0.2pt,nodesep=1pt,arcangle=-15]{xpa}{xm}
\ncput*{\scriptsize 1}
\ncarc[linewidth=0.2pt,nodesep=1pt,arcangle=-45]{xm}{x1p}
\ncput*{\scriptsize 2,6}
\ncarc[linewidth=0.2pt,nodesep=1pt,arcangle=-55]{x1p}{x1ma}
\ncput*{\scriptsize 3}
\ncarc[linewidth=0.2pt,nodesep=1pt,arcangle=25]{x1ma}{xpb}
\ncput*{\scriptsize 4}
\ncarc[linewidth=0.2pt,nodesep=1pt,arcangle=10]{xpb}{xm}
\ncput*{\scriptsize 5}
\ncarc[linewidth=0.2pt,nodesep=1pt,arcangle=-25]{x1p}{x1mb}
\ncput*{\scriptsize 7}
\ncarc[linewidth=0.2pt,nodesep=1pt,arcangle=45]{x1mb}{x11pa}
\ncput*{\scriptsize 8}
\ncarc[linewidth=0.2pt,nodesep=1pt,arcangle=45]{x11pa}{x11m}
\ncput*{\scriptsize 9}
\ncarc[linewidth=0.2pt,nodesep=1pt,arcangle=-45]{x11m}{daimon}
\ncput*{\scriptsize 10}

\ncline[nodesep=1pt]{-}{daimon}{x11m}
\ncline[nodesep=1pt]{-}{x1p}{x11m}
\ncline[nodesep=1pt]{-}{x1p}{xm}
\ncline[nodesep=1pt]{-}{x11pa}{x1mb}
\ncline[nodesep=1pt]{-}{xpb}{x1mb}
\ncline[nodesep=1pt]{-}{xpb}{x1ma}
\ncline[nodesep=1pt]{-}{xpa}{x1ma}

\ncbox[nodesep=.5cm,boxsize=2.2,linearc=.2,
linestyle=dotted]{x11pa}{xpa}

\ncbox[nodesep=.5cm,boxsize=2,linearc=.2,
linestyle=dotted]{daimon}{xm}
}
  \caption{Example of interaction with repetitions}
  \label{fig:intrep}
\end{figure}

\end{exa}

\begin{exa} \label{ex copies}
We now  check for $\D,\E$  in  Example ~\ref{ex intrep} that $\pl \D, \E \pr = \pl \rt{\D}, \E, \E[\sigma/\xi]  \pr$
 as pictured in Figure ~\ref{fig:intrepcop}.  Since
$\rt{\D}$ is linear in this example,  we no longer need to show
 the
pointers explicitly.

\begin{figure}[htbp]
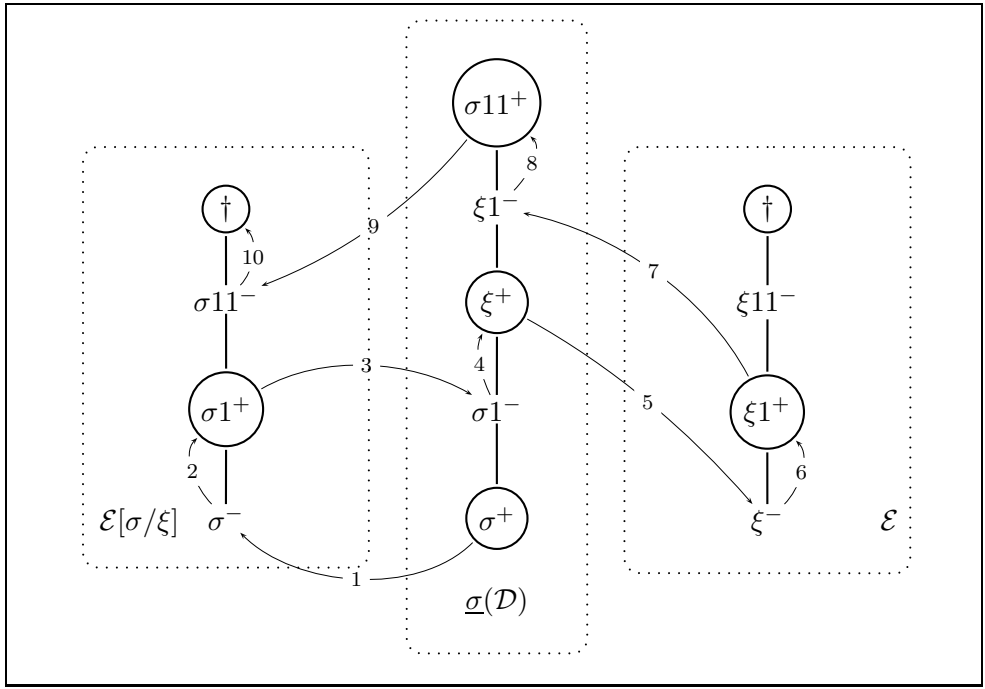

  \centering
\fbox{
\psmatrix[rowsep=13pt,colsep=0.5cm]
&&&&&&&&&&&&&&&&& \\
&&&&&&&&\circlenode{x11pas}{$\sigma 11^+$}&&&&&&& \\
&&&\circlenode{daimons}{$\dai$}&&&&&\rnode{x1mb}{$\xi 1^-$}&&&&& \circlenode{daimon}{$\dai$}&& &&\\
&&&\rnode{x11ms}{$\sigma 11^-$}&&&&&\circlenode{xpb}{$\xi^+$}&&&&& \rnode{x11m}{$\xi 11^-$} && &&\\
&&&\circlenode{x1ps}{$\sigma 1^+$}&&&&&\rnode{x1ma}{$\sigma 1^-$} &&&&& \circlenode{x1p}{$\xi 1^+$}&& &&\\
&&$\E[\sigma/\xi] \!\!\!\!\!\! $&\rnode{xms}{$\sigma^-$}&&&&&\circlenode{xpa}{$\sigma^+$}&&&&&  \rnode{xm}{$\xi^-$} && $\E$ &&\\
&&&&&&&&\rnode{A}{$\rt\D$}&&&&&&& &&\\
&&&&&&&&&&&&&&&&& \\
\endpsmatrix
 \psset{shortput=nab,arrows=->,labelsep=6pt}
 \small

\ncarc[linewidth=0.2pt,nodesep=1pt,arcangle=45]{xpa}{xms}
\ncput*{\scriptsize 1}
\ncarc[linewidth=0.2pt,nodesep=1pt,arcangle=45]{xms}{x1ps}
\ncput*{\scriptsize 2}
\ncarc[linewidth=0.2pt,nodesep=1pt,arcangle=30]{x1ps}{x1ma}
\ncput*{\scriptsize 3}
\ncarc[linewidth=0.2pt,nodesep=1pt,arcangle=25]{x1ma}{xpb}
\ncput*{\scriptsize 4}
\ncarc[linewidth=0.2pt,nodesep=1pt,arcangle=10]{xpb}{xm}
\ncput*{\scriptsize 5}
\ncarc[linewidth=0.2pt,nodesep=1pt,arcangle=-45]{xm}{x1p}
\ncput*{\scriptsize 6}
\ncarc[linewidth=0.2pt,nodesep=1pt,arcangle=-25]{x1p}{x1mb}
\ncput*{\scriptsize 7}
\ncarc[linewidth=0.2pt,nodesep=1pt,arcangle=-45]{x1mb}{x11pas}
\ncput*{\scriptsize 8}
\ncarc[linewidth=0.2pt,nodesep=1pt,arcangle=15]{x11pas}{x11ms}
\ncput*{\scriptsize 9}
\ncarc[linewidth=0.2pt,nodesep=1pt,arcangle=-45]{x11ms}{daimons}
\ncput*{\scriptsize 10}

\ncline[nodesep=1pt]{-}{daimon}{x11m}
\ncline[nodesep=1pt]{-}{x1p}{x11m}
\ncline[nodesep=1pt]{-}{x1p}{xm}
\ncline[nodesep=1pt]{-}{daimons}{x11ms}
\ncline[nodesep=1pt]{-}{x1ps}{x11ms}
\ncline[nodesep=1pt]{-}{x1ps}{xms}
\ncline[nodesep=1pt]{-}{x11pas}{x1mb}
\ncline[nodesep=1pt]{-}{xpb}{x1mb}
\ncline[nodesep=1pt]{-}{xpb}{x1ma}
\ncline[nodesep=1pt]{-}{xpa}{x1ma}

\ncbox[nodesep=.5cm,boxsize=1.2,linearc=.2,
linestyle=dotted]{x11pas}{A}

\ncbox[nodesep=.5cm,boxsize=1.9,linearc=.2,
linestyle=dotted]{daimon}{xm}

\ncbox[nodesep=.5cm,boxsize=1.9,linearc=.2,
linestyle=dotted]{daimons}{xms}
}
\caption{Example of interaction with copies}
  \label{fig:intrepcop}
\end{figure}

\end{exa}

\subsection{What are the difficulties} \label{probl}
We are ready to discuss which are the difficulties in extending the approach of ludics to a setting where strategies are non-linear.

\subsubsection*{Problem 1: Separation} The first problem  when strategies have repetitions is with separation.
Let us  give a simple example of why separation fails if we allow repetitions.

\begin{exa}[\cite{MauTh}]
Let $\D_1,\D_2 : \xi^+$ and $\E  : \xi^-$ be strategies as in Figure ~\ref{fig:separation}, where $x= (\xi , I)$, $y=(\xi i , J)$.
 We
cannot
find a strategy orthogonal to $\D_1$ but
not orthogonal to $\D_2$. For example,
the interactions between $\D_1$ and $\E$ and
 $\D_2$ and $\E$ produce  the same normal form  $\Dai$.

\begin{figure}[h]
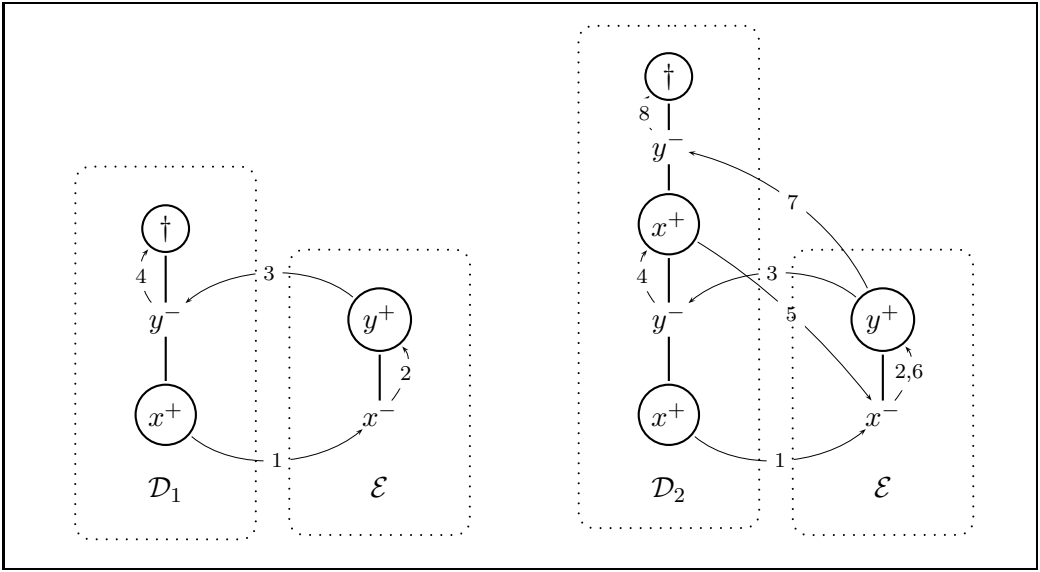

\centering
\fbox{

\psmatrix[rowsep=11pt,colsep=0.5cm]
&&&&&&&&&&&&& &&&& && &\\

& && &&&&&&&&&& \circlenode{daimon2}{$\dai$}&&&&  &&& \\
& && &&&& &&&&&& \rnode{ym3}{$y^-$}&&&&  &&& \\
& && \circlenode{daimon1}{$\dai$}&&&& &&&&&& \circlenode{xp3}{$x^+$}&&&& &&  &\\
& && \rnode{ym1}{$y^-$} &&&& \circlenode{yp1}{$y^+$}&&&&&& \rnode{ym2}{$y^-$} &&&& \circlenode{yp2}{$y^+$} && & \\
& && \circlenode{xp1}{$x^+$}&&&&  \rnode{xm1}{$x^-$} &&&&&& \circlenode{xp2}{$x^+$}&&&&  \rnode{xm2}{$x^-$} && &\\
& && \rnode{D1}{$\D_1$}&&&& \rnode{E1}{$\E$}&&&&&& \rnode{D2}{$\D_2$}&&&& \rnode{E2}{$\E$}  &&& \\
& &&&&&&&&&&&& &&&& && &\\
\endpsmatrix
\psset{shortput=nab,arrows=->,labelsep=6pt}
\small
\ncarc[nodesep=1pt,arcangle=-60]{x11pa}{x1ma}

\ncarc[linewidth=0.2pt,nodesep=1pt,arcangle=-40]{xp1}{xm1}
\ncput*{\scriptsize 1}
\ncarc[linewidth=0.2pt,nodesep=1pt,arcangle=-40]{xm1}{yp1}
\ncput*{\scriptsize 2}
\ncarc[linewidth=0.2pt,nodesep=1pt,arcangle=-40]{yp1}{ym1}
\ncput*{\scriptsize 3}
\ncarc[linewidth=0.2pt,nodesep=1pt,arcangle=40]{ym1}{daimon1}
\ncput*{\scriptsize 4}

\ncarc[linewidth=0.2pt,nodesep=1pt,arcangle=-40]{xp2}{xm2}
\ncput*{\scriptsize 1}
\ncarc[linewidth=0.2pt,nodesep=1pt,arcangle=-40]{xm2}{yp2}
\ncput*{\scriptsize 2,6}
\ncarc[linewidth=0.2pt,nodesep=1pt,arcangle=-40]{yp2}{ym2}
\ncput*{\scriptsize 3}
\ncarc[linewidth=0.2pt,nodesep=1pt,arcangle=40]{ym2}{xp3}
\ncput*{\scriptsize 4}
\ncarc[linewidth=0.2pt,nodesep=1pt,arcangle=10]{xp3}{xm2}
\ncput*{\scriptsize 5}
\ncarc[linewidth=0.2pt,nodesep=1pt,arcangle=-25]{yp2}{ym3}
\ncput*{\scriptsize 7}
\ncarc[linewidth=0.2pt,nodesep=1pt,arcangle=45]{ym3}{daimon2}
\ncput*{\scriptsize 8}

\ncline[nodesep=1pt]{-}{daimon1}{ym1}
\ncline[nodesep=1pt]{-}{ym1}{xp1}
\ncline[nodesep=1pt]{-}{x1p}{xm}

\ncline[nodesep=1pt]{-}{yp1}{xm1}
\ncline[nodesep=1pt]{-}{yp2}{xm2}

\ncline[nodesep=1pt]{-}{daimon2}{ym3}
\ncline[nodesep=1pt]{-}{xp3}{ym3}
\ncline[nodesep=1pt]{-}{xp3}{ym2}
\ncline[nodesep=1pt]{-}{xp2}{ym2}

\ncbox[nodesep=.5cm,boxsize=1.2,linearc=.2,
linestyle=dotted]{daimon1}{D1}

\ncbox[nodesep=.5cm,boxsize=1.2,linearc=.2,
linestyle=dotted]{yp1}{E1}

\ncbox[nodesep=.5cm,boxsize=1.2,linearc=.2,
linestyle=dotted]{yp2}{E2}

\ncbox[nodesep=.35cm,boxsize=1.2,linearc=.2,
linestyle=dotted]{daimon2}{D2}
}

 \caption{Non-separation}
 \label{fig:separation}
\end{figure}
\end{exa}

In this work, we ignore separation all together.
As we discussed in  Section ~\ref{int_compl}, even if separation is an important property,
 we don't need it in order to have
interactive types and internal completeness.
In future work, it may be possible to refine our setting using Maurel's techniques   ~\cite{MauTh}.

\subsubsection*{Problem 2: Enough tests (counter-strategies)}

The second problem | which we believe being the deeper one| has to do with having enough tests, \ie enough counter-strategies.\\
As in ~\cite{GirLoc}, we have defined an interactive type
 to be any  set of strategies closed
by biorthogonal.
Assume we have defined how to interpret formulas
of $\HS$, like $P = \posm (N_1 \otimes \cdots \otimes N_n )$ and $P\b = \negm (N_1^\bot \parr \cdots \parr N_n^\bot)$.

We would like to associate  to each ``good'' strategy in the interpretation of
a formula, say the positive formula $P$ above, in a behaviour that we indicate with $\bP$,
a syntactical proof of $\vdash P$ (full completeness).
 If $\D:\xi^+ \in \bP$, we would like to  transform it into
a strategy $\D'\in \ \vdash\bP_\xi,  \bP_{\sigma}$ (where distinct names indicate distinct copies). This
corresponds to the contraction rule (in its upwards reading).

The natural idea is to use the same technique as in  ~\cite{AbrJagMac}, and
to rename the root, and all the actions which are
hereditarily justified by it.
  We have already illustrated this operation in
Section ~\ref{renaming} (Figure ~\ref{fig:D'}).
 From $\D:\xi^+$, we obtain a new strategy $\D':\xi^+,\sigma^+$, where $\D'=\rt{\D}$.\\
We would like to prove that:
$$(\ast)~\D\in \mathbf{\vdash \bP_\xi} \ \ \Rightarrow \ \ (\ast \ast)~\rt{\D} \in
\mathbf{\vdash  \bP_\xi, \bP_\sigma}.
$$
To have $(\ast \ast)$, we need (see Definition ~\ref{seq_def}) to know that
$\rt{\D} \bot \{\E,\F \}$ for each $\E\in \bP_\xi^\bot$ and each $\F\in \bP_\sigma^\bot$.
Since $\bP_\sigma^\bot$ is a copy (renamed in $\sigma$) of  $\bP_\xi^\bot$, we
 can also write this condition as \begin{equation} \label{eq2}
 \rt{\D} \bot \{\E, \F[\sigma/\xi]\}, \end{equation}
where both $\F$ and $\E$ vary in $\bP_\xi^\bot$.

However, from Equation ~(\ref{Copies}) we \emph{only} have that
$\rt{\D} \bot \{\E, \E[\sigma/\xi]\}$:  two copies of \emph{the same}
 (up to renaming) strategy $\E$.
This  fact can be rephrased by saying that  in
our  ``HO
 setting'', strategies in the type $\bP^\bot$, which is, {
 roughly speaking,} ``of the form
$\negm \bC$"
are  {\em uniform:}
every time we find a repeated action in   $\bP$, Opponent
$\bP^\bot$ reacts in the same way.

\subsection{A solution: non-uniform tests}
The need for having enough tests appears  similar to the one which has  led Girard to the
introduction of the {\em daimon rule}:
in ludics, one typically opposes to an abstract ``proof of $A$''
 an abstract ``counter-proof of $A$.'' To have enough tests (that is,
to have both proofs of $A$ and proofs of $A\b$) there is a new rule
which allow us to justify any premise.

Similarly here, when we oppose to a proof of $\bP$  a proof of $\bP\b$, we need
enough counter-strategies. We are led to enlarge the universe of tests by introducing {\em non-uniform counter-strategies}. This is extremely natural to realize in an AJM setting ~\cite{AbrJagMac,thesisbas},
where a strategy of type ``$\negm\bC$" is a sort of infinite tensor
of strategies on
``$\bC$", each one with its {\em index of copy}.  To have HO non-uniform counter-strategies, we introduce a  non-deterministic sum of strategies.
Let us illustrate the idea, which we will formalize in the next section.

\subsubsection*{Non-uniform counter-strategies} \label{NUtests}
The idea is to allow  a kind of {\em  ``non-deterministic sum''}
of {\em negative} strategies $\E,\F$. Let us, for now, {\em informally}  write such a  sum of $\E$ and $\F$ in the following way, which is reminiscent of non-deterministic sum in $\pi$-calculus, where the two possible (non-deterministic) choices are prefixed by $\tau$ actions:
$$\tau.\E + \tau.\F$$

\begin{enumerate}[$\bullet$]
\item
During the composition with other strategies, we might have to use several times this strategy, hence ``entering'' it several times.
Every time is presented with this choice, normalization will non-deterministically  choose one of the two possible continuations.
The choice could  be different at each repetition.

\item    To define {\em orthogonality}, we essentially set:
\begin{quote}
\emph{$\D\bot (\tau.\E + \tau.\F)$ if and only if
$\pl\D,\tau.\E + \tau.\F \pr$  is total for each possible choice among the $\tau$'s}.
\end{quote}
It is immediate that:
\begin{equation} \label{eq star}
\D\bot (\tau.\E + \tau.\F) \ \Rightarrow \ \D\bot \E \mbox{ and } \D\bot \F.
\end{equation}

\end{enumerate}\medskip

\noindent As we will see, if $\E\in \bG$ and $\F\in \bG$ for $\bG$ \emph{interpreting a formula} of $\HS$,
 we have that
$(\tau.\E + \tau.\F) \in \bG$, and vice-versa.
Hence:
\begin{enumerate}[$\bullet$]
\item
 if $\D\in \bP$, for each $\E,\F\in \bP\b$  we have $\D \bot  (\tau.\E + \tau.\F)$.
\item By using  Equation (\ref{Copies}) we  have  that
$\rt{\D} \bot \{(\tau.\E + \tau.\F),(\tau.\E + \tau.\F)[\sigma/\xi]\}$.
\item
By using Equation (\ref{eq star}), we deduce that
$ \rt{\D} \bot \{\E ,\F[\sigma/\xi]\}$, as we wanted.
\end{enumerate}

\subsubsection*{Linearity of the root} Observe that by construction, in $\rt{\D}$
the action at the root is positive and it is the only action on the name  $\sigma$.
 We can hence apply
the same argument we have already given in Section ~\ref{compl+}
for the internal completeness of tensor.

As a consequence, if $P= \posm(N_1\otimes N_2)$, given
a ``good" strategy $\D\in {\bP_\xi}$,
we have that $\rt{\D}$ actually belongs to ${\mathbf \vdash \bP_\sigma,\bP_\xi}$
and  $(\sigma,\{1,2\})^+$ occurs linearly in  $\rt{\D}$
(only at the root).
Hence, $\rt{\D}$  can be decomposed
in strategies $\rt{\D}_1  \in \ \vdash \bP_\xi, \bN_{\sigma 1}$ and
 $\rt{\D}_2  \in \ \vdash \bP_\xi, \bN_{\sigma 2}$.

This allows us to associate to $\D\in {\mathbf \vdash  \bPP,\bP}$ a proof which essentially has this form:

\begin{prooftree}
\AxiomC{$\vdots$}
\noLine
\UnaryInfC{$\vdash \Pi, \posm (N_1\otimes N_2), \ N_1$}
\AxiomC{$\vdots$}
\noLine
\UnaryInfC{$\vdash \Pi, \posm (N_1\otimes N_2), \ N_2$}
\RightLabel{$\otimes$ + contractions}
\BinaryInfC{$\vdash \Pi, \posm (N_1\otimes N_2), \ N_1\otimes N_2$}
\RightLabel{dereliction + contraction}
\UnaryInfC{$\vdash \Pi, \ \posm (N_1\otimes N_2)$}
\end{prooftree}



\section{Ludics with repetitions: non-uniform strategies} \label{lwR section II}
In this section we technically implement     the ideas which we have
presented in  Section ~\ref{NUtests}.
In particular, we revise the definition of universal arena and strategy so to accommodate actions which corresponds to the $\tau$ actions we have informally introduced, and
a notion of strategy  which correspond to the $\tau$-sum. In this section we use ideas which have been developed to bridge between ludics and concurrency in ~\cite{FP09}, but here we choose a more standard presentation.

\subsubsection*{The silent arena}
We extend the
set of actions with a set $M_\tau= \{t\}\cup \{t_i : i\in \mathbb{N}\}$ of actions which we call {\em silent}.
We define the arena $A_{\tau}=(M_\tau,\vdash, \lambda)$ as follows.
\begin{enumerate}[$\bullet$]
\item the set of moves is $M_\tau$;
\item  the polarity is $\lambda(t)=+$, $\lambda(t_i)=-$;
\item $t\vdash t_i$, for each $i$.
\end{enumerate}
We can represent $A_{\tau}$ as follows:
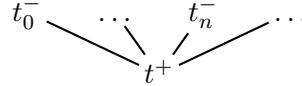
\begin{center}
\pstree[treemode=U,nodesep=2pt,levelsep=5ex]{\TR{$t^+$}}{
\TR{$t_0^-$} \TR{$\ldots$} \TR{$t_n^-$} \TR{$\ldots$}}
\end{center}

\begin{defi}[Universal arena on an interface]
Let $A(\xi,\epsilon)$ and $A_\dai$ be as in Definition ~\ref{A_name}.
We extend the construction of universal arena on an interface
(Definition ~\ref{Uarenaonint}) as follows.

Let $\Pi = \xi_1,\ldots,\xi_n$ be a (possibly empty) {\em positive} interface.

The {\bf extended universal arena on the interface $\Pi$} is the arena
 $$U^*(\Pi) := \  A(\xi_1,+) \parallel \cdots \parallel A(\xi_n,+)\parallel A_\dai \parallel A_\tau.$$

The  {\bf universal arena
on a (negative) interface $\{\sigma^-\} \cup \Pi$} is the arena
$$U^*(\Gamma) := \ A(\sigma,-)\freccia  U^*(\Pi).$$

\end{defi}

The silent actions play a role similar to that of $\dai$ in  Definition ~\ref{untyped}.
 The actions in both $A_{\dai}$ and $A_{\tau}$ are ``special actions'' which are not localized on a name.
Observe that even when the interface is empty, the universal arena still contains $A_{\dai} \parallel A_{\tau}$.

\subsubsection*{Non uniform strategies}
As one may expected, we are now going to take strategies on  $U^*(\GG)$. We reformulate the definition of strategy,

\begin{defi}[Non-uniform strategies]  \label{N.U. def} \label{nust}
Let $\Gamma$ be an interface.  A \textbf{non-uniform strategy} (n.u. strategy for short) $\D$ on $\Gamma$, written $\D:\Gamma$,
is a prefix-closed set of {\em non-empty} views (as in Definition ~\ref{viewdef}) on the arena  $U^*(\GG) $ , such that:
\begin{enumerate}[(1)]
\item {\em Coherence.}  If $s.m,s.n\in \D$ and $m \not= n$ then $m,n$ are
negative.

\item {\em $\tau$-Positivity.}  If $s.m$ is maximal in $\D$ (\ie no other view extends it), and  $m$ is a \emph{proper} action (\ie and action on a name),  then
$m$ is positive.

\end{enumerate}\medskip

\noindent We will call {\bf deterministic} a n.u. strategy which has no silent actions.
(Observe that in such a case  the only special action is $\dai$, and the definition coincides with the usual one, as in Definition ~\ref{untyped}.)
\end{defi}

The new Positivity condition says that if a maximal view  terminates with a proper action, that action must be positive.
However, a maximal view may terminate with a silent action $t_i$ (negative).
This is necessary to our construction, and more precisely to the definition of orthogonality
(see Section  ~\ref{orthog_sec} and Example ~\ref{ex nondetorm}).

\begin{rem}Definition ~\ref{nust} makes explicit the difference in role between proper actions and  ``special'' actions, those
in $A_\dai \cup A_\tau$. This will be apparent in normalization. Since communication propagates along the names, the interaction between strategies  only takes place on the proper actions. The $\dai$ and the silent actions are never internal | as they have no names.
But they have a fundamental role  in the definition of orthogonality.
\end{rem}

As a notational convention, from
now on,\emph{ by strategy we always mean n.u. strategy}, otherwise we specify ``deterministic'' strategy.

\paragraph{\bf Notation $\tau_i$}  In the pictures, we will write $\tau_i$ for the segment of sequence $t.t_i$; this to convey the intuition that a pair $t.t_i$ represent a ``$\tau$-action."

\begin{exa}
The following is an example of  non-uniform strategy:
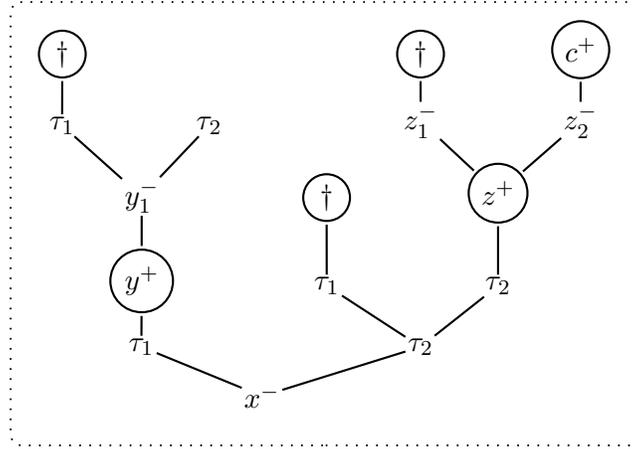
\begin{center}
\psmatrix[rowsep=8.5pt,colsep=0.3cm]
 &  && &  & &  &  & \\
\circlenode{dai1}{$\dai$} & & & &\rnode{up1}{$\mathstrut$} & & \circlenode{dai2}{$\dai$}
& & \circlenode{c}{$c^+$} \\
\rnode{tau1}{$\tau_1$} & & \rnode{tau2}{$\tau_2$}
& & & & \rnode{z1}{$z_1^-$} & & \rnode{z2}{$z_2^-$} \\

 & \rnode{y1}{$y_1^-$} &
& & \circlenode{dai3}{$\dai$} & &  & \circlenode{z}{$z^+$} & \\

 & \circlenode{y}{$y^+$} &
& & \rnode{tau3}{$\tau_1$}& &  & \rnode{tau4}{$\tau_2$} & \\

 & \rnode{tau5}{$\tau_1$} &
& & & & \rnode{tau6}{$\tau_2$} &  & \\

 &  &
& \rnode{root}{$x^-$}& \rnode{name1}{}& &  &  & \\


\endpsmatrix
 \psset{shortput=nab,arrows=->,labelsep=6pt}
 \small
\ncline[nodesep=1pt]{-}{root}{tau5}
\ncline[nodesep=1pt]{-}{root}{tau6}
\ncline[nodesep=1pt]{-}{y}{tau5}
\ncline[nodesep=1pt]{-}{y}{y1}
\ncline[nodesep=1pt]{-}{y1}{tau2}
\ncline[nodesep=1pt]{-}{y1}{tau1}
\ncline[nodesep=1pt]{-}{dai1}{tau1}
\ncline[nodesep=1pt]{-}{tau3}{tau6}
\ncline[nodesep=1pt]{-}{tau4}{tau6}
\ncline[nodesep=1pt]{-}{tau3}{dai3}
\ncline[nodesep=1pt]{-}{tau4}{z}
\ncline[nodesep=1pt]{-}{z1}{z}
\ncline[nodesep=1pt]{-}{z2}{z}
\ncline[nodesep=1pt]{-}{z1}{dai2}
\ncline[nodesep=1pt]{-}{z2}{c}
\ncbox[nodesep=.5cm,boxsize=4.2,linearc=.15,
linestyle=dotted]{name1}{up1}
\end{center}
\end{exa}\vspace{-15 pt}

\subsubsection*{Sum of strategies} We  use non-uniform strategies to
capture the idea of ``non-uniform'' tests.  As anticipated in Section ~\ref{NUtests},
a non-uniform strategies can be seen as a non-deterministic sum of ``standard'' strategies.

\begin{defi}[$\tau$-sum] \label{tau sum} Given a family of strategies on the same interface (and hence all with the same polarity),
we define their sum.
 Let $S$ be a non-empty subset of $\mathbb{N}$.

\begin{enumerate}[$\bullet$]
\item If  $\{\D_i:\Pi\}_{i \in S}$
  is a family of positive  strategies, we
  define their {positive sum}: $$ {\bigoplus_{i \in S}}^\tau \D_i :=  \displaystyle \bigcup_{i \in S} \{t^+.t_i^-.\D_i\}.$$
 When $S$ is a finite set, say $\{1,\dots,k\}$, we  write
  $\D_1 \oplus^\tau \dots \oplus^\tau \D_k$.\\

\item If $\{x^-.\D_i:\Gamma\}_{i \in S}$
  is a family of negative strategies which have the  same  root $x^-$,
 we define their {negative sum}:
  $$ \sumN{i \in S} x^-.\D_i :=   x^-.\sumP{i \in S}\D_i.$$
 In the finite case, we also write
  $x^-.\D_1 +^\tau \dots +^\tau x^-.\D_k$.
\end{enumerate}

 \end{defi}\medskip

\noindent The following is easy to check.
\begin{prop}
Let $\D_i$ be a family of positive strategies, and $\F_i$ a family of negative strategies. We have that  $ \bigoplus_{i \in S}^\tau \D_i $
is a  positive  strategy, and $ \sum_{i \in S}^\tau \F_i $ is a negative  strategy
(in the sense of Definition \emph{~\ref{N.U. def}}). \qed
\end{prop}
A $\tau$-sum of strategies can be seen as a \emph{superposition}
 of strategies in a way that they do not overlap.

\subsubsection*{Totality}
Roughly speaking, a non-uniform strategy is total if it is not $\Fid$, but also not obtained via $\tau$-sum with $\Fid$. Precisely:

\begin{defi}[Totality, non-uniform strategies]\label{nutot}
 A strategy $\D$ is total if $\D\not=\Fid$ and for each $s\in \D$, there
are pointing sequences $p$ and $q$ such that:
 \begin{center}$p.a.q\in \D $, \ \  $s\pref p.a.q$ \ and \  $a\not\in A_{\tau}$.
\end{center}
\end{defi}\medskip

\noindent In words, each path from the root of $\D$ has to take to at least an action which is not silent.
So for example, $\Fid \oplus^\tau \Fid$ or $\bigoplus_{1}^\tau \Fid$ are partial strategies, as well as $\D \oplus^\tau \Fid$, whatever positive strategy is $\D$.
Notice also that any negative strategy is total, since
it is either empty | and we have already discussed
the reason of its totality in Section ~\ref{tot section} | or it is not empty and
each root is by definition a  negative proper action (non-silent, because negative silent actions are never initial).

 As in the linear case, in an untyped setting, partial strategies emerge naturally in case of
 unsuccessful interaction. In fact, they  have a key role in the definition of orthogonality.
Before discussing orthogonality and behaviours in the case of non-uniform strategies, we first make precise the definition of normalization.

\subsubsection*{Strategies on the empty interface}
In the linear case, a strategy on the empty interface is a strategy on
$A_\dai$, hence, as we have already seen, there are only two possibilities: $\Dai$ or $\Fid$.
The introduction of silent actions give more possibilities; a {\em non-uniform strategy} on the empty interface is a non-uniform strategy on the arena $A_{\dai} \cup A_{\tau}$. Beside $\Dai$ or $\Fid$, there are other strategies, those which have as root  $t^+\in A_\tau$. Each view can then contain several repetitions of silent actions, and terminate or not  with a $\dai$  action.


\section{Strategies with repetitions: normalization via the VAM abstract machine}\label{VAM} \label{VAM section}

The basic  notions we use in this section (cut-net, visible action, \ldots)
are exactly the same as defined in Section ~\ref{lin_norm}.

Composition of strategies  in our setting works accordingly to the standard paradigm  of
``\emph{parallel composition plus hiding}.''

\begin{enumerate}[(1)]

\item Given a cut-net $\R$, we calculate the set of its \emph{interaction} $I(\R)$ (this is the parallel composition), which is a set of pointing strings, calculated via an abstract machine, the VAM ~\cite{CurHerb}.

\item  From the interactions $I(\R)$,  we obtain the strategy which corresponds to   the \emph{normal form} $\pl \R \pr$ (Definition ~\ref{normal form})
by hiding the internal communication.

\end{enumerate}\medskip

\noindent Since we allow for the repetition of actions,  there might be several occurrence of the same action $a$ in a strategy.
To define the VAM, we need the following notion.

\begin{defi}[View extraction] \label{View}
Let $s=x_1\dots x_n$ be  a   pointing string of actions. We define the  \textbf{view} of $s$ denoted
by $\ulcorner s \urcorner$  as
follows:
\begin{enumerate}[$\bullet$]
\item $\ulcorner s \urcorner:= s$ if $s$ is empty;
\item $\ulcorner s.x^+ \urcorner := \ulcorner s \urcorner.x^+$  ($x$ is positive);

\item $\ulcorner s.x^- \urcorner= x^-$, if $x$ is initial (\ie it does not point to any previous action).
\item $\ulcorner s.x^- \urcorner := \ulcorner q \urcorner.x^-$ if
  $s= q.r.x$ and $x$ points to
 the last action of $q$.
\end{enumerate}
Given a  pointing string of actions $s=x_1\dots x_n$ we obtain
a subsequence $\ulcorner s \urcorner = x_{k_1}\dots x_{k_m}$, where
$1 \leq {k_1} < \dots < {k_m} \leq n$.
We say that the element at position $j$ in $\ulcorner s \urcorner$ ($1\leq  j \leq m$) \textbf{corresponds} to the element at position
$k_j$ in $s$  ($1 \leq k_j \leq n$). We will use this in the following.

As for the pointers,   the operation $\ulcorner \ \urcorner$ is
pointer preserving in the sense that if $x_{k_y}, x_{k_z}$ in $\view s$  respectively correspond to $x_i,x_j$ in $s$,
and $x_j$ points to $x_i$ in $s$, then $x_{k_z}$  points to  $x_{k_y}$ in $\view s$.

\end{defi}

In words, we trace back from the end of $s$: (i) following the pointers
of negative actions of $s$ and
erasing all actions under such pointers, (ii) bypassing  positive
actions, (iii) stopping the process when  we
reach an initial negative action.
\footnote{In general, the procedure of view extraction
may delete some pointers. This will never happen
for the pointing strings we consider in this paper.}

\begin{exa}

Given $s=x_1.x_2\dots x_6$ and $t=y_1.y_2\dots y_5$ as follows:

\begin{center}
\begin{tabular}{ccc}
\psmatrix[rowsep=10pt,colsep=0.2cm]

 \rnode{p}{$s \ = $} &  \rnode{p2}{$a^+$}
 & \rnode{p3}{$a^-$}
 & \rnode{p5}{$b^+$} & \rnode{p6}{$b_0^-$}
 &  \rnode{p8}{$a_0^+$}&
  & \rnode{p9}{$a_0^-$}
\endpsmatrix
 \psset{shortput=nab,arrows=->,labelsep=4pt}
 \small
\ncarc[linewidth=0.2pt,nodesep=1pt,arcangle=45]{p6}{p5}
\ncarc[linewidth=0.2pt,nodesep=1pt,arcangle=45]{p8}{p3}
\ncarc[linewidth=0.2pt,nodesep=1pt,arcangle=45]{p9}{p2}

& $\qquad$ &

\psmatrix[rowsep=10pt,colsep=0.2cm]

 \rnode{p}{$t \ = $} & \rnode{p2}{$a^+$}
 & \rnode{p3}{$a^-$}
 & \rnode{p5}{$b^+$} & \rnode{p6}{$b_0^-$}
 &  \rnode{p8}{$a_0^+$}&
\endpsmatrix
 \psset{shortput=nab,arrows=->,labelsep=4pt}
 \small
\ncarc[linewidth=0.2pt,nodesep=1pt,arcangle=45]{p6}{p5}
\ncarc[linewidth=0.2pt,nodesep=1pt,arcangle=45]{p8}{p3}
\end{tabular}
\end{center}
\vspace{.8 cm}
\noindent we get $\ulcorner s \urcorner = x_1.x_6$ and $\ulcorner t \urcorner
=y_2.y_3.y_4.y_5$:
\begin{center}
\begin{tabular}{ccc}
\psmatrix[rowsep=10pt,colsep=0.2cm]

 \rnode{pp}{$\ulcorner s \urcorner \ = $}
&  \rnode{p6}{$a^+$} & \rnode{p7}{$a_0^-$}
\endpsmatrix
 \psset{shortput=nab,arrows=->,labelsep=4pt}
 \small

\ncarc[linewidth=0.2pt,nodesep=1pt,arcangle=45]{p7}{p6}

& $\qquad$ &

\psmatrix[rowsep=10pt,colsep=0.2cm]

 \rnode{p}{$\ulcorner t \urcorner \ = $}
 & \rnode{p3}{$a^-$}
 & \rnode{p5}{$b^+$} & \rnode{p6}{$b_0^-$}
 &  \rnode{p8}{$a_0^+$}&
\endpsmatrix
 \psset{shortput=nab,arrows=->,labelsep=4pt}
 \small
\ncarc[linewidth=0.2pt,nodesep=1pt,arcangle=45]{p6}{p5}
\ncarc[linewidth=0.2pt,nodesep=1pt,arcangle=45]{p8}{p3}
\end{tabular}
\end{center}
\vspace{.5 cm}
\end{exa}\medskip

\noindent We will also rely on the following Lemma.
\begin{lem}\label{view_prop}
Let $\R=\{\D_1, \dots,\D_n\}$ be a cut-net. The following properties hold:
\begin{enumerate}[\em(1)]
\item Given an address $\xi$, and a polarity $\epsilon$, $\xi^{\epsilon}$ occurs in at most one of the interfaces.
\item Each view    $s\in \bigcup \{\D_i : 1 \leq i \leq n\}$  belongs exactly to one $\D_i$.
\end{enumerate}
\end{lem}

\proof \hfill \begin{enumerate}[(1)]
\item
By definition of cut-net (Definition ~\ref{cut-net}).

 \item The first action $x$ in the view $s$ is enough to determine to which $\D_i$ it belongs. If $x$ is a proper action
 $(\xi, I)^{\epsilon}$, the conclusion is immediate from the previous point ($x$ belongs to the same strategy to which $\xi^{\epsilon}$ belongs).
If $x$ is  in $A_{\dai} \cup A_{\tau}$,  the interface must be positive; we already observed that a cut-net
contains at most one positive interface. \qed
\end{enumerate}

\subsection{VAM (View-Abstract-Machine)}
Let us first  informally explain how the  abstract machine
calculates the interaction of a cut-net $\R$. The machine visits actions of the strategies in $\R$
and  collects the sequences of visited actions, proceeding as follows:

\begin{enumerate}[$\bullet$]

\item
We start on
 the roots of
the main strategy of a cut-net $\R$.

 \item If we visit a visible action $a$ occurring in some  $\D \in \R$,
we continue to explore the current strategy $\D$. The process
  branches when $a$ is a branching node of $\D$.

\item
If we visit an
 internal action $a^+$ occurring in $\D$ we match it
with its opposite
 $a^-$ occurring in $\E \in \R$, then we continue to collect actions  in $\E$ (this is a \emph{jump}
of the machine). Since there could be several occurrences
 of
 $a^-$ in $\E$, we use $\ulcorner \ \urcorner$
 to determine the  correct occurrence of action to which we have to move.

\item We may eventually stop when either we reach a maximal action
or  an internal action which has no match.
\end{enumerate}\medskip

\noindent We now give the formal definition of the VAM.
The definition below is a reformulation of the machine defined in ~\cite{CurHerb}.

\begin{defi}[VAM] \label{interaction}\label{VAMdef} Let $\R = \{\D_1,\dots,\D_n\}$ be a cut-net.
 $VAM(\R)$
 is the
 set of  pointing strings  defined as
follows. The construction preserves the following invariant:

\begin{center}
(*) \quad  if $p\in VAM(\R)$ then $\ulcorner p \urcorner$ is a view which belongs to one of the $\D\in \R$.
\end{center}

\begin{enumerate}[(1)]

\item {\bf (Initialization)}
If the main strategy
of $\R$ is  empty, we set $VAM(\R):= \emptyset$. Otherwise, for each  $a$ root
 of the main strategy,
$a \in VAM(\R)$.

\item
 Let $ p=x_1\dots x_n \in VAM(\R)$. We have the following two cases.
\begin{enumerate}[a.]
\item[(a)] {\bf (Continuation)}

  The action $x_n$ is either a {\bf negative action} or a {\bf
    positive visible action}.  We consider $\ulcorner p \urcorner$.
  There exists a unique strategy $\D \in \R$ such that $\ulcorner p
  \urcorner \in \D$ (because of the invariant (*) and Lemma
  ~\ref{view_prop}).  For each action $a$ which extends $\ulcorner p
  \urcorner$ in $\D$ (\ie such that $\ulcorner p \urcorner.a \in \D$)
  we { set} $p.a \in VAM(\R)$ where the pointer for $a$ is given by
  the equation $\ulcorner p.a \urcorner = \ulcorner p \urcorner.a$.

\item[(b)] {\bf (Jump)} The action $x_n= a^+$ is an \textbf{internal positive}
action. We consider
the sequence $p.a^-$ obtained
by adding the action $a^-$ to $p$ and possibly a pointer as follows. If $x_n =(\xi i,J)^+$ points to $x_i =(\xi,I)^-$
 we add
a pointer from $a^-$ to $x_{i-1}$
in $p$.
If there is  $\D \in \R$ such that $ \ulcorner p.a^- \urcorner \in \D$, we
set $ p.a^- \in VAM(\R)$.  That occurrence of $a^-$ is the \emph{match} of $a^+$.
\end{enumerate}
\end{enumerate}
\end{defi}

\begin{rem}[The pointers in case (2)] \label{rem pointers}
In the case (2)(a), the equation $\ulcorner p.a \urcorner = \ulcorner p \urcorner.a$ summarizes
the following conditions.
\begin{enumerate}[$\bullet$]
\item Assume $a$ is negative. $a$ must point
to $x_n$.
\item Assume   $a$ is positive. Let  $\ulcorner p \urcorner =x_{k_1} \ldots x_{k_m}$, where
each $x_{k_j}$ corresponds to en element $x_i$ in $p$. If in the view $\ulcorner p \urcorner.a\in \D$ we have that
 $a$ points to $x_{k_j}$,
in $p.a$, we have that $a$ points to the corresponding element.
\end{enumerate}
In the case (2)(b),
 we have the following distinct situations.
\begin{enumerate}[$\bullet$]
\item
If  $x_n =(\xi i,J)^+$ points to $x_i =(\xi,I)^-$, then $i >1$, because if $x_i$ is a negative hidden action
and by definition of cut-net it cannot be a root of the main strategy of $\R$. (Moreover, it is easily seen that $x_{i-1}=(\xi,I)^+$
so that $x_i$ has been introduced by a Jump case.)

\item
If $x_n =(\xi ,J)^+$ points to a negative
action $x_i=(\sigma,K)^-$ with $\sigma$ and $\xi$ disjoint,
it means  that both occurrences of actions belongs to the same
strategy $\D\in \R$ of interface $\sigma^-,\xi^+,\Pi$.
We do not need to add any pointer,
because the match for $x_n$ (if any) must be a root
of a strategy $\E : \xi^-,\Pi'$ of $\R$.

\item
Last, if  $x_n =(\xi ,J)^+$ does not point to any action,
then $x_n$ occurs in a strategy  $\D\in \R$ of interface $\xi^+,\Pi$.
Again, we do not need to add any pointer,
because the match for $x_n$ (if any) must be a root
of a strategy $\E : \xi^-,\Pi'$ of $\R$.
\end{enumerate}
\end{rem}

\begin{rem}
Observe that the interface $\Gamma$ of the cut-net $\R$  is negative  if and only if the main strategy is negative. In such a case, the
first action in each $p\in VAM(\R)$ is a root $a^-$ of the main strategy. Such an action is visible, and never occurs again in $p$.
\end{rem}

\subsection{Normal form}

\begin{defi}[Hiding]
Given a cut-net $\R$,  and   $p=x_1\ldots x_n \in VAM(\R)$, we define $hide(p)$ as the pointing string obtained as follows:
\begin{enumerate}[(1)]
\item If  $x_1=a^-$ is negative (see the remark above), the pointers are updated as follows: for each $c^+$ {\em visible} action pointing to an internal action $b^-$, we make $c^+$ point to $x_1$.

\item We hide all the internal actions.\\  All pointers between actions which are visible are preserved, \ie if $p=x_1.\dots.x_n$ and $x_i$ points to $x_j$ ($1\leq j < i \leq n $) in $p$, and both $x_i,x_j$ are visible, then  in $hide(p)$ the action (which corresponds to) $x_i$ still points to the action  (which corresponds to) $x_j$.
\end{enumerate}\medskip

\noindent What remains after hiding  is the subsequence of  visible actions of $p$, written ${hide}(p)$, with the  inherited pointer structure.
 If $X\subseteq VAM(\R)$, we will also write $hide(X)$ for the set $\{hide(p) :  p\in X\}$.
\end{defi}

\begin{rem}\label{hide_rem}
Observe that after  step (1) of hiding,  each visible action in $p$
either  does not point to any previous action (but only because it is initial in the arena induced by the interface of the cut-net)
or
it points to a visible action. The latter point deserves some explanations.
If  an action  $c^+$  is {\em visible} and points
 to an {\em internal} action (say $b^-$),
 it means
  that   both $b^-$ and $c^+$ belong to the same strategy $\D: \xi^-,\Pi$. Moreover, by definition of internal action and of
   the universal arena
 $U^\ast(\xi^-,\Pi)$, the action $b^-$ must be an action  $(\xi,I)^-$, initial in $ A(\xi^-)$, the action $c^+$ must be initial in $U^\ast(\Pi)$,  and the pointer form  $c^+$ to $b^-$
must correspond to the
 enabling $b^-\vdash c^+$
 introduced by the construction $ A(\xi,-)\freccia  U^\ast(\Pi)$.

\end{rem}

By using the $VAM$ machine, we are now going to define the  set $I(\R)$ of  the interactions of a cut-net  $\R$,
and from this its normal form $\pl \R \pr$.

\begin{defi}[Normal form] \label{normal form}\label{norm_new}
Let $\R$ be a cut-net.
We define the set  $I(\R)$   of the {\bf interactions} of $\R$ as
  the closure under non-empty prefix of
\begin{center}
 $\{q.c\in VAM(\R) :$   $c$ is  visible and    not a proper negative action $\}.$
  \end{center}

\noindent The \textbf{normal form}
 of  $\R$, denoted by $\pl \R \pr$ is defined as
\begin{center}
 $\pl \R \pr = \{hide(p):  p\in I(\R) \mbox{ and } hide(p) \mbox{ non-empty}\}$.
 \end{center}

\end{defi}\medskip

\noindent Observe that since $VAM(\R)$ is closed by non-empty prefix, we have that
$I(\R) \subseteq  VAM(\R)$. \smallskip

The normal form of a cut-net is a strategy.
We show this fact in Proposition ~\ref{norm_is_st}.

\begin{lem}\label{coh_p}
If $p.m,p.n\in VAM(\R)$ and $m\not=n$, then $m,n$ are {\em negative} and {\em visible}.
\end{lem}
\proof
First of all, we observe that the polarity of $m$ and $n$ is the same. Assume $m,n$ are positive.
If $p$ is empty, then $m,n$ are both roots of the main strategy  (Case (1) in the VAM). By Coherence condition
(Definition ~\ref{N.U. def}), we have that $m=n$.
 The case of $p$  non-empty is similar, with  $m,n$ which extend  $\ulcorner p \urcorner $ (Case (2)(a) in the VAM).

We have established that if $m\not=n$, then $m,n$ are negative. They must also be visible, because
 the only way to extend $p$ with a negative, internal  action is Case (2)(b). However,  in such a case since
the  extension of  $\ulcorner p\urcorner $ is uniquely given by the construction, we would have that $m=n$.
\qed

\begin{cor}\label{cor_p}
Let $p,q\in VAM(\R)$. If $hide(p)=hide(q)$ then $p\pref q$ or $q\pref p$.
\end{cor}
\proof
Assume that neither $p\pref q$ nor $q\pref p$ hold, and let  $s$ be the longest common prefix of $p$ and $q$, \ie $s.a\pref p, s.b\pref q$, and $a\not=b$. By Proposition ~\ref{coh_p},
$a,b$ are negative and visible, hence $hide(s.a) \not= nide(s.b)$ which contradicts the fact that  $hide(p)=hide(q)$.
\qed

\begin{prop}\label{norm_is_st}
If $\R$ is a cut-net of interface $\GG$, then $\pl\R\pr$ is a strategy on $\GG$.
\end{prop}
\proof
We show the following:
\begin{enumerate}[(1)]
\item if $p\in VAM(p)$, then $hide(p)$ is a view (on the arena $U^{\ast}(\Gamma)$);
\item the set  $hide(VAM(\R))$ satisfies the Coherence condition (as given in Definition ~\ref{N.U. def});
\item $I(\R)$ and $\cut \R$ satisfy $\tau$-Positivity
(as given in Definition ~\ref{N.U. def}).\
\end{enumerate}\smallskip

\begin{enumerate}[(1)]
\item \begin{enumerate}[a.]
\item {\bf $hide(p)$ is a justified sequence} (on the appropriate arena). It is  immediate to check that the pointers in $hide(p)$ satisfy the
 justification condition, as this fact is inherited from the same property of  the strategies which take part
 in the construction of $p$.  The only delicate aspect are the pointers
 in the case where the interface of $\R$ is negative, \ie $\GG=\sigma^-,\Pi$. More precisely, the pointers
  which in the arena $U^*(\ss^-,\Pi) = A(\sigma,-)\freccia  U^*(\Pi)$ correspond to the  enabling $a\vdash c$, with $a$ initial in $ A(\sigma,-)$ and $c$
  initial in $U^\ast(\Pi)$. However, this situation is taken care by  step (1) in the hiding (the updating of the pointers, see Remark ~\ref{hide_rem}).

\item
{\bf $hide(p)$ is alternating}. First of all, we observe that if  $p\in VAM(\R)$, then $p$ is alternating (by construction). As a consequence, $hide(p)$ is also alternating, because all occurrences of  internal actions appear (and are deleted) as pairs  $p=p'.a^+.a^-.p''$, or {they occur} in the very last position.
\item
{\bf $hide(p)$ is a view}.
Each visible negative action in $p$ is either initial (Case (1)), or points to its immediate predecessor (Case (2)(a)).
\end{enumerate}

\item
Assume $s.m\not=s.n$.
 We have that   $s.m=hide(p.m)$, $s.n=hide(q.n)$ and $hide(p)=hide(q)=s$.
Using Corollary ~\ref{cor_p},  let assume
 $p\pref q$.
Let us consider $p.x\pref q.n$. By Lemma ~\ref{coh_p}, we have that if $m\not= x$, then $m,x$ are negative.
 The execution of the VAM introduces both $p.m$ and $p.x$ by checking how  $\view{p}$ is extended in $\R$. Since $m$ is visible,
 $x$ must be as well. From this  we deduce that $x=n$ and then that  $p=q$. Hence, $m,n$ are negative, by Lemma ~\ref{coh_p}.

\item
 We first observe that  the $\tau$-Positivity condition holds in $ I(\R)$. Assume that  $p=q.c \in I(\R)$ and has no extensions in  $I(\R)$; then $c$ is not  a proper negative action. Moreover, from the fact that $c$ is visible,  it follows that $\cut \R$ satisfies $\tau$-Positivity. \qed
\end{enumerate}

Like for the strategies in ~\cite{GirLoc}, our strategies are a variant of abstract B\"ohm trees (see ~\cite{CurLLL2,MauTh} for a { description of ludics strategies in terms of B\"ohm trees}).
Abstract B\"ohm trees normalize via the VAM abstract machine  ~\cite{CurAbs,CurHerb}, which we have described in this section;
a fundamental property that normalization  satisfies is associativity.

\begin{thm}[Associativity]\label{associativity}
Let $\R$ be a cut-net which can be partitioned into cut-nets
$\R = \R_1,\ldots,\R_n$. We have:
$$
\pl \R \pr = \pl \pl \R_1 \pr,\ldots, \pl \R_n \pr \pr\eqno{\qEd}
$$
\end{thm}\medskip

\noindent We omit the formal proof of this fact, which is established in ~\cite{CurAbs,CurHerb}.
We must observe that in our setting, the only essential
 difference (\wrt abstract B\"ohm trees)  are
 the conditions on the polarity of the maximal actions, but this is irrelevant  to establish  associativity.


\subsection{Renamings and normalization}
We now state a property which relates
renamings (in the sense of Section ~\ref{renaming})
and normalization.
 We first generalize
the renaming
operator defined in Section ~\ref{renaming}. We now
allow  simultaneous renamings in  arbitrary
interfaces.\smallskip

Let $\Gamma = \{\alpha_1^{\epsilon_1},\ldots, \alpha_k^{\epsilon_k},\alpha_{k+1}^{\epsilon_{k +1}}, \ldots,
\alpha_n^{\epsilon_n}\}$ be an interface,
 $\Delta= \{\alpha_1^{\epsilon_1},\ldots, \alpha_k^{\epsilon_k}\}$
and  $\Delta'=\beta_1,\ldots,\beta_k$
names such that $\Gamma' := \{\beta_1^{\epsilon_1},\ldots,\beta_k^{\epsilon_k}\}
\cup \{\alpha_{k+1}^{\epsilon_{k +1}}, \ldots,
\alpha_n^{\epsilon_n}\}$
forms an interface.

Let $\D$ be a strategy  on $\Gamma$.
By $\D[\Delta'/\Delta]$ we denote the strategy on interface
$\Gamma' $ obtained
from $\D$ by \textbf{renaming}, in all occurrences  of action,
the prefix $\alpha_l$ into $\beta_l$ for any
 $1 \leq l \leq k$.\\
We observe that $\D[\Delta'/\Delta]$ is indeed a strategy
(in the sense of Definition ~\ref{N.U. def})  because:
\begin{enumerate}[(1)] \item
we explicitly request that
$\Gamma'$ forms an interface;
\item
the conditions of  being a strategy
are inherited from $\D$, since the operation
of renaming preserves the polarity
of the actions,  their nature (proper, silent, \ldots), and
the pointer structure of the strategy (our operation
acts only on names of proper actions, all the existing  pointers are preserved).
\end{enumerate}\medskip

\noindent The main property we need in the sequel is  following one.
Let us  consider the following data:
\begin{enumerate}[$\bullet$]
\item
  $\D: \Gamma,\xi^+,\sigma^+$;
  \item
  $\E : \xi^-,\Delta$, where $\Delta = \alpha_1^+,\ldots,\alpha_k^+$;
  \item
$\E[\sigma/\xi,\Delta'/\Delta]: \sigma^-,\Delta'$,
where  $\Delta'=\beta_1^+,\ldots,\beta_k^+$ (an ``isomorphic" copy of $\E$);
\item  $\D[\xi/\sigma] : \Gamma,\xi^+$;
\item $(\F_j) = \F_1,\ldots,\F_n$ a list of strategies.
\end{enumerate}
Let us assume that  $\R := \{\D,\E,\E[\sigma/\xi,\Delta'/\Delta],(\F_j) \}$
forms a cut-net of interface $\Lambda,\Delta,\Delta'$,
where names in $\Lambda$ comes from $\Gamma$
and the names of the interfaces of $(\F_j)$,
so that
$\R' := \{\D[\xi/\sigma], \E,(\F_j)\}$ is a
cut-net  on interface $\Lambda,\Delta$.
We have:

\begin{prop} \label{prop copies vera}
$$\pl \D,\E,\E[\sigma/\xi,\Delta'/\Delta],(\F_j) \pr \ [\Delta/\Delta'] = \pl \D[\xi/\sigma], \E,(\F_j) \pr.$$
\end{prop}
\proof (Sketch.)
By reasoning on the names of the  interfaces, we can easily deduce:
\begin{enumerate}[$\bullet$]
\item The names in $\Delta$ and $\Delta'$ (resp.
$\Delta$) are not cuts in $\R$ (resp.\ $\R'$).
\item
 $\D$ (resp.\ $\F_j$) is the main strategy
of $\R$ if and only if $\D[\xi/\sigma]$ (resp.\ $\F_j$) is the main
strategy of $\R'$.
\item
Neither $\E$ nor $\E[\sigma/\xi,\Delta'/\Delta]$ (resp.\ $\E$)
is the main strategy of $\R$ (resp.\ $\R'$).

\end{enumerate}
If we normalize $\R$ and $\R'$, they share a very similar dynamics:  the only difference is that  the
part of interaction between $\D$, $\E$
and the ``isomorphic" copy of $\E$ in $\R$
is reproduced
in $\R'$ by
$\D[\sigma/\xi]$ and $\E$.
But this creates no
relevant differences on the pointing
string in $VAM(\R)$ and
$VAM(\R')$, since   we have a pointing string \eg
$$\begin{array}{lccc}
p & = & \ldots (\sigma,I) \ldots (\xi.\gamma,K) \ldots (\alpha_i.L)
\ldots (\beta_j.M) \ldots & \in VAM(\R) \\
\end{array}
$$
if and only if we have a ``corresponding" pointing string of the same length
$$\begin{array}{lccc}
 p' & = & \ldots (\xi,I) \ldots (\xi.\gamma,K) \ldots (\alpha_i.L)
\ldots (\alpha_j.M) \ldots & \in VAM(\R') \\
\end{array}
$$
where, for $p=x_1\ldots x_n$ and
$p'= y_1 \ldots y_n$:
\begin{enumerate}[$\bullet$]
\item the polarity of $x_i$ and $y_i$ is the same, for $1 \leq i \leq n$;
\item
 the pointer structures of $p$
and $p'$
are exactly the same in the sense
that  $x_i$ points
to $x_j$ if and only if  $y_i$ points to $y_j$;

\item
for any $1 \leq i \leq n$,   either
$x_i=y_i$ or $x_i=(\sigma.\gamma,I)$ and
$y_i=(\xi.\gamma,I)$ or
$x_i=(\alpha_l.\gamma,I)$ and
$y_i=(\beta_l.\gamma,I)$, for $1 \leq l \leq k$.
\end{enumerate}\medskip

\noindent This  follows from the fact that
using the views extraction operation on pointing strings
given by the VAM, we can univocally determine a specific view of strategy. In particular, given $p$ (resp.\ $p'$) we can univocally
reconstruct $p'$ (resp.\ $p$).

Since  the nature (visible, hidden, proper, silent,\ldots) of the action in $p$ and $p'$ corresponds
elementwise too,
it  follows that  $p\in I(\R)$   if and only if
there is a ``corresponding"  $p' \in I(\R')$.
 The difference between $p$ and $p'$ then
is only on names occurring in proper actions,
but  names generated by  cuts $\xi$ and $\sigma$
will be erased by hiding, thus
obtaining  a view $s \in \pl \R \pr$
and a ``corresponding" view $s' \in \pl \R'\pr$,
and names generated
by $\Delta'$ which are not
in $s'$ are renamed by using
the renaming
$[\Delta/\Delta']$ which is only needed to this aim.

A formal proof can be carried out by induction
on the length of the pointing strings.\qed\medskip

\noindent As a special instance of the previous
proposition we have:

\begin{prop}[Copies] \label{propo copies1}
Let $\D: \xi^+$ and $\E: \xi^-$ be  strategies.
We have:
\begin{equation} \label{copies equation1}
\pl \D , \E \pr = \pl \rt\D, \E, \E[\sigma/\xi] \pr.
\end{equation}
\end{prop}

\section{Orthogonality and interactive types} \label{orthog_sec}
\subsection{Orthogonality}

The definition of orthogonality is the same as for linear strategies (Definition \ref{ortho def});
  we repeat it for convenience.

 Like in the linear case (and with a similar meaning)
   orthogonality is a relation which  is defined   on   {\em total} strategies (with the notion of totality being now  that in Definition
 ~\ref{nutot}).

\begin{defi}[Orthogonality, orthogonal set]  Let $\D:\GG$ be a total strategy and
$(\E_\xi)_{\xi\in \GG}$ be
a family of  counter-strategies (Definition \ref{counter-strategies}).
$\D$ and  $(\E_\xi)$
 are said to be \textbf{orthogonal}, written  $\D \bot (\E_\xi)$
 if $\pl \D, (\E_\xi) \pr$ is total.

Given  a set $\bS$ of total strategies on the same  interface $\Gamma$, its \textbf{orthogonal set} is defined as $$\bS\b:=\{(\E_\xi) : ~ (\E_\xi)  \mbox{ is a family of counter-strategies \wrt $\Gamma$ and } (\E_\xi)\bot \D \mbox{ for any } \D \in \bS \}.$$
Similarly, given a set $\bC$ of families of counter-strategies \wrt $\Gamma$,
its \textbf{orthogonal set} is defined as $$\bC\b:=\{\D :   \D \mbox{ is total  on interface } \Gamma \mbox{ and } \D \bot (\E_\xi)  \mbox{ for any } (\E_\xi) \in \bC \}.$$
\end{defi}\medskip

\noindent Observe that $\pl \D, (\E_\xi) \pr$ is a strategy on the empty interface. By Definition ~\ref{nutot}, we have
orthogonality if
$\pl \D, (\E_\xi) \pr \not=\Fid$ and for each $s\in \pl \D, (\E_\xi) \pr$ there is $p.\dai\in \pl \D, (\E_\xi) \pr$ such that $s\pref p.\dai$.
The intuition is that each  interaction should lead to $\dai$. Coherently with this intuition, we will also see that  two distinct views in the normal form of a closed net always branch on silent actions.\\

The following lemma is a direct consequence of the definition of orthogonality:

\begin{lem} \label{ortho_cor}\label{prop per lemma1} Let $\E :\xi^-$ be a negative strategy such that
 $\E=\sumN{i \in S} \E_i$.
 Let $\{\D_1,\dots,\D_n,\E\}$ be a closed cut-net.
  We have that:
 $$ \pl \D_1,\dots,\D_n, \E \pr  \mbox{ is total }   \ \Rightarrow  \ \pl \D_1,\dots,\D_n, \E_i \pr \ \mbox{ is total, for each } i \in S.$$
 Hence in particular, for any strategy $ \D \bot \E $, we have
 $$ \D \bot \E   \ \Rightarrow  \ \D \bot \E_i, \ \mbox{ for each } i \in S.\eqno{\qEd}$$
\end{lem}\medskip

\noindent The converse of Lemma ~\ref{ortho_cor} does not hold in general. We now give a concrete example,
which is also useful to better understand composition and orthogonality.

\begin{exa} \label{ex nondetorm}
Let us consider the following strategies.

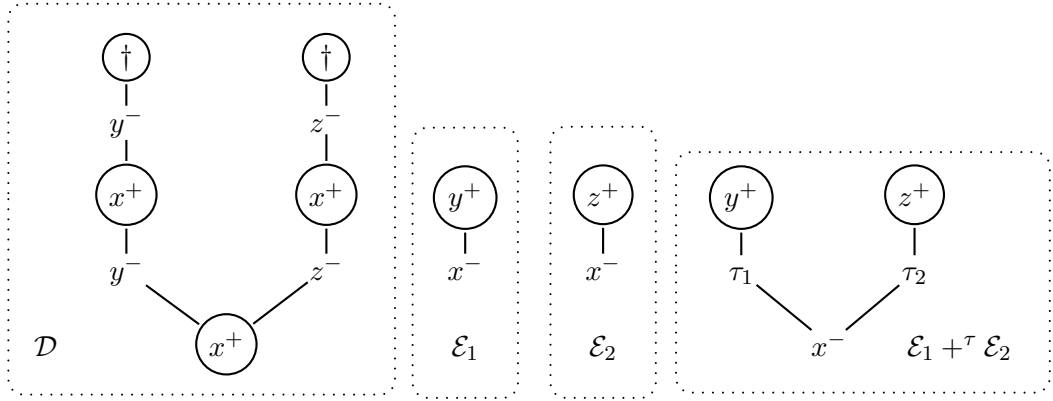
\begin{center}
\psmatrix[rowsep=9pt,colsep=0.5cm]
&&&&&&&&&&&&\\
&\circlenode{dai1}{$\dai$}&  &
\circlenode{dai2}{$\dai$}
& &&&&&& &&\\
&\rnode{ym2}{$y^-$}&  &
\rnode{zm2}{$z^-$}
& &&&&&& &&\\
\rnode{x0}{$\mathstrut$}&\circlenode{x1}{$x^+$}&  &
\circlenode{x2}{$x^+$}\rnode{x3}{$\mathstrut$}
 &&\circlenode{yp}{$y^+$}&&\circlenode{zp}{$z^+$}&&
\circlenode{y1}{$y^+$} &&
\circlenode{z1}{$z^+$}\\
&\rnode{ym1}{$y^-$}&  & \rnode{zm1}{$z^-$}
 &&  \rnode{root2}{$x^-$}
&&  \rnode{root3}{$x^-$}
&& \rnode{tau1}{$\mathstrut\tau_1$} &
& \rnode{tau2}{$\mathstrut\tau_2$}&\rnode{z2}{$\mathstrut$}  \\
\rnode{name1}{$\D$}&
& \circlenode{root1}{$x^+$} &
 &&\rnode{name2}{$\E_1$}
&& \rnode{name3}{$\E_2$}
&& &  \rnode{root4}{$x^-$}
&& \rnode{name4}{\quad\llap{$\E_1 +^\tau \E_2$}}\\
\endpsmatrix
\psset{shortput=nab,arrows=->,labelsep=6pt}
\small
\ncline[nodesep=1pt]{-}{root1}{ym1}
\ncline[nodesep=1pt]{-}{x1}{ym1}
\ncline[nodesep=1pt]{-}{x1}{ym2}
\ncline[nodesep=1pt]{-}{dai1}{ym2}
\ncline[nodesep=1pt]{-}{root1}{zm1}
\ncline[nodesep=1pt]{-}{x2}{zm1}
\ncline[nodesep=1pt]{-}{x2}{zm2}
\ncline[nodesep=1pt]{-}{dai2}{zm2}
\ncline[nodesep=1pt]{-}{root2}{yp}
\ncline[nodesep=1pt]{-}{root3}{zp}
\ncline[nodesep=1pt]{-}{root4}{tau1}
\ncline[nodesep=1pt]{-}{root4}{tau2}
\ncline[nodesep=1pt]{-}{y1}{tau1}
\ncline[nodesep=1pt]{-}{z1}{tau2}

\ncbox[nodesep=.5cm,boxsize=0.7,linearc=.2,
linestyle=dotted]{name2}{yp}
\ncbox[nodesep=.5cm,boxsize=0.7,linearc=.2,
linestyle=dotted]{name3}{zp}
\ncbox[nodesep=.7cm,boxsize=1.6,linearc=.2,
linestyle=dotted]{tau1}{z2}
\ncbox[nodesep=.5cm,boxsize=2.6,linearc=.2,
linestyle=dotted]{x0}{x3}
\end{center}\vspace{12 pt}
If we compose $\D$ with $\E_i$, it is rather clear that we always reach  $\dai$,
hence $\D \bot \E_1$ and $\D \bot \E_2$.
On the other hand, if we compose
$\D$ with $\E_1 +^\tau \E_2$, we  have the
interaction as (partially) described below.
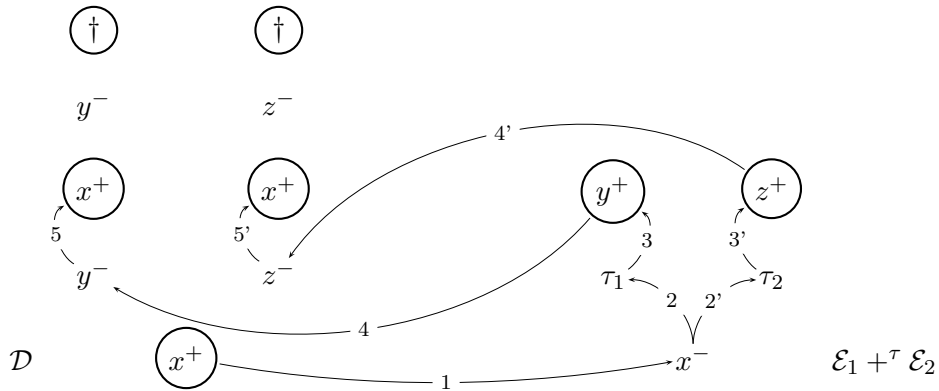
\begin{center}
\psmatrix[rowsep=13pt,colsep=0.4cm]
&&&&&&&&&&&&&&& \\
&\circlenode{daimon1}{$\dai$}&\rnode{upD}{$\phantom{a}$}&\circlenode{daimon2}{$\dai$}&&&&&&&&&&&& \\
&\rnode{ym2}{$y^-$}&&\rnode{zm2}{$z^-$}&&&&&&&&&& \rnode{upE}{$\phantom{a}$}&& \\
&\circlenode{xpl}{$x^+$}&&\circlenode{xpr}{$x^+$}&&&&&&&&&\circlenode{yp}{$y^+$}&  &\circlenode{zp}{$z^+$}& \\
&\rnode{ym1}{$y^-$} &&\rnode{zm1}{$z^-$} &&&&&&&&&\rnode{tau1}{$\tau_1$}& &\rnode{tau2}{$\tau_2$}& \\
$\D$&&\circlenode{xp}{$x^+$}&&&&&&&&&&&  \rnode{xm}{$x^-$} && $\E_1 +^\tau \E_2$ \\
\endpsmatrix
 \psset{shortput=nab,arrows=->,labelsep=6pt}
 \small
 \ncarc[nodesep=1pt,arcangle=-60]{x11pa}{x1ma}

\ncarc[linewidth=0.2pt,nodesep=1pt,arcangle=-10]{xp}{xm}
\ncput*{\scriptsize 1}
\ncarc[linewidth=0.2pt,nodesep=1pt,arcangle=-45]{xm}{tau1}
\ncput*{\scriptsize 2}
\ncarc[linewidth=0.2pt,nodesep=1pt,arcangle=45]{xm}{tau2}
\ncput*{\scriptsize 2'}
\ncarc[linewidth=0.2pt,nodesep=1pt,arcangle=-55]{tau1}{yp}
\ncput*{\scriptsize 3}
\ncarc[linewidth=0.2pt,nodesep=1pt,arcangle=55]{tau2}{zp}
\ncput*{\scriptsize 3'}
\ncarc[linewidth=0.2pt,nodesep=1pt,arcangle=40]{yp}{ym1}
\ncput*{\scriptsize 4}
\ncarc[linewidth=0.2pt,nodesep=1pt,arcangle=-45]{zp}{zm1}
\ncput*{\scriptsize 4'}
\ncarc[linewidth=0.2pt,nodesep=1pt,arcangle=60]{ym1}{xpl}
\ncput*{\scriptsize 5}
\ncarc[linewidth=0.2pt,nodesep=1pt,arcangle=60]{zm1}{xpr}
\ncput*{\scriptsize 5'}

\ncline[nodesep=1pt]{-}{daimon1}{ym2}
\ncline[nodesep=1pt]{-}{xpl}{ym2}
\ncline[nodesep=1pt]{-}{xpl}{ym1}
\ncline[nodesep=1pt]{-}{xp}{ym1}
\ncline[nodesep=1pt]{-}{daimon2}{zm2}
\ncline[nodesep=1pt]{-}{xpr}{zm2}
\ncline[nodesep=1pt]{-}{xpr}{zm1}
\ncline[nodesep=1pt]{-}{xp}{zm1}

\ncline[nodesep=1pt]{-}{yp}{tau1}
\ncline[nodesep=1pt]{-}{tau1}{xm}
\ncline[nodesep=1pt]{-}{zp}{tau2}
\ncline[nodesep=1pt]{-}{tau2}{xm}

\ncbox[nodesep=.5cm,boxsize=2.9,linearc=.2,
linestyle=dotted]{upD}{xp}
\ncbox[nodesep=.5cm,boxsize=3.6,linearc=.2,
linestyle=dotted]{upE}{xm}
\end{center}\vspace{-21 pt}
After the steps tagged by $5$ and $5'$ the interaction ``re-enters''
in $\E_1 +^\tau \E_2$. The steps which follow
$5$ are described below (for the steps which follow
$5'$ the situation is symmetric).

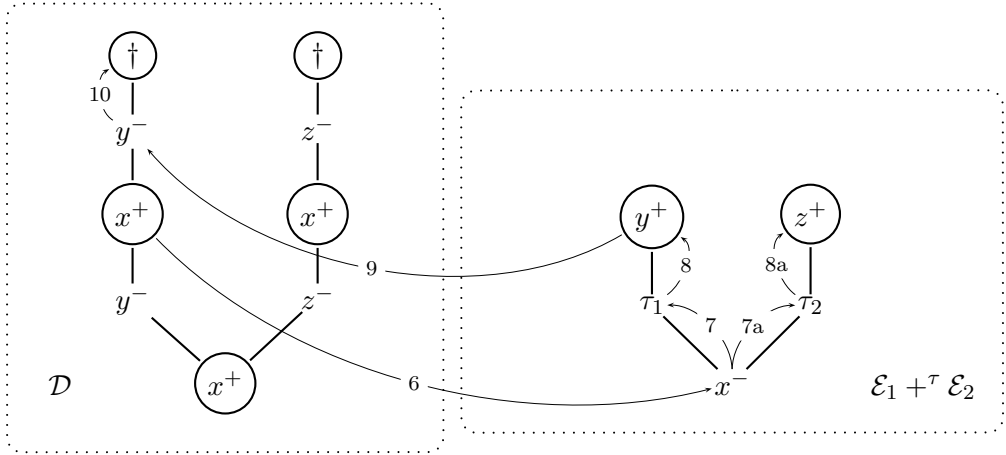
\begin{center}
\psmatrix[rowsep=13pt,colsep=0.4cm]
&&&&&&&&&&&&& && \\
&\circlenode{Daimon1}{$\dai$}&\rnode{UpD}{$\mathstrut$}&\circlenode{Daimon2}{$\dai$}&&&&&&&&&&&& \\
&\rnode{Ym2}{$y^-$}&&\rnode{Zm2}{$z^-$}&&&&&&&&&& \rnode{UpE}{$\phantom{a}$}&& \\
&\circlenode{Xpl}{$x^+$}&&\circlenode{Xpr}{$x^+$}&&&&&&&&&\circlenode{Yp}{$y^+$}&  &\circlenode{Zp}{$z^+$}& \\
&\rnode{Ym1}{$y^-$} &&\rnode{Zm1}{$z^-$} &&&&&&&&&\rnode{Tau1}{$\tau_1$}& &\rnode{Tau2}{$\tau_2$}& \\
$\D$&&\circlenode{Xp}{$x^+$}&&&&&&&&&&&  \rnode{Xm}{$x^-$} && $\E_1 +^\tau \E_2$ \\
&&&&&&&&&&&&& && \\
\endpsmatrix
 \psset{shortput=nab,arrows=->,labelsep=6pt}
 \small
 \ncarc[nodesep=1pt,arcangle=-60]{x11pa}{x1ma}

\ncarc[linewidth=0.2pt,nodesep=1pt,arcangle=-30]{Xpl}{Xm}
\ncput*{\scriptsize 6}
\ncarc[linewidth=0.2pt,nodesep=1pt,arcangle=-45]{Xm}{Tau1}
\ncput*{\scriptsize 7}
\ncarc[linewidth=0.2pt,nodesep=1pt,arcangle=45]{Xm}{Tau2}
\ncput*{\scriptsize 7a}
\ncarc[linewidth=0.2pt,nodesep=1pt,arcangle=-55]{Tau1}{Yp}
\ncput*{\scriptsize 8}
\ncarc[linewidth=0.2pt,nodesep=1pt,arcangle=55]{Tau2}{Zp}
\ncput*{\scriptsize 8a}
\ncarc[linewidth=0.2pt,nodesep=1pt,arcangle=40]{Yp}{Ym2}
\ncput*{\scriptsize 9}
\ncarc[linewidth=0.2pt,nodesep=1pt,arcangle=60]{Ym2}{Daimon1}
\ncput*{\scriptsize 10}
\ncline[nodesep=1pt]{-}{Daimon1}{Ym2}
\ncline[nodesep=1pt]{-}{Xpl}{Ym2}
\ncline[nodesep=1pt]{-}{Xpl}{Ym1}
\ncline[nodesep=1pt]{-}{Xp}{Ym1}
\ncline[nodesep=1pt]{-}{Daimon2}{Zm2}
\ncline[nodesep=1pt]{-}{Xpr}{Zm2}
\ncline[nodesep=1pt]{-}{Xpr}{Zm1}
\ncline[nodesep=1pt]{-}{Xp}{Zm1}
\ncline[nodesep=1pt]{-}{Yp}{Tau1}
\ncline[nodesep=1pt]{-}{Tau1}{Xm}
\ncline[nodesep=1pt]{-}{Zp}{Tau2}
\ncline[nodesep=1pt]{-}{Tau2}{Xm}
\ncbox[nodesep=.5cm,boxsize=2.9,linearc=.2,
linestyle=dotted]{UpD}{Xp}
\ncbox[nodesep=.5cm,boxsize=3.6,linearc=.2,
linestyle=dotted]{UpE}{Xm}
\end{center}
Notice that after the step tagged by $8a$ we have a \emph{deadlock}:
the action $z^+$ should match an action $z^-$ above
(\ie justified) by
the  \emph{last visited} occurrence of  $x^+$ (the leftmost one), but
there is no such an action since we only have $y^-$.

The result of composition is:
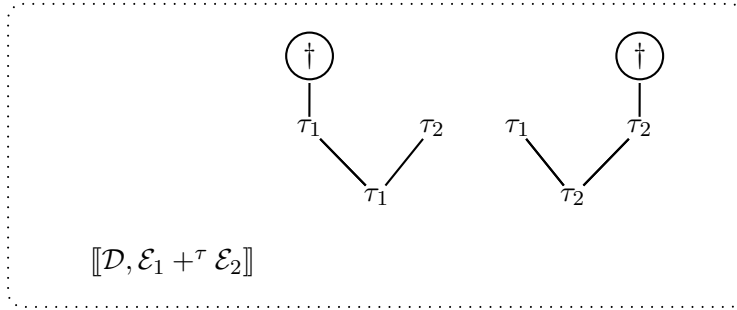
\begin{center}
\psmatrix[rowsep=13pt,colsep=0.4cm]
&&&&&&&\\

&\circlenode{dai1}{$\dai$} &\rnode{up}{$\mathstrut$}&&& &&\circlenode{dai2}{$\dai$}\\
&\rnode{tau1}{$\tau_1$} &&\rnode{tau2}{$\tau_2$} &&\rnode{tau3}{$\tau_1$} &&\rnode{tau4}{$\tau_2$}\\
&&\rnode{root1}{$\tau_1$}& && &\rnode{root2}{$\tau_2$} &\\
\rnode{name}{$\pl \D, \E_1 +^\tau \E_2 \pr $}& &\rnode{a}{}&& &&&\\
&&&&&&&\\
\endpsmatrix
 \psset{shortput=nab,arrows=->,labelsep=6pt}
 \small
\ncline[nodesep=1pt]{-}{dai1}{tau1}
\ncline[nodesep=1pt]{-}{dai2}{tau4}
\ncline[nodesep=1pt]{-}{root1}{tau2}
\ncline[nodesep=1pt]{-}{root2}{tau4}
\ncline[nodesep=1pt]{-}{root1}{tau1}
\ncline[nodesep=1pt]{-}{root2}{tau3}
\ncbox[nodesep=.5cm,boxsize=4.9,linearc=.2,
linestyle=dotted]{up}{a}

\end{center}
which has four maximal views:
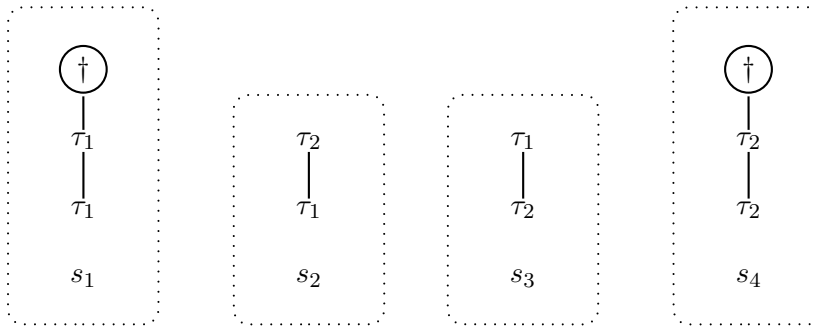
\begin{center}
\psmatrix[rowsep=13pt,colsep=2.5cm]
&&&\\
\circlenode{tdai1}{$\dai$} &   &
 & \circlenode{tdai2}{$\dai$}  \\
\rnode{ttau11}{$\tau_1$} & \rnode{ttau22}{$\tau_2$}  &
\rnode{ttau33}{$\tau_1$}  & \rnode{ttau44}{$\tau_2$}  \\
\rnode{ttau1}{$\tau_1$} & \rnode{ttau2}{$\tau_1$}  &
\rnode{ttau3}{$\tau_2$}  & \rnode{ttau4}{$\tau_2$}  \\
\rnode{tname1}{$s_1$} & \rnode{tname2}{$s_2$} &
\rnode{tname3}{$s_3$} & \rnode{tname4}{$s_4$}  \\
&&&\\
\endpsmatrix
 \psset{shortput=nab,arrows=->,labelsep=6pt}
 \small
\ncline[nodesep=1pt]{-}{tdai1}{ttau11}
\ncline[nodesep=1pt]{-}{tdai2}{ttau44}
\ncline[nodesep=1pt]{-}{ttau1}{ttau11}
\ncline[nodesep=1pt]{-}{ttau2}{ttau22}
\ncline[nodesep=1pt]{-}{ttau3}{ttau33}
\ncline[nodesep=1pt]{-}{ttau4}{ttau44}
\ncline[nodesep=1pt]{-}{root2}{tau4}
\ncline[nodesep=1pt]{-}{root1}{tau1}
\ncline[nodesep=1pt]{-}{root2}{tau3}
\ncbox[nodesep=.5cm,boxsize=1,linearc=.2,
linestyle=dotted]{tname1}{tdai1}
\ncbox[nodesep=.5cm,boxsize=1,linearc=.2,
linestyle=dotted]{tname2}{ttau22}
\ncbox[nodesep=.5cm,boxsize=1,linearc=.2,
linestyle=dotted]{tname3}{ttau33}
\ncbox[nodesep=.5cm,boxsize=1,linearc=.2,
linestyle=dotted]{tname4}{tdai2}

\end{center}
From this we conclude that $\D \not\!\!\bot \! \ \ \E_1 +^\tau \E_2$
because $s_2$ and $s_3$ do not satisfy the totality condition
of Definition ~\ref{nutot}.
\end{exa}

We will use also  the following properties of normalization.

\begin{lem}\label{norm_lem} Let $\D,\D_1,\D_2$
be strategies on the same positive  interface
$ \Pi=\xi 1^+,\dots,\xi n^+$.
Let
 $(\E_i)_{\xi i  \in \Pi} = (\E_i) = \{\E_1,\ldots,\E_n\}$
be a  family of (negative) counter-strategies \emph{\wrt}$\Pi$.

We have the following:

\begin{enumerate}[\em(a)]

\item
 $\{\E_1,\ldots,\E_n\} \ \bot \ \D$ \ if and only if \
 $x^+.\{\E_1,\dots,\E_n\} \ \bot \ x^-.\D$, where $x=(\xi,I_n)$;

\item $(\E_i)\ \bot \  \D_1 \oplus^\tau \D_2$
\ if and only if \  $(\E_i) \ \bot \ \D_1 $
and $(\E_i) \ \bot \ \D_2$.

\end{enumerate}
\end{lem}

\proof \hfill
\begin{enumerate}[(a)]
\item By construction, the root action  $x$ is never repeated
in $x^+.\{\E_1,\dots,\E_n\}$. By definition of  VAM, the interaction in the cut-net  $\{x^+.\{\E_1,\dots,\E_n\},  x^-.\D  \}$ starts by matching $x^+$ with
$x^-$, and then continue as in $\{\E_1,\dots,\E_n, \D  \}$. More precisely,
$$VAM(x^+.\{\E_1,\dots,\E_n\},  x^-.\D  ) =\{x^+,x^+.x^-\} \cup
\{x^+.x^-.q: q\in VAM(\E_1,\dots,\E_n, \D  )\}.$$
If $I(\E_1,\dots,\E_n, \D  )\not=\emptyset$, then $$I(x^+.\{\E_1,\dots,\E_n\},  x^-.\D  )=\{x^+,x^+.x^-\} \cup
\{x^+.x^-.q: q\in I(\E_1,\dots,\E_n, \D  )\}.$$ If $I(\E_1,\dots,\E_n, \D  )=\emptyset$, then $I(x^+.\{\E_1,\dots,\E_n\},  x^-.\D  )$ is also empty.  By hiding, we have  the conclusion.

\item  By definition of VAM,
$$\begin{array}{rcl}
VAM(\E_1,\dots,\E_n, \D_1 \oplus^\tau \D_2) & = &
\{t.t_1.q: q\in VAM(\E_1,\dots,\E_n, \D_1)\}  \\ &  \cup &
\{t.t_2.q: q\in VAM(\E_1,\dots,\E_n, \D_2)\}. \\
\end{array}$$
Since
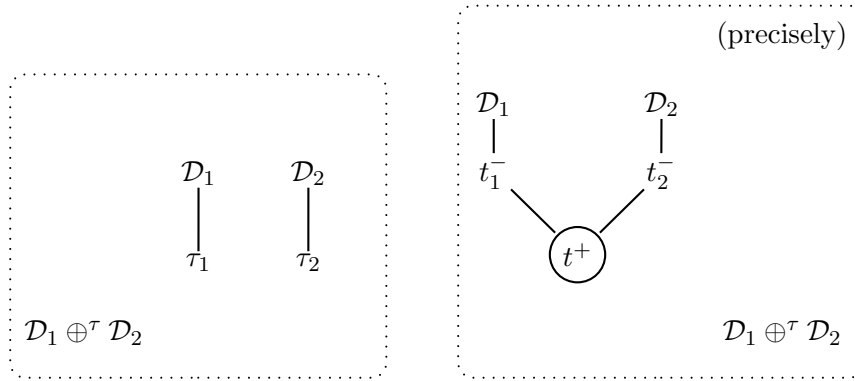
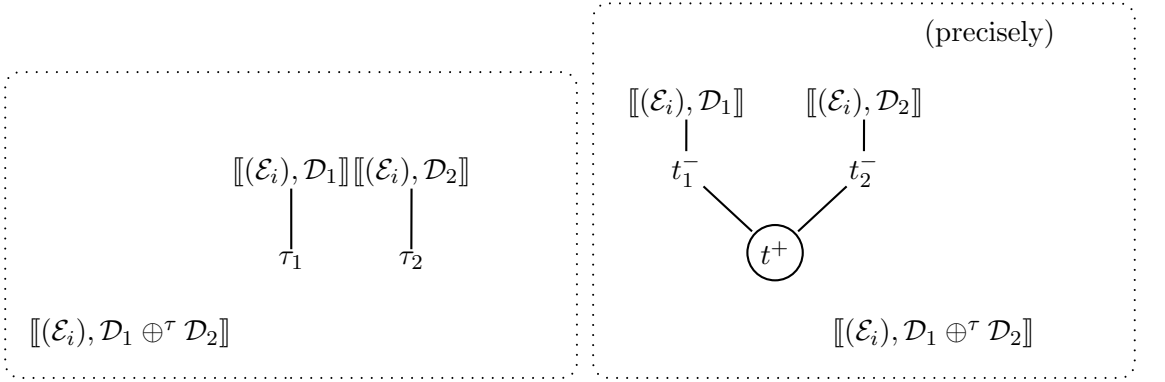
\begin{center}
\psmatrix[rowsep=13pt,colsep=0.5cm]

&&&&&&&&&&\\
&& & & &
& & & & \rnode{up2}{} &(precisely)\\
&\rnode{up1}{}& & & &
& & \rnode{ll1}{$\D_1$}& & \rnode{ll2}{$\D_2$} & \\
&\rnode{leaf1}{$\D_1$}& &\rnode{leaf2}{$\D_2$} & &
& & \rnode{br1}{$t_1^-$}& & \rnode{br2}{$t_2^-$} &\\
&\rnode{tau1}{$\tau_1$}& &\rnode{tau2}{$\tau_2$} & &
& & & \circlenode{root}{$t^+$}& &\\
$\D_1 \oplus^\tau \D_2$ &\rnode{name1}{} && & &
& & & & \rnode{name2}{}&$\D_1 \oplus^\tau \D_2$\\

\endpsmatrix
 \psset{shortput=nab,arrows=->,labelsep=6pt}
 \small
\ncline[nodesep=1pt]{-}{leaf1}{tau1}
\ncline[nodesep=1pt]{-}{leaf2}{tau2}
\ncline[nodesep=1pt]{-}{root}{br1}
\ncline[nodesep=1pt]{-}{root}{br2}
\ncline[nodesep=1pt]{-}{ll1}{br1}
\ncline[nodesep=1pt]{-}{ll2}{br2}

\ncbox[nodesep=.5cm,boxsize=2.5,linearc=.2,
linestyle=dotted]{name1}{up1}
\ncbox[nodesep=.5cm,boxsize=2.7,linearc=.2,
linestyle=dotted]{name2}{up2}
\end{center}
we have
\begin{center}
\psmatrix[rowsep=13pt,colsep=0.01cm]

&&&&&&&&&&\\
&& & & &
& & & & \rnode{up2}{} & (precisely)\\
&\rnode{up1}{}& & & &
& & \rnode{ll1}{$\pl  (\E_i), \D_1\pr$}& & \rnode{ll2}{$\pl  (\E_i), \D_2\pr$} & \\
&\rnode{leaf1}{$\pl  (\E_i), \D_1\pr$}& &\rnode{leaf2}{$\pl  (\E_i), \D_2\pr$} & & $\phantom{ll}$
& & \rnode{br1}{$t_1^-$}& & \rnode{br2}{$t_2^-$} &\\
&\rnode{tau1}{$\tau_1$}& &\rnode{tau2}{$\tau_2$} & & $\phantom{aaaaaaaaaa}$
& & & \circlenode{root}{$t^+$}& &\\
\hspace{-2cm}$\pl(\E_i) ,  \D_1 \oplus^\tau \D_2 \pr $& \rnode{name3}{}&& & &
& & & & \rnode{name2}{}& \hspace{-1.5cm}$\pl(\E_i) ,  \D_1 \oplus^\tau \D_2 \pr $\\

\endpsmatrix
 \psset{shortput=nab,arrows=->,labelsep=6pt}
 \small
\ncline[nodesep=1pt]{-}{leaf1}{tau1}
\ncline[nodesep=1pt]{-}{leaf2}{tau2}
\ncline[nodesep=1pt]{-}{root}{br1}
\ncline[nodesep=1pt]{-}{root}{br2}
\ncline[nodesep=1pt]{-}{ll1}{br1}
\ncline[nodesep=1pt]{-}{ll2}{br2}

\ncbox[nodesep=.5cm,boxsize=3.8,linearc=.2,
linestyle=dotted]{name3}{up1}
\ncbox[nodesep=.5cm,boxsize=3.6,linearc=.2,
linestyle=dotted]{name2}{up2}
\end{center}\vspace{-12 pt}
\qed
\end{enumerate}

\subsection{Interactive types}
As already defined in Section ~\ref{types}, a behaviour $\bG$ is a set of strategies closed
by biorthogonal. Since we have abandoned the limitation of ``strict linearity'' which we had in Section \ref{linearity} , now behaviours are never empty.

\begin{defi}
A {\ behaviour} on the interface $\GG$
is a set $\bG$ of   strategies $\D:\GG$ such that $\bG\b\b=\bG$.
A behaviour is positive or negative according to the polarity of the interface.
\end{defi}

\begin{prop}
A behaviour is always non-empty.
\end{prop}
\proof
A positive behaviour always contains at least  $\Dai$. A negative behaviour on the interface $\sigma^-,\xi_1^+,\ldots,\xi_n^+$, always contains at least the following
strategy $\Dai^-$ (called  negative daimon in ~\cite{GirLoc}):

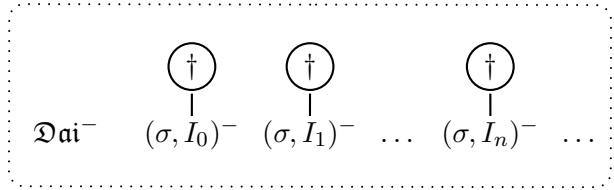
\begin{center}
\psmatrix[rowsep=9pt,colsep=0.3cm]

& &  &
&
&& \\

& & \circlenode{dai0}{$\dai$} &
\circlenode{dai1}{$\dai$} &
& \circlenode{dain}{$\dai$} & \\

$\Dai^-$ & & \rnode{a0}{$(\sigma,I_0)^-$} &
\rnode{a1}{$(\sigma,I_1)^-$} &
\ldots & \rnode{an}{$(\sigma,I_n)^-$} & \ldots \\

\endpsmatrix
 \psset{shortput=nab,arrows=->,labelsep=6pt}
 \small
\ncline[nodesep=1pt]{-}{dai0}{a0}
\ncline[nodesep=1pt]{-}{dai1}{a1}
\ncline[nodesep=1pt]{-}{dain}{an}

\ncbox[nodesep=.5cm,boxsize=4.0,linearc=.2,
linestyle=dotted]{dai1}{a1}
\end{center}
Indeed, let $ \D : \sigma^+$ and
$\E_1 : \xi_1^-$, \ldots,
$\E_n : \xi_n^-$ be total counter-strategies.
Let us  consider then the closed cut-net
$\{\Dai^-,\D,\E_1,\ldots,\E_n\}$
and calculate its normal form.
By definition of the abstract machine,
we start by collecting
actions in $\D$ (because it is the main strategy).
Since $\D$ is total,
after a sequence
of silent actions we reach $\dai$ or
we reach a positive action $(\sigma,I_n)^+$
(both cases are possible ``in parallel", since $\D$ is non-uniform).
In the latter case this proper action matches
its opposite $(\sigma,I_n)^-$
and  we eventually reach $\dai$.
The normal form is then a total strategy
on the empty interface, since each path from
the root of $\pl\Dai^-,\D,\E_1,\ldots,\E_n\pr$
leads to $\dai$.
 \qed

\begin{exa}[$\posm \Zero$, $\negm \Toppo$] \label{zero,top2} Let us see what happens to  the behaviour generated from  $\Dai:\xi^+$
(\cf Example ~\ref{zero,top}) in the non-linear setting.
$$
 \posm \Zero := \{\Dai\} \b \b \mbox{ on interface } \xi^+; \qquad
\negm \Toppo := \{\Dai\}\b \mbox{ on interface } \xi^-.
$$
We have that $\posm \Zero$ contains a unique deterministic strategy, which is $\Dai$. $\posm \Zero$ also contains strategies which are set of views of the form $s=t.t_i.\dots.t.t_j.\dai$.
On the other side, $\negm \Toppo$ contains  all negative strategies which have interface $\xi$, including the empty one.

When we take  $\posm \Zero$ on the empty interface, it consists exactly of  all the total strategies on the empty interface.
\end{exa}

\section{Ludics with repetitions: types and internal completeness}
\label{type sez}
In this section, we give constructions for the behaviours which correspond to the construction of  $\HS$ formulas,
and prove that they enjoy internal completeness.

\subsection{\texorpdfstring{$\HS$}{HS} types} \label{compound types}
In this section, we use the same
 constructions on strategies  as  in Section ~\ref{typecons}. The resulting behaviours are different, because normalization is different (\ie  non-linear).

\subsubsection{Constructions on strategies.} Let
$\D_i:\xi i^-,\Pi \ (1\leq i \leq n)$   be $n\geq 0$  negative strategies (which can possibly also be empty). We obtain a new
positive strategy on the interface $\xi^+,\Pi$, denoted by $ \D_1\bullet \dots \bullet \D_n$,
by adding to the union of the strategies the positive root $(\xi,I_n)^+$,
\ie
$$ \D_1\bullet \dots\bullet \D_n:= (\xi,I_n)^+. \{\D_1, \dots, \D_n\}.$$

\noindent We observe that the root of the resulting strategy is linear. A converse construction also exists.

\begin{lem}\label{projections}
Let $\D: \xi^+,\Pi$ be a positive strategy having as root a linear occurrence of the proper action $(\xi, I_n)^+$.
 We can write  $\D$ as $\D_1\bullet \dots\bullet \D_n$, with   $\D_i:\xi i^-,\Pi$ for each $1 \leq i \leq n$.
\end{lem}
\proof
Let us write $\D$ as $(\xi, I_n)^+.\E$.
 All views in $\E$ have a first action  of the form  $(\xi i, K)^-$, with  $1 \leq i \leq n$.
 We partition $\E$ into maximal subsets of views which start with an action on the same name,    that is, we set
$\D_i := \{s \in \E : s= (\xi i, K)^-.s' $ for some $ K   \}$.
Then it is immediate to verify that  each $\D_i$ is in fact a negative strategy
on interface $\xi i^-,\Pi$ and that $\E= \D_1 \cup \ldots \cup \D_n$. We finally conclude
  $\D= \D_1\bullet\dots\bullet \D_n$.
\qed

\begin{lem} \label{conversion}
 Let  $\D_1:\xi1^-,\Pi$, \dots, \ $\D_n:\xi n^- , \Pi$   be  negative strategies.
Let $(\F_{\alpha})_{\alpha\in \Pi}$ be a family of counter-strategies (\emph{\wrt} $\Pi$). We have that
$$\pl {\D_1\bullet\dots\bullet \D_n,(\F_\alpha) }\pr = \pl \D_1,(\F_\alpha)\pr \bullet \dots \bullet  \pl\D_n,(\F_\alpha) \pr .$$
That is,
$$\pl (\xi,I_n)^+.\{\D_1,\dots, \D_n\},(\F_\alpha) \pr = (\xi,I_n)^+.\{\pl\D_1,(\F_\alpha)\pr, \dots,  \pl \D_n,(\F_\alpha)\pr \} .$$
\end{lem}
\proof
   It easily follows  from the definition of VAM. We first observe that by construction, the root of $\D_1\bullet\dots\bullet \D_n$ is $(\xi,I_n)^+$ and that it is  a visible action. This implies that it is also the root of $\cut {\D_1\bullet\dots\bullet \D_n,(\F_\alpha)} $.
Since $(\xi,I_n)^+$ occurs linearly, the interaction never uses (occurrences of) $(\xi,I_n)^+$ again, and we can write the equations above.
\qed


From now until the end of Section ~\ref{type sez}, let us
 fix  a family of $n$ ($n\geq 0$) negative behaviours $\bN_{\xi i}:\xi_i^-$, with $1\leq i \leq n$.
  We
 define
$$\bN_{\xi 1}\bullet \dots \bullet \bN_{\xi n}:=  \{\D_1\bullet \dots\bullet \D_n: \D_i\in \bN_{\xi i} \}.$$
We define a new positive (resp.\ negative) behaviour on the interface  $\xi^+$ (resp.\ $\xi^-$) as follows:
$${\bF^+_\xi(\bN_{\xi 1},\dots,\bN_{\xi n})} := (\bN_{\xi 1}\bullet \dots \bullet \bN_{\xi n})\b\b; \qquad
 \bF^-_\xi(\bN^\bot_{\xi 1},\dots,\bN^\bot_{\xi n}) := (\bN_{\xi 1}\bullet \dots \bullet \bN_{\xi n})\b.$$

  \begin{rem}\label{root_rem}
We stress once more that, by construction, all strategies in
$\bN_{\xi 1}\bullet \dots \bullet \bN_{\xi n}$ have as root  $x=(\xi, I_n )^+$, which is
 \emph{linear}.  The repetitions of occurrences of $x$ are obtained via the closure
by biorthogonality, and hence only belong to $(\bN_{\xi 1}\bullet \dots \bullet \bN_{\xi n})\b\b$.
\end{rem}

When $n=0$ we write
$\semone$ and  $\sembot$ for
the positive behaviour and the negative one
given by the previous constructions respectively.
Precisely, $\semone = \{\D\}\b\b$
and  $\sembot = \{\D\}\b$,
where $\D= (\xi,\emptyset)^+$.


\begin{lem}\label{pos_root}  If the root $x$ of $\D\in \bP_\xi=  (\bN_{\xi 1}\bullet \dots \bullet \bN_{\xi n})\b\b$ is a proper action, then  $x=(\xi,I_n)^+$.
\end{lem}
\proof    By construction, all the strategies in $\bN_{\xi 1}\bullet \dots \bullet \bN_{\xi n}$ have root $x=(\xi,I_n)^+$.
Hence, $(\bN_{\xi 1}\bullet \dots \bullet \bN_{\xi n})\b$ contains the strategy $\E:= (\xi,I_n)^-.\dai$. Since any $\D \in
 (\bN_{\xi 1}\bullet \dots \bullet \bN_{\xi n})\b\b$ has to be orthogonal to $\E$, when its root is a proper action  it must be $(\xi,I_n)^+$.
\qed\vspace{-6 pt}

\subsection{Sequent of behaviours}
The definition of sequent of behaviours remains  the same as in the linear case, and Proposition ~\ref{closure} still holds.
For the reader's convenience we repeat them below.

\begin{defi}[Sequent of behaviours]\label{seq_def}
Let   $\Gamma=\xi_1^{\epsilon_1}, \dots,\xi_n^{\epsilon_n} $ ($n\geq 0$) be an interface, and  let
 $\bGG = \bG_{\xi_1},\dots,
\bG_{\xi_n}$    ($n \geq 0$)   behaviours of respective polarities $\epsilon_1, \dots, \epsilon_n$.

We define a new behaviour  on the same interface $\Gamma$,
which we call {\bf sequent of behaviours} and denote by $\vdash \bGG$,
as follows:
$$ \vdash \bGG := \{ \D :\D \mbox{ is total on interface }
\Gamma \mbox{ and }
 \D \bot \{\E_1,\ldots,\E_n\}  \mbox{ for all  } \E_1 \in \bG_{\xi_1}^\bot, \ldots,\E_n \in \bG_{\xi_n}^\bot  \}.$$
\end{defi}

\noindent Observe  that
 if $\Gamma$ is empty, then  $\vdash\bGG$ consists
 of those  strategies
on the empty interface which are total
(\cf Section ~\ref{lwR section II}). Hence,
 $\vdash\bGG$ is $\{\Dai\}\b\b$.

We will use the following two results.

\begin{prop} \label{closure2} Let $\bA, \bG_1, \dots \bG_n$ ($n\geq 0$) be a sequence of
behaviours, and $\bGG= \bG_1, \dots \bG_n$.  We have that:
\begin{enumerate}[$\bullet$]
\item
$\D\in \ \vdash \bGG,\bA $ if and only if  for each
$\F \in \bA\b$, $\pl \D,\F \pr \in \ \vdash \bGG$.

\item $\D\in \ \vdash \bGG,\bA $ if and only if
 $\pl \D,(\E_i)  \pr \in \ \vdash \bA$, for each family
$(\E_i)$ such that
$\E_1 \in \bG_{1}^\bot,\ldots,$ $\E_n \in \bG_{n}^\bot$. \qed
\end{enumerate}
\end{prop}\smallskip

\noindent The proof is the same as for Proposition ~\ref{closure} (but now using
Theorem ~\ref{associativity}).  Observe again that if $\GG=\emptyset$,
the condition in the first claim is simply a reformulation of the
definition of orthogonality: $\D\bot\F$ if $\cut{\D,\F}$ is a total
strategy on the empty interface.\vfill\eject

\begin{lem}\label{neg_root} Let $\bN= (\bN_{\xi 1}\bullet \dots \bullet \bN_{\xi n})\b$.   We have that
\begin{enumerate}[\em (1)]
\item If $\F \in \bN$, then $\F \neq \emptyset$.
\item  If $\F \in \ \vdash \bN, \bPP$, then $\F \neq \emptyset$.
\end{enumerate}
\end{lem}
\proof \hfill
\begin{enumerate}[(1)] \item
 $\F\not=\emptyset$ because each $\D\in \bN_{\xi 1}\bullet \dots \bullet \bN_{\xi n}$ has root $(\xi,I_n)^+$,
and $\F$ has to be orthogonal  to $\D$.
 \item The second point is proven with a similar argument.
\qed
\end{enumerate}

\subsection{Internal completeness}

The following two propositions are the core of internal completeness.
Internal completeness for a negative behaviour is the same as in the linear case.
For the positive case, we also first state internal completeness for those strategies whose root is linear.
The key element which allows us to reduce the general case to this one is Lemma ~\ref{singleroot}.

\begin{prop}[Internal completeness of $\bF^-$]\label{neg_comp} Let $\bP_{\xi i}=\bN_{\xi i}^\bot$ and $x= (\xi, I_n)$.
$$ (1) ~x^-.\D\in \bF^- (\bP_{\xi_1},\dots,\bP_{\xi_n}) \ \Leftrightarrow \
(2)~ \D\in \ \vdash \bP_{\xi_1},\dots,\bP_{\xi_n}.$$
\end{prop}
\proof
The proof follows immediately from the definitions. Expanding the definition, we obtain the two following properties, which are equivalent
by using Lemma ~\ref{norm_lem} (a):

\begin{enumerate}[(1)]
\item    $x^-.\D \ \bot \  x^+.\{\E_1,\dots,\E_n\}$, for any
$\E_1\in \bP_{\xi_1}^\bot,\dots, \E_n \in \bP_{\xi_n}^\bot$;
\item    $ \D \ \bot  \ \{\E_1,\dots, \E_n\}$,
for any $\E_1\in \bP_{\xi_1}^\bot,\dots, \E_n \in \bP_{\xi_n}^\bot$.
\qed
\end{enumerate}

\begin{prop}[Linear internal completeness of $\bF^+$] \label{lin comp F+}
Let $\D \in \bF^+(\bN_{\sigma 1},\dots,\bN_{\sigma n})$ have a root
that is
a linear occurrence of a proper action.  Then $\D = \D_1
\bullet\dots\bullet \D_n$, with $\D_i \in \bN_{\sigma i}$, for any $1
\leq i \leq n$.
\end{prop}
\proof
  By Lemma ~\ref{pos_root} we know that the root
 of $\D$ must be $(\sigma,I_n)^+$. Since it occurs linearly, we can apply  Lemma ~\ref{projections} and then write
 $\D$   as $\D_1\bullet\dots\bullet \D_n$ .
We already observed (Remark ~\ref{importante}), that the argument in
Proposition ~\ref{compl+}
only relies on the fact that the root is a linear occurrence of action.
We can hence repeat the same proof as in ~\ref{compl+}.
\qed

The essential property which allows us to use the argument we sketched in Section ~\ref{NUtests} is the following  Lemma ~\ref{singleroot} (1).
 Point (2) guarantees that all tests we are interested in have the form requested by point (1).

\begin{lem} \label{poscomp_lem} \label{singleroot}
Let
$\bN_\xi= (\bN_{\xi 1}\bullet \dots \bullet \bN_{\xi n})\b$,   $\bP_\xi= (\bN_{\xi 1}\bullet \dots \bullet \bN_{\xi n})\b\b$,
and $x=(\xi,I_n)$.

  \begin{enumerate}[\em(1)]

\item Let $\F_1,\F_2\in \bN_\xi$. Assume $\F_1=x^-.\D_1,\F_2=x^-.\D_2$. We have that
 $\F_1 +^\tau \F_2\in \bN_\xi$ (\cf Definition \emph{~\ref{tau sum}}).

  \item Let us denote by $\bN_\xi\single{x}$ the set of all $\F\in \bN_\xi$ such that
$\F$ has $x$ as  {\em unique} root.
We have that
$(\bN_\xi\single{x})\b= \bN_\xi^\bot$ i.e., $$\bP_\xi = (\bN_\xi\single{x})\b.$$
\end{enumerate}
\end{lem}

\proof \hfill
\begin{enumerate}[(1)]
\item

By internal completeness (Proposition ~\ref{neg_comp}),
$\D_1,\D_2 \in \ \vdash \bN_{\xi 1}^\bot,\dots, \bN_{\xi n}^\bot$.
By Lemma ~\ref{norm_lem}, $\D_1 \oplus ^\tau \D_2\in \  \vdash \bN_{\xi 1}^\bot,\dots, \bN_{\xi n}^\bot$.
 By Proposition ~\ref{neg_comp} again,
we conclude $\F_1+^\tau \F_2 = x^-. (\D_1 \oplus ^\tau \D_2) \in \bN$.

\item  Let  $\F\in (\bN_{\xi 1}\bullet \dots \bullet \bN_{\xi n})\b$. Being  on interface $\xi$, a negative strategy $\F$
 may have several roots, all of the form $(\xi,K)^-$
 (all views in $\F$ start with such an action).   We define $\F_x \subseteq \F$ as the subset of views whose first action
   is $x=(\xi,I_n)$. $\F_x$ is clearly  a strategy. We proceed in two steps.

 (i) We prove that for each $\F\in \bN_\xi$,   then $\F_x\in \bN_\xi$.

  Let  $\D :\xi^+ \in \bN_{\xi 1}\bullet \dots \bullet \bN_{\xi n}$. A positive strategy has a unique root.
 By construction, the root of $\D$ is the action $x^+=(\xi,I_n)^+$.  For each $\F\in  \bN_\xi$, it must be $\D \bot \F$.
Since $\D$ is the main strategy in the cut-net $\{\D,\F\}$, $\F$
  need to have $x^-=(\xi,I_n)^-$ as a root, in order to match $x^+$, otherwise
  those strategies would not be orthogonal.
  We observe that the normalization with any $\D \in \bN_{\xi 1}\bullet \dots \bullet \bN_{\xi n}$ only uses views in $\F_x$, because $\D$ by construction does not contain any other occurrence of action on the name $\xi$ but the root.
 As a consequence, $\cut{\D,\F} = \cut{\D,\F_x} $.

  (ii) We can now prove that $(\bN_\xi\single{x})^\bot \subseteq \bN_\xi^\bot$ (the other inclusion is obvious, since $\bN_\xi\single{x} \subseteq \bN_\xi$). Let $\D\in (\bN_\xi\single{x})\b $. We prove that  for each $
    \F\in \bN_\xi$, $\D\bot \F$, {  \ie $\D \in \bN_\xi^\bot$.} More precisely, we prove that $\cut{\D,\F}=\cut{\D,\F_x}$, by showing that
    the part of $\F$ used in the interaction against $\D$ is all
  contained in $\F_x$.

  Assume there exists  $p\in I(\D,\F)$ such that $\view p \not\in \F_x $, and let us take it minimal, \ie $\view p =y^-$, where $y=(\xi,J), J\not=I_n$ (\ie the interaction enters the root of another subtree of $\F_y$  of $\F$).
  In this case $p=p'.y^+.y^-$, with $\view {t}\in \{\D,\F_x\}$, for each $t\pref p'.y^+$. We observe that
  $p'.y^+$ has no extensions in $VAM(\D,\F_x)$ (because there is no match for $y^+$), and that it contains no $\dai$.
  Let $s$ be the maximal prefix of $p'.y^+$ which has an extension $q.\dai$ in $ VAM(\D,\F_x)$ (\ie an extension which terminates with a $\dai$ action).
  We have $s.a\pref p'.y^+$,  $s.b\pref q.\dai$, and
  by Lemma ~\ref{coh_p}, we have that $a,b$ are negative silent actions. Hence $s.a\in I(\D,\F_x)$, and $hide(s.a)\in \cut{\D,\F_x}$,
  but $hide(s.a)$ has no extension terminating with a $\dai$ action, against the hypothesis that $\D\bot\F_x$.

Observe that, putting all elements together,  we have also established that
\begin{center}

$\cut{\D,\F}=\cut{\D,\F_x}$ for each $\D\in \bN_\xi^\bot, \F\in \bN_\xi$.

\end{center}
We will use this in the sequel.
\qed
\end{enumerate}\medskip

\noindent Observe that the property at the point (1) above does not hold in general, for arbitrary behaviours
 (see Example
~\ref{ex nondetorm}  and take $\bG = \{\D\}^\bot$).

\begin{lem}\label{seq}Let $
  \D\in \bP_\xi=(\bN_{\xi 1}\bullet \dots \bullet \bN_{\xi n})\b\b$, such that
the root is a proper action.
 We have that  $\rt{\D}\in \mathbf{\vdash \bP_{\xi},\bP_{\sigma}}$. Moreover, the new root on $\sigma$ is linear.
\end{lem}
\proof
By Lemma ~\ref{poscomp_lem} (2),
 $ \bP_\xi =(\bN_\xi\single{x})\b$. For all pairs $\E,\F \in \bN_\xi\single{x}$,  we have $\D \bot \E$ and $\D \bot \F$.
By Lemma~\ref{poscomp_lem} (1),
we have that $\D \bot \ (\E +^\tau \F)$.
Using
Proposition ~\ref{propo copies1}, we have that
$\rt{\D} \bot \{\E +^\tau \F , (\E +^\tau \F)[\sigma/\xi]\}$
and {by using twice} Lemma ~\ref{ortho_cor}
 we have that
$\rt{\D} \bot \{ \E , \F[\sigma/\xi] \}$, that is
$\rt{\D}\in \mathbf{\vdash \bP_{\xi},\bP_{\sigma}}$.
{The linearity of $\rt \D$ is given by the construction ``$\rt \_$."}
\qed

\begin{cor}[Internal completeness of $\bF^+$] \label{int comp F+}
Let $\D\in \bP_\xi=\bF^+(\bN_{\xi 1},\dots,\bN_{\xi n})$ be as in Lemma ~\emph{~\ref{seq}}.
 Then
$\rt \D= \D_1'\bullet\dots\bullet \D_n'$ where each
$\D_i' \in \ \vdash \bN_{\sigma i},\bP_\xi.$
\end{cor}

\proof
 By Lemma ~\ref{pos_root}, the root $x$ of $\D \in \bP_\xi$ is $x=(\xi,I_n)$.
By Lemma  ~\ref{seq}, we have that
$\rt {\D}\in \  \vdash \bP_\xi,\bP_\sigma$. The root  $(\sigma, I_n)$ is the
 only occurrence of action on $\sigma$.
By Lemma ~\ref{projections}, we  can write $\rt {\D}$  as
$\rt {\D}=\E_1\bullet\dots\bullet \E_n$.
Now we use Proposition ~\ref{closure2}.  For each $\A_\xi\in \bP_\xi^\bot $, we have $\cut {\rt {\D},\A}\in \bP_\sigma$.
By Lemma ~\ref{conversion}
we  have that:
$\cut {\E_1\bullet\dots\bullet \E_n,\A} = \cut{\E_1,\A} \bullet \dots \bullet  \cut{\E_1,\A} $.
By Proposition ~\ref{lin comp F+}, $\cut{\E_i,\A}\in \bN_{\sigma i}$.
 By using again Proposition ~\ref{closure2}, we have that $\E_i \in \ \vdash \bN_{\sigma i},\bP_\xi$. \qed




\section{Ludics with repetitions: full completeness} \label{completeness}
In this section, we show that our model
is fully complete with respect to  $\HS$
(Section ~\ref{HS}).\smallskip

As usual in game semantics (\eg ~\cite{LauSvS,MelTab}), not all strategies are suitable
to be
interpretation of a proof. In general, strategies which are interpretations
of a proof have to
satisfy some \emph{winning conditions} which describe  a strategy with  "good properties."
Our winning strategies are those that are finite,
deterministic, daimon-free and \emph{material} (see below).

We now introduce
the notion of materiality.

 \subsection{Materiality} \label{materiality}
It is important to have in mind that normalization does not
   necessarily visit all the actions of a strategy. This is exactly what underlies the notion of \emph{materiality}.
 Let us first examine an example, to understand materiality.

By definition of normalization (Section ~\ref{VAM section}),
 at each step, the machine examines  (\ie \emph{visits}) an occurrence of
 action $a$ in a  view $s.a$ belonging to a strategy of the cut-net.   We say that  the view $s.a$ is \emph{used} or \emph{visited}.

 \begin{exa} \label{matesempio}
 Let $\D,\E,\F$ be the  strategies in Figure ~\ref{fig:mat}.

 \begin{figure}[h]
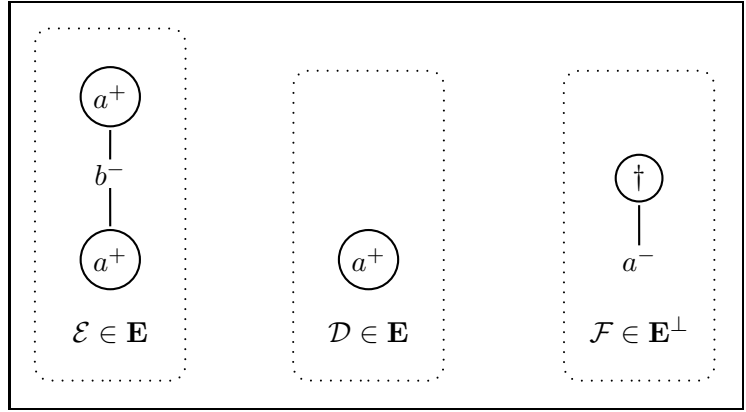

   \centering
\fbox{

\psmatrix[rowsep=9pt,colsep=0.3cm]

&&$\phantom{[a/b]}$&&&&
&&&&&&&&&&&&&&
\\
&&
\circlenode{leaf1}{$a^+$}&&&&&&
&&\rnode{fake2}{}&&&&&&&&\rnode{fake3}{}&&
\\
&&\rnode{node1}{$b^-$}&&&&&&
&&&&&&&&
&&\circlenode{node3}{$\dai$}&&
\\
&&\circlenode{root1}{$a^+$}&&&&&&
&&\circlenode{root2}{$a^+$} &&&&&&
&&\rnode{root3}{$a^-$}&&
\\
&&\rnode{name1}{$\E \in \bE$}&&&&&&
&&\rnode{name2}{$\D \in \bE$} &&&&&&
&&\rnode{name3}{$\F \in \bE^\bot$}&&
\\
&&&&
&&&&&&&&&&&&&&&&
\\
\endpsmatrix
 \psset{shortput=nab,arrows=->,labelsep=6pt}
 \small
\ncline[nodesep=1pt]{-}{root1}{node1}
\ncline[nodesep=1pt]{-}{leaf1}{node1}
\ncline[nodesep=1pt]{-}{root3}{node3}

\ncbox[nodesep=.5cm,boxsize=1,linearc=.2,
linestyle=dotted]{name1}{leaf1}
\ncbox[nodesep=.5cm,boxsize=1,linearc=.2,
linestyle=dotted]{name2}{fake2}
\ncbox[nodesep=.5cm,boxsize=1,linearc=.2,
linestyle=dotted]{name3}{fake3}

}

   \caption{Materiality}
   \label{fig:mat}
 \end{figure}

 \noindent Consider the behaviour $\bE = \{\D\}\b\b$ and notice that $\F \in \bE\b$.
 Observe also that $\E \in \bE$, but  the normalization
 between $\E$ and $\F$ uses only the first  action
 $a^+$; the action $b^-$ is never visited through the
 interaction between $\E$ and $\F$.

 The notion of {\em materiality} exactly captures the significant part of a strategy $\D$ in a behaviour, \ie   the part that is really used to react to the tests in $\bE\b$.
  \end{exa}\medskip

\noindent To make these ideas formal, we proceed in two steps.

  First, in Section ~\ref{used_sec},  we consider a strategy $\D$, and a family of counter-strategies $(\E_i)$.   We define
  the restriction of $\D$ to the part $\D'\subseteq \D$ which is really used (visited) to produce the normal form $\cut{\D, (\E_i)}$.
  We have that $\D'$ is a strategy, and that $\cut{\D', (\E_i)}= \cut{\D, (\E_i)}$. Any occurrence of action in $\D$ which is outside $\D'$
  will never be accessed during this normalization and in fact those | not visited | actions are not significant in the calculation.

  Then, in Section ~\ref{mat_sec}, we consider $\D$ inside a behaviour $\bG$. From the point of view of the behaviour $\bG$, the part of $\D$ which is significant is  {\em the part that is really used to react to the tests}
  ({\em strategies of $\bG\b$}). This leads to the definition of material part of a strategy in a behaviour.


\subsubsection{Part of a cut-net used by the normalization}\label{used_sec}
When normalizing two strategies $\D,\E$, only a part $\D'\subseteq\D, \E'\subseteq \E$ is used to produce the normal form; this is the  set of views which are examined by the machine in Definition ~\ref{VAMdef}.
If we run the machine on $\{\D,\E\}$ or on $\{\D',\E'\}$, the result will be the same.

\begin{defi}\label{used_part}
Let $\R$ be a cut-net.
For each  $\D\in \R$,  we define  the \textbf{restriction}  of $\D$ to those of its views which are \emph{used} (or \emph{visited}) in the process of interaction
 to produce $\cut \R$ as follows:
$$\D^\R  = \D \cap \{\ulcorner p \urcorner: p\in I(\R) \}.$$
If
$\R  = \{\D, \E_1, \dots, \E_n\}$, we will also  write $\D[\E_1, \dots, \E_n]$ for $\D^\R$.

\end{defi}

\begin{lem}\label{bounded_prop}
Let $\R$ be a cut-net.
For each  $\D\in \R$, $\D^\R$ is a strategy.
\end{lem}
\proof
Let $\D\in \R$.
Since $\D^\R\subseteq\D$, to show that it is a strategy, we only need to show that it satisfies $\tau$-Positivity, in the sense of
Definition ~\ref{N.U. def}. Let $s.a$ be maximal in $\D^\R$. By construction, $s.a=\view{p}$, for some $p=x_1\dots x_n\in I(\R)$. Moreover, by definition of view extraction $\view{}$ (Definition ~\ref{View})  $x_n=a$. If $a$ is a proper negative action, by definition of  $I(\R)$, $p $ must have an extension.
Let us consider a minimal one, $p.c^+$. It has to  be that
$\view{p}.c^+\in \D$ (by Definition ~\ref{VAMdef} of VAM).  We have that $\view{p.c^+}=\view{p}.c^+=sa.c^+\in \D^\R$, against the hypothesis that $s.a$ is maximal.
\qed

The part of a strategy which we have defined in Definition ~\ref{used_part} is the part of the cut-net in which normalization takes place.

\begin{lem}\label{interno} Let $\R = \{\D_1, \dots,\D_n\}$   be
a cut-net. We have that
$$\pl \D_1, \dots,\D_n  \pr = \pl \D_1^\R,\dots,\D_n^\R\pr.$$
\end{lem}
\proof It is a straightforward consequence of Definition ~\ref{VAMdef}, since  all the views of $ \D_1, \dots,\D_n $ which are used for the
  construction of $I(\R)$  are contained in
 $ \D_1^\R,\dots,\D_n^\R$.
\qed


\subsubsection{Material part of a strategy in a behaviour}\label{mat_sec}
If $\D$ is a strategy in a behaviour $\bG$, the part of $\D$ which is used to interact with all the tests (\ie the strategies in $\bG\b$)
is the only significant part of $\D$
from the point of view of the behaviour.

To make the notion of materiality easier to grasp, we first give he definition in the case of behaviours on unary interfaces,
and then generalize it to sequents of behaviours.

\begin{defi}[Materiality] \label{mater_def}
Let $\bG$ be a behaviour and $\D$   a strategy in $\bG$. We define the \textbf{material part}
of $\D$ in $\bG$ as  $$|\D|_\bG := \displaystyle \bigcup \big\{ \D[\E] : \E \in \bG\b \big\}.$$
A strategy $\D \in \bG$ is said to be \textbf{material} in $\bG$ if  $\D = |\D|_\bG$.

For $\bGG = \bG_1,\ldots,\bG_n$, let  $\vdash \bGG$ be a sequent of behaviours,   $\D \in \ \vdash \bGG$, and $(\E_i)$ a family of counter-strategies
$\E_1 \in \bG_1^\bot,\ldots,\E_n \in \bG_n^\bot$.
The material part of $\D$ in $\vdash \bGG$
is  defined as
$$ |\D|_{\vdash \bGG} := \displaystyle \bigcup \big\{ \D[(\E_i)] : \E_1 \in \bG_1^\bot,\ldots,\E_n \in \bG_n^\bot \big\}.$$
 A strategy $\D \in \ \vdash \bGG$ is said material in $\vdash \bGG$ if  $\D = |\D|_{\vdash \bGG}$.
\end{defi}

As an immediate consequence of Lemma ~\ref{bounded_prop}, we have that:

\begin{lem} $|\D|_\bG$ and
 $|\D|_{\vdash\bGG}$
 are strategies. \qed
\end{lem}

\begin{exa}
Let us consider $\D$ and $\E$ as in  Example
 ~\ref{matesempio}. We  see that $\D$ is material in $\bE$ whereas $\E$ is not.
 \end{exa}

The content of the definition of materiality is made explicit by the properties below.

\begin{prop} Let $\bG$ be a behaviour {on a unary interface} and $\D \in \bG$. We have:
  \begin{enumerate}[\em(1)]

 \item $\pl |\D|_\bG, \E \pr = \pl \D,\E \pr$,  for each $\E\in \bG\b$.

 \item  $|\D|_\bG \bot \E$,   for each $\E\in \bG\b$. In particular,
$|\D|_\bG \in \bG$.
\end{enumerate}
\end{prop}
\proof

By Lemma ~\ref{interno}, we have that
 $\pl \D[\E],\E \pr=\pl\D,\E\pr$. From this we have (1) which implies  (2).
\qed
The previous proposition obviously extends to sequents of behaviours.

\begin{lem}\label{neg_root_mat} Let $\bN= (\bN_{\xi 1}\bullet \dots \bullet \bN_{\xi n})\b$ be a behaviour on interface $\xi^-$.   We have that if $\F$ is material in $\bN$,
then it has a single root which is  $x=(\xi,I_n)^-$.\\
 In other words, with the notations introduced in Lemma \emph{~\ref{poscomp_lem}},
 we have $|\F|_\bN \in \bN\single{x}$, for each $\F\in \bN$.
\end{lem}

\proof
The proof  is actually part of the proof of Lemma ~\ref{poscomp_lem} (2), because we have proven that given $\F\in \bN$, for each $\D\in \bN\b$,
we have
 $\cut{\D,\F}=\cut{\D,\F_x}$,   which in turn implies $\F[\D] \subseteq \F_x$.
\qed

\begin{exa}[Materiality with constant types] \label{ex mat top} Let us fix a name $\xi$.
The  material and deterministic strategies in the constant type $ \semone$
are  $\D=(\xi,\emptyset)^+$ and $\Dai$. Moreover,
$\D$ is \emph{winning} in $\semone$ (see Definition  ~\ref{winning strat}).

The only material and deterministic strategy which inhabits $\sembot$   is $\E = (\xi,\emptyset)^-.\Dai$. In particular,
there are no winning strategies in $\sembot$, \ie
`` no syntactical derivations of $\vdash \negm \ult$."
\end{exa}

\subsection{Completeness theorems} \label{completenessA}
In Section ~\ref{intF}, we describe the interpretation of a formula $F$  of $\HS$ into a
behaviour $\bF$ and similarly the interpretation of a syntactical sequent of $\HS$ into a sequent of behaviours. A derivation
of a sequent in $\HS$ will be interpreted by a \emph{winning strategy} {  which belongs to the interpretation of the sequent}.

\begin{defi}[Winning strategy] \label{winning strat}
A strategy $\D \in \ \vdash \bGG$  is said \textbf{winning} if
it is finite, deterministic, daimon-free and material in $\vdash \bGG$.
\end{defi}

In  the sequel,  finiteness, determinism, daimon-freeness and materiality
are also called \emph{winning conditions}.

\begin{rem}[Finiteness condition] We here assume {\em finiteness} among the winning conditions.
However, recent work by
Basaldella and Terui  ~\cite{BasalTLCA09}  shows
an exciting property of  interactive types:
any material, deterministic and
daimon free strategy in a
behaviour
which is interpretation
of  a logical formula is finite.
We are confident that this result is also
valid our setting;   we need to  check this
in detail and   we postpone
it to a subsequent work.
\end{rem}

The rest of this article is then devoted to proving the following
 theorems.

\begin{enumerate}[\hbox to8 pt{\hfill}]
\item\noindent{\hskip-12 pt\bf Soundness:}\ (Theorem ~\ref{interpretation st})
Let $\pi$ be a derivation of a sequent $\vdash \Gamma$ in $\HS$
interpreted by a sequent of behaviours $\vdash \bGG$.
There exists a winning strategy  $\D(\pi) \in \ \vdash \bGG$.
\item\noindent{\hskip-12 pt\bf Full Completeness:}\ (Theorem ~\ref{full compl st}) Let $\vdash \bGG$ be the interpretation of
a sequent $\vdash \Gamma$ of $\HS$ and
let  $\D\in \ \vdash \bGG$.  If  $\D$ is  winning, then
$\D$ is the interpretation of a cut-free derivation $\pi$ of the sequent $\vdash \Gamma$ in $\HS$.
\end{enumerate}

\subsection{Interpretation of formulas, sequents, derivations}\label{intF}
In the rest of this work, \emph{we only consider
behaviours
inductively defined as follows},
using the types constructors introduced in Section ~\ref{type sez}:
$$\bP_\xi::= \bF^+(\bN_{\xi1},\ldots,\bN_{\xi n}) \ \qquad \ \bN_\xi::=
 \bF^-(\bP_{\xi 1},\ldots,\bP_{\xi n}) $$
 where $n \geq 0$ and $\xi$ is an arbitrary name.

We now interpret formulas and sequents of $\HS$.
Intuitively, given a $\HS$ formula $F$ we associate to it
a behaviour $\bF$ of the same polarity. Given a sequent $\vdash F_1,\ldots,F_n$,
we  associate to the occurrences of formula
$F_1,\ldots,F_n$  behaviours $\bF_1,\ldots,\bF_n$  to form
a sequent of behaviours $\vdash \bF_1,\ldots,\bF_n$.

\begin{defi}[Interpretation of formulas and sequents of $\HS$] \label{def_int seq} \hfill
\begin{enumerate}[(1)]
\item[(1)] Given a formula $F$ and an arbitrary name
$\xi$ we associate a behaviour of the same polarity $\bF$ on interface $\xi$
inductively as follows.
$$
 \begin{array}{rcl}
\posm(N_1\otimes \cdots \otimes N_n) \ ^\xi & := & \bF^+( N_1 \ ^{\xi1},\ldots, N_n \ ^{\xi n}); \\ \negm(P_1\parr \cdots \parr P_n) \ ^\xi & := & \bF^-( P_1 \ ^{\xi1},\ldots, P_n \ ^{\xi n}).  \\
 \end{array}
$$

\item[(2a)]
Given a  positive sequent  $\vdash P_1,\ldots,P_n$,
and a positive interface $\xi_1^+ \ldots,\xi_n^+$
we associate the
 sequent of behaviours $\vdash P_1 \ ^{\xi_1},\ldots,P_{n}
 \ ^{\xi_n}$.

\item[(2b)]
Given a  negative sequent  $\vdash N, P_1,\ldots,P_n$
and a negative interface $\sigma^-,\xi_1^+ \ldots,\xi_n^+$
we associate the
 sequent of behaviours $\vdash N \ ^{\sigma}, P_1 \ ^{\xi_1},\ldots,P_{n}
 \ ^{\xi_n}$.
\end{enumerate}
\end{defi}\medskip

\noindent For instance, we have $\posm \uno  \ ^\xi = \semone$ and  $\negm \ult \ ^\xi =
\sembot$ on interface $\xi^+$ and $\xi^-$ respectively.

In the sequel, we  indicate the behaviour $ F \ ^\xi$
by $\bF_\xi$  or just by $\bF$ when the name $\xi$ is  clear from the context or irrelevant for our purposes.
Similarly, we write $\vdash \bP_{\xi_1},\ldots,\bP_{\xi_n}$
and $\vdash \bN_{\sigma}, \bP_{\xi_1},\ldots,\bP_{\xi_n}$
or just $\vdash \bPP$ and   $\vdash \bN,\bPP$.\smallskip

We are now ready to define the interpretation of a derivation $\pi$
of a sequent $\vdash \Gamma$ in $\HS$.

\begin{defi}[Interpretation of derivations of $\HS$] \label{def interpretation}
Let $\pi$ be a derivation
of a sequent $\vdash F_1,\ldots,F_n$ in $\HS$
and $\vdash \bF_{\xi_1},\ldots,\bF_{\xi_n}$ its interpretation
on an  arbitrary interface $\xi_1,\ldots,\xi_n$.
The \textbf{interpretation of $\pi$ in} $\vdash \bF_{\xi_1},\ldots,\bF_{\xi_n}$ is a strategy $\D(\pi)$ on interface
$\xi_1,\ldots,\xi_n$
inductively defined
as follows.

\begin{enumerate}[\hbox to8 pt{\hfill}]
\item\noindent{\hskip-12 pt\bf Positive rule:}\  $\pi$ ends with a positive rule \textrm{Pos}$_n$ and the principal formula   is $P = \posm(N_1 \otimes \cdots \otimes N_n)$.
\begin{center}
\AxiomC{$\vdots$ $\pi_1$}\noLine\UnaryInfC{$\vdash \Pi,P,N_1$}
\AxiomC{$\ldots$}
\AxiomC{$\vdots$ $\pi_n$}\noLine\UnaryInfC{$\vdash \Pi,P,N_n$}
\RightLabel{\scriptsize{Pos$_n$}}
\TrinaryInfC{$ \vdash  \Pi, P$} \DisplayProof
\end{center}
Let $\D(\pi_i)$ be the interpretation of $\pi_i$ in
 $\vdash \bPP, \bP_\xi, \bN_{\sigma i}$, for  $1 \leq i \leq n$. The strategy
$$\D(\pi) :=
(\sigma,I_n)^+.\{\D_1, \ldots,\D_n\}[\xi/\sigma]
$$
is the interpretation of $\pi$ in $\vdash  \mathbf{\Pi}, \bP_\xi$.
\item\noindent{\hskip-12 pt\bf Negative rule:}\  $\pi$ ends with a negative rule \textrm{Neg}$_n$, where $N = \negm(P_1 \parr \cdots \parr P_n)$.
\begin{center}
\AxiomC{$\vdots$ $\pi_0$}\noLine\UnaryInfC{$\vdash  \Pi,P_1,\ldots,P_n$}
\RightLabel{\scriptsize{Neg$_n$}}
\UnaryInfC{$\vdash \Pi,N$} \DisplayProof
\end{center}

Let $\D(\pi_0)$ be the interpretation of $\pi_0$ in
 $\vdash \bPP, \bP_{\xi 1},\ldots, \bP_{\xi n}$. The strategy
$$\D(\pi) :=
(\xi,I_n)^-.\D(\pi_0)
$$
is the interpretation of $\pi$ in
$\vdash \mathbf{\Pi}, \bN_\xi$.

\item\noindent{\hskip-12 pt\bf Cut:}\ $\pi$ ends with a \textrm{Cut} rule:
\begin{center}
\AxiomC{$\vdots$ $\pi_1$}
\noLine \UnaryInfC{$\vdash \Xi,\Pi,P$}

\AxiomC{$\vdots$ $\pi_2$}
\noLine
\UnaryInfC{$\vdash \Delta,P\b$}
\RightLabel{\scriptsize{Cut}}
\BinaryInfC{$\vdash \Xi,\Pi,\Delta$}
\DisplayProof
\end{center}
Let $\D(\pi_1)$ and $\D(\pi_2)$ be the interpretation
of $\pi_1$ and $\pi_2$ in
$\vdash \mathbf{\Xi},\bPP,\bP_\xi$
and $\vdash \bDD,\bP^\bot_\xi$ respectively. The strategy
 $$\D(\pi) :=
\pl \D(\pi_1),\D(\pi_2) \pr
$$
is the interpretation of $\pi$ in $\vdash \mathbf{\Xi},\bPP, \bDD$.   (Here we are  assuming that the names in $\bPP$ and $\bDD$ are disjoint.)

\end{enumerate}
\end{defi}

\begin{exa}
Let us consider the case $n=0$ in the interpretation
of the positive and negative rule respectively.
Following the previous definition (with the same notation) we have:
\begin{enumerate}[$\bullet$]
\item The strategy $(\xi,\emptyset)^+$
is the interpretation in $\vdash \bPP,\semone_\xi$ of the derivation
\AxiomC{}
\RightLabel{\scriptsize{Pos$_0$}}
\UnaryInfC{$\vdash \Pi,\posm \uno$}
\DisplayProof \ (where the behaviour $\semone_\xi$ interprets the occurrence of the principal formula
 $\posm \uno$).
\item
The strategy $(\xi,\emptyset)^-.\D(\rho)$
 is the interpretation in $\vdash \bPP,\sembot_\xi$ of the derivation
\begin{center}
\AxiomC{$\vdots$ $\rho$} \noLine
\UnaryInfC{$\vdash \Pi$}
\RightLabel{\scriptsize{Neg$_0$}}
\UnaryInfC{$\vdash \Pi,\negm \ult$}
\DisplayProof
\end{center}
\end{enumerate}

\end{exa}

\subsection{Soundness}
Having fixed the interpretations of formulas, sequents
and derivations we are ready to show:

\begin{thm}[Soundness]  \label{interpretation st}
Let $\pi$ be a derivation  of a sequent $\vdash \Gamma$ in $\HS$  and
 $\D(\pi)$ be the interpretation of $\pi$
in a sequent of behaviours  $\vdash \bGG$. We have that:
\begin{center}$\D(\pi)$ is a winning strategy  in $\vdash \bGG$.
\end{center}
   Moreover, the interpretation is invariant under cut-elimination.
 \qed
\end{thm}

The proof consists of several lemmas
and it is given in Appendix ~\ref{soundness appendice}.


\subsection{Full completeness} \label{full completeness}

   We can finally prove the following:

\begin{thm}[Full Completeness] \label{full compl st}
Let $\vdash \bGG$ be a  sequent of behaviours which is interpretation of the sequent $\vdash \Gamma$ in $\HS$.
If $\D$ is a  winning strategy
in   $\vdash \bGG$
then it
 is the interpretation of a  cut-free derivation $\pi$ of the sequent $\vdash \Gamma$ in $\HS$.

\end{thm}
\proof
Since  $\D$ is finite, we can reason by induction on the number $\nu$ of its actions.
Let us examine the other  implications of the winning conditions on $\D$.
Since  $\D$ is a daimon-free and deterministic strategy, all occurrences of actions in $\D$ are proper actions. Moreover, we have that $\D$ is non-empty, and has a single root (both properties always hold for positive strategies, and here also in case $\D$ is negative by Lemma ~\ref{neg_root} and ~\ref{neg_root_mat}, because $\D$ is material).

Like in the linear case, we will use | back and forth | the definition  of sequent of behaviours and more precisely Proposition ~\ref{closure2}.
Let $\vdash \bGG$ be the interpretation of the sequent $\vdash \Gamma$, and $\D\in \ \vdash \bGG$
 a winning strategy.
Our purpose is
 to associate to $\D$ a derivation $\D^\star$
of   $\vdash \Gamma$
 in
$\HS$, by progressively decomposing $\D$, \ie
inductively writing ``the last rule.''

The formula on which the last rule is applied is indicated by the name of the root action.
The argument is the same as in the linear case.  For example, assume that the root of $\D$ is $(\xi,I)$;
then if  $\D\in \ \vdash \bF_\xi,\bQ_\sigma$,
the behaviour which corresponds to the last  rule is the one on $\xi$, \ie $\bF_\xi$.\\

In the sequel,  we  consider sequents
of behaviours of the form $\vdash \bPP,\bF_\xi$, which are interpretations of a sequent $\vdash  \Pi, F$ in $\HS$.
$\bF$ is the interpretation of a formula $F$
and $\bPP$ is a sequence of $m$ behaviours  $\bQ_{\alpha_1},\ldots,\bQ_{\alpha_m}$  which respectively interpret formulas
$Q_1, \dots, Q_m$.
Observe that
  $\bPP$  always consists of positive behaviours only.

  We use the following convention:
  we write $(\E)$ and
  $(\E) \in \bC$ for $\E_1,\ldots,\E_m$
  and
   $\E_1 \in \bQ_{\alpha_1}^\bot,\ldots,
   \E_m \in \bQ_{\alpha_m}^\bot$
   respectively.

We have two cases.
\subsubsection*{Positive case}
Let  $\D = (\xi,I_n)^+.\{\D_1,\dots,\D_n\}$ ($n\geq 0$) be a positive winning
strategy which belongs to $\vdash \bPP, \bP_\xi$,
 where $\bP_\xi=\bF^+(\bN_{\xi 1}, \dots, \bN_{\xi k})$
 is the  interpretation of the formula of $\HS$
$P = \posm (N_1 \otimes \cdots \otimes N_k)$.

 By  Proposition ~\ref{closure2},  for any $(\E)\in \bC$, we have that
$\C=\cut{\D,(\E)}\in \bP_\xi$.  The root of
$\pl \D,(\E) \pr$
is still $(\xi,I_n)^+$ (by the definition of the abstract machine, since it is a visible action which is root of the main). Hence $k=n$.

We  now  use internal completeness.

\begin{enumerate}[(1)]
\item By  internal completeness of positive connectives (Proposition
~\ref{int comp F+}), we have that the strategy
$\rt{\pl \D,(\E) \pr}$ can be written as $\C_1\bullet \dots \bullet \C_n$,
where  $\C_i\in \ \vdash \bP_\xi,\bN_{\sigma i}$.
\item It is immediate
 that $\rt {\cut{\D,(\E)}} = \cut {\rt \D,(\E)} $. By Lemma ~\ref{projections} , we can write $\rt \D=(\sigma,I_n)^+.\{\D_1',\dots,\D_n'\}$ as
$\D_1' \bullet \dots \bullet \D_n'$.  By Lemma ~\ref{conversion}, we have that
$ \cut{  \D_1' \bullet \dots \bullet \D_n', (\E)}  = \cut{\D_1',(\E)} \bullet \dots \bullet \cut{\D_n',(\E)}$.

From (1), we conclude that $\cut{ \D'_i,(\E)}\in \ \vdash
\bP_\xi, \bN_{\sigma i}$.
\end{enumerate}
By applying
Proposition ~\ref{closure2} again, we have that
$\D'_i\in \ \vdash \bPP, \bP_\xi,\bN_{\sigma i}$.

If $\nu = 1$, $\D$ consists of a single action  $(\xi,I_n)^+$, and we must have  $n=0$.
Otherwise, we would  have empty $\D'_i$'s, and we already observed that this cannot be the case.
Hence, $\nu = 1$ gives $\bP_\xi= \semone$, and we have that     $\D =  (\xi,\emptyset)^+$ is the interpretation of the  rule \rm{Pos$_0$}.

 If $\nu >1$, we note that the
  winning conditions are preserved for each $\D'_i$ (by Lemma ~\ref{matpos1} and Lemma ~\ref{matpos2} below)
  and
  the number of actions  decreases.
 Hence, the inductive hypothesis applies, and
 we can  write the derivation:
\begin{center}
\AxiomC{$\vdots \ \D_1^{'\star}$}\noLine
\UnaryInfC{$\vdash \Pi,P,N_1$}
\AxiomC{$\ldots$}
\AxiomC{$\vdots \ \D_n^{'\star}$}\noLine
\UnaryInfC{$\vdash \Pi,P,N_n$}
\RightLabel{\scriptsize{Pos$_n$}}
\TrinaryInfC{$ \vdash   \Pi,P$} \DisplayProof
\end{center}

\subsubsection*{Negative case}
Let us  consider a negative winning strategy
$\D \in \ \vdash \bPP, \bN_\xi$, where the behaviour $\bN_\xi=\bF^-(\bP_1,\dots,\bP_n)$ is the interpretation of a formula
 $N = \negm(P_1 \parr \cdots \parr P_n)$ of $\HS$.
Let
$\D=x^-.\D'$, with $x=(\xi,I_n)$.

For any  family $(\E) \in \bC$, we have
\begin{enumerate}[(1)]
\item
$\pl \D,(\E)\pr \in \bN_\xi$, and the root is still $x^-$. This
allows us
to use internal completeness.

\item By internal completeness of negative connectives (Proposition~\ref{neg_comp}), we conclude that
    $\pl \D,(\E)\pr$ is of the form $x^-.\D''$
    with $\D'' \in \ \vdash \bP_{\xi 1},\dots, \bP_{\xi n}$.

    \item By the definition of normalization,
    $$ \pl \D,(\E) \pr = \pl x^-.\D', (\E) \pr = x^-.\pl \D',(\E)\pr.$$
From this, we have that
 $\D'' =
\pl \D',(\E)\pr$ and hence
 $\pl \D',(\E)\pr \in \ \vdash \bP_{\xi 1},\dots, \bP_{\xi n}$.

  \end{enumerate}
By applying
Proposition ~\ref{closure2} again, we have that
 $ \D'\in
 \ \vdash \bPP, \bP_{\xi 1},\dots, \bP_{\xi n}$.
The  winning conditions are preserved for $\D'$ (Lemma ~\ref{matneg} below) and
  the number of actions  decreases.
 Hence, the inductive hypothesis applies. Then, we can  write the derivation:
\begin{center}
\hfill
\AxiomC{$\vdots \ \D^{'\star}$}\noLine
\UnaryInfC{$\vdash\Pi,  P_1,\dots,P_n$}
\RightLabel{\scriptsize{Neg$_n$}}
\UnaryInfC{$ \vdash   \Pi,N$} \DisplayProof\qed
\end{center}\medskip

\noindent We still have to prove that the winning conditions are preserved by the deconstructions
we have used in the last proof. The only condition which is not obvious is materiality.

To show that a strategy $\D$ is material in $\vdash \bGG$, we check that for each $s\in \D $ there is a certain family of counter-strategies $(\E) \in \bC$  such that   $s=\view p$ and $ p\in I(\D,(\E))$. In other words,
$s$ is used to produce the normal form $\cut{\D,(\E) }$ (see Definitions ~\ref{mater_def} and ~\ref{used_part}).

\begin{lem}\label{matpos1} Let $\bP_\xi=\bF^+(\bN_{\xi 1}, \dots, \bN_{\xi n})$,  $\D\in \ \vdash  \bPP,\bP_\xi$ and $\rt \D \in \ \vdash \bPP, \bP_\xi,\bP_\sigma$.
If $\D$ is material in $\vdash \bPP,\bP_\xi$ then  $\rt \D$ is material in $\vdash \bPP,\bP_\xi,\bP_\sigma$.

\end{lem}

\proof \wloge, we  assume $\bPP=\bQ_{\alpha}$.
If $\D:\alpha^+, \xi^+$ is material, it means that for each $t\in \D$, we have $t=\view q$, with $q\in I(\D,\E,\A)$ for a certain
$\E \in \bP_\xi^\bot$ and a certain $\A \in \bQ_\alpha^\bot$.

  Let us consider $\R=\{ \rt \D,\E, \E[\sigma/\xi],\A  \}$ and $\R'=  \{ \D,\E,\A  \}$.
As discussed in the proof of Proposition ~\ref{prop copies vera},  $p\in I(\R)$   if and only if
  there is a ``corresponding"  $p' \in I(\R')$,
  where $p'$ is obtained from $p$ by the operation of renaming of pointing strings described in  the proof of Proposition
~\ref{prop copies vera}.

Now, let us consider a view $s\in \rt \D:\alpha^+,\xi^+,\sigma^+ $.
We have that  $s'\in \D$ (where $s'$ is obtained from $s$ by the same operation of renaming), and since $\D$ is material,
$s'=\view{q}$ for $q\in I(\D,\E,\A)$, where  $\E$ and $\A$ are those provided by the materiality of $\D$. Let    $p$ be the
pointing string  such that $q=p'$.
 We conclude that  $s=\view p$, for
  $p\in I(\rt \D,\E, \E[\sigma/\xi],\A)$.
\qed

\begin{lem}\label{matpos2}Let $\bP_\sigma=\bF^+(\bN_{\sigma 1}, \dots, \bN_{\sigma n})$, $\D=\D_1\bullet \dots \bullet \D_n\in \ \vdash \bPP,\bP_\sigma$, and  $\D_i\in \ \vdash \bPP,\bN_{\sigma i}$.
If $\D_1\bullet \dots \bullet \D_n$ is material in $\vdash \bPP,\bP_\sigma$, then each $\D_i$ is material in $\vdash \bPP,\bN_{\sigma i}$.
\end{lem}
\proof  For concreteness, we only discuss the case $\bPP=\bQ$, and $n=2$ (the general case being a simple generalization). Let $\D=x^+.\{\D_1,\D_2\}$ (we recall that
$x$ occurs linearly in $\D$).
  For each $x^-.\E\in \bP_\sigma^\bot $,  we have  by internal completeness (Proposition ~\ref{neg_comp}) that
   $\E\in \ \vdash \bN_{\sigma 1}^\bot,\bN_{\sigma 2}^\bot$; moreover
 for each  $\A\in \bQ\b$,
   $\cut{\D_i,\A}\in \bN_{\sigma_i}$.

Let $s\in \D_1$. By construction,
  $x^+.s$ is a view in $\D$, and since $\D$ is material,  $x^+.s$
 belongs to $\D[x^-.\E,\A]$, for a
 certain $x^-.\E\in \bP_\sigma^\bot $ and a certain $\A\in \bQ\b$ (we restrict our attention to single-rooted strategies in $\bP_\sigma^\bot $,
 thanks to  Lemma ~\ref{singleroot}). It is then straightforward to check that
if $x^+.s$ is used in the interaction which  produces
 $\cut{x^+.\{\D_1,\D_2\},x^-.\E, \A}$,
  then  $s$ is used in the interaction which produces  $\cut{\D_1, \F,\A}$, where   $\F:=\cut{\E,\D_2,\A}\in \bN_{\sigma 1}^\bot$.
\qed

\begin{lem}\label{matneg} Let $\bN_\xi = \bF^-(\bP_{\xi 1},\dots,\bP_{\xi n})$. If $x^-.\D\in \ \vdash \bPP,\bN_\xi$ is material, then $\D\in \ \vdash \bPP, \bP_{\xi 1},\dots,\bP_{\xi n}$ is material.

\end{lem}

\proof The argument is  similar to the one for Lemma ~\ref{logical}(2), or rather it is its inverse.
The only subtlety is that to ensure that the root of the counter-strategy we fix in $\bN_\xi^\bot $ is linear, we  need to
  go through some lengthy but straightforward steps of renaming (which  we omit here) by repeated use of Proposition \ref{prop copies vera}.
\qed

\section{Conclusion} \label{conclusion}
In this work, we started by recalling the standard
notion of
HO strategy and we have
shown how ludics strategies can be expressed
in term HO strategies by giving an universal arena.
We have revised the main results
of the higher-level part of ludics (namely, \emph{internal
completeness}) giving direct proofs of them using
basic properties of the dynamics only.
We have motivated and introduced the notion
of non-uniform strategy and shown
that we still have   a suitable form of  internal completeness
when strategies are non-linear and
non-uniform.
From this,
we finally have shown a full completeness
result with respect to  $\HS$.

\subsection*{Related and future work}
\subsubsection*{Maurel's exponentials}
Maurel  ~\cite{MauTh} has built a sophisticated setting to recover a form of separation when having repetitions in ludics;
however, the complexity of the setting prevented him from going further and studying
 interpretation and full
completeness issues.
In this paper, we ignore separation all together, and in fact we
 show that
we don't need it in order to have
interactive types and internal completeness.
In future work, we hope it may be possible to refine our setting by using Maurel's techniques.
 In Maurel's setting, strategies have
 a quantitative information carried by probabilistic values (called {\em coefficients}).
The values in the coefficients
  have a central role, and must satisfy a  set of ``quantitative conditions'' inspired by measure theory.
This is  fundamentally different from our ``indexed silent actions", as the specific natural number which is
chosen as index for a silent action is irrelevant
 (in particular, all the indexes  can be interchanged, and this does not affect
orthogonality), and there are no condition attached.
 Our indexed silent actions have the same role as in ~\cite{varyosh}.
 However,  in a way, we think that our approach could be seen as a simplification | or  rather a     kind of  quotient |
on Maurel's coefficients;
on this grounds, we hope
it  may be possible to refine our silent actions by attaching probabilities to them,  without
losing our high-level results.

\subsubsection*{AJM style exponentials for ludics}
A different solution that uses
AJM  style exponentials has been studied  by
the first of the two authors in ~\cite{thesisbas}.
Essentially, the strategies which inhabit a semantical type $\negm \bA$ are those of the form $(\D,1) \cup (\E,2) \cup \dots$ \ :
an \emph{indexed (and disjoint) superimposition}
of strategies $\D,\E,\dots$ of $\bA$.
However,  the approach we use  in this paper, which exploits similar ideas | namely the disjoint superimposition which is technically implemented by using silent actions here |
 is considerably simpler,   as we do not need
 to consider further (rather complex) ``uniformity" conditions
 to discriminate those  strategies which are interpretations of syntactical derivations; here the ``uniform strategies"  are simply the deterministic strategies.
We think that the approach to repetitions we implement in this paper is
more suitable for more applicative uses of ludics ~\cite{FagPic,alexis,Terui}.

\subsubsection*{Computational ludics}
By using the  approach we present in this paper,
Basaldella and Terui  ~\cite{BasalTLCA09}
have recently  extended Terui's computational ludics
\cite{Terui}  in order to accommodate
exponentials.

Their paper is aimed at analyzing  the
traditional logical duality between
proofs and models from
the point of view of ludics and  they get an alternative proof
of full completeness
based on a direct construction of a counter-model. Very interestingly, that work
  also reveals
an exciting property of the  ``\emph{interactive types}.''
Unlike in standard HO game semantics, finiteness does not need to be requested as a condition
for strategies to be winning; it is rather an outcome of the closure by orthogonality.
In fact, Basaldella and Terui show that
any material, deterministic strategy in a
behaviour
which is interpretation
of  logical formula is finite.
We are confident that this result is also
valid our setting.    However,  a careful verification
of all the details is needed.

\subsubsection*{Non-deterministic innocent strategies}

They have been introduced by Harmer in ~\cite{harthesis}, with
 the purpose
of modeling  non-determinism (in  PCF with
erratic choice).\\
In this paper we introduce non-uniform strategies, which are realized by means of
 {\em non-deterministic sums}.
However, the purpose of our {\em non-deterministic sums} is  to implement {\em non-uniformity} via ``formal sums'' of strategies, in order
 to provide enough tests to make possible the
interactive approach of ludics. The different purpose is reflected in the composition, which is simpler in our setting, where is in fact reduced to deterministic composition.\\
Our strategies could be seen as a
 ``concrete'' implementation of Harmer's solution, in a simplified setting.
Harmer overcomes the problems with composition
 moving from naive non-deterministic strategies $S: A\rightarrow  B$
to an ``indirect'' definition of strategies of the kind $S: A \times \mathbb{N} \rightarrow  B$.
We have instead silent actions,  which can be seen as $\tau$ actions carrying an index $i\in \mathbb{N}$. These actions have
a two-fold role: they guard the sum (as in ~\cite{varyosh}), and provide an ``index of  copy'' (as in AJM  game semantics,
but here the index is unfold only when needed), but do not go so far as to model non-determinism. In particular, we do not introduce any quotient on the strategies.

\subsubsection*{Game semantics for linear logic}
When we work with innocent strategies, in this paper we consider the variant of HO strategies introduced  in  ~\cite{LauSvS}.
Having now enriched the setting of ludics with duplicative features, we want to make comparisons with
other   kinds of  game semantics for linear  logic.
In particular, we are interested in finding
connections  between our treatment of duplication and the resource modalities of games semantics
introduced by Melli{\`e}s and   Tabareau in ~\cite{MelTab}.

\subsubsection*{Abstract machines}
Curien and Herbelin in ~\cite{CurHerb} have studied
composition of strategies as sets of views.
In particular they have developed the  View-Abstract-Machine (VAM)
(see also ~\cite{CurAbs})
which is the device we use in this paper.

\section*{Acknowledgement}
We are in debt with Olivier Laurent for his  sharp remarks and detailed suggestions which
significantly contributed to improve an earlier version of our work.
Many thanks to Pierre-Louis Curien,   Kazushige Terui, Mauro Piccolo, for fruitful discussions and helpful suggestions.
Finally, we
gratefully
 acknowledge  the
 anonymous referees whose
 in depth revision and many detailed comments have given an invaluable aid in improving the quality of this paper.



\appendix

\section{Admissibility of the \textrm{Cut}-rule
and expressivity of \texorpdfstring{$\HS$}{HS}} \label{calcolo_app}

In  Section ~\ref{cut adm_app} we prove that the cut-rule is admissible in $\HS$.

We then   discuss the expressivity of
 $\HS$ by relating it to more standard systems (in Sections ~\ref{intuition} and ~\ref{corrispondenza}).
Since we are interested
in clarifying the nature of the logical rules and sequents of $\HS$, we omit the cut-rule in all system.

In the sequel we use  extensively
 the  following notions.

\begin{defi}[Depth, height] \label{depth height}
We define the
 \textbf{depth} of a formula $F$, noted by $d(F)$
 as the length of the longest branch of $F$ in the obvious
 tree representation of $F$.
More precisely,   let  $F$ be a formula in some language and   $G_1,\ldots,G_n$
its immediate subformulas. We define   $d(F)$
as the natural number inductively given by:
$$
 d(F)  :=   \sup\{d(G_1),\ldots, d(G_n) \} +1
.$$

Let $\pi$ be a syntactical derivation of a sequent in
some system.
The \textbf{height} of $\pi$ , noted by $h(\pi)$,
is inductively given as follows:
if $\pi$ ends with a rule whose premises are derived by
$\pi_1,\ldots,\pi_n$, then $$h(\pi) := \sup \{h(\pi_1),\ldots, h(\pi_n) \} +1.$$
\end{defi}\medskip

\noindent For instance, if $F$ (resp.\ $\pi$)  is a formula (resp.\ derivation) in $\HS$, we have that:
\begin{enumerate}[$\bullet$]
\item
  $d(F) =1$ if and only if
 either $F= \posm \uno$ or $F=\negm \ult$;
\item
 $h(\pi) = 1$
if and only if $\pi$ is of the form
 \AxiomC{}
\RightLabel{\scriptsize{Pos$_0$}}
\UnaryInfC{$\vdash \Pi, \posm \uno$}
\DisplayProof \ for some  $\Pi$.
\end{enumerate}

The following lemma is also useful in the sequel.

\begin{prop}[Structural rules] \label{struct} \hfill
\begin{enumerate}[\em(1)]
\item \emph{Weakening}:  if $\vdash \Gamma$ is cut-free derivable then $\vdash \Gamma,Q$ is cut-free derivable.

\item  \emph{Contraction}: if $\vdash \Gamma,Q,Q$ is cut-free derivable then $\vdash \Gamma,Q$ is cut-free derivable.
\end{enumerate}
\end{prop}
\proof
\hfill
\begin{enumerate}[(1)]
\item Let $\pi$ a be derivation of $\vdash \Gamma$.
By induction on the height of $\pi$ we  now construct a derivation $\pi'$
of $\vdash \Gamma,Q$.
\begin{enumerate}[a.]
\item
$\pi$ ends with a positive rule, where $P=\posm(N_1 \otimes \cdots \otimes N_n)$:

\begin{prooftree}
\AxiomC{$\vdots$ $\pi_1$}\noLine \UnaryInfC{$\vdash \Pi,P,N_1$}
\AxiomC{$\ldots$} \noLine
\AxiomC{$\vdots$ $\pi_n$}\UnaryInfC{$\vdash \Pi,P,N_n$}
\RightLabel{\scriptsize{Pos}$_n$}
\TrinaryInfC{$\vdash \Pi,P$}
\end{prooftree}

By inductive hypothesis, $\pi_i'$ is a derivation
of $\vdash \Pi,Q,P,N_i$ for any $1 \leq i \leq n$.
We can then apply a  rule \textrm{Pos}$_n$  and obtain

\begin{prooftree}
\AxiomC{$\vdots$ $\pi_1'$}\noLine \UnaryInfC{$\vdash \Pi,Q,P,N_1$}
\AxiomC{$\ldots$} \noLine
\AxiomC{$\vdots$ $\pi_n'$}\UnaryInfC{$\vdash \Pi,Q,P,N_n$}
\RightLabel{\scriptsize{Pos}$_n$}
\TrinaryInfC{$\vdash \Pi,Q,P$}
\end{prooftree}

\item
$\pi$ ends with a negative rule, where $N=\negm(P_1 \parr \cdots \parr P_n)$:

\begin{prooftree}
\AxiomC{$\vdots$ $\rho$} \noLine
\UnaryInfC{$\vdash \Pi, P_1,\ldots,P_n$}
\RightLabel{\scriptsize{Neg$_n$}}
\UnaryInfC{$\vdash \Pi, N$}
\end{prooftree}

By inductive hypothesis, $\rho'$ is a derivation of
$\vdash \Pi,Q, P_1,\ldots,P_n$.
We can then apply a  rule \textrm{Neg}$_n$   and obtain

\begin{prooftree}
\AxiomC{$\vdots$ $\rho'$} \noLine
\UnaryInfC{$\vdash \Pi,Q, P_1,\ldots,P_n$}
\RightLabel{\scriptsize{Neg$_n$}}
\UnaryInfC{$\vdash \Pi,Q, N$}
\end{prooftree}

\end{enumerate}

\item
Let $\pi$ a be derivation of $\vdash \Gamma,Q,Q$.
By induction on the height of $\pi$ we  now construct a derivation $\pi'$
of $\vdash \Gamma,Q$.
\begin{enumerate}[a.]
\item
$\pi$ ends with a positive rule,where the principal formula  of \textrm{Pos}$_n$  is
(an occurrence of) some formula  $P = \posm(N_1 \otimes \cdots \otimes N_n)$ in $\Pi,Q,Q$:\vspace{-12 pt}
\begin{prooftree}
\AxiomC{$\vdots$ $\pi_1$} \noLine
\UnaryInfC{$\vdash \Pi,Q,Q,N_1$}
\AxiomC{$\ldots$}
\AxiomC{$\vdots$ $\pi_n$} \noLine
\UnaryInfC{$\vdash \Pi,Q,Q,N_n$}
\RightLabel{\scriptsize{Pos}$_n$}
\TrinaryInfC{$\vdash \Pi,Q,Q$}
\end{prooftree}
By inductive hypothesis, $\pi_i'$ is a derivation of
$\vdash \Pi,Q,N_i$
for any $1 \leq i \leq n$.
Since     $P$ also occurs in
 $\Pi,Q$,  we can then apply \textrm{Pos}$_n$
and  obtain\vspace{-12 pt}
\begin{prooftree}
\AxiomC{$\vdots$ $\pi_1'$} \noLine
\UnaryInfC{$\vdash \Pi,Q,N_1$}
\AxiomC{$\ldots$}
\AxiomC{$\vdots$ $\pi_n'$} \noLine
\UnaryInfC{$\vdash \Pi,Q,N_n$}
\RightLabel{\scriptsize{Pos}$_n$}
\TrinaryInfC{$\vdash \Pi,Q$}
\end{prooftree}

\item
$\pi$ ends with
a negative rule, where $N=\negm(P_1 \parr \cdots \parr P_n)$:

\begin{prooftree}
\AxiomC{$\vdots$ $\rho$} \noLine
\UnaryInfC{$\vdash \Pi,Q,Q, P_1,\ldots,P_n$}
\RightLabel{\scriptsize{Neg$_n$}}
\UnaryInfC{$\vdash \Pi,Q,Q,N$}
\end{prooftree}
 By inductive hypothesis, $\rho'$ is a derivation of
$\vdash \Pi,Q, P_1,\ldots,P_n$.
We can then apply  \textrm{Neg}$_n$ rule
and obtain\vspace{-12 pt}
\begin{prooftree}
\AxiomC{$\vdots$ $\rho'$} \noLine
\UnaryInfC{$\vdash \Pi,Q, P_1,\ldots,P_n$}
\RightLabel{\scriptsize{Neg$_n$}}
\UnaryInfC{$\vdash \Pi,Q,N$}
\end{prooftree}
\qed
\end{enumerate}
\end{enumerate}

\subsection{ Admissibility of the \textrm{Cut}-rule} \label{cut adm_app}

In this section we show
that the cut-rule

\begin{prooftree}
\AxiomC{$\vdash \Xi,\Pi,P$}
\AxiomC{$\vdash \Delta, P\b$}
\RightLabel{\scriptsize{Cut}}
\BinaryInfC{$\vdash \Xi,\Pi,\Delta$}
\end{prooftree}

\noindent is admissible in $\HS$  (more precisely, the cut-rule is admissible
in $\HS$ without \textrm{Cut}).

\begin{thm}[Cut-elimination] \label{principal}
Let $\pi$ and $\rho$ be cut-free derivations of
$\vdash \Xi,\Pi,P$ and $\vdash \Delta, P\b$ respectively.
The sequent $\vdash \Xi,\Pi,\Delta$ is derivable with
a cut-free derivation $\theta$.
\end{thm}
\proof
The proof  by induction on the
 pair $( d(P),h(\pi)+h(\rho))$ | where $d(P)$ and $h(\pi)$
 denote the depth of the formula $P$ and height of the derivation
$\pi$ respectively (Definition  ~\ref{depth height}) |  ordered
 lexicographically:
$(d(P),h(\pi)+h(\rho)) < (d(P'),h(\pi')+h(\rho'))$
if and only if  either $d(P) < d(P')$ or
$d(P) = d(P')$ and $h(\pi)+h(\rho) < h(\pi')+h(\rho')$.

Observe that we are assuming that $\pi$ and $\rho$ are cut-free
derivations, this implies that
the last rule of $\rho$ is always a negative rule
having    $P\b$ as principal formula.

We now give a procedure to obtain a cut-free derivation
$\theta$  of $\vdash \Xi,\Pi,\Delta$.
We have  three distinct cases, depending on the last rule of $\pi$.\smallskip

\noindent {\textbf{(a)}}
The last rule of $\pi$ is a positive rule \textrm{Pos}$_n$  and   $P$
is  principal in \textrm{Pos}$_n$.

For readability, we only consider the cases $n=0$ and $n=2$.
The general $n$-ary case straightforwardly follows.\smallskip

If $n=0$, then
$P= \posm \uno$, $P\b = \negm \ult$ and $d(P)=1$. We have derivations:

\begin{prooftree}
\AxiomC{}
\RightLabel{\scriptsize{Pos$_0$}}
\UnaryInfC{$\vdash \Pi,\posm \uno$}

\AxiomC{$\vdots$ $\rho_0$}
\noLine \UnaryInfC{$\vdash \Delta$}
\RightLabel{\scriptsize{Neg$_0$}}
\UnaryInfC{$\vdash \Delta, \negm \ult$}
\noLine
\BinaryInfC{}
\end{prooftree}
The cut-free derivation
$\theta$ of
$\vdash \Pi,\Delta$,
is obtained from $\rho_0$ by
weakening, in the sense of Proposition
 ~\ref{struct} (1) (weakening on positive formulas).\smallskip

\noindent If $n=2$, then
$P = \posm(N_1 \otimes  N_2)$ and
$P\b = \negm(N_1^\bot \parr  N_2^\bot)$. We have
derivations:\vspace{-12 pt}
\begin{prooftree}
\AxiomC{$\vdots$ $\pi_1$}
\noLine \UnaryInfC{$\vdash \Pi,P,N_1$}
\AxiomC{$\vdots$ $\pi_2$}
\noLine \UnaryInfC{$\vdash \Pi,P,N_2$}
\RightLabel{\scriptsize{Pos$_2$}}
\BinaryInfC{$\vdash \Pi,P$}

\AxiomC{$\vdots$ $\rho_0$}
\noLine \UnaryInfC{$\vdash \Delta, N_1^\bot,N_2^\bot$}
\RightLabel{\scriptsize{Neg$_2$}}
\UnaryInfC{$\vdash \Delta, P\b$}
\noLine
\BinaryInfC{}
\end{prooftree}
We now construct  a cut-free derivation $\theta$
of $\vdash \Pi,\Delta$ as follows.

Consider for $i \in \{1,2\}$ the following
derivations:\vspace{-12 pt}
\begin{prooftree}
\AxiomC{$\vdots$ $\pi_i$}
\noLine \UnaryInfC{$\vdash \Pi,P,N_i$}

\AxiomC{$\vdots$ $\rho$}
\noLine
\UnaryInfC{$\vdash \Delta,P\b$}
\noLine
\BinaryInfC{}
\end{prooftree}
 Since
$h(\pi_i) < h(\pi)$, we have $
(d(P),h(\pi_i)+h(\rho)) < (d(P),h(\pi)+h(\rho)) $ and
the induction hypothesis yields a cut-free
derivation  $\psi_i$ of  $\vdash \Pi,\Delta,N_i$.
Consider now\vspace{-12 pt}
\begin{prooftree}
\AxiomC{$\vdots$ $\rho_0$}
\noLine \UnaryInfC{$\vdash \Delta, N_1^\bot,N_2^\bot$}

\AxiomC{$\vdots$ $\psi_1$}
\noLine
\UnaryInfC{$\vdash \Pi,\Delta,N_1$}
\noLine
\BinaryInfC{}
\end{prooftree}
Since $N_1^\bot $ is an immediate subformula of $P$,
we have $d(N_1^\bot) < d(P)$ and
by induction we have
 a cut-free derivation $\theta_1$ of $\vdash \Pi,\Delta,\Delta, N_2^\bot$.
 Similarly, from $\theta_1$ and $\psi_2$ we
 get
 a cut-free derivation $\theta_2$
of $\vdash \Pi,\Pi,\Delta,\Delta,\Delta$.
From $\theta_2$ we  finally obtain a cut-free derivation $\theta$ of $\vdash \Pi,\Delta$
by repeatedly applying Proposition  ~\ref{struct} (2)
(contraction on positive formulas).\smallskip

\noindent\textbf{(b)}
The last rule of $\pi$ is a positive rule  \textrm{Pos}$_n$ and  $P$
is \emph{not}  principal in \textrm{Pos}$_n$:\vspace{-6 pt}
\begin{prooftree}
\AxiomC{$\vdots$ $\pi_1$}
\noLine \UnaryInfC{$\vdash \Pi,P,Q,N_1$}
\AxiomC{$\ldots$}
\AxiomC{$\vdots$ $\pi_n$}
\noLine \UnaryInfC{$\vdash\Pi,P,Q,N_n$}
\RightLabel{\scriptsize{Pos$_n$}}
\TrinaryInfC{$\vdash \Pi,P,Q$}

\AxiomC{$\vdots$ $\rho$}
\noLine
\UnaryInfC{$\vdash \Delta, P\b$}
\noLine
\BinaryInfC{}
\end{prooftree}
and the principal formula of \textrm{Pos}$_n$  is
the occurrence of formula $Q=\posm(N_1 \otimes \cdots \otimes N_n)$. We now define a cut-free derivation $\theta$ of $\vdash \Pi,Q,\Delta$ as follows.
For $1 \leq i \leq n$ consider pairs of derivations\vspace{-6 pt}
\begin{prooftree}
\AxiomC{$\vdots$ $\pi_i$}
\noLine \UnaryInfC{$\vdash \Pi,P,Q,N_i$}

\AxiomC{$\vdots$ $\rho$}
\noLine
\UnaryInfC{$\vdash \Delta, P\b$}
\noLine
\BinaryInfC{}
\end{prooftree}
 Since
$h(\pi_i) < h(\pi)$, we have $
(d(P),h(\pi_i)+h(\rho)) < (d(P),h(\pi)+h(\rho)) $ and
by induction we get
 a cut-free
derivation  $\psi_i$ of  $\vdash \Pi,Q,\Delta,N_i$.
We now apply \textrm{Pos}$_n$  to the sequents
derived by $\psi_1,\ldots,\psi_n$, to obtain the
cut-free derivation $\theta$ of $\vdash \Pi,Q,\Delta$:\vspace{-6 pt}
\begin{prooftree}
\AxiomC{$\vdots$ $\psi_1$}
\noLine \UnaryInfC{$\vdash \Pi,Q,\Delta,N_1$}
\AxiomC{$\ldots$}
\AxiomC{$\vdots$ $\psi_n$}
\noLine \UnaryInfC{$\vdash\Pi,Q,\Delta,N_n$}
\RightLabel{\scriptsize{Pos$_n$}}
\TrinaryInfC{$\vdash \Pi,Q,\Delta$}
\end{prooftree}
Notice that when $n=0$  and hence
$Q= \posm \uno$ and $h(\pi) = 1$,  we have that\vspace{-6 pt}
\begin{prooftree}
\AxiomC{}
\RightLabel{\scriptsize{Pos$_0$}}
\UnaryInfC{$\vdash \Pi,P,\posm\uno$}
\AxiomC{$\vdots$ $\rho$}
\noLine
\UnaryInfC{$\vdash \Delta, P\b$}
\noLine
\BinaryInfC{}
\end{prooftree}
  The procedure described above gives the cut-free derivation
\AxiomC{}
\RightLabel{\scriptsize{Pos$_0$}}
\UnaryInfC{$\vdash \Pi,\Delta,\posm\uno$}
\DisplayProof.  \smallskip

\noindent\textbf{(c)}
The last rule of $\pi$ is a negative rule \textrm{Neg}$_n$,
having  $N=\negm(P_1 \parr \cdots \parr P_n)$ as principal formula.\vspace{-6 pt}
\begin{prooftree}
\AxiomC{$\vdots$ $\pi_0$}
\noLine \UnaryInfC{$\vdash \Pi,P,P_1,\ldots,P_n$}
\RightLabel{\scriptsize{Neg$_n$}}
\UnaryInfC{$\vdash \Pi,P,N$}
\AxiomC{$\vdots$ $\rho$}
\noLine
\UnaryInfC{$\vdash \Delta, P\b$}
\noLine
\BinaryInfC{}
\end{prooftree}
To build a cut-free derivation of $\vdash N,\Pi,\Delta$,
we first consider\vspace{-6 pt}
\begin{prooftree}
\AxiomC{$\vdots$ $\pi_0$}
\noLine \UnaryInfC{$\vdash \Pi,P,P_1,\ldots,P_n$}
\AxiomC{$\vdots$ $\rho$}
\noLine
\UnaryInfC{$\vdash \Delta, P\b$}
\noLine
\BinaryInfC{}
\end{prooftree}
 Since
$h(\pi_0) < h(\pi)$, we have $
(d(P),h(\pi_0)+h(\rho)) < (d(P),h(\pi)+h(\rho)) $ and
by induction we get
 a cut-free
derivation  $\psi$ of  $\vdash \Pi,P_1,\ldots,P_n,\Delta$.
Applying \textrm{Neg}$_n$, we finally get a cut-free derivation
$\theta$ of $\vdash N,\Pi,\Delta$\vspace{-6 pt}
\begin{prooftree}
\AxiomC{$\vdots$ $\psi$}
\noLine \UnaryInfC{$\vdash \Pi,P_1,\ldots,P_n,\Delta$}
\RightLabel{\scriptsize{Neg$_n$}}
\UnaryInfC{$\vdash N,\Pi,\Delta$}\rlap{\lower 12 pt\hbox to 272 pt{\hfill\qEd}}
\end{prooftree}\medskip

\noindent By using the previous theorem, we have:

\begin{cor}\label{cut elimination}
If  $\vdash \Gamma$ is derivable in $\HS$
then $\vdash \Gamma$ is derivable in  $\HS$ without \emph{\textrm{Cut}}.
\end{cor}
\proof
By induction on the height of a derivation
$\pi$ of  $\vdash \Gamma$.
Suppose that $\pi$ ends with:

\begin{prooftree}
\AxiomC{$\vdots$ $\pi_1$} \noLine \UnaryInfC{$\vdash \Xi_1,\Pi_1$}
\AxiomC{$\ldots$}
\AxiomC{$\vdots$ $\pi_k$} \noLine \UnaryInfC{$\vdash \Xi_k,\Pi_k$}
\RightLabel{\scriptsize{Rule}}
\TrinaryInfC{$\vdash \Xi,\Pi$}
\end{prooftree}

By inductive hypothesis,
we have cut-free derivations $\psi_i$ of   $\vdash \Xi_i,\Pi_i$,
for any $1 \leq i \leq k$.
If \textrm{Rule} is not  \textrm{Cut},
we apply it to the sequents $\vdash \Xi_i,\Pi_i$ derived from $\psi_i$ and get a cut-free derivation of the conclusion.
If  \textrm{Rule} is \textrm{Cut} (and hence $k=2$), from the cut-free derivations $\psi_1$ and $\psi_2$ we get a cut-free derivation
of the conclusion, by means of
Theorem  ~\ref{principal}.
\qed

\subsection{\texorpdfstring{$\HS$}{MELLS} related to the intuitionistic sequent calculus \texorpdfstring{$\mathbf{LJ}$}{LJ}} \label{intuition}

We show a correspondence between the calculus $\HS$  and a fragment of intuitionistic  logic sequent calculus $\mathbf{LJ}$, that we call $\mathbf{LJ}_0$,
in which any formula is (hereditarily) a negation of a (possibly empty) conjunction of formulas.\vfill\eject

 The main motivation for doing this is that in
$\mathbf{LJ}_0$ we can employ the
 bilateral presentation of intuitionistic sequents | with at most
one formula on the right side of the entailment symbol |
to
represent, in a more traditional way, the asymmetry between  the polarized
formulas of $\HS$.\smallskip

Formulas of $\mathbf{LJ}_0$  are given by the following grammar:
$$\sA \ :: =   \  \neg (\sA_1 \wedge \cdots \wedge \sA_n) \qquad (n \ge 0) .$$
We write $\neg \coslj$ when $n=0$.\footnote{We chose the symbol
 $\coslj$ because  $\coslj$   usually denotes the nullary version of $\wedge$.
Consistently, we write $\neg \coslj$ for the nullary version of $\neg (\sA_1 \wedge \cdots \wedge \sA_n)$.}

Analogously to the standard calculus
 $\mathbf{LJ}$, a sequent of $\mathbf{LJ}_0$ is a pair of  (possibly empty) multi-sets of formulas
$\sPi,\mathsf{\Xi}$,
written $\sPi \vdash_{int} \mathsf{\Xi}$,
such that
 $\mathsf{\Xi}$ contains  at most one (occurrence of) formula.

The calculus $\mathbf{LJ}_0$ consists of two kinds of rules.
They are given in Table  ~\ref{LJ_0_fig}.

\begin{table}[h]
  \centering

\fbox{
\begin{tabular}{cc}
&\\
Left rules : &
\AxiomC{$\sPi, \sA   \vdash_{int} \sA_1$} \AxiomC{$\ldots$}   \AxiomC{$\sPi, \sA  \vdash_{int} \sA_n $} \RightLabel{\scriptsize{$\mathsf{L}_n$}}
  \TrinaryInfC{$\sPi,\sA \vdash_{int} $}
\DisplayProof

\\
$\sA = \neg (\sA_1 \wedge \cdots \wedge \sA_n)$ and $n\geq 0 $ & \\
& \\

Right rules : &

  \AxiomC{$  \sPi,\sA_1,\ldots, \sA_n \vdash_{int}  $}
   \RightLabel{\scriptsize{$\mathsf{R}_n$}}
  \UnaryInfC{$\sPi \vdash_{int}  \sA  $} \DisplayProof

\\

$\sA = \neg (\sA_1 \wedge \cdots \wedge \sA_n)$ and $n\geq 0 $ & \\

&\\
\end{tabular}
}
 \caption{$\mathbf{LJ}_0$}
  \label{LJ_0_fig}
\end{table}
\noindent In particular, when $n=0$ we have:
\begin{center}
\AxiomC{} \RightLabel{\scriptsize{$\mathsf{L}_0$}}
  \UnaryInfC{$\sPi,\neg \coslj  \vdash_{int} $}
\DisplayProof
\qquad
  \AxiomC{$  \sPi \vdash_{int} $}
   \RightLabel{\scriptsize{$\mathsf{R}_0$}}
  \UnaryInfC{$\sPi \vdash_{int}  \neg \coslj $} \DisplayProof
\end{center}

\subsubsection{From $\HS$ to $\mathbf{LJ}_0$}

The translation  $^\ast$ of formulas  of $\HS$ into formulas  of  $\mathbf{LJ}_0$ is given as follows.
$$
\begin{array}{rclrcl}
\posm(N_1  \otimes \cdots \otimes N_n) \ ^\ast &\!:=\!& \neg
(N_1^\ast \wedge \cdots \wedge N_n^\ast); &  \negm(P_1  \parr \cdots
\parr P_n) \ ^\ast &\!:=\!&\neg (P_1^\ast \wedge \cdots \wedge P_n^\ast).
\end{array}
$$
Given a multi-set $\Gamma= F_1,\ldots,F_n$  of formulas of
 of $\HS$ we denote
by $\Gamma^\ast$ the multi-set $F_1^\ast,\ldots,F_n^\ast$
of formulas of $\mathbf{LJ}_0$.

Given a sequent $\vdash \Pi,\Xi$ of $\HS$,
where $\Pi$ is a multi-set of positive formulas and $\Xi$ is either empty of it consists of
exactly of one (occurrence of) negative formula, we define
$\vdash \Pi,\Xi \ ^\ast := \Pi^\ast \vdash_{int} \Xi^\ast $.

We have the following:

\begin{prop} \label{LJ0}
If $\vdash \Pi,\Xi$  is derivable in $\HS$ then
$\vdash \Pi,\Xi \ ^\ast$  is derivable in $\mathbf{LJ}_0$.
\end{prop}
\proof
By induction on the height of the derivation $\pi$ of
$\vdash \Pi,\Xi$ in $\HS$.

\begin{enumerate}[$\bullet$]
\item
$\pi$ ends with a positive rule, where $P=\posm(N_1 \otimes \cdots \otimes N_n)$.

\begin{prooftree}
\AxiomC{$\vdash \Pi,P,N_1$}
\AxiomC{$\ldots$}
\AxiomC{$\vdash \Pi,P,N_n$}
\RightLabel{\scriptsize{Pos}$_n$}
\TrinaryInfC{$\vdash \Pi,P$}
\end{prooftree}
By inductive hypothesis, we have derivable sequents
 $$\vdash \Pi,P,N_i \ ^\ast \ = \ \Pi^\ast,\neg(N_1^\ast \wedge  \cdots \wedge N_n^\ast) \vdash_{int} N_i^\ast,$$
for any $1 \leq i \leq n$.
We can then apply  $\mathsf{L}_n$-rule and obtain

\begin{prooftree}
\AxiomC{$\Pi^\ast,\neg(N_1^\ast \wedge \cdots \wedge N_n^\ast) \vdash_{int} N_1^\ast$}  \AxiomC{$\ldots$}  \AxiomC{$\Pi^\ast,\neg(N_1^\ast \wedge \cdots \wedge N_n^\ast) \vdash_{int} N_n^\ast $}
     \RightLabel{\scriptsize{$\mathsf{L}_n$}}
  \TrinaryInfC{$ \Pi^\ast, \neg(N_1^\ast \wedge \cdots \wedge N_n^\ast) \vdash_{int}  $}
\end{prooftree}
Since $\vdash \Pi, P \ ^\ast \ = \
\Pi^\ast,\neg(N_1^\ast \wedge \cdots \wedge N_n^\ast) \vdash_{int}$, we are done.
\item
$\pi$ ends with a negative rule, where $N=\negm(P_1 \parr \cdots \parr P_n)$.
\begin{prooftree}
\AxiomC{$\vdash \Pi, P_1,\ldots,P_n$}
\RightLabel{\scriptsize{Neg$_n$}}
\UnaryInfC{$\vdash \Pi, N$}
\end{prooftree}
By inductive hypothesis,  the sequent
$$\vdash \Pi, P_1,\ldots,P_n \ ^\ast \ = \ \Pi^\ast,P_1^\ast,\ldots,P_n^\ast  \vdash_{int}$$ is derivable.
We can then apply  $\mathsf{R}_n$-rule and obtain
\begin{prooftree}
\AxiomC{$\Pi^\ast,P_1^\ast,\ldots,P_n^\ast  \vdash_{int}$}
     \RightLabel{\scriptsize{$\mathsf{R}_n$}}
  \UnaryInfC{$\Pi^\ast   \vdash_{int} \neg(P_1^\ast \wedge \cdots \wedge P_n^\ast) $}
\end{prooftree}
Since $ \vdash \Pi, N \ ^\ast \ = \
\Pi^\ast   \vdash_{int} \neg(P_1^\ast \wedge \cdots \wedge P_n^\ast) $, we are done. \qed
\end{enumerate}

\subsubsection{From  $\mathbf{LJ}_0$ to $\HS$}
We also
define an inverse translation $ ^\diamondsuit$
as follows.

We first define two translations, noted
by $ ^{\sPl}$ and $ ^{\sNl}$, from
formulas of $\mathbf{LJ}_0$ to positive and negative formulas of $\HS$ respectively:
$$
\begin{array}{rclrcl}
\neg (\sA_1 \wedge \cdots \wedge \sA_n ) \ ^{\sPl} & := & \posm(\sA_1^{\sNl} \otimes  \cdots \otimes \sA_n^{\sNl} ); &  \neg (\sA_1 \wedge \cdots \wedge \sA_n ) \ ^{\sNl} & := & \negm(\sA_1^{\sPl} \parr \cdots \parr \sA_n^{\sPl} ).  \\
\end{array}
$$
 In particular, $\neg \coslj  \ ^{\sPl}  =  \posm \uno$. Notice also that  $\sA^{\sPl} = (\sA^{\sNl})\b$.

Given a multi-set $\sGamma=\sA_1,\ldots,\sA_n$ of formulas
of $\mathbf{LJ}_0$ we write
$\sGamma^{\sPl}$ (resp.\ $\sGamma^{\sNl}$)
for $\sA_1^{\sPl},\ldots,\sA_n^{\sPl}$
(resp.\ $\sA_1^{\sNl},\ldots,\sA_n^{\sNl}$).
Given a  sequent $\sPi \vdash_{int} \mathsf{\Xi}$  of $\mathbf{LJ}_0$
we define $\sPi \vdash_{int} \mathsf{\Xi} \ ^\diamondsuit \ := \ \vdash \sPi^{\sPl}, \mathsf{\Xi}^{\sNl}$.

Notice that $\sPi \vdash_{int} \mathsf{\Xi} \ ^\diamondsuit$
is always a sequent of $\HS$, since it contains
at most one (occurrence of) negative formula $\mathsf{\Xi}^{\sNl}$.

We have:

\begin{prop} \label{L to M}
If $\sPi \vdash_{int} \mathsf{\Xi}$ is derivable in $\mathbf{LJ}_0$ then
 $\sPi \vdash_{int} \mathsf{\Xi} \ ^\diamondsuit$  is derivable in $\HS$.
\end{prop}
\proof
By induction on the height of a derivation $\pi$ of
$\sPi \vdash_{int} \mathsf{\Xi}$ in $\mathbf{LJ}_0$.

\begin{enumerate}[$\bullet$]
\item
$\pi$ ends with a left rule, where  $\sA = \neg (\sA_1 \wedge \cdots \wedge \sA_n)$.

\begin{prooftree}
\AxiomC{$\sPi, \sA   \vdash_{int} \sA_1$} \AxiomC{$\ldots$}   \AxiomC{$\sPi, \sA  \vdash_{int} \sA_n $} \RightLabel{\scriptsize{$\mathsf{L}_n$}}
  \TrinaryInfC{$\sPi,\sA \vdash_{int} $}
\end{prooftree}
By inductive hypothesis, we have derivable sequents
 $$\sPi, \sA   \vdash_{int} \sA_i \ ^\diamondsuit \ = \ \vdash \sPi^{\sPl},\posm(\sA_1^{\sNl} \otimes \cdots \otimes \sA_n^{\sNl}), \sA_i^{\sNl},$$
for any $1 \leq i \leq n$.
We can then apply  \textrm{Pos}$_n$-rule and obtain
\begin{prooftree}
 \AxiomC{$\vdash \sPi^{\sPl},\posm(\sA_1^{\sNl} \otimes \cdots \otimes \sA_n^{\sNl}), \sA_1^{\sNl}$}  \AxiomC{$\ldots$}  \AxiomC{$\vdash \sPi^{\sPl},\posm(\sA_1^{\sNl} \otimes \cdots \otimes \sA_n^{\sNl}), \sA_n^{\sNl}$}
     \RightLabel{\scriptsize{Pos$_n$}}
  \TrinaryInfC{$ \vdash \sPi^{\sPl},\posm(\sA_1^{\sNl} \otimes \cdots \otimes \sA_n^{\sNl})  $}
\end{prooftree}
Since $\sPi,\sA \vdash_{int} \ ^\diamondsuit \ = \
\vdash \sPi^{\sPl},\posm(\sA_1^{\sNl} \otimes \cdots \otimes
\sA_n^{\sNl})$, we conclude the argument.
\item
$\pi$ ends with a right rule, where $\sA = \neg (\sA_1 \wedge \cdots \wedge \sA_n)$.
\begin{prooftree}
  \AxiomC{$  \sPi,\sA_1,\ldots, \sA_n \vdash_{int}  $}
   \RightLabel{\scriptsize{$\mathsf{R}_n$}}
  \UnaryInfC{$\sPi \vdash_{int}  \sA  $}
\end{prooftree}
By inductive hypothesis,  the sequent
$$  \sPi,\sA_1,\ldots, \sA_n \vdash_{int}   \ ^\diamondsuit \ = \ \vdash  \sPi^{\sPl},\sA_1^{\sPl},\ldots, \sA_n^{\sPl}$$ is derivable.
Applying  \textrm{Neg}$_n$-rule we obtain
\begin{prooftree}
  \AxiomC{$\vdash  \sPi^{\sPl},\sA_1^{\sPl},\ldots, \sA_n^{\sPl}$}
     \RightLabel{\scriptsize{Neg$_n$}}
  \UnaryInfC{$\vdash  \sPi^{\sPl}, \negm(\sA_1^{\sPl} \parr \cdots \parr \sA_n^{\sPl}) $}
\end{prooftree}
Since $ \sPi \vdash_{int}  \sA  \ ^\diamondsuit \ = \
\vdash  \sPi^{\sPl}, \negm(\sA_1^{\sPl} \parr \cdots \parr
\sA_n^{\sPl}) $, we conclude the argument. \qed
\end{enumerate}

\subsubsection{Composing  $^\ast$ and $^\diamondsuit$} \label{composing}

We can finally show that the translations
$^\ast$ and $^\diamondsuit$ are the inverse of each other,
in the sense we are now going to make precise.

We first show the following lemma.

\begin{lem} \label{lemma per inverse} \hfill
\begin{enumerate}[\em(1)]
\item
For any positive formula $P$ of $\HS$, we have
$(P^{\ast})^\sPl  = P$. Similarly, for any negative
formula $N$, we have $(N^{\ast})^\sNl  = N$.
\item For any formula $\sA$ of $\mathbf{LJ}_0$, we
have $(\sA^{\sPl})^\ast = (\sA^{\sNl})^\ast = \sA$.
\item For any sequent $\vdash \Gamma$ of $\HS$, we have
$(\vdash \Gamma \ ^\ast)^\diamondsuit \ = \ \vdash \Gamma$.
\item For any sequent $\sPi \vdash_{int} \mathsf{\Xi}$ of $\mathbf{LJ}_0$, we have
$(\sPi \vdash_{int} \mathsf{\Xi} \ ^\diamondsuit)^\ast \ = \
\sPi \vdash_{int} \mathsf{\Xi}$.
\end{enumerate}
\end{lem}

\proof \hfill
\begin{enumerate}[(1)]
\item
By induction on the depth of $F$.

Let $P = \posm(N_1  \otimes \cdots \otimes N_n)$. We have
$
(P^{\ast})^{\sPl}  =   \neg (N_1^\ast \wedge \cdots \wedge N_n^\ast)^{\sPl}
 =  \posm( (N_1^\ast)^{\sNl}  \otimes \cdots \otimes (N_n^\ast)^{\sNl}) $,
and by inductive hypothesis we conclude the argument. Similarly,
let $N = \negm(P_1  \parr \cdots \parr P_n)$. We have
$
(N^{\ast})^\mathsf{n}  =   \neg (P_1^\ast \wedge \cdots \wedge P_n^\ast)^{\sNl}
 =  \negm( (P_1^\ast)^{\sPl}  \parr \cdots \parr (P_n^\ast)^{\sPl}) $,
and again by inductive hypothesis we conclude the argument.

\item
By induction on the depth of $\sA=\neg (\sA_1 \wedge \cdots \wedge \sA_n )$.

We have
$
(A^{\sPl})^\ast  =   \posm(\sA_1^{\sNl} \otimes  \cdots \otimes \sA_n^{\sNl} )^\ast  =
\neg ((\sA_1^{\sNl})^\ast \wedge \cdots \wedge (\sA_n^{\sNl})^\ast )$,
and by inductive hypothesis we conclude the argument. Similarly,
$(A^{\sNl})^\ast  =   \negm(\sA_1^{\sPl} \parr  \cdots \parr \sA_n^{\sPl} )^\ast  =
\neg ((\sA_1^{\sPl})^\ast \wedge \cdots \wedge (\sA_n^{\sPl})^\ast )$
and again by inductive hypothesis we conclude the argument.

\item Let  $\vdash \Gamma = \vdash P_1,\ldots,P_n, \Xi$.
We have $(\vdash P_1,\ldots,P_n,\Xi \ ^\ast)^\diamondsuit = 
 ( P_1^\ast,\ldots,P_n^\ast \vdash_{int} \Xi^\ast )^\diamondsuit =$
$\vdash (P_1^\ast)^{\sPl},\ldots, (P_n^\ast)^{\sPl},
(\Xi^\ast)^{\sNl}$, and by point (1) above we conclude the argument.
\item If $\sPi \vdash_{int} \mathsf{\Xi} \ = \ \sA_1,\ldots,\sA_n \vdash_{int} \mathsf{\Xi}$, we get
$(\sA_1,\ldots,\sA_n \vdash_{int} \mathsf{\Xi} \ ^\diamondsuit)^\ast =
  ( \vdash \sA_1^{\sPl},\ldots,\sA_n^{\sPl} , \mathsf{\Xi}^{\sNl}
  )^\ast =  (\sA_1^{\sPl})^\ast,\ldots,(\sA_n^{\sPl})^\ast
  \vdash_{int} (\mathsf{\Xi}^{\sNl})^\ast $, and by point (2) above we
  conclude the argument.
\qed
\end{enumerate}\medskip

\noindent Given a derivation $\pi$ of a sequent $\vdash \Pi,\Xi$  of $\HS$
we denote by $\pi^\ast$ the derivation of the sequent
$\vdash \Pi,\Xi \ ^\ast$  of $\mathbf{LJ}_0$  given
 by Proposition ~\ref{LJ0}.
Similarly, given
a derivation $\pi$ of a sequent
$\sPi \vdash_{int} \mathsf{\Xi}$ of  $\mathbf{LJ}_0$
we denote by $\pi^\diamondsuit$  the derivation of the sequent
$\sPi \vdash_{int} \mathsf{\Xi} \ ^\diamondsuit$  of $\HS$
given by Proposition ~\ref{L to M}.

We can finally show the following:

\begin{thm} \hfill
\begin{enumerate}[\em(1)]
\item $(\pi^\ast)^\diamondsuit = \pi$;
\item $(\pi^\diamondsuit)^\ast = \pi$.
\end{enumerate}
\end{thm}
\proof \hfill
\begin{enumerate}[(1)]
\item
By induction on the height of the derivation $\pi$
of a sequent $\vdash \Pi,\Xi$ of $\HS$.

Suppose that $\pi$ ends with
\begin{prooftree}
\AxiomC{$\vdots \ \pi_1$} \noLine
\UnaryInfC{$\vdash \Pi_1,\Xi_1$}
\AxiomC{$\ldots$}
\AxiomC{$\vdots \ \pi_n$} \noLine
\UnaryInfC{$\vdash \Pi_n,\Xi_n$}
\RightLabel{\scriptsize{Rule}}
\TrinaryInfC{$\vdash \Pi,\Xi$}
\end{prooftree}
By Proposition ~\ref{LJ0}, we get the derivation $\pi^\ast$ ending with
\begin{prooftree}
\AxiomC{$\vdots \ \pi_1^\ast$} \noLine
\UnaryInfC{$\Pi_1^\ast \vdash_{int} \Xi_1^\ast$}  \AxiomC{$\ldots$}  \AxiomC{$\vdots \ \pi_n^\ast$} \noLine
\UnaryInfC{$\Pi_n^\ast  \vdash_{int} \Xi_n^\ast $}
     \RightLabel{\scriptsize{\scriptsize{Rule}$^\ast$}}
  \TrinaryInfC{$ \Pi^\ast \vdash_{int} \Xi^\ast  $}
\end{prooftree}
and by Proposition  ~\ref{L to M} we finally get the derivation
$(\pi^\ast)^\diamondsuit$ ending with
\begin{prooftree}
\AxiomC{$\vdots \ (\pi_1^\ast)^\diamondsuit$} \noLine
\UnaryInfC{$\vdash (\Pi_1^\ast )^{\sPl},(\Xi_1^\ast )^{\sNl}$}
\AxiomC{$\ldots$}
\AxiomC{$\vdots \ (\pi_n^\ast)^\diamondsuit$} \noLine
\UnaryInfC{$\vdash (\Pi_n^\ast )^{\sPl},(\Xi_n^\ast )^{\sNl}$}
\RightLabel{$($\scriptsize{Rule}$^\ast)^\diamondsuit$}
\TrinaryInfC{$\vdash (\Pi^\ast )^{\sPl},(\Xi^\ast )^{\sNl}$} \end{prooftree}
 By Lemma ~\ref{lemma per inverse} (3),
$\vdash (\Pi_1^\ast )^{\sPl},(\Xi_1^\ast )^{\sNl} \ =  \ \vdash \Pi_1, \Xi_1$, \ldots,
$\vdash (\Pi_n^\ast )^{\sPl},(\Xi_n^\ast )^{\sNl} \ =  \ \vdash \Pi_n, \Xi_n$ and
$\vdash (\Pi^\ast )^{\sPl},(\Xi^\ast )^{\sNl} \ =  \ \vdash \Pi, \Xi$. It is immediate to verify  that the principal and the auxiliary (occurrences of) formulas of  \textrm{Rule}
are exactly  the same of  $($\textrm{Rule}$^\ast)^\diamondsuit$,
and hence
the two  expressions  denote  the same rule.
By inductive hypothesis $(\pi_1^\ast)^\diamondsuit = \pi_1$,\ldots,$(\pi_n^\ast)^\diamondsuit = \pi_n$.
We can finally conclude $(\pi^\ast)^\diamondsuit = \pi$.

\item
By induction on the height of the derivation $\pi$
of a sequent $\sPi \vdash_{int} \mathsf{\Xi}$ of $\mathbf{LJ}_0$.

Suppose that $\pi$ ends with
\begin{prooftree}
\AxiomC{$\vdots \ \pi_1$} \noLine
\UnaryInfC{$\sPi_1 \vdash_{int} \mathsf{\Xi}_1$}
\AxiomC{$\ldots$}
\AxiomC{$\vdots \ \pi_n$} \noLine
\UnaryInfC{$\sPi_1 \vdash_{int} \mathsf{\Xi}_n$}
\RightLabel{\scriptsize{$\mathsf{Rule}$}}
\TrinaryInfC{$\sPi \vdash_{int} \mathsf{\Xi}$}
\end{prooftree}
By Proposition ~\ref{L to M}, we get the derivation $\pi^\diamondsuit$ ending with
\begin{prooftree}
\AxiomC{$\vdots \ \pi_1^\diamondsuit$} \noLine
\UnaryInfC{$ \vdash \sPi_1^{\sPl},\mathsf{\Xi}_1^{\sNl}$}  \AxiomC{$\ldots$}  \AxiomC{$\vdots \ \pi_n^\diamondsuit$} \noLine
\UnaryInfC{$ \vdash \sPi_n^{\sPl},\mathsf{\Xi}_n^{\sNl}$}
\RightLabel{\scriptsize{$\mathsf{Rule}^\diamondsuit$}}
  \TrinaryInfC{$ \vdash \sPi^{\sPl},\mathsf{\Xi}^{\sNl} $}
\end{prooftree}
and by Proposition  ~\ref{LJ0} we finally get the derivation
$(\pi^\diamondsuit)^\ast$ ending with
\begin{prooftree}
\AxiomC{$\vdots \ (\pi_1^\diamondsuit)^\ast$} \noLine
\UnaryInfC{$(\sPi_1^{\sPl})^\ast \vdash_{int} (\mathsf{\Xi}_1^{\sNl})^\ast$}
\AxiomC{$\ldots$}
\AxiomC{$\vdots \ (\pi_n^\diamondsuit)^\ast$} \noLine
\UnaryInfC{$(\sPi_n^{\sPl})^\ast \vdash_{int} (\mathsf{\Xi}_n^{\sNl})^\ast$}
\RightLabel{\scriptsize{$(\mathsf{Rule}^\diamondsuit)^\ast$}}
\TrinaryInfC{$(\sPi^{\sPl})^\ast \vdash_{int} (\mathsf{\Xi}^{\sNl})^\ast$}
\end{prooftree}
 By Lemma ~\ref{lemma per inverse} (4),
$(\sPi_1^{\sPl})^\ast \vdash_{int} (\mathsf{\Xi}_1^{\sNl})^\ast \ =  \ \sPi_1 \vdash_{int} \mathsf{\Xi}_1$, \ldots,
$(\sPi_n^{\sPl})^\ast \vdash_{int} (\mathsf{\Xi}_n^{\sNl})^\ast \ =  \ \sPi_n \vdash_{int} \mathsf{\Xi}_n$ and
$(\sPi^{\sPl})^\ast \vdash_{int} (\mathsf{\Xi}^{\sNl})^\ast \ =  \ \sPi \vdash_{int} \mathsf{\Xi}$. As before, it is immediate to verify  that the principal and the auxiliary (occurrences of) formulas of  $\mathsf{Rule}$
are exactly  the same of  $(\mathsf{Rule}^\diamondsuit)^\ast$, and hence
they denote  the same rule.
By inductive hypothesis $(\pi_1^\diamondsuit)^\ast = \pi_1$,\ldots,$(\pi_n^\diamondsuit)^\ast = \pi_n$.
We finally conclude $(\pi^\diamondsuit)^\ast = \pi$.
\qed
\end{enumerate}

\subsection{On the expressivity of \texorpdfstring{$\HS$}{HS}} \label{corrispondenza}
In this part, we discuss the relation between $\HS$ and some other
(more standard) polarized variants of $\mathbf{MELL}$ (see also
\cite{BasalTLCA09}).

\subsubsection{$\mathbf{MELL}_{\mathsf{pol}}$}
We first recall the syntax of $\mathbf{MELL}_{\mathsf{pol}}$ ~\cite{LaurentThesis}, the fragment of $\mathbf{MELL}$
given by the following data.

Formulas of $\mathbf{MELL}_{\mathsf{pol}}$  are the \emph{polarized} formulas given
by the following grammar:
$$
\begin{array}{rrclcccc}
\mbox{Positive formulas :} & \sP & ::= & \mathsf{1} & |  &    \sP \otimes \sP & |  &  \oc \sN \\
\mbox{Negative formulas :} &  \sN & ::= & \mathsf{\bot} & |  & \sN \parr \sN & |  &  \wn \sP \\

\end{array}
$$
Rules  of $\mathbf{MELL}_{\mathsf{pol}}$ are the standard rules of $\mathbf{MELL}$
(Table ~\ref{mell_fig}) applied to sequents containing polarized formulas.\\

For our purposes, it is  convenient to redefine
the syntax of
$\mathbf{MELL}_{\mathsf{pol}}$ by considering formulas in a certain canonical form,
using the syntactical isomorphisms of linear logic
$$(\sA \otimes \sB) \otimes \sC \cong \sA \otimes (\sB \otimes \sC),
\qquad \sA \otimes \mathsf{1} \cong \sA,$$
$$
(\sA \parr \sB) \parr \sC \cong \sA \parr (\sB \parr \sC), \qquad
 \sA \parr \mathsf{\bot} \cong \sA.$$
We \emph{redefine} the formulas of
$\mathbf{MELL}_{\mathsf{pol}}$  as follows:
$$
\begin{array}{rrclcrclc}
\mbox{Positive formulas :} \qquad & \sP & ::= & \oc \sM & \qquad  & \sQ &
::= &    \sP_1 \otimes \dots\otimes \sP_n & (n \geq 0) \\
\mbox{Negative formulas :} \qquad &  \sN & ::= & \wn \sQ & \qquad & \sM
& ::=  & \sN_1 \parr \dots  \parr \sN_n & (n \geq 0) \\

\end{array}
$$
The constant $\mathsf{1}$ (resp.\ $\mathsf{\bot}$)
is given by $\sP_1 \otimes \dots\otimes \sP_n$  (resp.
$\sN_1 \parr \dots  \parr \sN_n$) with  $n=0$.

Notice that in this reformulation
we now allow the
 unary  tensor ``$\otimes \oc \sM$" and par
 ``$\parr \wn \sQ$."
 We also point out that
 $\otimes \oc \sM$ is different from $\oc \sM$
 as they have different outermost connectives.
 Similarly, $\parr \wn \sQ$ is different
 from $\wn \sQ$.
On the other hand,
 we do not have formulas like $\wn(\oc(\wn(\ldots)))$.
 This is not a big loss, since we can consider
 formulas of the form
  $\wn(\otimes\oc(\parr\wn(\ldots)))$ in  place of them.

The rules
are
 almost the same  we gave for   $\mathbf{MELL}$. The only difference
is that
 we here consider tensor and par rules of any arity $n \geq 0$.
 They are given in Table ~\ref{mellpol_fig}.\\

\begin{table}[h]
  \centering

\fbox{
\begin{tabular}{cc}
&\\
Multiplicative rules &
\AxiomC{$\vdash \sGamma_1,\sP_1$}
\AxiomC{$\ldots$}
\AxiomC{$\vdash \sGamma_n, \sP_n$}
\RightLabel{\scriptsize{$\otimes_n$}}
\TrinaryInfC{$\vdash \sGamma_1,\ldots,\sGamma_n, \sP_1 \otimes \cdots \otimes \sP_n$}
\DisplayProof
\\ &
\\ &
\AxiomC{$\vdash \sGamma, \sN_1,\ldots, \sN_n$}
\RightLabel{\scriptsize{$\parr_n$}}
\UnaryInfC{$\vdash \sGamma,  \sN_1 \parr \cdots \parr \sN_n $}
\DisplayProof
\\
&\\

& \\
Exponential and structural rules  &
\AxiomC{$\vdash  \sN_1,\ldots,\sN_n, \sM$}
\RightLabel{\scriptsize{$\oc$}}
\UnaryInfC{$\vdash \sN_1,\ldots,\sN_n, \oc \sM$}
\DisplayProof

\quad
\AxiomC{$\vdash  \sGamma, \sQ$}
\RightLabel{\scriptsize{$\wn$}}
\UnaryInfC{$\vdash  \sGamma,\wn \sQ$}
\DisplayProof
\\
& \\ &
\AxiomC{$\vdash  \sGamma$}
\RightLabel{\scriptsize{W}}
\UnaryInfC{$\vdash  \sGamma,\sN$}
\DisplayProof
\quad
\AxiomC{$\vdash  \sGamma,\sN,\sN$}
\RightLabel{\scriptsize{C}}
\UnaryInfC{$\vdash \sGamma,\sN$}
\DisplayProof \\
&\\
\end{tabular}
}
 \caption{$\mathbf{MELL}_{\mathsf{pol}}$}
  \label{mellpol_fig}
\end{table}

\noindent The following lemmas are useful in the sequel.

\begin{lem} \label{at most one +}
If the sequent $\vdash \sGamma$ is derivable in
$\mathbf{MELL}_{\mathsf{pol}}$
then $\sGamma$ contains at most one occurrence
of positive formulas.
\end{lem}
\proof
By induction on the height of a derivation $\pi$ of $\vdash \sGamma$.
\begin{enumerate}[$\bullet$]
\item Suppose that $\pi$ ends with
a $\otimes_n$-rule.
By inductive hypothesis, since
$\sP_i$ is positive, any $\sGamma_i$
consists of negative formulas only.
Hence, in final sequent,  $\sP_1 \otimes \cdots \otimes \sP_n$ is the only occurrence of positive formula.

\item Suppose that $\pi$ ends with a $\parr_n$-rule.
In this case, the number of occurrence of positive formulas
in the premise is the same as in the conclusion.

\item Suppose that $\pi$ ends with the
$\oc$-rule.
In this case, there is no  occurrence of positive formula
 in the premise and exactly one
 in the conclusion.

\item Suppose that $\pi$ ends with the
$\wn$-rule.
In this case,
there exactly one  occurrence of  positive formula
 in the premise and none
 in the conclusion.
\item Suppose that $\pi$ ends with a structural rule
\textrm{W} or \text{C}.
In this case the number of occurrence of positive formulas
in the premise is the same as in the conclusion.
\qed

\end{enumerate}

\begin{lem} \label{lemma 2}
If   $\vdash \sGamma, \sN_1 \parr \cdots \parr \sN_m$ is derivable then
$\vdash \sGamma, \sN_1, \ldots,\sN_m$
is derivable.
\end{lem}
\proof
By induction on the height of the derivation $\pi$ of $\vdash \sGamma, \sN_1 \parr \cdots \parr \sN_m$,
we construct a derivation
 $\pi'$ of $\vdash \sGamma, \sN_1, \ldots,\sN_m$ as follows.
\begin{enumerate}[$\bullet$]
\item Suppose that $\pi$ ends with
\begin{prooftree}
\AxiomC{$ \vdots \  \pi_1$} \noLine
\UnaryInfC{$\vdash \sGamma_1,\sP_1$}
\AxiomC{$ \vdots \  \pi_i$} \noLine
\UnaryInfC{$\ldots \qquad \vdash \sGamma_i,\sN_1 \parr \cdots \parr \sN_m,\sP_i \qquad \ldots $}
\AxiomC{$ \vdots \  \pi_n$} \noLine
\UnaryInfC{$\vdash \sGamma_n, \sP_n$}
\RightLabel{\scriptsize{$\otimes_n$}}
\TrinaryInfC{$\vdash \sGamma_1,\ldots,
\sGamma_i,\sN_1 \parr \cdots \parr \sN_m,\ldots,
\sGamma_n, \sP_1 \otimes \cdots \otimes \sP_n$}
\end{prooftree}
By inductive hypothesis,  $\pi_i'$ derives $\vdash \sGamma_i,\sN_1, \ldots,  \sN_m,\sP_i$.
We take
\begin{prooftree}
\AxiomC{$ \vdots \  \pi_1$} \noLine
\UnaryInfC{$\vdash \sGamma_1,\sP_1$}
\AxiomC{$ \vdots \  \pi_1'$} \noLine
\UnaryInfC{$\ldots \qquad \vdash \sGamma_i,\sN_1, \ldots,  \sN_m,\sP_i \qquad \ldots $}
\AxiomC{$ \vdots \  \pi_n$} \noLine
\UnaryInfC{$\vdash \sGamma_n, \sP_n$}
\RightLabel{\scriptsize{$\otimes_n$}}
\TrinaryInfC{$\vdash \sGamma_1,\ldots,
\sGamma_i,\sN_1, \ldots,  \sN_m,\ldots,
\sGamma_n, \sP_1 \otimes \cdots \otimes \sP_n$}
\end{prooftree}

\item Suppose that $\pi$ ends with a $\parr_k$-rule.
We distinguish two subcases.

If $\sN_1 \parr \cdots \parr \sN_m$
is the principal formula in the last rule
\begin{prooftree}
\AxiomC{$ \vdots \  \pi_0$} \noLine
\UnaryInfC{$\vdash \sGamma, \sN_1,\ldots, \sN_m$}
\RightLabel{\scriptsize{$\parr_m$}}
\UnaryInfC{$\vdash \sGamma,  \sN_1 \parr \cdots \parr \sN_m $}
\end{prooftree}
we take  the derivation $\pi'=\pi_0$.

Otherwise,  $\sN_1 \parr \cdots \parr \sN_m$
is not the principal formula in the last rule
\begin{prooftree}
\AxiomC{$ \vdots \  \pi_0$} \noLine
\UnaryInfC{$\vdash \sGamma,\sN_1 \parr \cdots \parr \sN_m,  \ \sN_1',\ldots, \sN_n'$}
\RightLabel{\scriptsize{$\parr_n$}}
\UnaryInfC{$\vdash \sGamma, \sN_1 \parr \cdots \parr \sN_m, \  \sN_1' \parr \cdots \parr \sN_n' $}

\end{prooftree}
By inductive hypothesis, $\pi_0'$ derives $\vdash \sGamma,\sN_1,
\ldots,  \sN_m, \ \sN_1',\ldots, \sN_n'$. We  take 
\begin{prooftree}
\AxiomC{$ \vdots \  \pi_0'$} \noLine
\UnaryInfC{$\vdash \sGamma,\sN_1, \ldots,  \sN_m, \ \sN_1',\ldots, \sN_n'$}
\RightLabel{\scriptsize{$\parr_n$}}
\UnaryInfC{$\vdash \sGamma, \sN_1, \ldots,  \sN_m, \  \sN_1' \parr \cdots \parr \sN_n' $}

\end{prooftree}

\item Suppose that $\pi$ ends with
\textrm{Rule} $\in\{\wn,$\textrm{W}$,$\textrm{C}$\}$
\begin{prooftree}

\AxiomC{$ \vdots \  \pi_0$} \noLine
\UnaryInfC{$\vdash  \sGamma',\sN_1 \parr \cdots \parr \sN_m$}
\RightLabel{\scriptsize{Rule}}
\UnaryInfC{$\vdash  \sGamma,\sN_1 \parr \cdots \parr \sN_m$}

\end{prooftree}
By inductive hypothesis, $\pi_0'$ derives $\vdash \sGamma',\sN_1, \ldots,  \sN_m$. We  take
\begin{prooftree}

\AxiomC{$ \vdots \  \pi_0'$} \noLine
\UnaryInfC{$\vdash  \sGamma',\sN_1, \ldots,  \sN_m$}
\RightLabel{\scriptsize{Rule}}
\UnaryInfC{$\vdash  \sGamma,\sN_1, \ldots,  \sN_m$}
\end{prooftree}
\end{enumerate}
We finally observe that there are
no other cases: due to the presence of
$\sN_1 \parr \cdots \parr \sN_m$ in the final
sequent, $\pi$ cannot end with the $\oc$-rule.
\qed

\subsubsection{$\mathbf{MELL}_{\mathsf{pol}}^\ast$}

The next step is to consider
the following subsystem of $\mathbf{MELL}_{\mathsf{pol}}$
that we call
$\mathbf{MELL}_{\mathsf{pol}}^\ast$.

Formulas of $\mathbf{MELL}_{\mathsf{pol}}^\ast$
are the same of $\mathbf{MELL}_{\mathsf{pol}}$ but
sequents are now multi-set of formulas
of the form $ \vdash  \sDelta \ = \ \vdash \sN_1,\ldots,\sN_{k-1},\sF_k$, for some $k \geq 0$. In other words,
a sequent of $\mathbf{MELL}_{\mathsf{pol}}^\ast$
contains at most
one occurrence of formulas which is not a $\wn$-formula.
In some cases we  also denote a sequent of  $\mathbf{MELL}_{\mathsf{pol}}^\ast$ by $\vdash \wn\sGamma,\sF$.

The rules of  $\mathbf{MELL}_{\mathsf{pol}}^\ast$ are the same
of  $\mathbf{MELL}_{\mathsf{pol}}$, with the obvious
modifications due to the constraint on sequents.
They are given in Table ~\ref{mellpolast_fig}.

\begin{table}[h]
  \centering

\fbox{
\begin{tabular}{cc}
&\\
Multiplicative rules &
\AxiomC{$\vdash \wn\sGamma_1,\sP_1$}
\AxiomC{$\ldots$}
\AxiomC{$\vdash \wn\sGamma_n, \sP_n$}
\RightLabel{\scriptsize{$\otimes_n$}}
\TrinaryInfC{$\vdash \wn\sGamma_1,\ldots,\wn\sGamma_n, \sP_1 \otimes \cdots \otimes \sP_n$}
\DisplayProof
\\ &
\\ &
\AxiomC{$\vdash  \wn\sGamma, \sN_1,\ldots, \sN_n$}
\RightLabel{\scriptsize{$\parr_n$}}
\UnaryInfC{$\vdash  \wn\sGamma,  \sN_1 \parr \cdots \parr \sN_n $}
\DisplayProof
\\
&\\

& \\
Exponential and structural  rules  &
\AxiomC{$\vdash   \wn\sGamma, \sM$}
\RightLabel{\scriptsize{$\oc$}}
\UnaryInfC{$\vdash \wn\sGamma, \oc \sM$}
\DisplayProof

\quad
\AxiomC{$\vdash  \wn\sGamma, \sQ$}
\RightLabel{\scriptsize{$\wn$}}
\UnaryInfC{$\vdash  \wn\sGamma,\wn \sQ$}
\DisplayProof
\\
& \\  &
\AxiomC{$\vdash  \sDelta$}
\RightLabel{\scriptsize{W}}
\UnaryInfC{$\vdash \sDelta,\sN$}
\DisplayProof
\quad
\AxiomC{$\vdash  \sDelta,\sN,\sN$}
\RightLabel{\scriptsize{C}}
\UnaryInfC{$\vdash \sDelta,\sN$}
\DisplayProof \\
&\\
\end{tabular}
}

 \caption{$\mathbf{MELL}_{\mathsf{pol}}^\ast$}
  \label{mellpolast_fig}
\end{table}

\noindent The main consequence of  restriction on the shape of sequents
is that a sequent of the form

$$\vdash  \wn\sGamma, \wn \sQ,  \sN_1 \parr \cdots \parr \sN_n $$
cannot be inferred using the $\wn$-rule on $\sQ$,
as the resulting premise
$$\vdash  \wn\sGamma,  \sQ,  \sN_1 \parr \cdots \parr \sN_n $$
which is a sequent in $\mathbf{MELL}_{\mathsf{pol}}$,
would not be a sequent of $\mathbf{MELL}_{\mathsf{pol}}^\ast$.\smallskip

Clearly, if $\vdash \sDelta$
is derivable in $\mathbf{MELL}_{\mathsf{pol}}^\ast$,
then  $\vdash \sDelta$ is derivable in
$\mathbf{MELL}_{\mathsf{pol}}$.
But we also have the converse.

\begin{prop} \label{pol to ast}
If $\vdash \sDelta$ is derivable in
 $\mathbf{MELL}_{\mathsf{pol}}$ then $\vdash \sDelta$ is derivable
in  $\mathbf{MELL}_{\mathsf{pol}}^\ast$.
\end{prop}

By induction on the height of a derivation $\pi$ of $\vdash \sDelta$ in  $\mathbf{MELL}_{\mathsf{pol}}$
we construct a derivation $\pi^\ast$
of $\vdash \sDelta$ in  $\mathbf{MELL}_{\mathsf{pol}}^\ast$. There are several cases
to analyze.
\begin{enumerate}[\hbox to8 pt{\hfill}]

\item\noindent{\hskip-12 pt\bf Structural rules:}\
$\pi$ ends with a structural rule \textrm{Rule} $\in \{$\textrm{W}$,$\textrm{C}$\}$:
\begin{prooftree}

\AxiomC{$\vdots \ \pi_0$} \noLine
\UnaryInfC{$\vdash  \sDelta'$}
\RightLabel{\scriptsize{Rule}}
\UnaryInfC{$\vdash  \sDelta$}
\end{prooftree}
 Since structural rules only affect
$\wn$-formulas, it is clear that
 $\vdash  \sDelta'$ is
  a sequent of $\mathbf{MELL}_{\mathsf{pol}}^\ast$.
The inductive hypothesis yields
a  derivation $\pi_0^\ast$ of the premise.
We can then apply \textrm{Rule} and conclude.

\item\noindent{\hskip-12 pt\bf Positive case:}\
Suppose that $\vdash \Delta$ contains an (occurrence of) positive formula $\sF$.
We have the following subcases.
\begin{enumerate}[$\bullet$]

\item $\sF =
\sP_1 \otimes \cdots \otimes \sP_n$ and
 $\pi$ ends with a $\otimes_n$-rule:
\begin{prooftree}
\AxiomC{$\vdots \ \pi_1$} \noLine
\UnaryInfC{$\vdash \wn\sGamma_1,\sP_1$}
\AxiomC{$\ldots$}
\AxiomC{$\vdots \ \pi_n$} \noLine
\UnaryInfC{$\vdash \wn\sGamma_n, \sP_n$}
\RightLabel{\scriptsize{$\otimes_n$}}
\TrinaryInfC{$\vdash \wn\sGamma_1,\ldots,\wn\sGamma_n, \sP_1 \otimes \cdots \otimes \sP_n$}
\end{prooftree}
 Since  the premises are sequent of $\mathbf{MELL}_{\mathsf{pol}}^\ast$,
the inductive hypothesis yields the
  derivations $\pi_1^\ast,\ldots,\pi_n^\ast$
  of the premises
in $\mathbf{MELL}_{\mathsf{pol}}^\ast$.
We can then apply $\otimes_n$ and conclude.

\item $\sF =
\oc \sM$ and
 $\pi$ ends with the $\oc$-rule:
\begin{prooftree}
\AxiomC{$\vdots \ \pi_0$} \noLine
\UnaryInfC{$\vdash \wn\sGamma,\sM$}
\RightLabel{\scriptsize{$\oc$}}
\UnaryInfC{$\vdash \wn\sGamma,\oc\sM$}
\end{prooftree}
 Since $\vdash \wn\sGamma,\sM$
is a sequent of $\mathbf{MELL}_{\mathsf{pol}}^\ast$,
the inductive hypothesis yields
a derivation $\pi_0^\ast$ of the premise.
We can then apply $\oc$ and conclude.
\end{enumerate}
There are no other cases,
since the only possibility left out would be
an inference
of the form
\begin{prooftree}

\AxiomC{$\vdash  \wn\sGamma', \sQ,\sF$}
\RightLabel{\scriptsize{$\wn$}}
\UnaryInfC{$\vdash  \wn\sGamma',\wn \sQ,\sF$}

\end{prooftree}
but by Lemma ~\ref{at most one +}, $\vdash  \wn\sGamma', \sQ,\sF$ is not derivable
in $\mathbf{MELL}_{\mathsf{pol}}$,
as it contains two occurrences of positive formula.

\item\noindent{\hskip-12 pt\bf Negative case:}\
Suppose that
$\vdash \Delta$ contains no positive formula,
 so that $\vdash \Delta \ = \ \vdash \wn \sGamma,\sF$, where $\sF$
is an  (occurrence of)
negative formula.
We have the following subcases.
\begin{enumerate}[$\bullet$]

\item $\sF =
\sN_1 \parr \cdots \parr \sN_n$.
In such a case $\pi$ does not necessarily end with
a $\parr_n$-rule. For instance,
the last rules of $\pi$ could be
\begin{prooftree}
\AxiomC{$\vdash  \wn\sGamma', \sQ, \sN_1 \parr \cdots \parr \sN_n$}
\RightLabel{\scriptsize{$\parr_n$}}
\UnaryInfC{$\vdash  \wn\sGamma', \wn\sQ, \sN_1 \parr \cdots \parr \sN_n $}

\end{prooftree}
and $\vdash  \wn\sGamma', \sQ, \sN_1 \parr \cdots \parr \sN_n$
is not a sequent of $\mathbf{MELL}_{\mathsf{pol}}^\ast$,
as we have already discussed.
We then proceed as follows.

Observe that the formula $\sN_1 \parr \cdots \parr \sN_n$ cannot be
affected by means of structural rules, and
the contexts of the $\otimes_n$-rules are splitting. Hence,
there is a unique branch in $\pi$ where
at some stage the formula
$\sN_1 \parr \cdots \parr \sN_n$ is decomposed
by means of a $\parr_n$-rule:
\begin{prooftree}

\AxiomC{$\vdots \ \pi_0$}
\noLine
\UnaryInfC{
$\vdash  \sGamma', \sN_1,\ldots, \sN_n$}
\RightLabel{\scriptsize{$\parr_n$}}
\UnaryInfC{$\vdash   \sGamma',  \sN_1 \parr \cdots \parr \sN_n $}
\noLine
\UnaryInfC{$\vdots   $}

\noLine
\UnaryInfC{$\phantom{^{[\sN_1 \parr \cdots \parr \sN_n]}} \vdots   ^{[\sN_1 \parr \cdots \parr \sN_n]} $}
\noLine
\UnaryInfC{$\vdash  \wn\sGamma,  \sN_1 \parr \cdots \parr \sN_n $}
\end{prooftree}
We also observe that $\vdash   \sGamma', \sN_1,\ldots, \sN_n$
might not be a sequent of
$\mathbf{MELL}_{\mathsf{pol}}^\ast$
and that there is no application of the $\oc$-rule in the branch.

Let $\rho$ be
the proof-tree obtained from $\pi$
by replacing the previous branch with
\begin{prooftree}

\AxiomC{$\vdots \ \pi_0$}
\noLine
\UnaryInfC{
$\vdash   \sGamma', \sN_1,\ldots, \sN_n$}
\noLine
\UnaryInfC{$\vdots   $}

\noLine
\UnaryInfC{$\phantom{^{ [\sN_1, \ldots,\sN_n]}} \vdots  ^{ [\sN_1, \ldots,\sN_n]} $}
\noLine
\UnaryInfC{$\vdash  \wn\sGamma,  \sN_1, \ldots,\sN_n $}
\RightLabel{\scriptsize{$\parr_n$}}
\UnaryInfC{$\vdash  \wn\sGamma,  \sN_1 \parr \cdots \parr \sN_n $}
\end{prooftree}

By Lemma ~\ref{lemma 2},  $\rho$ is a correct derivation
of $\vdash  \wn\sGamma,  \sN_1 \parr \cdots \parr \sN_n$
in $\mathbf{MELL}_{\mathsf{pol}}$ ending with a $\parr_n$-rule.
Moreover, the height of $\rho$ is the same of $\pi$
because the new branch has the same height of the previous one. We can then apply the inductive hypothesis
to the derivation, say $\rho_0$, of the premise $\vdash  \wn\sGamma,  \sN_1, \ldots,\sN_n $
of the last inference rule of $\rho$.
We obtain a derivation  $\rho_0^\ast$ of
$\vdash  \wn\sGamma,  \sN_1, \ldots,\sN_n $
in $\mathbf{MELL}_{\mathsf{pol}}^\ast$.
To conclude, we apply $\parr_n$.

\item $\sF =
\wn \sQ$ and
 $\pi$ ends with the $\wn$-rule:
\begin{prooftree}
\AxiomC{$\vdots \ \pi_0$} \noLine
\UnaryInfC{$\vdash \wn\sGamma,\sQ$}
\RightLabel{\scriptsize{$\wn$}}
\UnaryInfC{$\vdash \wn\sGamma,\wn\sQ$}
\end{prooftree}
 Since $\vdash \wn\sGamma,\sQ$
is a sequent of $\mathbf{MELL}_{\mathsf{pol}}^\ast$,
the inductive hypothesis yields
a  derivation $\pi_0^\ast$.
We can then apply $\wn$ and conclude the argument
\qed
\end{enumerate}
\end{enumerate}\medskip

\noindent It is now possible to give a correspondence
between $\HS$ and $\mathbf{MELL}_{\mathsf{pol}}^\ast$.
We do this in the next section.

\subsubsection{Correspondence with $\mathbf{MELL}_{\mathsf{pol}}^\ast$}

We first observe that any \emph{exponential} formula
(that is, a $\wn$-formula or a $\oc$-formula) of $\mathbf{MELL}_{\mathsf{pol}}^\ast$
can be generated by  the following grammar:
$$
\begin{array}{rrclc}
\mbox{Positive exponential formulas :} & \sP & ::= & \oc (
  \sN_1 \parr \dots  \parr  \sN_n)    & (n \geq 0) \\
\mbox{Negative exponential formulas :} &  \sN & ::= & \wn(\sP_1 \otimes \dots\otimes  \sP_n) & (n \geq 0) \\

\end{array}
$$
We can then define a translation $^\circ$ from formulas
of $\HS$ to \emph{exponential} formulas of $\mathbf{MELL}_{\mathsf{pol}}^\ast$ recursively as follows:
$$
\begin{array}{rclrcl}
P \ ^\circ  & = & \posm (N_1 \otimes \cdots \otimes N_n)  \ ^\circ  & := & \wn (N_1  ^\circ \otimes \cdots \otimes N_n   ^\circ); \\
N \ ^\circ  & = &
   \negm (P_1 \parr \cdots \parr P_n) \ ^\circ & := & \oc(P_1^\circ \parr \cdots \parr P_n^\circ).\\
\end{array}
$$

Notice that the translation $^\circ$ \emph{inverts} the polarity.
Given a multi-set of formulas $\Gamma = F_1,\ldots,F_n$  of $\HS$
we write $\Gamma^\circ$ for the multi-set $F_1^\circ,\ldots,F_n^\circ$ of formulas
of $\mathbf{MELL}_{\mathsf{pol}}^\ast$. Notice that, as a consequence of
the restriction on the polarities for
 sequents of  $\HS$,
$\vdash \Gamma^\circ$ is always a sequent of $\mathbf{MELL}_{\mathsf{pol}}^\ast$, as it contains
at most one occurrence of $\oc$-formula
(all the remaining ones are $\wn$-formulas).

\begin{prop} \label{HS to POL}
If $\vdash \Gamma$ is derivable in $\HS$
then $\vdash \Gamma ^\circ$ is derivable in $\mathbf{MELL}_{\mathsf{pol}}^\ast$.
\end{prop}

\proof
By induction on the height of a derivation $\pi$ of  $\vdash \Gamma$ in $\HS$.
\begin{enumerate}[$\bullet$]
\item $\pi$ is \AxiomC{}
\RightLabel{\scriptsize{Pos$_0$}}
 \UnaryInfC{$\vdash  \Pi,\posm \uno$} \DisplayProof.
We set:
 \begin{prooftree}

\AxiomC{}
\RightLabel{\scriptsize{$\mathsf{1}$}}
\UnaryInfC{$\vdash \mathsf{1}$}
\RightLabel{\scriptsize{$\wn$}}
\UnaryInfC{$\vdash \wn \mathsf{1}$}
\UnaryInfC{$\vdots$ weakenings $\vdots$}
\UnaryInfC{$\vdash  \Pi^\circ, \wn \mathsf{1}$}
\end{prooftree}

\item $\pi$ ends with a positive rule on
$P = \posm (N_1 \otimes \cdots \otimes  N_n)$  (with $n \geq 1$):\\

\begin{center}
\AxiomC{$\vdash \Pi,P,N_1$}
\AxiomC{$\ldots$}
\AxiomC{$\vdash \Pi,P,N_n$}
\RightLabel{\scriptsize{Pos$_n$}}
\TrinaryInfC{$\vdash \Pi,P$} \DisplayProof
\end{center}
By inductive hypothesis,
$\vdash \Pi^\circ,P^\circ,N_i^\circ$ is derivable for any $1 \leq i \leq n$.
We set:

\begin{center}
\AxiomC{$\vdash \Pi^\circ,P^\circ,N_1^\circ$}
\AxiomC{$\ldots$}
\AxiomC{$\vdash \Pi^\circ,P^\circ,N_n^\circ$}
\RightLabel{\scriptsize{$\otimes_n$}}
\TrinaryInfC{$\vdash \Pi^\circ,\ldots,\Pi^\circ,P^\circ,\ldots,P^\circ, N_1^\circ \otimes \cdots \otimes N_n^\circ $}
\RightLabel{\scriptsize{$\wn$}}
\UnaryInfC{$\vdash \Pi^\circ,\ldots,\Pi^\circ,P^\circ,\ldots,P^\circ, \wn(N_1^\circ \otimes \cdots \otimes N_n^\circ)$}
\UnaryInfC{$\vdots$ contractions $\vdots$}
\UnaryInfC{$\vdash \Pi^\circ,P^\circ$}
\DisplayProof
\end{center}

\item $\pi$ ends with a negative rule on
$N = \negm (P_1 \parr \cdots \parr  P_n)$ (with $n \geq 0$):\\

\begin{center}
\AxiomC{$\vdash \Pi,P_1,\ldots,P_n$}
\RightLabel{\scriptsize{Neg$_n$}}
\UnaryInfC{$\vdash \Pi,\negm (P_1 \parr \cdots \parr  P_n)$} \DisplayProof
\end{center}

\noindent By inductive hypothesis,
$\vdash \Pi^\circ,P_1^\circ,\ldots,P_n^\circ$ is derivable  and we set:\\

\begin{center}
\AxiomC{$\vdash \Pi^\circ,P_1^\circ,\ldots,P_n^\circ$}
\RightLabel{\scriptsize{$\parr_n$}}
\UnaryInfC{$\vdash \Pi^\circ,P_1^\circ \parr \cdots \parr  P_n^\circ$}
\RightLabel{\scriptsize{$\oc$}}
\UnaryInfC{$\vdash \Pi^\circ,\oc (P_1^\circ \parr \cdots \parr
  P_n^\circ)\rlap{\hbox to164 pt{\hfill\qEd}}$}
\DisplayProof
\end{center}
\end{enumerate}\medskip

\noindent To show the converse,
we define a  translation $^\bullet$
from formulas of $\mathbf{MELL}_{\mathsf{pol}}^\ast$ to
formulas of $\HS$ recursively as follows:
$$
\begin{array}{rclrclrclcrclc}
\sP  \ ^\bullet & = & \oc \sM  \ ^\bullet & := & \sM^\bullet; & \qquad  & \sQ \ ^\bullet & = &
\sP_1 \otimes \cdots \otimes \sP_n \ ^\bullet  &
:= &   \posm (\sP_1^\bullet \otimes \cdots \otimes \sP_n^\bullet );  \\
\sN \ ^\bullet & = &  \wn \sQ  \ ^\bullet & := & \sQ^\bullet; & \qquad & \sM \ ^\bullet & = &
\sN_1 \parr \cdots \parr \sN_n \ ^\bullet
& :=  & \negm (\sN_1^\bullet \parr \cdots \parr \sN_n^\bullet). \\

\end{array}
$$
We observe that the translation $^\bullet$
\emph{inverts} the polarity of the exponential formulas
whereas it
\emph{preserves}
 the polarity
of the other formulas.
Given a multi-set of formulas $\sGamma = \sF_1,\ldots,\sF_k$  of $\mathbf{MELL}_{\mathsf{pol}}^\ast$
we write $\sGamma^\bullet$ for the multi-set $\sF_1^\bullet,\ldots,\sF_k^\bullet$ of formulas
of $\HS$.

Consider now a sequent $\vdash \sDelta \ = \ \vdash \sN_1,\ldots, \sN_{k-1},\sF_k $ of
$\mathbf{MELL}_{\mathsf{pol}}^\ast$.
Since $\sN_1,\ldots, \sN_{k-1}$ are
$\wn$-formula, the multi-set
$\sN_1^\bullet,\ldots, \sN_{k-1}^\bullet,\sF_k^\bullet$
contains at most one occurrence of negative formula.
Hence,  every sequent
 $\vdash \sDelta$ of
$\mathbf{MELL}_{\mathsf{pol}}^\ast$ is sent to a sequent
$\vdash \sDelta  ^\bullet$  of $\HS$.

We are now ready to show the converse correspondence.

\begin{prop} \label{pol to HS}
If $\vdash  \sDelta$ is derivable in $\mathbf{MELL}_{\mathsf{pol}}^\ast$
then $\vdash  \sDelta  ^\bullet$  is derivable in
$\HS$.
\end{prop}

\proof

By induction on the height of a derivation $\pi$ of a sequent
$\vdash  \sDelta$ in $\mathbf{MELL}_{\mathsf{pol}}^\ast$.
\begin{enumerate}[WW :]
\item[$\otimes_n$ : ]  Suppose that $\pi$ ends with
\begin{prooftree}
\AxiomC{$\vdash \wn\sGamma_1,\sP_1$}
\AxiomC{$\ldots$}
\AxiomC{$\vdash \wn\sGamma_n, \sP_n$}
\RightLabel{\scriptsize{$\otimes_n$}}
\TrinaryInfC{$\vdash \wn\sGamma_1,\ldots,\wn\sGamma_n, \sP_1 \otimes \cdots \otimes \sP_n$}
\end{prooftree}

We have to show a derivation of
$$\vdash \wn\sGamma_1,\ldots,\wn\sGamma_n, \sP_1 \otimes \cdots \otimes \sP_n \ ^\bullet \  = \ \
\vdash \wn\sGamma_1^\bullet,\ldots,\wn\sGamma_n^\bullet, \posm(\sP_1^\bullet \otimes \cdots \otimes \sP_n^\bullet).
 $$
By inductive hypothesis, the sequent
$\vdash \wn\sGamma_i^\bullet,\sP_i^\bullet$
is derivable for any $1 \leq i \leq n$.
By Proposition
 ~\ref{struct} (1) $\vdash \wn\sGamma_1^\bullet,\ldots,\wn\sGamma_n^\bullet,\sQ^\bullet,\sP_i^\bullet$ is also derivable
for any $1 \leq i \leq n$, where  $\sQ^\bullet = \posm  (\sP_1^\bullet \otimes \cdots \otimes \sP_n^\bullet)$.
We take:

\begin{prooftree}
\AxiomC{$\vdash \wn\sGamma_1^\bullet,\ldots,\wn\sGamma_n^\bullet,\sQ^\bullet,\sP_1^\bullet$}
\AxiomC{$\ldots$}
\AxiomC{$\vdash \wn\sGamma_1^\bullet,\ldots,\wn\sGamma_n^\bullet,\sQ^\bullet,\sP_n^\bullet$}
\RightLabel{\scriptsize{Pos$_n$}}
\TrinaryInfC{$\vdash  \wn\sGamma_1^\bullet,\ldots,\wn\sGamma_n^\bullet, \sQ^\bullet$}
\end{prooftree}

\item[$\parr_n$ : ] Suppose that $\pi$ ends with

\begin{prooftree}
\AxiomC{$\vdash \wn \sGamma, \sN_1,\ldots, \sN_n$}
\RightLabel{\scriptsize{$\parr_n$}}
\UnaryInfC{$\vdash \wn \sGamma,  \sN_1 \parr \cdots \parr \sN_n $}
\end{prooftree}

We have to show a derivation of
$$\vdash \wn \sGamma,  \sN_1 \parr \cdots \parr \sN_n  \ ^\bullet \  = \ \
\vdash \wn \sGamma^\bullet,  \negm(\sN_1^\bullet \parr \cdots \parr \sN_n^\bullet).
 $$
By inductive hypothesis, the sequent
$\vdash \wn \sGamma^\bullet, \sN_1^\bullet,\ldots, \sN_n^\bullet$ is derivable. We  take:

\begin{prooftree}
\AxiomC{$\vdash \wn \sGamma^\bullet, \sN_1^\bullet,\ldots, \sN_n^\bullet$}
\RightLabel{\scriptsize{$\mathsf{Neg}_n$}}
\UnaryInfC{$\vdash \wn \sGamma^\bullet,  \negm(\sN_1^\bullet \parr \cdots \parr \sN_n^\bullet) $}
\end{prooftree}

\item[$\oc$ : ]  Suppose that $\pi$ ends with
\begin{prooftree}

\AxiomC{$\vdash \wn \sGamma,\sM$}
\RightLabel{\scriptsize{$\oc$}}
\UnaryInfC{$\vdash  \wn \sGamma,\oc\sM $}
\end{prooftree}
By inductive hypothesis, the sequent
$\vdash \wn \sGamma^\bullet,\sM^\bullet$ is derivable. Since
$ \oc  \sM \ ^\bullet = \sM^\bullet$, we conclude the argument.

\item[$\wn$ : ] Suppose that $\pi$ ends with

\begin{prooftree}
\AxiomC{$\vdash \wn \sGamma,\sQ$}
\RightLabel{\scriptsize{$\wn$}}
\UnaryInfC{$\vdash \wn\sGamma, \wn \sQ $}
\end{prooftree}
 By inductive hypothesis, the sequent
$\vdash \wn \sGamma^\bullet,\sQ^\bullet$ is derivable.
Since $ \wn  \sQ \ ^\bullet = \sQ^\bullet$, we conclude the argument.

\item[\textrm{W} : ] Suppose that $\pi$ ends with

\begin{prooftree}
\AxiomC{$\vdash \sDelta$}
\RightLabel{\scriptsize{W}}
\UnaryInfC{$\vdash  \sDelta, \sN $}
\end{prooftree}
By inductive hypothesis,  the sequent
$\vdash \sDelta^\bullet$ is derivable.
By Proposition
~\ref{struct} (1), the sequent $\vdash  \sDelta^\bullet, \sN^\bullet $
is also derivable.

\item[\textrm{C} : ] Suppose that $\pi$ ends with

\begin{prooftree}
\AxiomC{$\vdash  \sDelta,\sN,\sN$}
\RightLabel{\scriptsize{C}}
\UnaryInfC{$\vdash   \sDelta,\sN $}
\end{prooftree}
By inductive hypothesis,  the sequent
$\vdash  \sDelta^\bullet, \sN^\bullet,\sN^\bullet$ is derivable. By
Proposition
~\ref{struct} (2), the sequent $\vdash  \sDelta^\bullet, \sN^\bullet$
is also derivable.
\qed
\end{enumerate}

\noindent Regarding  the composition of the translations $\ ^\circ$ and $\ ^\bullet$,
we  observe the following properties.

\begin{lem} \label{lemma per inverse due} \hfill
\begin{enumerate}[\em(1)]
\item
For any formula $F$ of $\HS$, we have
$(F^{\circ})^\bullet  = F$.
\item For any \emph{exponential} formula $\sF$ of $\mathbf{MELL}_{\mathsf{pol}}^\ast$,
we have $(\sF^{\bullet})^\circ = \sF$.

\item For any sequent $\vdash \Gamma$ of $\HS$, we have
$(\vdash \Gamma \ ^\circ)^\bullet \ = \ \vdash \Gamma$.

\item Let $ \sDelta=\sF_1,\ldots, \sF_n$. We get
$(\vdash \sF_1,\ldots, \sF_n \, ^\bullet)^\circ  = \, (\vdash \sF_1
^\bullet,\ldots, \sF_n^\bullet)^\circ = \     \vdash (\sF_1
^\bullet)^\circ,\ldots, (\sF_n^\bullet)^\circ$, and by point (2) above we
conclude the argument.

\item For any sequent $\vdash \sDelta$ of
$\mathbf{MELL}_{\mathsf{pol}}^\ast$ consisting of
exponential formulas only, we have
$(\vdash \sDelta \ ^\bullet)^\circ \ = \
\vdash \sDelta$.
\end{enumerate}
\end{lem}

\proof \hfill
\begin{enumerate}[(1)]
\item
By induction on the depth of $F$.
Let $P = \posm(N_1  \otimes \cdots \otimes N_n)$. We have:
$$ \begin{array}{rcl}
(P^{\circ})^{\bullet}
 & = &  \posm(N_1  \otimes \cdots \otimes N_n) \ ^\circ \ ^\bullet \\

 & = &  \wn (N_1^\circ  \otimes \cdots \otimes N_n^\circ) \ ^\bullet \\
& = &  N_1^\circ  \otimes \cdots \otimes N_n^\circ \ ^\bullet \\
 & = &  \posm((N_1^\circ)^\bullet  \otimes \cdots \otimes (N_n^\circ)^\bullet), \end{array}$$
and by inductive hypothesis we conclude the argument. The negative case is similar.
\item
By induction on the depth of $\sF$.
Let $\sP = \oc (\sN_1 \parr \cdots \parr \sN_n)$.
We have:
$$ \begin{array}{rcl}
(\sP^\bullet)^\circ & = &  \oc (\sN_1 \parr \cdots \parr \sN_n) \ ^\bullet \ ^\circ\\
& = &  \sN_1 \parr \cdots \parr \sN_n \ ^\bullet \ ^\circ\\
 & = &
 \negm(\sN_1^\bullet \parr \cdots \parr \sN_n^\bullet) \ ^\circ \\ &   = &
\oc ((\sN_1^\bullet)^\circ \parr \cdots \parr (\sN_n^\bullet)^\circ),  \end{array}
$$
and by inductive hypothesis we conclude the argument. The negative case is similar.

\item Let  $ \Gamma=F_1,\ldots, F_n$.
We get  $(\vdash F_1,\ldots,F_n \, ^\circ)^\bullet  = \,
(\vdash F_1^\circ,\ldots,F_n^\circ )^\bullet  = \ \vdash
(F_1^\circ)^\bullet,\ldots,(F_n^\circ)^\bullet$, and  by point (1) above
we conclude the argument.


\item Let $ \sDelta=\sF_1,\ldots, \sF_n$. We get
$(\vdash \sF_1,\ldots, \sF_n \, ^\bullet)^\circ  = \, (\vdash \sF_1
^\bullet,\ldots, \sF_n^\bullet)^\circ = \     \vdash (\sF_1
^\bullet)^\circ,\ldots, (\sF_n^\bullet)^\circ$, and by point (2) above we
conclude the argument.\qed


\end{enumerate}\medskip

\noindent We can finally collect  the previous results in the next theorem.
\vfill\eject

\begin{thm} \hfill
\begin{enumerate}[\em(1)]
\item Let $\vdash \sDelta$ be a sequent of
$\mathbf{MELL}_{\mathsf{pol}}$ consisting
of exponential formulas only.\\
$\vdash \sDelta$ is derivable in
$\mathbf{MELL}_{\mathsf{pol}}$
if and only if $\vdash \sDelta^\bullet$
is derivable
in $\HS$.

\item $\vdash \Gamma$ is derivable in
$\HS$
if and only if $\vdash \Gamma^\circ$
is derivable
in $\mathbf{MELL}_{\mathsf{pol}}$.
\end{enumerate}
\end{thm}

\proof \hfill
\begin{enumerate}[(1)]
\item
If $\vdash \sDelta$ is a derivable sequent of $\mathbf{MELL}_{\mathsf{pol}}$,
then by Lemma ~\ref{at most one +}
it contains at most one occurrence of positive
formula. Since  $\vdash \sDelta$ is made of
exponential formulas only, it is of the
form $\vdash \wn\sGamma,\sF$, where
$\sF$ is either a $\wn$-formula or a $\oc$-formula.
In particular, $\vdash \sDelta$ is a sequent of
$\mathbf{MELL}_{\mathsf{pol}}^\ast$. By Proposition
 ~\ref{pol to ast}, $\vdash \sDelta$ is a
  derivable sequent
of  $\mathbf{MELL}_{\mathsf{pol}}^\ast$.
By Proposition ~\ref{pol to HS},
$\vdash \sDelta^\bullet$
is derivable in $\HS$.

Conversely, if $\vdash \sDelta^\bullet$
is derivable in $\HS$ then, by Proposition
~\ref{HS to POL},
$(\vdash \sDelta^\bullet)^\circ$ is derivable
in $\mathbf{MELL}_{\mathsf{pol}}^\ast$
and hence in of $\mathbf{MELL}_{\mathsf{pol}}$.
By Lemma ~\ref{lemma per inverse due} (4),
$(\vdash \sDelta^\bullet)^\circ \ = \ \vdash \sDelta$
and we conclude the argument.

\item
If $\vdash \Gamma$
is derivable in $\HS$ then, by Proposition
~\ref{HS to POL},
$\vdash \Gamma^\circ$ is derivable
in $\mathbf{MELL}_{\mathsf{pol}}^\ast$ and hence also in
$\mathbf{MELL}_{\mathsf{pol}}$.

For the converse, we observe that by the definition of $\ ^\circ$,
$\vdash \Gamma^\circ$ is a sequent of  $\mathbf{MELL}_{\mathsf{pol}}^\ast$.
Since we are assuming that it is derivable in $\mathbf{MELL}_{\mathsf{pol}}$,
by Proposition
 ~\ref{pol to ast}, it is also derivable
in
$\mathbf{MELL}_{\mathsf{pol}}^\ast$.
By Proposition ~\ref{pol to HS},
$(\vdash \Gamma^\circ)^\bullet$
is derivable in $\HS$.
By Lemma ~\ref{lemma per inverse due} (3),
$(\vdash \Gamma^\circ)^\bullet \ = \ \vdash \Gamma$
and we conclude the argument.
\qed
\end{enumerate}

\section{Soundness} \label{soundness appendice}

We first show some technical lemmas.

\begin{lem}[Logical rules] \label{logical}\hfill
\begin{enumerate}[\em(1)]
\item
 Let $\bP = \bF^+(\bN_1,\ldots,\bN_n)$  be a positive behaviour.\\
  If $\E_1 \in \ \vdash \bPP,\bN_{\sigma 1}$, \ldots, $\E_n
 \in \ \vdash \bPP,\bN_{\sigma n}$
then $\D := (\sigma,I_n)^+.\{\E_1,\ldots,\E_n\}
\in \ \vdash \bPP, \bP_\sigma$.\\
 Moreover, if $\E_1,\ldots,\E_n$
are winning, then $\D$ is winning.

\item

 Let
$\bN = \bF^-(\bP_1,\ldots,\bP_n)$ be a negative behaviour.\\
If $\E \in \ \vdash \bPP,\bP_{\xi 1},\ldots,\bP_{\xi n}$
then $\D := (\xi,I_n)^-.\E \in \ \vdash \bPP, \bN_{\xi}$.\\
Moreover, if $\E$
is winning, then $\D$ is winning.

\end{enumerate}
\end{lem}
\proof Let us fix $\bPP = \bQ_{\alpha_1},\ldots,\bQ_{\alpha_k}$
and  arbitrary strategies $\F_1 \in \bQ_{\alpha_1}^\bot,\ldots,\F_k \in \bQ_{\alpha_k}^\bot$. We will write
$(\F_j)$ and $(\F_j) \in \bC$  for $\F_1,\ldots,\F_k$
and $\F_1 \in \bQ_{\alpha_1}^\bot,\ldots,\F_k \in \bQ_{\alpha_k}^\bot$
respectively.
\begin{enumerate}[(1)]
\item For sake of clarity, we distinguish two subcases.

If $n =0$, then $\bP = \semone$ and $\D$ is just  $(\sigma,\emptyset)^+$.
It is then immediate to show (1), since for any $\A \in \bP^\bot_\sigma = \sembot_\sigma$, we have
$\pl \D, \A, (\F_j) \pr = \pl \D,\A \pr$
and $\D$ and $\A$ are obviously orthogonal.
It is also immediate to check that $\D$ is winning.

If $n > 0$, by Proposition ~\ref{closure2}, $\E_i \in \ \vdash \bPP,\bN_{\sigma i}$ if and only if
$\E_i' := \pl \E_i,(\F_j)\pr \in \bN_{\sigma i}$,
for any $1 \leq i \leq n$.
By construction of $\bF^+(\bN_1,\ldots,\bN_n)$, the strategy
$(\sigma,I_n)^+.\{\E_1',\ldots,\E_n'\} \in \bP_\sigma$.  By {using
Lemma ~\ref{conversion}} and  Proposition ~\ref{closure2} again, we conclude
 $\D \in \ \vdash \bPP, \bP_\sigma$.

As for winning conditions, the only one which is not immediate is materiality.
Let  $\A$ be a strategy in $\bN_{\sigma i}^\bot$ and
 $ \A^\ast := (\sigma,I_n)^-.\A \in \bP_\sigma^\bot$.
Observe that by construction $(\sigma,I_n)^+$ occurs linearly in $\D$ at the root. This makes
the interaction of the
cut-net $\{ \D,
\A^\ast,(\F_j) \}$ particularly simple to describe:
(i)
$(\sigma,I_n)^+$ and $(\sigma,I_n)^-$ match,
(ii) after  matching, the interaction is
 exactly as in the cut-net
 $\{\E_i,
\A,(\F_j)\}$. So we have (in the notation of Definition ~\ref{used_part})
$$\D[
\A^\ast,(\F_j)]
= (\sigma,I_n)^+.   \E_i[
\A,(\F_j)].$$
Since $\E_i$ is material, if we  let vary $\A \in \bN^\bot_{\sigma i}$ and $(\F_j) \in \bC$
and use the counter-strategies $\A^\ast,(\F_j)$,
we can visit  the whole subtree $(\sigma,I_n)^+.\E_i \subseteq \D$.
Applying the same reasoning to each $1 \leq i \leq n$ we have our claim.

\item

By Proposition ~\ref{closure2}, $\E \in \ \vdash \bPP,\bP_{\xi 1},\ldots,\bP_{\xi n }$ if and only if
$\E' := \pl \E,(\F_j)\pr \in  \ \vdash \bP_{\xi 1},\ldots,$ $ \bP_{\xi n}$.
By Proposition ~\ref{neg_comp},
$(\xi,I_n)^-.\E' \in \ \vdash  \bN_{\xi}$.
Since we have that
$ (\xi,I_n)^-.\pl \E,(\F_j)\pr = \pl(\xi,I_n)^-. \E,(\F_j)\pr$,
by using Proposition ~\ref{closure2} again, we conclude
 $\D \in \ \vdash \bPP, \bN_{\xi}$.

As for winning conditions, the only one which is not immediate is materiality.\\
For  $\A_1, \in \bP_{\xi 1}^\bot,\ldots,\A_n \in \bP_{\xi n}^\bot$,
let us set $\A^\ast := (\xi,I_n)^+.\{\A_1,\ldots,$ $ \A_n\} \in \bN_\xi^\bot$.
Since by construction $(\xi,I_n)^+$ occurs linearly in $\A^\ast$,
the interaction of the
cut-net $\{ \D,
\A^\ast,(\F_j) \}$ can  be simply described as follows:
(i) the actions
$(\xi,I_n)^+$ and $(\xi,I_n)^-$ match, (ii)
after  matching the interaction is the exactly as in the cut-net
 $\{\E,
\A_1,\ldots,\A_n, $ $ (\F_j) \}$. So we have
$$\D[
\A^\ast,(\F_j) ]
= (\xi,I_n)^-.   \E[
\A_1,\ldots,\A_n,(\F_j) ].$$
Since $\E$ is material,  if we  let vary $\A_1, \in \bP_{\xi 1}^\bot,\ldots,\A_n \in \bP_{\xi n}^\bot$ and $(\F_j) \in \bC$,
we can use the counter-strategies $\A^\ast,(\F_j)$ in order
to completely visit $\D$.
\qed
\end{enumerate}

\begin{lem}[Structural rules]\label{contraction} \hfill
\begin{enumerate}[\em(1)]
\item \emph{Weakening}:
if
$\D \in \ \vdash \bGG$  then  $\D
\in \ \vdash \bGG,\bQ_\xi$.

 Moreover, if $\D$ is winning (in $ \vdash \bGG$)
 then $\D$ is winning (in $\vdash \bGG,\bQ_\xi$).
\item \emph{Contraction}:
if $\D \in \ \vdash \bGG, \bQ_\xi,\bQ_\sigma$ then
 $\D[\xi/\sigma] \in \ \vdash \bGG, \bQ_\xi$.

 Moreover, if $\D$ is winning
 then $\D[\xi/\sigma]$ is winning.

\end{enumerate}
\end{lem}
\proof Let us fix $\bGG = \bF_{\alpha_1},\ldots,\bF_{\alpha_k}$
and  arbitrary strategies $\F_1 \in \bF_{\alpha_1}^\bot,\ldots,\F_k \in \bF_{\alpha_k}^\bot$. We will write
$(\F_j)$ and $(\F_j) \in \bC$  for $\F_1,\ldots,\F_k$
and $\F_1 \in \bF_{\alpha_1}^\bot,\ldots,\F_k \in \bF_{\alpha_k}^\bot$
respectively.
\begin{enumerate}[(1)]
\item  Since no action with name
$\xi$ occurs in $\D$, we have for any
$\E \in \bQ_\xi^\bot$:
$$\pl \D ,(\F_j) \pr = \pl \D ,\E,(\F_j) \pr,$$
 where $\D$ is taken on the interface $\alpha_1,\ldots,\alpha_k$ of $\vdash \bGG$ (resp.
  $\alpha_1,\ldots,\alpha_k,\xi^+$ of $\vdash \bGG,\bQ_\xi$)
 on the LHS (resp.\ RHS) of the equality above.
 Moreover it is easily seen that the interactions of
 the two cut-nets are exactly the same.
  The result then immediately follows.

\item We first prove that
 $\D \in \ \vdash \bGG, \bQ_\xi,\bQ_\sigma$ implies
 $\D[\xi/\sigma] \in \ \vdash \bGG,\bQ_\xi$.
By hypothesis, for any $\A,\B\in \bQ\b_\xi$, we have
that
$\D\bot \{\A,\B[\sigma/\xi],(\F_j)\}$.
When $\A= \B$, we have that
$\D \bot \{\A,  \A[\sigma/\xi],(\F_j)\}$
which  implies
$\D[\xi/\sigma] \bot \{\A,(\F_j)\}$
by Proposition ~\ref{prop copies vera}.
Hence, we have
$\D[\xi/\sigma] \in \ \vdash \bGG,\bQ_\xi$.

As for winning conditions, the only one which is not immediate to prove is materiality.\\
By Lemma ~\ref{poscomp_lem} (2), we have
that $\bQ_\xi = (\bQ_\xi^\bot\single{x})\b$, where
recall that $\bQ_\xi^\bot\single{x}$ consists of the negative strategies
of $\bQ\b$ which have a unique negative root $x=(\xi,I_n)$.
Since $\D$ is material in $ \vdash \bGG, \bQ_\xi,\bQ_\sigma$,
we have by
Definition ~\ref{mater_def}:
$$\D = \displaystyle \bigcup \big\{ \D[\A,\B[\sigma/\xi],(\F_j)] \ : \ \A,\B \in  \bQ_\xi^\bot\single{x} \mbox{ and } (\F_j) \in \bC \big\}.
$$

Since  $ \A,\B$ have the same root $x$,  by Lemma ~\ref{poscomp_lem} (1) the strategy
$\A +^\tau \B \in \bQ_\xi^\bot$.
The strategy $\A +^\tau \B$ has the same  unique root $x$ too, hence
$\A +^\tau \B \in \bQ_\xi^\bot\single{x}$.

Observe that
using the counter-strategy $\A+^\tau \B$ we are able to visit
the part of $\D$ which can be visited
interchanging $\A$ and $\B$ (and possibly visit new actions), \ie
$$
\begin{array}{rcl}
\D[\A,\B[\sigma/\xi], (\F_j)]
& \subseteq & \D[\A+^\tau \B, (\A +^\tau \B)[\sigma/\xi], (\F_j)] \\
\D[\B,\A[\sigma/\xi], (\F_j)]
& \subseteq & \D[\A+^\tau \B, (\A +^\tau \B)[\sigma/\xi], (\F_j)] \\
\end{array}$$
But then, we have that
$$\begin{array}{rcl}
\D
& = &  \displaystyle \bigcup \big\{ \D[\A+^\tau \B,(\A+^\tau \B)[\sigma/\xi],(\F_j)] \ : \ \A+^\tau \B \in  \bQ_\xi^\bot\single{x} \mbox{ and } (\F_j) \in \bC \big\} \\
 & = &  \displaystyle \bigcup \big\{ \D[\C,\C[\sigma/\xi],(\F_j)] \ : \ \C \in  \bQ_\xi^\bot\single{x} \mbox{ and } (\F_j) \in \bC \big\}.\\
\end{array}
$$
Similarly as in Proposition ~\ref{prop copies vera}, we can derive that
$$\D[\C, \C[\sigma/\xi], (\F_j)] [\xi/\sigma] = \D[\xi/\sigma] [\C,  (\F_j)],$$
because the interactions
of the cut-nets $\{\D,\C,\C[\sigma/\xi], (\F_j)\}$
and $\{\D[\xi/\sigma],\C,  (\F_j)\}$ differ
only in the names of some  hidden actions but have the same pointer structure.
Hence:
$$\begin{array}{rcl}
\D[\xi/\sigma]& =  &   \displaystyle \bigcup \big\{ \D[\C,\C[\sigma/\xi],(\F_j)] \ : \ \C \in  \bQ_\xi^\bot\single{x} \mbox{ and } (\F_j) \in \bC \big\}[\xi/\sigma]\\
& =  &   \displaystyle \bigcup \big\{ \D[\C,\C[\sigma/\xi],(\F_j)][\xi/\sigma] \ : \ \C \in  \bQ_\xi^\bot\single{x} \mbox{ and } (\F_j) \in \bC \big\}\\
& =  &   \displaystyle \bigcup \big\{ \D[\xi/\sigma] [\C,  (\F_j)] \ : \ \C \in  \bQ_\xi^\bot\single{x} \mbox{ and } (\F_j) \in \bC \big\},\\
\end{array}
$$
which shows the materiality of $\D[\xi/\sigma]$ in $\vdash \bGG,\bQ_\xi$.
\qed
\end{enumerate}\medskip

\noindent We can now prove the following:

\begin{prop} \label{prop per soundness}
Let $\pi$ be a cut-free derivation  of a sequent $\vdash \Gamma$ in $\HS$  and
 $\D(\pi)$ be the interpretation of $\pi$
in a sequent of behaviours  $\vdash \bGG$.
$\D(\pi)$ is a winning strategy  in $\vdash \bGG$.
\end{prop}
\proof
By induction on the height of $\pi$.
As for positive rules, we use Lemmas ~\ref{logical} (1)
and~\ref{contraction}.
As for negatives rules, we use Lemma ~\ref{logical} (2).
\qed

In order to expand the previous proposition
to derivation with cuts, we   need to study the relation
between composition of strategies and the procedure
of cut-elimination defined in the proof of admissibility
of the cut-rule of $\HS$ (Theorem ~\ref{principal}).\\

We first show the following:

\begin{lem}[Cut-rule] \label{cut rule lemma}
If $\D \in \ \vdash \mathbf{\Xi},\bPP,\bP_\xi$
and $\E \in \ \vdash \bDD,\bP^\bot_\xi$
then $\pl \D,\E \pr \in \  \vdash \mathbf{\Xi},\bPP,\bDD$
\end{lem}
\proof
We only consider the case in which $\mathbf{\Xi}$
is empty, the case $\mathbf{\Xi} =\bN$ is similar.

Let $\bPP = \bA_{\alpha_1},\ldots,\bA_{\alpha_n}$
and $\bDD = \bB_{\beta_1},\ldots,\bB_{\beta_k}$
on disjoint interfaces
$\alpha_1^+,\ldots,\alpha_n^+$ and
$\beta_1^+,\ldots,\beta_k^+$ respectively.
Let $\A_1 \in  \bA_{\alpha_1}^\bot,\ldots, \A_n \in \bA_{\alpha_n}^\bot$
and  $\B_1 \in  \bB_{\beta_1}^\bot,\ldots, \B_k \in \bB_{\beta_k}^\bot$
be arbitrary strategies and write
$(\A_j)$ and $(\B_l)$ for $\A_1,\ldots,\A_n$
and $\B_1,\ldots,\B_k$ respectively.

By Proposition
~\ref{closure2} we have that
$\pl \D,(\A_j) \pr \in \bP_\xi$ and $\pl \E ,(\B_l) \pr \in \bP_\xi^\bot$ which implies
that $\pl \pl \D,(\A_j) \pr,  \pl \E ,(\B_l) \pr \pr$
is total. By  associativity,
we have that
$\pl \pl \D,\E\pr,   (\A_j) ,(\B_l)  \pr$
is total, which shows that $\pl \D,\E \pr \in \  \vdash \bPP,\bDD$.
\qed

 We now relate our interpretation and Proposition ~\ref{struct},
 which deals with the structural rules of $\HS$.\vfill\eject

\begin{lem}\label{lemma per prop} \hfill
\begin{enumerate}[\em(1)]
\item Let $\D(\pi)$ be the interpretation of a cut-free derivation $\pi$ of $\vdash \Gamma$ in a sequent of behaviours $\vdash \bGG$ and $\D(\pi')$ be
the interpretation of the cut-free derivation $\pi'$ of
$\vdash \Gamma,Q$ in the sequent of behaviours $\vdash \bGG,\bQ_\xi$ as  given by Proposition \emph{~\ref{struct} (1)}.
We have $\D(\pi)= \D(\pi')$.
\item
Let $\D(\pi)$ be interpretation of a cut-free derivation $\pi$ of $\vdash \Gamma,Q,Q$ in a sequent of behaviours $\vdash \bGG,
\bQ_\xi,\bQ_\sigma$ and $\D(\pi')$ be
the interpretation of the cut-free derivation $\pi'$ of
$\vdash \Gamma,Q$ in a sequent of behaviours $\vdash \bGG,\bQ_\xi$ as  given by Proposition \emph{~\ref{struct} (2)}.
We have $\D(\pi') =\D(\pi)[\xi/\sigma]$.
\end{enumerate}
\end{lem}

\proof By induction on the height of $\pi$.
\qed

We can finally show the correspondence between
composition and cut-elimination.

\begin{lem}[Composition and cut-elimination]  \label{corr lemma}
Let $\pi$ and $\rho$ be cut-free derivations in $\HS$ of
$\vdash \Xi,\Pi,P$ and $\vdash \Delta, P\b$ respectively
 and $\D(\pi)$ and $\D(\rho)$ be the interpretation of $\pi$ and $\rho$
in
$\vdash \mathbf{\Xi},\bPP,\bP_\xi$
and $\vdash \bDD,\bP^\bot_\xi$ respectively.
Let $\theta$ be the cut-free derivation of
$\vdash \Xi,\Pi,\Delta$ as given by Theorem ~\ref{principal}
and $\D(\theta)$ its interpretation in
$\vdash \mathbf{\Xi},\bPP,\bDD$.

We have  $\pl \D(\pi),\D(\rho) \pr = \D(\theta)$.
\end{lem}
\proof
As in Theorem ~\ref{principal}, the proof is given by induction
on the pair $(d(P),h(\pi)+h(\rho))$,
where $d$ (resp.\ $h$) denotes
the depth of a formula (resp.\ the height of
a proof), as given in Definition ~\ref{depth height}.
We distinguish three subcases.\smallskip

\noindent {\textbf{(a)}}
The last rule of $\pi$ is a positive rule \textrm{Pos}$_n$ and   $P$
is  principal in \textrm{Pos}$_n$. As in Theorem ~\ref{principal}, we only consider the
cases $n=0$ and $n=2$.\smallskip

If $n =0$ and we have
\begin{center}
\AxiomC{}
\RightLabel{\scriptsize{Pos$_0$}}
\UnaryInfC{$\vdash \Pi, \posm \uno$}

\AxiomC{$\vdots$ $\rho_0$}
\noLine \UnaryInfC{$\vdash \Delta$}
\RightLabel{\scriptsize{Neg$_0$}}
\UnaryInfC{$\vdash \Delta, \negm \ult$}
\noLine
\BinaryInfC{}
\DisplayProof
\end{center}
the procedure described in the proof of  Theorem ~\ref{principal}
gives the cut free derivation $\theta$ of $ \vdash \Pi,\Delta$,
where $\theta$ is obtained from $\rho_0$ by
means of Proposition ~\ref{struct} (1) (weakening on positive formulas).

By our interpretation, we have
that $\D(\pi) = (\xi,\emptyset)^+$ and $\D(\rho) = (\xi,\emptyset)^-.\D(\rho_0)$. By normalization
and Lemma
~\ref{lemma per prop} (1), we have
$\pl \D(\pi),\D(\rho) \pr = \D(\rho_0) =\D(\theta)$.\smallskip

If $n=2$, for $P = \posm(N_1 \otimes N_2)$ and
$P\b = \negm(N_1^\bot \parr  N_2^\bot)$, we have
\begin{center}
\AxiomC{$\vdots$ $\pi_1$}
\noLine \UnaryInfC{$\vdash \Pi,P,N_1$}
\AxiomC{$\vdots$ $\pi_2$}
\noLine \UnaryInfC{$\vdash \Pi,P,N_2$}
\RightLabel{\scriptsize{Pos$_2$}}
\BinaryInfC{$\vdash \Pi,P$}

\AxiomC{$\vdots$ $\rho_0$}
\noLine \UnaryInfC{$\vdash \Delta, N_1^\bot,N_2^\bot$}
\RightLabel{\scriptsize{Neg$_2$}}
\UnaryInfC{$\vdash \Delta, P\b$}
\noLine
\BinaryInfC{}
\DisplayProof
\end{center}
Suppose that $\pi$ is interpreted by $\D(\pi)
=(\sigma,\{1,2\})^+.\big\{\D(\pi_1), \D(\pi_2)\big\}[\xi/\sigma]$
in the sequent of behaviours $\vdash \bPP,\bP$ on interface
$\Pi,\xi^+$ and  $\rho$ is interpreted by $\D(\rho) =(\xi,\{1,2\})^-.\D(\rho_0)$
in the sequent of behaviours $\vdash \bDD,\bP^\bot$
on interface
$\Delta,\xi^-$.
Since the construction
 of
 the cut-free derivation $\theta$
 of $\vdash \Pi,\Delta$ (as given in Theorem ~\ref{principal})
involves ``copies" of derivations and
contractions, we also consider:
\begin{enumerate}[$\bullet$]
\item $\A_1 := \D(\pi_1)[\Pi'/\Pi]$, the interpretation
of $\pi_1$ in  the sequent of behaviours
$\vdash \bPP,\bP,\bN_1$ on interface
$\Pi',\xi^+,\sigma_1^-$,
\item    $\A_2 := \D(\pi_2)[\Pi''/\Pi]$, the interpretation
of $\pi_2$ in  the sequent of behaviours
$\vdash \bPP,\bP,\bN_2$ on interface
$\Pi'',\xi^+,\sigma_2^-$,
\item    $\B := \D(\rho)[\sigma/\xi,\Delta'/\Delta]$, the interpretation
of $\rho$ in  the sequent of behaviours
$\vdash \bDD,\bP^\bot$ on interface
$\Delta',\sigma^-$,
\item    $\B_0 := \D(\rho_0)[\sigma 1/\xi 1,\sigma 2/\xi 2,\Delta'/\Delta]$, the interpretation
of $\rho_0$ in  the sequent of behaviours
$\vdash \bDD,\bN_1^\bot,\bN_2^\bot$ on interface
$\Delta',\sigma 1^-, \sigma 2^-$. Equivalently, $\B_0$
is obtained from $\B$ by removing its root $(\sigma,\{1,2\})^-$,\ie $\B = (\sigma,\{1,2\})^-.\B_0$,
\item    $\C_1 := \D(\rho)[\Delta''/\Delta]$, the interpretation
of $\rho$ in  the sequent of behaviours
$\vdash \bDD,\bP^\bot$ on interface
$\Delta'',\xi^-$,
\item    $\C_2 := \D(\rho)[\Delta'''/\Delta]$, the interpretation
of $\rho$ in  the sequent of behaviours
$\vdash \bDD,\bP^\bot$ on interface
$\Delta''',\xi^-$,
\end{enumerate}
 where the names $\Pi',\Pi'',\Delta',\Delta'',\Delta'''$
are all fresh and disjoint.

Recall that the procedure described in Theorem ~\ref{principal}
introduces  cut-free derivations
$\psi_1$ of $\vdash \Pi,\Delta,N_1$
(from $\pi_1$ and $\rho$),
$\psi_2$ of $\vdash \Pi,\Delta,N_2$ (from $\pi_2$ and $\rho$),
$\theta_1$ of $\vdash \Pi,\Delta,\Delta,N_2^\bot$
(from $\rho_0$ and $\psi_1$),
$\theta_2$ of $\vdash \Pi,\Pi,\Delta,\Delta,\Delta$
(from $\theta_1$ and $\psi_2$).
Finally, the cut-free derivation
 $\theta$ of $\vdash \Pi,\Delta$ is obtained from $\theta_2$ by means of contractions, in the sense of Proposition ~\ref{struct} (2)
(contraction on positive formulas).

In terms of our interpretation, we have:
$$\D(\theta) = \pl \pl \B_0, \pl \A_1,\C_1 \pr \pr, \pl \A_2,\C_2 \pr \pr [\Pi/\Pi',\Pi/\Pi'',\Delta/\Delta',\Delta/\Delta'',\Delta/\Delta''']$$
where  $\pl \A_1,\C_1 \pr$ (resp.
$\pl \A_2,\C_2 \pr$) interprets the derivation $\psi_1$ (resp.\ $\psi_2$),
$\pl \B_0, \pl \A_1,\C_1 \pr\pr$ interprets the derivation  $\theta_1$,
$\pl \pl \B_0, \pl \A_1,\C_1 \pr \pr, \pl \A_2,\C_2 \pr \pr$
interprets the derivation  $\theta_2$ and
finally, by  Lemma ~\ref{lemma per prop} (2), the renamings $[\Pi/\Pi',\Pi/\Pi'',$ $\Delta/\Delta',\Delta/\Delta'',\Delta/\Delta''']$
take care of the contractions.

We have to show that $\pl \D(\pi), \D(\rho) \pr = \D(\theta)$.

Writing $a$ (resp.\ $b$) for $(\sigma,\{1,2\})$ (resp.\ $(\xi,\{1,2\})$ and using
Proposition ~\ref{prop copies vera} and the associativity of the normalization,
we get:

 $$
\begin{array}{rcl}
\pl \D(\pi), \D(\rho)\pr & = &
 \pl   a^+.\big\{\D(\pi_1), \D(\pi_2)\big\}[\xi/\sigma] \ , \ b^-.\D(\rho_0)\pr  \\
& = & \pl a^+.\big\{\D(\pi_1), \D(\pi_2)\big\} \ , \ b^-.\D(\rho_0) \ , \ \B \pr [\Delta/\Delta']   \\
& = & \pl \pl a^+.\big\{\D(\pi_1) \ , \ \D(\pi_2)\big\},b^-.\D(\rho_0) \pr \ , \ \B \pr [\Delta/\Delta']   \\
\end{array}$$
In
$\pl a^+.\big\{\D(\pi_1) \ , \ \D(\pi_2)\big\},b^-.\D(\rho_0) \pr$ , the action $a^+$ is  visible and then we can ``push"
$\D(\rho)$ in both premises $\D(\pi_1)$ and $\D(\pi_2)$
using the strategies $\A_1,\A_2,\C_1,\C_2$ introduced above
as follows:
 $$
\begin{array}{rcl}
\pl a^+.\big\{\D(\pi_1) \ , \ \D(\pi_2)\big\},b^-.\D(\rho_0) \pr & \hspace{-1.5mm}= \hspace{-1.5mm} &
 a^+.\big\{\pl \A_1,\C_1 \pr \ , \ \pl \A_2, \C_2 \pr\big\}[\Pi/\Pi',\Pi/\Pi'',\Delta/\Delta'',\Delta/\Delta'''] \\
\end{array}$$
where the renamings $
[\Pi/\Pi',\Pi/\Pi'',\Delta/\Delta'',\Delta/\Delta''']$
ensure that the strategy on LHS and the strategy on RHS of the equality
are on the same interface $\Pi,\Delta,\sigma^+$.
We have that
 $$
\begin{array}{rcl}
\pl \D(\pi), \D(\rho)\pr & = &
 \pl    a^+.\big\{\pl \A_1,\C_1 \pr \ , \ \pl \A_2, \C_2 \pr\big\}[\Pi/\Pi',\Pi/\Pi'',\Delta/\Delta'',\Delta/\Delta''']  \ , \ \B \pr [\Delta/\Delta']   \\
& = &
 \pl    a^+.\big\{\pl \A_1,\C_1 \pr \ , \ \pl \A_2, \C_2 \pr\big\}  \ , \ \B \pr [\Pi/\Pi',\Pi/\Pi'',\Delta/\Delta',\Delta/\Delta'',\Delta/\Delta''']   \\
& = &
 \pl   \B_0  \ , \  \pl \A_1,\C_1 \pr \ , \ \pl \A_2, \C_2 \pr \pr [\Pi/\Pi',\Pi/\Pi'',\Delta/\Delta',\Delta/\Delta'',\Delta/\Delta'''] \\
& = &
 \pl  \pl \B_0  \ , \  \pl \A_1,\C_1 \pr \pr \ , \ \pl \A_2, \C_2 \pr \pr [\Pi/\Pi',\Pi/\Pi'',\Delta/\Delta',\Delta/\Delta'',\Delta/\Delta'''] \\ & = & \D(\theta),
  \\
\end{array}$$
where the second equality above is justified by the
fact that
those  renamings on ``contexts" of visible actions
do not modify the calculation of the normal form
(and we have correct cut-nets on both sides of the equality),
the third one by the fact that
by construction $a^+$ occurs linearly (recall that
  $\B = a^-.\B_0$) and the fourth one by associativity.\smallskip

\noindent\textbf{(b)}
The last rule of $\pi$ is a positive rule  \textrm{Pos}$_n$ and  $P$
is \emph{not}  principal in \textrm{Pos}$_n$:

\begin{center}
\AxiomC{$\vdots$ $\pi_1$}
\noLine \UnaryInfC{$\vdash \Pi,P,Q,N_1$}
\AxiomC{$\ldots$}
\AxiomC{$\vdots$ $\pi_n$}
\noLine \UnaryInfC{$\vdash\Pi,P,Q,N_n$}
\RightLabel{\scriptsize{Pos$_n$}}
\TrinaryInfC{$\vdash \Pi,P,Q$}

\AxiomC{$\vdots$ $\rho$}
\noLine
\UnaryInfC{$\vdash \Delta, P\b$}
\noLine
\BinaryInfC{}
\DisplayProof
\end{center}
and the principal formula of \textrm{Pos}$_n$  is
the occurrence of formula $Q=\posm(N_1 \otimes \cdots \otimes N_n)$.

Let $\D(\pi) = (\sigma,I_n)^+.\big\{\D(\pi_1),\ldots,\D(\pi_n)\big\} [\alpha/\sigma]$ be
the interpretation of $\pi$ in the sequent
of behaviours $\vdash \bPP,\bP_\xi,\bQ_\alpha$ on interface
$\Pi,\xi^+,\alpha^+$ and
$\D(\rho)$ the interpretation of $\rho$ in the sequent
of behaviours $\vdash \bDD,\bP_\xi^\bot$ on interface
$\Delta,\xi^-$.

The procedure described in Theorem ~\ref{principal}
gives a cut-free derivation $\theta$
which is interpre\-ted by
$\D(\theta)= (\sigma,I_n)^+.\big\{\pl \D(\pi_1),\D(\rho)\pr,\ldots,\pl\D(\pi_n),\D(\rho)\pr\big\} [\alpha/\sigma]$, where
each $\pl\D(\pi_i),\D(\rho)\pr$ is the strategy which is interpretation
 of $\vdash \Pi,Q,\Delta,N_i$ on interface  $\Pi,\alpha^+,\Delta,\sigma i^-$.

We have to show that $\pl \D(\pi), \D(\rho) \pr = \D(\theta)$,
but
this easily follows from the definition of normalization, observing that
the main strategy $\D(\pi)$ starts with a visible positive action.\smallskip

\noindent\textbf{(c)}
The last rule of $\pi$ is a negative rule \textrm{Neg}$_n$,
having  $N=\negm(P_1 \parr \cdots \parr P_n)$ as principal formula.

\begin{center}
\AxiomC{$\vdots$ $\pi_0$}
\noLine \UnaryInfC{$\vdash \Pi,P,P_1,\ldots,P_n$}
\RightLabel{\scriptsize{Neg$_n$}}
\UnaryInfC{$\vdash \Pi,P,N$}
\AxiomC{$\vdots$ $\rho$}
\noLine
\UnaryInfC{$\vdash \Delta, P\b$}
\noLine
\BinaryInfC{}
\DisplayProof
\end{center}
Suppose that $\D(\pi) = (\alpha,I_n)^-.\D(\rho)$ interprets $\pi$ in the sequent
of behaviours $\vdash \bPP,\bP_\xi,\bN_\alpha$ on interface
$\Pi,\xi^+,\alpha^-$ and
$\D(\rho)$ interprets $\rho$ in the sequent
of behaviours $\vdash \bDD,\bP_\xi^\bot$ on interface
$\Delta,\xi^-$.

The procedure described in Theorem ~\ref{principal}
gives a cut-free derivation $\theta$
which is interpreted by
$\D(\theta)= (\alpha,I_n)^-. \pl\D(\pi_0),\D(\rho)\pr$, where
 $\pl\D(\pi_0),\D(\rho)\pr$ is the strategy which is interpretation
 of $\vdash \Pi,P_1,\ldots,P_n,\Delta$ on interface  $\Pi,\alpha1^+,\ldots,\alpha n^+,\Delta$.

We have to show that $\pl \D(\pi), \D(\rho) \pr = \D(\theta)$,
but
this easily follows from the definition of normalization, observing that
the main strategy $\D(\pi)$  starts with a visible negative action.\qed

We can finally prove:

\begin{thm}[Soundness]
Let $\pi$ be a derivation  of a sequent $\vdash \Gamma$ in $\HS$  and
 $\D(\pi)$ be the interpretation of $\pi$
in a sequent of behaviours  $\vdash \bGG$.
\begin{center}$\D(\pi)$ is a winning strategy  in $\vdash \bGG$.
\end{center}
Moreover, the interpretation is invariant under cut-elimination.
\end{thm}
\proof
If $\pi$ is cut-free, the result is given
in Proposition ~\ref{prop per soundness}.
If $\pi$ contains cuts, then using Lemmas ~\ref{logical},
~\ref{contraction} and
~\ref{cut rule lemma} we can inductively show that
 $\D(\pi) \in \ \vdash \bGG$.

By repeatedly applying
Lemma ~\ref{corr lemma}, we eventually prove that
$\D(\pi)$ is also the interpretation
of a cut-free derivation interpreted in $\vdash \bGG$.
Finally, we  use Proposition ~\ref{prop per soundness} again.


\begin{thebibliography}{10}




\bibitem{Abr}
Abramsky, S.:
\newblock Axioms for definability and full completeness.
\newblock
In: Proof, Language, and Interaction (Essay in honor of Robin Milner) The MIT Press
\newblock (2000)
55--76.


\bibitem{AbrJagMac}
Abramsky, S.,   Jagadeesan, R.,  Malacaria, P.:
\newblock Full Abstraction for PCF.
\newblock  Inf. Comput. \textbf{163}(2) (2000) 409--470.
\bibitem{AndreoliLP}
 Andreoli, J.-M.:
\newblock Logic Programming with Focusing Proof in Linear Logic.
\newblock J. Log. Comput. \textbf{2}(3) (1992)
297--347.




\bibitem{thesisbas}
 Basaldella, M.:
\newblock On Exponentials in Ludics.
\newblock PhD Thesis (2008) University of Siena.


\bibitem{BasFag}
Basaldella, M., Faggian, C.:
\newblock Ludics with Repetitions (Exponentials, Interactive Types and Completeness).
\newblock  In: LICS. (2009) 375--384.

\bibitem{BasalTLCA09}
Basaldella, M., Terui, K.:
\newblock On the meaning of logical completeness.
\newblock Logical Methods in Computer Science
\textbf{6}(4:11) (2010) 1--35.

\bibitem{Coq}
Coquand, T.:
\newblock A semantics of evidence for classical arithmetic.
 \newblock J. Symb. Log.
\textbf{60}(1) (1995) 325-–337.

 \bibitem{CurAbs}
Curien, P.-L.:
\newblock Abstract {B}\"ohm trees.
\newblock Math. Struct. in Comp. Sci. \textbf{8}(6) (1998)
  559--591.

 \bibitem{CurLLL2}
Curien, P.-L.:
\newblock
Introduction to Linear Logic and Ludics, part  {II}.
\newblock  Advances of Mathematics (China) \textbf{35}(1) (2006) 1--44.


 \bibitem{CurGS}
Curien, P.-L.:
\newblock Notes on game semantics.
\newblock Manuscript (2006).


\bibitem{CurHerb}
Curien, P.-L., Herbelin, H.:
\newblock Abstract machines for dialogue games.
\newblock  Panoramas et Synth\`eses \textbf{27} (2009) 231--275.






\bibitem{FagTra}
Faggian, C.:
\newblock Travelling on designs.
\newblock In: CSL. (2002)  427--441.


\bibitem{introunif}
\newblock Faggian, C., Fleury, M.-R., Quatrini, M.: An introduction to uniformity in Ludics.
\newblock In: Linear logic in computer science,  London Math. Soc. Lecture Note Ser., \textbf{316}, Cambridge Univ. Press, Cambridge, (2004) 236--246.


\bibitem{FagHyl}
 Faggian, C.,   Hyland, J.M.E.:
\newblock Designs, disputes and strategies.
\newblock In : CSL. (2002)
442--457.




\bibitem{FagPic}
Faggian, C., Piccolo, M.:
\newblock Ludics is a model for the finitary linear pi-calculus.
\newblock In: TLCA. (2007) 148--162.

\bibitem{FP09}
Faggian, C., Piccolo, M.:
\newblock Partial Orders, Event Structures, and Linear Strategies.
\newblock In: TLCA. (2009) 95--111.







\bibitem{LinearLogic}
Girard, J.-Y.:
\newblock Linear Logic.
\newblock Theor. Comput. Sci.
\textbf{50}(1) (1987)
1--102.



 \bibitem{GirGoI1}
Girard, J.-Y.:
\newblock Geometry of interaction {I}: Interpretation of System {F}.
\newblock Logic Colloquium 88,
 In R. Ferro et al., (1989)
 221--260.

 \bibitem{GirLC}
Girard, J.-Y.:
\newblock A New Constructive Logic: Classical Logic.
\newblock
 Math. Struct. in Comp. Sci. \textbf{1}(3) (1991)
  255--296.



 \bibitem{meaning1}
Girard, J.-Y.:
\newblock On the meaning of logical rules {I}: syntax vs.
               semantics.
\newblock Computational Logic (U. Berger and H. Schwichtenberg eds)
Heidelberg Springer-Verlag
(1999)
215 -- 272.

 \bibitem{GirMeanII}
Girard, J.-Y.:
\newblock On the meaning of logical rules {II} : multiplicatives and additives.
\newblock Foundation of Secure Computation  (2000)
 183--212.


\bibitem{GirLoc}
Girard, J.-Y.:
\newblock Locus solum: From the rules of logic to the logic of rules.
\newblock Math. Struct. in Comp. Sci. \textbf{11}(3) (2001)
  301--506.

\bibitem{GirBook}
 Girard, J.-Y.:
\newblock
Le Point Aveugle, Cours de logique, Tome {II}:
Vers l'imperfection.
\newblock Visions des Sciences. Hermann (2007).

\bibitem{harthesis}
 Harmer, R. S.:
 \newblock Games and Full Abstraction for Nondeterministic
Languages. PhD Thesis (1999) University of London.

 \bibitem{harGS}
 Harmer, R. S.:
  \newblock Innocent game semantics.
  \newblock Manuscript (2006).

\bibitem{Hy-On}
Hyland, J.M.E., Ong, C.H.L.:
\newblock On full abstraction for {PCF}: I, {II}, and {III}.
\newblock Inf. Comput. \textbf{163}(2) (2000)  285--408.


\bibitem{HylScha}
Hyland, J.M.E., Schalk, A.:
\newblock Glueing and orthogonality for models of linear logic.
\newblock Theor. Comput. Sci. \textbf{294}(1-2) (2003)
183--231.

\bibitem{KriRea}
Krivine, J.-L.:
\newblock Realizability in classical logic.
\newblock  Panoramas et Synth\`eses \textbf{27} (2009) 197--229.



\bibitem{LaurentThesis}
Laurent, O.:
\newblock \'Etude de la polarization en logique.
\newblock PhD thesis, Universit\'{e} Aix-Marseille {II} (2002).


 \bibitem{LauPol}
Laurent, O.:
\newblock Polarized games.
\newblock Ann. Pure Appl. Logic \textbf{130}(1-3) (2004)  79--123.


\bibitem{LauSvS}
 Laurent, O.:
 \newblock  Syntax vs. semantics: A polarized approach.
 \newblock
Theor. Comput. Sci.
\textbf{343}(1-2)
(2005) 177--206.


\bibitem{MauTh}
 Maurel, F.:
\newblock Un cadre quantitatif pour la Ludique.
\newblock PhD Thesis (2004) Universit\'{e} Paris VII.



\bibitem{MelTab}
Melli{\`e}s, P.-A.,  Tabareau, N.:
\newblock Resource modalities in game semantics.
\newblock In: LICS. (2007)
389--398.

\bibitem{MelvouRea}
Melli{\`e}s, P.-A.,
Vouillon, J.:
\newblock Recursive Polymorphic Types and Parametricity in an Operational Framework.
\newblock
In: LICS. (2005)
82--91.

\bibitem{Nickau}
Nickau, H.:
\newblock Hereditarily Sequential Functionals: A Game-Theoretic Approach to Sequentiality.
\newblock PhD thesis,
Universit\"{a}t GH Siegen (1996).

\bibitem{paolini08tcs}
Paolini, L.:
\newblock Parametric $\lambda$-Theories.
\newblock Theor. Comput. Sci. \textbf{398}(1-3)
  (2008) 51--62.


\bibitem{Pittspoly}
Pitts, A.M.:
\newblock Parametric polymorphism and operational equivalence.
\newblock Math. Struct. in  Comp. Sci. \textbf{10}(3)
(2000) 321--359.


\bibitem{alexis}
Saurin, A.:
\newblock Towards Ludics Programming: Interactive Proof Search.
\newblock In: ICLP.
(2008)
253--268.


\bibitem{Terui}
Terui, K.:
\newblock Computational ludics.
 \newblock To appear in Theor. Comput. Sci. (2008).

\bibitem{yobeho}
 Yoshida, N.,   Berger, M.,   Honda, K.:
\newblock Strong Normalisation in the pi-Calculus.
\newblock In:  LICS. (2001)
311--322.

\bibitem{varyosh}
 Varacca, D.,   Yoshida, N.:
\newblock Typed Event Structures and the {\it pi}-Calculus: Extended
               Abstract.
\newblock In: MFPS.
(2006) 373--397.


\end{thebibliography}
\end{document}